\documentclass[a4paper,usenatbib,useAMS]{mnras}

\usepackage{graphicx}	
\usepackage{amsmath}	
\usepackage{amssymb}	
\usepackage{bm}		    
\usepackage{pdflscape}	
\usepackage{float}
\usepackage{subcaption}
\usepackage{enumitem}
\usepackage[below]{placeins}
\usepackage{afterpage}

\usepackage{booktabs}
\usepackage{multicol}   
\usepackage{multirow}
\usepackage{caption}

\usepackage[T1]{fontenc}
\usepackage{ae,aecompl}

\usepackage{newtxtext,newtxmath}

\title[DESI imaging systematics \& clustering]{Imaging Systematics and Clustering of DESI Main Targets}

\author[E. Kitanidis et al.]{Ellie Kitanidis$^{1}$, Martin White$^{1}$, Yu Feng$^{1}$, David Schlegel$^{2}$, Julien Guy$^{3}$, Arjun Dey$^{4}$, \and Martin Landriau$^{2}$, David Brooks$^{5}$, Michael Levi$^{2}$, John Moustakas$^{6}$, Francisco Prada$^{7}$, \and Gregory Tarle$^{8}$, Benjamin Alan Weaver$^{4}$ \\ \\ 
$^{1}$ Department of Physics, University of California, Berkeley, 366 LeConte Hall, Berkeley, CA 94720, USA\\
$^{2}$ Lawrence Berkeley National Laboratory, 1 Cyclotron Road, Berkeley, CA 93720, USA\\
$^{3}$ Sorbonne Universit\'{e}s, UPMC Universit\'{e} Paris 06, Universit\'{e} Paris-Diderot, CNRS-IN2P3 LPNHE 4 Place Jussieu, F-75252, Paris Cedex 05, France\\
$^{4}$ NSF's National Optical-Infrared Astronomy Research Laboratory, 950 N. Cherry Avenue, Tucson, AZ 85719, USA\\
$^{5}$ Department of Physics \& Astronomy, University College London, Gower Street, London, WC1E 6BT, UK\\
$^{6}$ Department of Physics and Astronomy, Siena College, 515 Loudon Road, Loudonville, NY 12211, USA\\
$^{7}$ Instituto de Astrofisica de Andaluc\'{i}a, Glorieta de la Astronom\'{i}a, s/n, E-18008 Granada, Spain\\
$^{8}$ Physics Department, University of Michigan Ann Arbor, MI 48109, USA}

\date{Last updated 2020 June 10; in original form 2019 November 13}

\pubyear{2019}

\begin{document}
\label{firstpage}
\pagerange{\pageref{firstpage}--\pageref{lastpage}}
\maketitle

\begin{abstract}
We evaluate the impact of imaging systematics on the clustering of luminous red galaxies (LRG), emission-line galaxies (ELG) and quasars (QSO) targeted for the upcoming Dark Energy Spectroscopic Instrument (DESI) survey. Using Data Release 7 of the DECam Legacy Survey, we study the effects of astrophysical foregrounds, stellar contamination, differences between north galactic cap and south galactic cap measurements, and variations in imaging depth, stellar density, galactic extinction, seeing, airmass, sky brightness, and exposure time before presenting survey masks and weights to mitigate these effects. With our sanitized samples in hand, we conduct a preliminary analysis of the clustering amplitude and evolution of the DESI main targets. From measurements of the angular correlation functions, we determine power law fits $r_0 = 7.78 \pm 0.26$ $h^{-1}$Mpc, $\gamma = 1.98 \pm 0.02$ for LRGs and $r_0 = 5.45 \pm 0.1$ $h^{-1}$Mpc, $\gamma = 1.54 \pm 0.01$ for ELGs. Additionally, from the angular power spectra, we measure the linear biases and model the scale dependent biases in the weakly nonlinear regime. Both sets of clustering measurements show good agreement with survey requirements for LRGs and ELGs, attesting that these samples will enable DESI to achieve precise cosmological constraints. We also present clustering as a function of magnitude, use cross-correlations with external spectroscopy to infer $dN/dz$ and measure clustering as a function of luminosity, and probe higher order clustering statistics through counts-in-cells moments.
\end{abstract}





\section{INTRODUCTION}
\label{sec:introduction}
The Dark Energy Spectroscopic Instrument (DESI; \citealt{DESI16}) is a ground-based dark energy experiment whose mission is to produce the largest three-dimensional map of the universe to date. This map will enable unprecedented constraints on dark energy (for a comprehensive review, refer to \citealt{Weinberg13} or \citealt{Amendola13}) by charting the expansion history of the universe through studies of baryon acoustic oscillations (BAO; see \citealt{Eisenstein05} or \citealt{Bassett10} reviews) and constraining the growth of structure through redshift-space distortion measurements (RSD; see e.g. \citealt{Ruggeri18} for a recent study). In addition, it will provide a means to precisely measure the sum of neutrino masses (\citealt{FontRibera14}), and to investigate theories of inflation (\citealt{Gariazzo15}, \citealt{Tellarini16}) and modified gravity (\citealt{Jain10}, \citealt{Joyce15}, \citealt{Casas17}, \citealt{Amendola18}). Installed on the Mayall 4-meter telescope at Kitt Peak, DESI is a Stage IV dark energy project\footnote{As defined in the Dark Energy Task Force report \citep{DarkEnergyTaskForce}.} consisting of a highly multiplexed fiber-fed spectrograph that can measure as many as 5000 spectra in parallel using robot fiber positioners. DESI will obtain spectra for four main target classes selected from imaging, including approximately 6 million luminous red galaxies (LRG) up to $z = 1.0$, 17 million [OII] emission-line galaxies (ELG) up to $z=1.6$, and 2.5 million quasars (QSO). QSOs with $z < 2.1$ will serve as tracers of the underlying dark matter distribution, while a high redshift sample of QSOs ($2.1 < z < 3.5$) will be used for their Lyman-$\alpha$ absorption features to probe the distribution of neutral hydrogen in the intergalactic medium. During ``bright time,'' when the position of the moon above the horizon impacts the observation of faint, high redshift targets, DESI will conduct the Bright Galaxy Survey (BGS), observing over 10 million galaxies up to $z \sim 0.4$, and also the Milky Way Survey (MWS) of local stars.

DESI is currently in its commissioning phase, with survey validation scheduled for the spring of 2020. At this stage of the project, it is vital to ensure that the targets selected from imaging will satisfy the science requirements of the collaboration, which demand meticulous control over all possible systematics. Since it is not obvious how systematics in the distribution of targets selected from imaging will translate to systematics in the 3D clustering of the spectroscopic samples, it is prudent to identify and mitigate them to the greatest extent possible. Furthermore, while acquiring spectra will allow us to remove any low redshift contaminants and deproject any purely angular systematics, the targeting efficiency of the survey will be adversely affected. In this paper, we analyze the impact of potential systematics from imaging and target selection on the observed clustering of the main DESI samples and develop methods to ameliorate those effects. Using the resulting value-added large-scale structure catalogs, we begin to characterize the properties of these tracer samples, both as a first step for analysis in cosmological studies and to aid in generating accurate mock catalogs.

The paper is organized as follows: Section~\ref{sec:methods} outlines our techniques for measuring and modeling clustering. Section~\ref{sec:catalogs} describes the imaging data and explains how we build our large-scale structure catalogs. In Section~\ref{sec:masks}, we implement and test some preliminary survey masks; Section~\ref{sec:masks/complete} deals with imaging completeness and its sensitivity to depth variations and color cuts, while Section~\ref{sec:masks/foregrounds} covers masking around bright foregrounds. In Section~\ref{sec:spatial}, we explore spatial variations in the clustering of the targets and identify additional problematic or anomalous regions. In Section~\ref{sec:systematics}, we investigate the effects of varying survey properties and instrument characteristics on the densities of targets, obtaining photometric weights for the most dominant systematics, with Section~\ref{sec:stellar_contam} focusing on the impact of stellar contamination in the QSO sample. Section~\ref{sec:characterization_angular} presents our preliminary characterization of the angular distribution of of the tracer samples, including mean surface densities, angular correlation functions, angular power spectra, linear biases, counts-in-cells moments, and clustering as a function of magnitude. In Section~\ref{sec:characterization_cross}, we use external spectroscopic catalogs to measure real-space projected cross-correlations, clustering $dN/dz$, and clustering as a function of luminosity. Section~\ref{sec:conclusions} summarizes our results and conclusions. 

Throughout, we will work in co-moving coordinates and assume a flat $\Lambda$CDM cosmology with $h = 0.676$, $\Omega_m h^2 =  0.142$, $\Omega_b h^2 = 0.022$, $n_s = 0.962$, and $\sigma_8 = 0.848$ (the default parameters in \texttt{CLASS}, see e.g. \citealt{Blas++11}). Additionally, all magnitudes are quoted as $AB$ magnitudes, unless otherwise specified.

\section{CLUSTERING MEASUREMENT AND THEORY}
\label{sec:methods}
To characterize the properties of DESI main samples and understand how systematics impact their observed distributions, we must be able to accurately quantify clustering as well as compare our results to theoretical predictions and survey expectations. To convert between angular clustering measurements and 3D clustering theory in Section~\ref{sec:characterization_angular}, we assume the fiducial  redshift distributions in the DESI Science Final Design Report (\citealt{DESI16}, henceforth FDR), calculated from cross-matching and photometric methods (private communications: Rongpu Zhou, Anand Raichoor, and Nathalie Palanque-Delabrouille for the DR7 LRG, ELG, and QSO $dN/dz$, respectively). These redshift distributions are plotted in Figure~\ref{fig:spec_overlapping_z}. In Section~\ref{sec:characterization_cross}, we use cross-correlations with external spectroscopy to obtain clustering $dN/dz$.

\subsection{Angular correlation functions}
\label{sec:methods/corrfuns}
One of the simplest and most powerful measurements of clustering is the two-point correlation function $\xi(r)$, which measures the excess probability, compared to a random Poisson distribution, that a pair of objects lie at a given separation (see e.g. \citealt{Peebles80}, \citealt{Peacock99}). The two-point correlation function and its Fourier transform, the power spectrum, fully characterize a Gaussian random field. For samples that lack redshifts, the 2D angular correlation function $w(\theta)$, representing the probability in excess of random of finding two objects separated by a given angle, may be used instead.

\subsubsection{Pair-count estimators}
\label{sec:methods/corrfunds/paircount}

We measure $w(\theta)$ of the targets with direct pair-count estimators, namely the Landy-Szalay estimator \citep{LandySzalay93}\footnote{Normalization factors are not included here.},
\begin{equation}\label{eqn:LSestimator}
\hat{w}_{LS}(\theta) = \frac{D_1D_2 - D_1R_2 - D_2R_1 + R_1R_2}{R_1R_2}
\end{equation}
where $DD$, $DR$, and $RR$ respectively refer to counts of data-data, data-random, and random-random pairs at average separation $\theta$ (within annular bins $\theta \pm \delta\theta$). For auto-correlations, this simplifies to $\hat{w}_{LS}(\theta) = (DD - 2DR + RR)/RR$. When cross-correlating with external data sets that don't have a corresponding random catalog readily available, we instead use the Davis-Peebles estimator \citep{DavisPeebles83}, 
\begin{equation} \label{eqn:DPestimator}
\hat{w}_{DP}(\theta) = \frac{D_1D_2 - D_2R_1}{D_2R_1}
\end{equation}
which has a slightly larger variance (see \citealt{LandySzalay93} for a comparison of pair-count estimators) but only requires one set of randoms. We count pairs within 16 logarithmically spaced angular bins between $\theta = 0.001^{\circ}$ and $\theta = 1^{\circ}$.

\subsubsection{Limber approximation}
\label{sec:methods/corrfuns/modelling}

We wish to generate a prediction for the observed angular clustering of objects in the sky, $w(\theta)$, given an assumed model for the full three-dimensional clustering, $\xi(r)$, and redshift distribution, $dN/dz$. An approximation first introduced by \citet{Limber53} and \citet{Rubin54} is frequently employed for this purpose. Briefly, the Limber approximation assumes redshift distributions that do not vary appreciably over the coherence length of the structures defined by $\xi(r)$. Though not a requirement, it is often further assumed that the sky may be treated as flat. In this section, we state the general result, as well as the simplified expression for the case of a power-law $\xi(r)$. A more detailed derivation can be found in e.g. \citet{Simon07} or \citet{LoverdeAfshordi08}.

Written in center-of-mass and relative coordinates, $\bar{r} = (r_1 + r_2)/2$ and $\Delta r = r_2 - r_1$, the Limber approximation straightforwardly relates the angular and spatial correlations between two samples, 
\begin{equation}
w_{1,2}(\theta) = \int_{0}^{\infty} d\bar{r} \ f_1(\bar{r}) \ f_2(\bar{r}) \int_{-\infty}^{\infty} d\Delta r \ \xi_{1,2}(R,\bar{r})
\end{equation}
where $f_1$ and $f_2$ are the normalized radial distributions, and $ R = \ \sqrt[]{\bar{r}^2\theta^2 + \Delta r^2}$. If we further presume a power-law for the correlation function, with correlation length $r_0$,
\begin{equation}
\xi(r) = \bigg( \frac{r}{r_0} \bigg) ^{-\gamma}
\end{equation}
then we can evaluate the $\Delta r$ integral directly, and Limber's approximation gives a particularly simple result, 
\begin{align}
\label{eqn:limber-pl}
w_{1,2}(\theta) &= \theta^{1-\gamma} \ r_0^\gamma \ \sqrt[]{\pi} \ \frac{\Gamma(\gamma/2 - 1/2)}{\Gamma(\gamma/2)} \int_{0}^{\infty} d\bar{r} \ f_1(\bar{r}) \ f_2(\bar{r}) \ \bar{r}^{1-\gamma} \nonumber \\
&\equiv A_w \theta^{1-\gamma} 
\end{align}
Using this equation with tabulated $dN/dz$'s, we can compare observation to theory by determining the clustering length $r_0$ and slope $\gamma$ that best fit the observed $w(\theta)$.

\subsection{Projected real-space cross-correlations}
\label{sec:methods/xcorrs}
We can extract additional clustering information by cross-correlating the samples with other samples of known redshift. We begin by deriving a relation between the angular correlation function $w(\theta)$ and the projected real-space correlation function $w_p(r_p)$, the latter being defined as the integral of the 2D spatial correlation function $\xi(r_{\pi}, r_p)$ over the line of sight $r_{\pi}$ \citep{DavisPeebles83}.

Starting with the simple case in which the spectroscopic sample lies at $\chi = \chi_0$, the flat-sky approximation yields
\begin{equation}
    \begin{aligned}
    w(\theta) = \int d\chi f(\chi) \xi \Big( \sqrt{\chi_0^2\theta^2 + (\chi-\chi_0)^2} \Big)
    \end{aligned}
\end{equation}
where $f(\chi)$ is the normalized radial distribution of the photometric sample, and the second integral over the radial distribution of the spectroscopic sample, a delta function, has been performed. Applying the Limber approximation (Section~\ref{sec:methods/corrfuns/modelling}), this simplifies to
\begin{equation}\label{eqn:wp}
    \begin{aligned}
        w(\theta) &\simeq f(\chi_0) \int dr_{\pi} \xi(r_p, r_{\pi}) \\
        &= f(\chi_0) w_p(r_p)
    \end{aligned}
\end{equation}
where $w_p(r_p)$ is
a real-space measurement, since redshift space distortions only affect $r_{\pi}$. 

Generalizing to a narrow spectroscopic redshift slice, such that the clustering can still be treated as constant over the slice, we adopt the approach of \cite{Padmanabhan++09}: for each pair in a given bin, we assume the photometric object lies at the same redshift as the spectroscopic object it is being correlated with, allowing us to re-bin the pair counts in transverse separation, $w_{\theta}(r_p)$, such that Equation~\ref{eqn:wp} becomes
\begin{equation}
    \begin{aligned}
        w_{\theta}(r_p) = \big\langle f(\chi) \big\rangle w_p(r_p)
    \end{aligned}
\end{equation}
where $\big\langle f(\chi) \big\rangle$ is averaged over the spectroscopic redshift bin in question and $w_{\theta}(r_p)$ is the angular correlation function but binned in physical distance instead of angle using the redshifts of the spectroscopic objects for conversion.

To compare with theory, we note also the form $w_p(r_p)$ takes for a power-law correlation function model $\xi(r) = ( \frac{r}{r_0}) ^{-\gamma}$:
\begin{equation}
    w_p(r_p) = r_p^{1-\gamma}r_0^{\gamma} \sqrt[]{\pi} \frac{\Gamma(\gamma/2 - 1/2)}{\Gamma(\gamma/2)}
\end{equation}

\subsection{Bootstrap errors}
\label{sec:methods/errors}
\label{sec:methods/errors/bootstrap}

As an internal error estimate, we use the bootstrap technique of \cite{Efron79}, splitting the sample into multiple subsamples and then randomly selecting with replacement to obtain many different realizations of the underlying distribution. Since resampling on individual objects has been shown to lead to unreliable errors (\citealt{Mo92}, \citealt{Fisher94}), with variance underestimated in underdense regions and overestimated in overdense regions, we instead partition the sky into equal area pixels and resample these. 

The choice of the bootstrap over similar methods such as the jackknife is motivated by comparative studies (e.g. \citealt{Norberg++09}) suggesting that, though the bootstrap tends to overestimate variance on all scales, it recovers the principal eigenvectors of the true covariance matrix in an unbiased fashion. As such, we caution that our bootstrap error bars are likely overestimated in some cases.

In detail, we use the \texttt{HEALPix} package\footnote{\url{http://healpix.sf.net}} \citep{Gorski++05}  with $N_{\text{SIDE}} = 4$ to divide the surface of a sphere into $192$ equal area pixels of approximate size ${\sim} 215 \text{ sq deg}$, then throw away any pixels that do not overlap with the footprint, leaving 83 pixels. We then randomly select pixels with replacement until the number of randoms in each bootstrap realization is similar to the number of randoms in the footprint. We use 500 bootstrap realizations to obtain an estimate of the variance. Our results are robust to variations in the $N_{\text{SIDE}}$ resolution and the number of bootstrap realizations.


\subsection{Angular power spectra}
\label{sec:methods/cls}
The angular power spectrum, $C_{\ell}$, is another powerful tool for quantifying clustering, allowing us to study the Fourier modes of the angular distribution of galaxies. It complements the statistical information derived from the angular correlation function, to which it is related via a Legendre transform: 
\begin{equation}
    w(\theta) = \sum_{\ell} \frac{2\ell+1}{4\pi} P_\ell(\cos\theta)C_\ell
\end{equation}
Large-scale systematics are more clearly visible in the power spectrum than in the correlation function, which potentially has long-wavelength modes affecting all angular scales. On the other hand, the correlation function is more sensitive to small scales, where nonlinear evolution dominates and introduces correlations between different $C_{\ell}$'s at large $\ell$. Additionally, it is faster to compute small-scale clustering in configuration space and large-scale clustering in Fourier space. Thus, we focus our analysis of the angular power spectrum on large scales $\ell \leq 500$, corresponding to angular scales greater than $\theta \sim 180^{\circ} / \ell \approx 0.4^{\circ}$, or spatial scales greater than a few $h^{-1}$ Mpc at the characteristic survey depth of 1 $h^{-1}$ Gpc.

\subsubsection{Measurement}

 We use \texttt{HEALPix} with $N_{\text{SIDE}} = 512$ and estimate the angular power spectrum from harmonic analysis of the pixelised map of density contrast $\delta_g = n/\bar{n} - 1$, where $n$ is the number of galaxies in a given pixel and $\bar{n}$ is the average density over the entire masked sample multiplied by the given pixel's effective area. We mask out pixels whose effective area is less than 25\% of its full area, such that only pixels which are fully (or mostly) inside the survey geometry are considered\footnote{Since effective area is calculated using a set of uniformly distributed random points, there is some natural Poisson variance, hence why we do not use a more restrictive threshold.}. Using \texttt{anafast}, we obtain the estimated angular power spectrum $C_{\ell}$, which is the sum of the signal and the shot noise. To first-order correct the effects of partial sky, we divide by a factor of $f_{\text{sky}}$, the fraction of sky covered by the masked footprint; full deconvolution of the mask is deferred to a future work. We find that the angular power spectrum of the mask has its power concentrated in the large-scale modes, with the mask dropping to half power at $\ell \sim 10$ and falling below 10\% power beyond $\ell \sim 20$. We also divide out the pixel window function. The variance of the estimator can be modelled analytically as
\begin{align}
    \sigma_{\ell}^2 = \frac{1}{f_{\text{sky}}}\frac{2C_{\ell}^2}{2\ell + 1}
\end{align}
%
%
On a small section of sky $\phi$, multipole resolution is limited by $\Delta \ell \approx 180^{\circ} / \phi$, with the $\ell_{\rm min}$ mode constrained to be the wavelength that fits into the angular patch \citep{Peebles80}. We bin $C_{\ell}$ using 10 linearly spaced bins from $\ell_{\rm min} = 30$ to $\ell_{\rm max} = 500$ and take the weighted arithmetic mean and variance for each bin.
\begin{align}\label{eqn:binnedCls}
\bar{C}_{\text{bin}} &= \frac{\sum\limits_{\ell \text{ in bin}}^{} \frac{C_{\ell}}{\sigma_{\ell}^2}}{\sum\limits_{\ell \text{ in bin}}^{} \frac{1}{\sigma_{\ell}^2}} \nonumber \\ 
\frac{1}{\sigma_{\text{bin}}^2} &= \sum\limits_{\ell \text{ in bin}}^{} \frac{1}{\sigma_{\ell}^2}
\end{align}

\subsubsection{Theory}

In the first-order correction to the Limber approximation \citep{LoverdeAfshordi08}, the multipole expansion of the galaxy angular power spectrum is given by
\begin{align}\label{eqn:Cls}
    C_{\ell} &= \int d\chi \ f(\chi)^2 \frac{1}{\chi^2} P_{g}(k = (\ell + 1/2)/\chi, z) \nonumber \\
    &= \int d\chi \ f(\chi)^2 \frac{b(z)^2}{\chi^2} P_{m}(k = (\ell + 1/2)/\chi, z)
\end{align}
where $f(\chi) \equiv dN/d\chi = dN/dz \frac{H(z)}{c}$ is the normalized radial distribution, $P_{m}$ is the linear dark matter power spectrum, and $b(z)$ is the large-scale bias, which we assume takes the form\footnote{$D(z)$ is the linear growth function with normalization $D(z=0)=1$. Thus the approximation $b\propto D^{-1}(z)$ assumes that the clustering is constant, since the evolution of $P(k,z)$ is cancelled by the evolution of $b^2(z)$.} $b(z) = b_0/D(z)$ as per the DESI FDR. In Section~\ref{sec:characterization/cls}, we fit the linear bias $b_0$ and also explore the scale-dependence of the bias in the weakly nonlinear regime.

\subsection{Clustering $dN/dz$}
\label{sec:methods/dndz}
\subsubsection{Overview}

The idea of using cross-correlations to infer redshift information about objects in the night sky has been circulating for decades (e.g. \citealt{SeldnerPeebles79}, \citealt{PhillippsShanks87}, \citealt{Landy++96}, \citealt{Ho08}), but has received renewed attention in the context of modern astronomical surveys, which are probing deeper than ever and imaging far more objects than can feasibly be targeted for spectroscopic observation. In recent years, a number of $dN/dz$ cross-correlation estimators have been proposed and studied (\citealt{Newman08}, \citealt{MatthewsNewman10}, \citealt{Schulz10}, \citealt{MatthewsNewman12}, \citealt{McQuinnWhite13}, \citealt{Menard13}) and applied to real or simulated data (\citealt{Schmidt13}, \citealt{Scottez++16}, \citealt{Hildebrandt++17}, \citealt{Scottez++18}, \citealt{Davis++18}, \citealt{Gatti18}, \citealt{Chiang18}, \citealt{Krolewski19}). Following \cite{Menard13}, which presents a simple and practical method for estimating the clustering $dN/dz$ of a sample, we probe the redshift distributions of objects targeted for DESI in Section~\ref{sec:characterization/dndz}. Unlike other methods, the M{\'e}nard method takes advantage of small-scale clustering information and reduces the impact of systematics by sidestepping autocorrelation functions. We briefly describe the formalism of M{\'e}nard method and the details of our implementation below.

\subsubsection{Method}

Consider two populations. Let the reference (spectroscopic) sample have a redshift distribution $dN_r/dz$, a mean surface density $\bar{n}_r$, and a total number of objects $N_r$. The corresponding properties for the unknown (photometric) sample will be labeled $dN_u/dz$, $\bar{n}_u$, and $N_u$, respectively. The angular cross-correlation between the reference sample and the unknown sample is estimated by 
\begin{equation}
w_{ur}(\theta,z) = \frac{\langle n_u(\theta,z) \rangle}{\bar{n}_u} - 1
\end{equation}
where $\langle n_u(\theta,z) \rangle$ is the mean surface density of objects from the unknown sample lying within an angular distance $\theta$ of objects in the reference sample at redshift $z$. To calculate this, we bin the reference sample into narrow redshift bins $\delta z_i$. Then, for each $\delta z_i$, we estimate $w_{ur}(\theta,z_i)$ by pair counting with the Davis-Peebles estimator Equation~\ref{eqn:DPestimator}.

In practice, we actually integrate over an annulus around each reference object, from $\theta_{\text{min}}$ to $\theta_{\text{max}}$, because the sensitivity of the estimator is improved by encoding information from many clustering scales \citep{Menard13}. In order to maximize the SNR, we weight each point by $\theta^{-1}$, which gives equal amounts of clustering information per logarithmic scale ($d\theta/\theta = d\log{\theta}$). 
\begin{equation}
\bar{w}_{ur}(z) =  \int_{\theta_{\text{min}}}^{\theta_{\text{max}}}{d\theta \ \frac{w_{ur}(\theta, z)}{\theta}}
\end{equation}
%
To avoid excess signal from cross-correlations between duplicate objects that appear in both catalogs, it is necessary to impose a minimum radius, $\theta_{\text{min}}$ which is at least as large as the astrometric uncertainties in the survey. Furthermore, as we go to smaller scales $<$ 1 Mpc, clustering becomes increasingly nonlinear and bias becomes increasingly scale-dependent, so the assumptions underpinning the estimator break down, potentially affecting the accuracy of the result. Finally, we note that as the scale falls below the mean separation of spectroscopic objects, cross-correlations between redshift bins become more significant. Meanwhile, at larger scales, the advantage of a linear bias\footnote{The bias measured by these angular cross-correlations is dominated by scales of hundreds of kpc to a few Mpc, and thus should be distinguished from the large-scale ($>$10 Mpc) bias, which may evolve differently.} must be balanced against the cost of degraded signal-to-noise since the clustering signal decreases with radius and the noise due to systematics increases as more background sources are included in the counts. Thus, for samples which have little to no bias evolution, small scales are ideal for recovering $dN/dz$; for samples with some bias evolution, intermediate scales are recommended; and for samples with extensive bias evolution, restricting to large scales may be the best strategy. 

\subsection{Counts-in-cells}
\label{sec:methods/cic}
To aid in simulations, we wish to also provide some measurement of the higher order clustering statistics of the DESI samples. In particular, for regions of high density, spectroscopic incompleteness due to the physical limitation of fiber allocation is expected to introduce systematic effects on the observed clustering (\citealt{Cahn17}, \citealt{Burden17}, \citealt{Pinol17}) which must be included in any realistic mock catalog. Thus, rich clustering information down to the scale of the DESI fiber patrol radius of $1.4^\prime$ is invaluable for the purpose of mock calibration and validation.

Since a discrete map of galaxies or quasars samples the continuous density field of matter, the number of galaxies within a randomly placed cell (counts-in-cells: \citealt{Hubble34}; \citealt{White79}; \citealt{Peebles80}) provides a window into the higher-order correlations. Let $P(N)$ be the probability that a cell with area $\Omega$ contains exactly $N$ galaxies. The factorial moments of this distribution, $F_p = \langle N(N-1)...(N-p+1) \rangle $, are related to the corresponding moments of the spatially smoothed underlying density field,  $\overline{\mu}_p = \langle (1 + \delta)^p \rangle$, via the simple relation $\overline{\mu}_p = \frac{F_p}{\langle N \rangle ^p}$, which tidily includes shot noise corrections arising from the fact that we are dealing with a discrete, locally Poissonian representation of the continuous field  (\citealt{SzapudiSzalay93b}, \citealt{Szapudi++96}). From these moments, we can extract the correlation functions of corresponding order (also smoothed over the characteristic cell size), $\overline{w}_p(\Omega) = \frac{1}{\Omega^p} \int_{\Omega}^{}d\Omega_1... d\Omega_p w_p(\theta_1,...,\theta_p)$. The first few are listed below (e.g. \citealt{Fry85}, \citealt{Fry++11}):
\begin{align}
  \overline{\mu}_2 &= 1 + \overline{w}_2 \nonumber \\
  \overline{\mu}_3 &= 1 + 3\overline{w}_2 + \overline{w}_3  \\
  \overline{\mu}_4 &= 1 + 6\overline{w}_2 + 3\overline{w}_2^2 + 4\overline{w}_3 + \overline{w}_4 \nonumber
\end{align}
%
%
\cite{SzapudiColombi96} classifies theoretical errors on counts-in-cells statistics as either cosmic errors (due primarily to shot noise, edge effects, and finite volume) or  measurement errors (due to the finite number of sampling cells), the latter scaling as the inverse of the number of sampling cells used. The counts-in-cells distribution and its moments are usually determined by throwing random cells over the region of interest, with massive oversampling required to control the measurement errors. However, \cite{Szapudi98} describes a method of implementing counts-in-cells that is essentially equivalent to using an infinite number of random cells, thereby eliminating the measurement error entirely. We implement this method using our publicly available code \texttt{infcic}\footnote{\url{ https://github.com/ekitanidis/infcic}}. Therefore, we present only the uncertainty associated with cosmic errors, which we approximate by calculating counts-in-cells over two large fields, one in each galactic hemisphere, and measuring the mean and dispersion, weighted by effective area.

\section{IMAGING CATALOGS}
\label{sec:catalogs}
\subsection{Imaging data}
\label{sec:catalogs/data}

The DECam Legacy Survey (DECaLS) is a wide-field photometric survey amassing deep multicolor imaging within the footprints of ongoing and future spectroscopic surveys. Using the DECam instrument \citep{Flaugher15} at the Blanco 4m telescope, DECaLS observed in three optical/near-infrared bands ($g$, $r$, $z$), complemented by four mid-infrared bands from the Wide-field Infrared Survey Explorer (WISE; \citealt{Wright10}). DECaLS aims to obtain images up to $5\sigma$ point-source depths $g = 24.7$, $r = 23.9$, and $z = 23.0$ AB mag, and is designed to boost the science power of spectroscopic observations by providing publicly available imaging with superior depth (1-2 magnitudes fainter) and enhanced image quality compared to existing photometry from SDSS, ATLAS, and Pan-STARRS. DECaLS covers a large equatorial region (bounded by $\delta < 32^\circ$ in galactic coordinates), corresponding to roughly two-thirds of the optical imaging used for DESI targeting, and is a key piece of the Legacy Survey project, which has imaged the full 14000 square degrees of extragalactic sky making up the DESI survey \citep{Dey18}.

\begin{table}
\centering\begin{tabular*}{0.97\columnwidth}{c p{0.3cm} p{0.3cm} p{0.3cm} p{0.3cm} p{0cm} p{0.5cm} p{0.5cm} p{0.5cm} p{0.5cm}}
\toprule
N$_{\text{EXP}}$ & \multicolumn{4}{c}{f$_{\text{sky}}$} & & \multicolumn{4}{c}{Area (deg$^2$)}
\\
\midrule
          & $g$ & $r$ & $z$ & all & & $g$ & $r$ & $z$ & all \\
          \cmidrule{2-5}
          \cmidrule{7-10}
0  & 0.14 & 0.11 & 0.05 & 1.0 & & 1572.5 & 1263.8 & 541.6 & 11243.6 \\
1  & 0.17 & 0.14 & 0.11 & 0.82 & & 1865.1 & 1529.1 & 1292.5 & 9273.5 \\
2  & 0.23 & 0.21 & 0.18 & 0.63 & & 2620.5 & 2329.9 & 2035.4 & 7114.6 \\
3  & 0.2 & 0.22 & 0.24 & 0.39 & & 2279.5 & 2430.0 & 2665.8 & 4380.5 \\
4  & 0.09 & 0.11 & 0.16 & 0.19 & & 1032.8 & 1274.3 & 1846.3 & 2185.9 \\
5+ & 0.17 & 0.21 & 0.25 & 0.11 & & 1873.2 & 2416.5 & 2861.9 & 1220.6 \\
\bottomrule
\end{tabular*}
\caption{Area and sky fraction covered by \textit{exactly} 0,1,2,3,4,5+ exposures in each optical band in DECaLS DR7. The ``all'' columns are cumulative, such that the $N_{\text{EXP}} = 1$ row refers to area and sky fraction covered by \textit{at least} one exposure in all bands. Areas are estimated by sampling the footprint with randoms. In this context, the sky fraction is defined relative to the total DECaLS footprint, whereas elsewhere in the paper, it is defined relative to the area where imaging exists (N$_{\text{EXP}} > 0 $) in all three optical bands.}
\label{table:passes}
\end{table}

\begin{figure}
\centering
\includegraphics[width=1\linewidth]{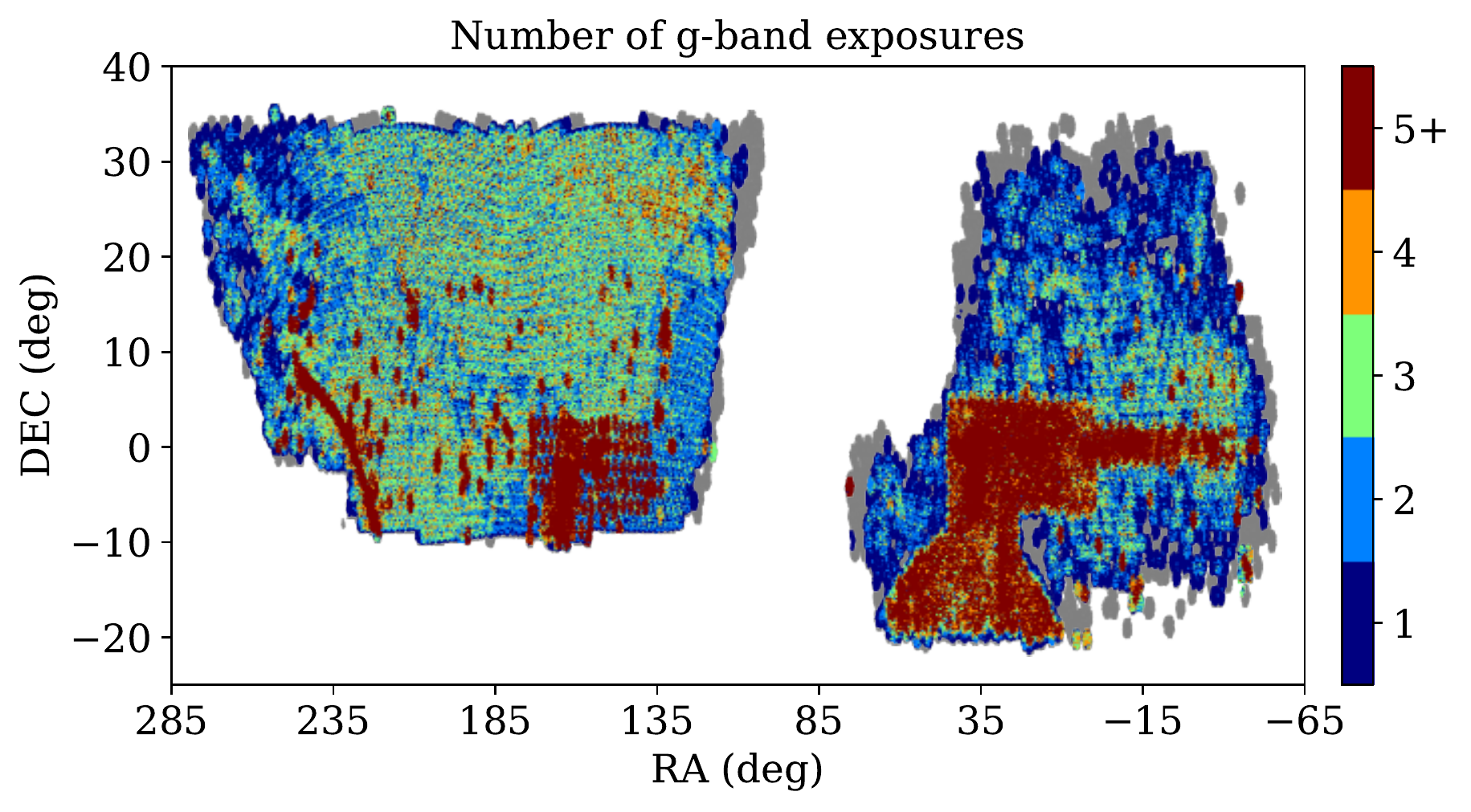}
\includegraphics[width=1\linewidth]{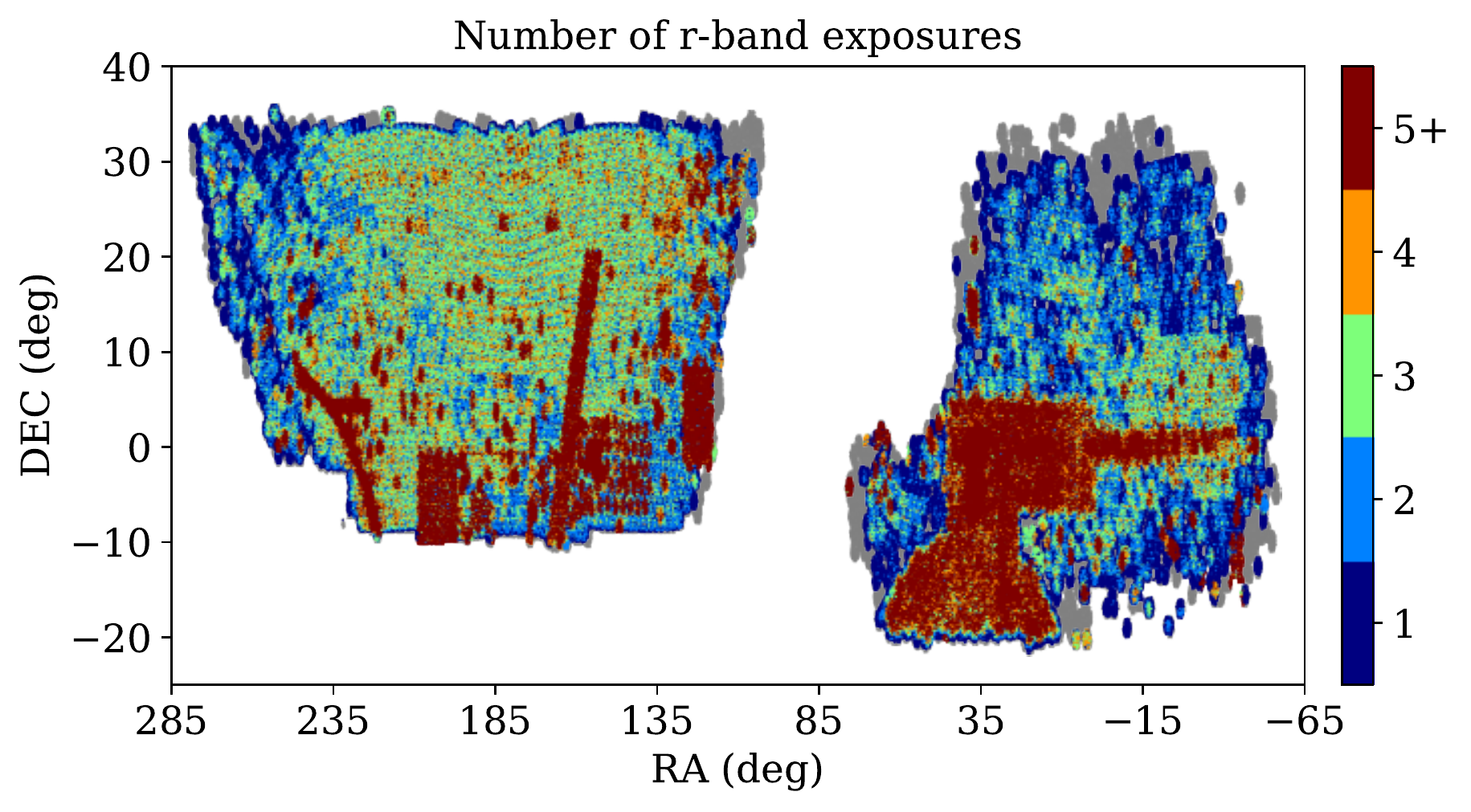}
\includegraphics[width=1\linewidth]{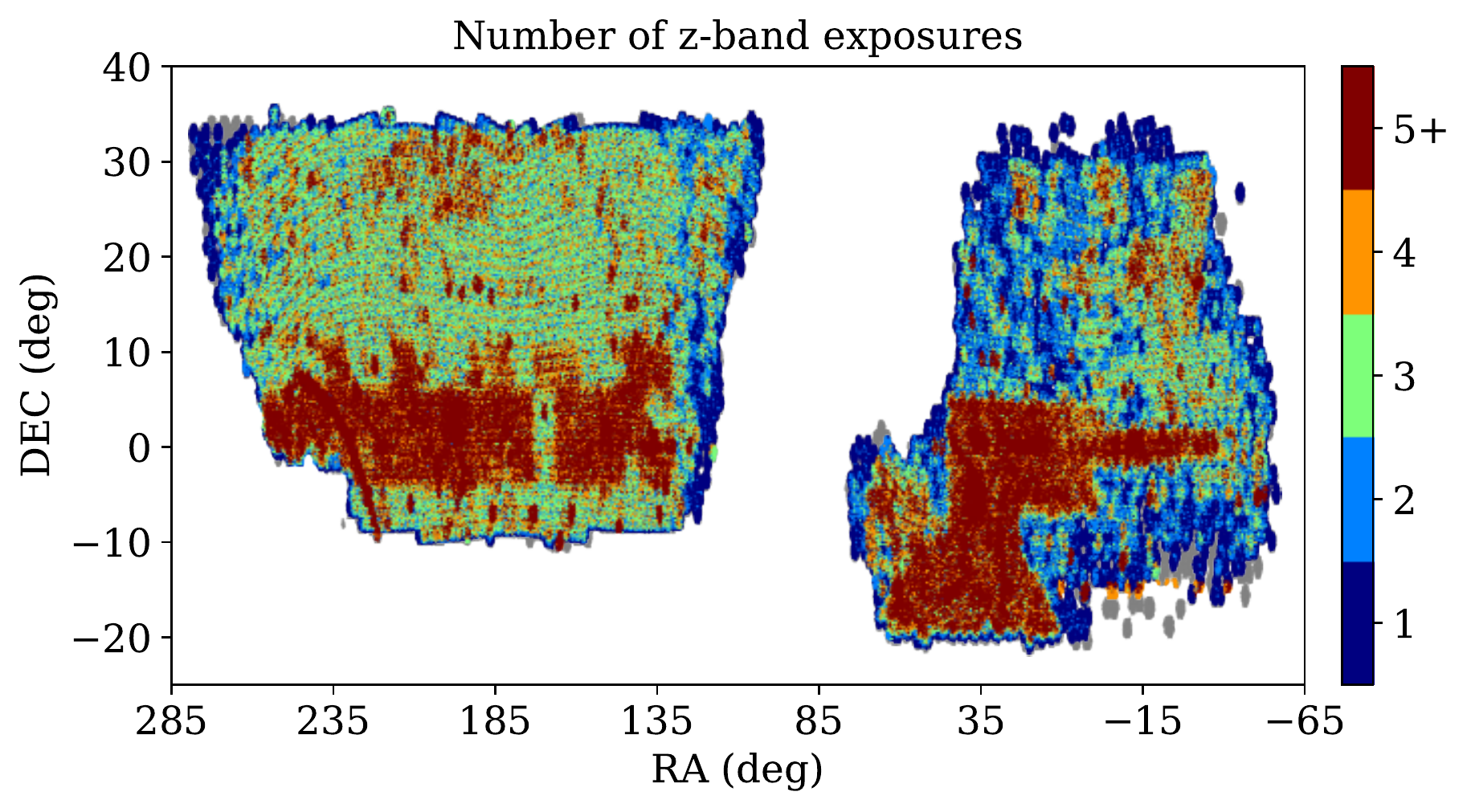}
\caption{Number of exposures in each DECaLS band in DR7, estimated by sampling the footprint with randoms. No map projection is applied here.}
\label{fig:map_nexp}
\end{figure}

Raw DECam images are processed through the NOAO Community Pipelines, with astrometric calibration and photometric characterization based on Pan-STARRS-1 measurements. The calibrated images are then run through \textit{The Tractor}\footnote{\url{https://github.com/dstndstn/tractor}} \citep{Lang16a}, which produces an inference-based catalog by optimizing the likelihood for source properties, given the data and a noise model. We use Data Release 7 (DR7), the seventh public data release of the Legacy Survey, which is the last DECaLS-only data release, including observations from August 2013 through March 2018 (NOAO survey program 0404). It also uses non-DECaLS observations from DECam conducted between August 2013 and  March 2018, including some data from the Dark Energy Survey (DES; \citealt{DES05}). Together, these cover approximately 9766 square degrees in the $g$-band, 9853 square degrees in the $r$-band, and 10610 square degrees in the $z$-band, with 9298 square degrees observed in all three optical bands (see also Table~\ref{table:passes}, Figure~\ref{fig:map_nexp}).

\subsection{Target selection}
\label{sec:catalogs/targets}

Table~\ref{table:targets} summarizes the primary target types evaluated in this paper. These targets are defined in great detail in the FDR and their selection algorithms are briefly outlined below.

\begin{table}
\begin{tabular}{cccc}
  \hline
  Target & Redshift range & \multicolumn{2}{c}{Selection bands} \\
    & & Primary & Other \\
   \hline
  LRG & 0.4 - 1.0 & $z$ & $g$, $r$, $W1$ \\
  ELG & 0.6 - 1.6 & $g$ &  $r$, $z$  \\
  QSO (tracers) & < 2.1 & $r$ & $g$, $z$, $W1$, $W2$   \\
  QSO (Ly-$\alpha$) & 2.1 - 3.5 & $r$ & $g$, $z$, $W1$, $W2$ \\
  \hline
\end{tabular}

\caption{Summary of selection properties for each of the dark time DESI target classes. $W1$ and $W2$ denote WISE bands. ``Primary'' refers to the band used to define the limiting magnitude, which is relevant for the completeness mask (see Section~\ref{sec:masks/complete}).}
\label{table:targets}
\end{table}

\subsubsection{LRG}
LRG targets are selected from the $g$, $r$, $z$, and $W1$ bands by applying a series of color cuts using extinction-corrected magnitudes. No morphology cut is applied. 
\begin{equation}
\begin{split}
18.01 < &z < 20.41 \\
0.75 < r &- z < 2.45 \\
-0.6 < (z - W1&) - 0.8 \ (r - z) \\
(z - 17.18) \ / \ 2 < r \ - \ &z < (z - 15.11) \ / \ 2 \\
(r - z > 1.15) \ \ &\text{||} \ \ (g - r > 1.65)
\end{split}
\end{equation}

\subsubsection{ELG}
ELG targets are selected from the $g$, $r$, and $z$ bands by applying a series of color cuts using extinction-corrected magnitudes. No morphology cut is applied. 
\begin{equation}
\begin{split}
21.00 < \ &g < 23.45 \\ 
0.3 < (r \ &- \ z) < 1.6 \\
(g \ - \ r) < 1.15 \ &(r \ - \ z) - 0.15 \\
(g \ - \ r) < 1.6 &- 1.2 \ (r \ - \ z)
\end{split}
\end{equation}

\subsubsection{QSO}
QSO targets are selected by applying a machine learning method based on Random Forests (RF) which relies only on extinction-corrected object colors in the $g$, $r$, $z$, $W_1$, $W_2$ bands (Christophe Y\`{e}che, private communication). The algorithm is trained using all known QSOs in the footprint with an initial cut of $r < 23$ against a sample of unresolved objects from Stripe 82 without known QSOs and objects exhibiting QSO-like variations in their light curves. In the target selection itself, a tighter initial cut of $r < 22.7$ is applied.

\subsection{Catalogs and randoms}
\label{sec:catalogs/lss}

DESI target catalogs and uniform random catalogs are created with our public code \texttt{ImagingLSS}\footnote{\url{https://github.com/desihub/imaginglss}}. \texttt{ImagingLSS} processes the outputs of the DECaLS pipeline and selects DESI targets from it, as well providing the option for auxiliary, user-defined targets. Uniform unclustered randoms are sampled from the imaging survey footprint. Geometric survey masks (for example, vetoing by proximity to bright objects) can be applied to both catalogs and randoms in a consistent manner. In addition to DECaLS data, \texttt{ImagingLSS} uses the SFD98 dust map \citep*{Schlegel98} to correct for extinction, as well as the Tycho-2 star catalog \citep{Hog00} and the AllWISE catalog \citep{AllWISE} to mask out bright stars. 

\FloatBarrier

\section{IMAGING MASKS}
\label{sec:masks}
We develop two types of initial masks, completeness and bright star, to reject possibly problematic regions of the imaging data. The effects of these masks on the survey efficiency and effective area are summarized in Table~\ref{table:masks}. Our baseline sample is selected from regions where imaging exists in all three optical bands used for targeting (``no mask''), and all sky fractions are quoted relative to this sample. In subsequent sections, we will review the purpose and implementation of each mask, with Section~\ref{sec:masks/complete} focused on imaging completeness, and Section~\ref{sec:masks/foregrounds} describing our bright star mask. We also investigate whether there is need for a mask around extended sources such as large galaxies.

\begin{table}
{\centering
\noindent\begin{tabular}{p{1.0cm}p{1.9cm}p{1.3cm}p{1.5cm}p{0.5cm}}
\toprule
Target & Mask & Number & Area (deg$^2$) & $f_{\rm sky}$ \\
\midrule
\multirow{4}{*}{LRG} & no mask & 4882206 & 9298.91 & 1.0 \\
                     & complete only      & 4872537 & 9281.0 & 1.0 \\
                     & bright star only   & 4304375 & 8584.68 & 0.92 \\
                     \cmidrule{2-5}
                     & \textbf{all masks} & 4296486 & 8568.65 & 0.92 \\
\hline
\multirow{4}{*}{ELG} & no mask & 23224353 & 9298.91 & 1.0 \\
                     & complete only      & 23032874 & 9227.78 & 0.99 \\
                     & bright star only   & 20329143 & 8638.84 & 0.93 \\
                     \cmidrule{2-5}
                     & \textbf{all masks} & 20166887 & 8575.15 & 0.92 \\
\hline
\multirow{4}{*}{QSO} & no mask & 3125148 & 9298.91 & 1.0 \\
                     & complete only      & 3116976 & 9288.35 & 1.0 \\
                     & bright star only   & 1814359 & 6655.11 & 0.72 \\
                     \cmidrule{2-5}
                     & \textbf{all masks} & 1810801 & 6648.68 & 0.71 \\
\bottomrule
\end{tabular}}
\caption{Summary of masks and how each affects the number of targets and the effective area and sky fraction. Here, sky fractions are quoted relative to the ``no mask'' case, which is simply the joint regions of the footprint where imaging is available in all three optical bands.}
\label{table:masks}
\end{table}

\subsection{Survey depth and completeness}
\label{sec:masks/complete}

Tractor catalogs contain an estimation of the imaging depth at each observed pixel in the footprint. This depth is affected by the number of exposures, exposure times, observational conditions, and instrument effects\footnote{Note that we do not explicitly apply the \texttt{ALLMASK} flag, which uses the NOAO Community Pipeline's data quality map to mask out bad pixels on the CCD and pixels affected by bleed trail, transients such as cosmic rays, and saturation. These effects are accounted for in the estimation of the depths \citep{Dey18} and thus are perforce included in our completeness mask.}. Due to the multi-pass nature of the imaging survey and the fact that it is ongoing, variations in depth across the footprint are substantial. In DR7, some of the sky has been covered just once, while the deepest regions have received five or more passes (see Table~\ref{table:passes} and Figure~\ref{fig:map_nexp}).

\begin{figure}
\centering
\includegraphics[width=1\linewidth]{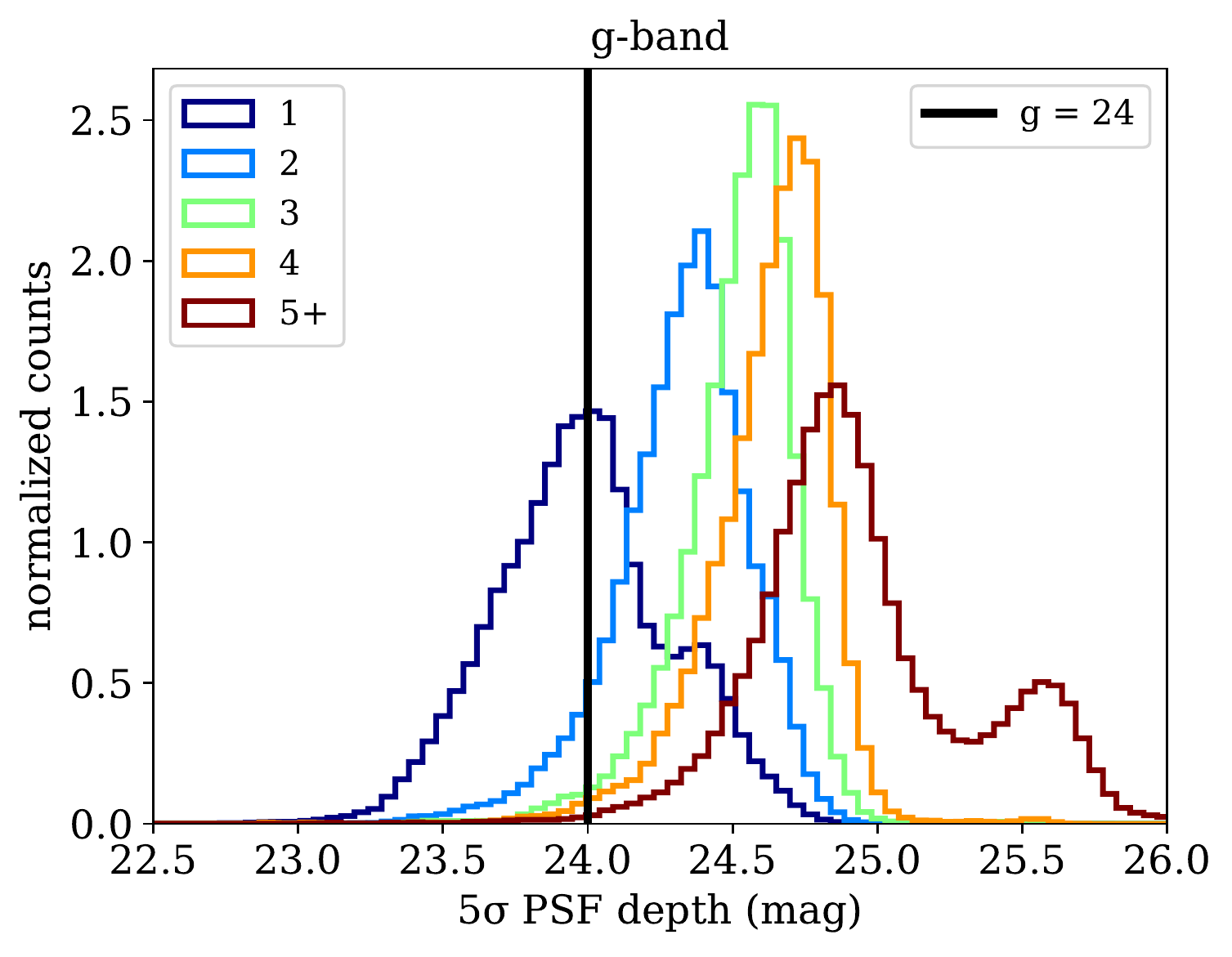}
\includegraphics[width=1\linewidth]{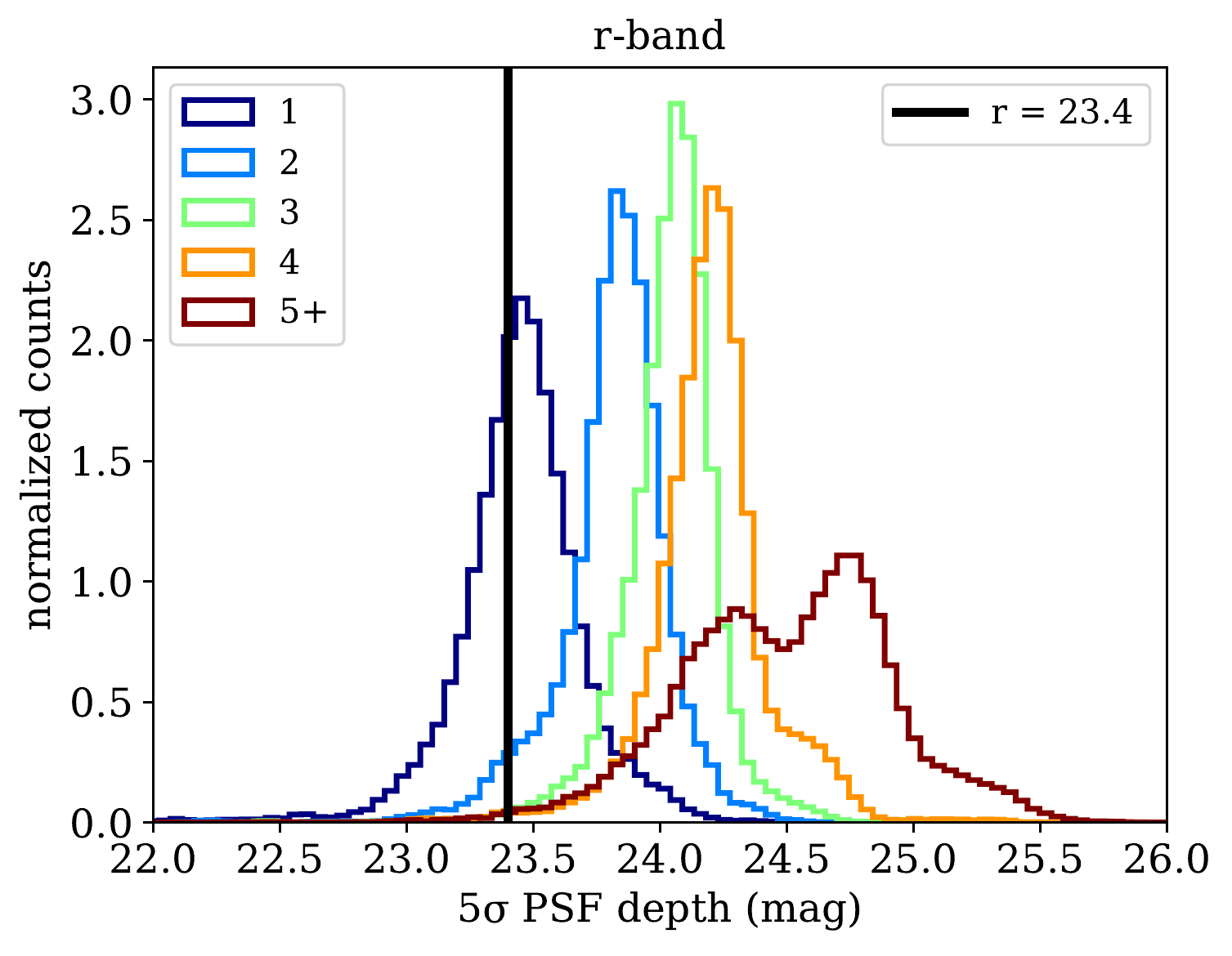}
\includegraphics[width=1\linewidth]{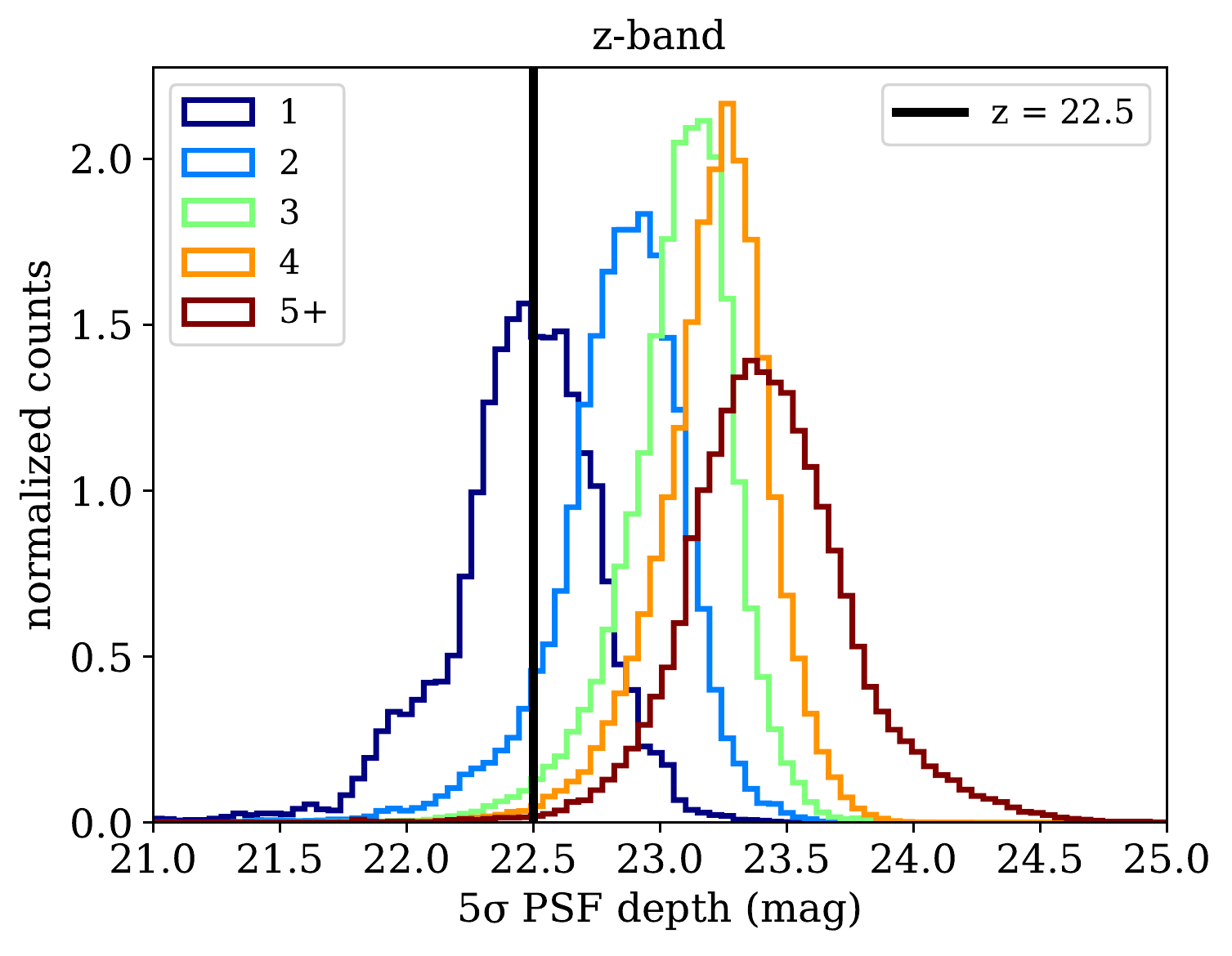}
\caption{Histograms of $5\sigma$ point-source depths of randoms in each band, normalized as probability densities, with the colored curves corresponding to different numbers of exposures. The solid vertical lines are the DESI nominal $5\sigma$ depth requirements $g = 24.0$, $r = 23.4$, $z = 22.5$ for an ELG galaxy with half-light radius of 0.45 arcsec.}
\label{fig:depths_nexp}
\end{figure}

\begin{figure}
\centering
\includegraphics[width=\linewidth]{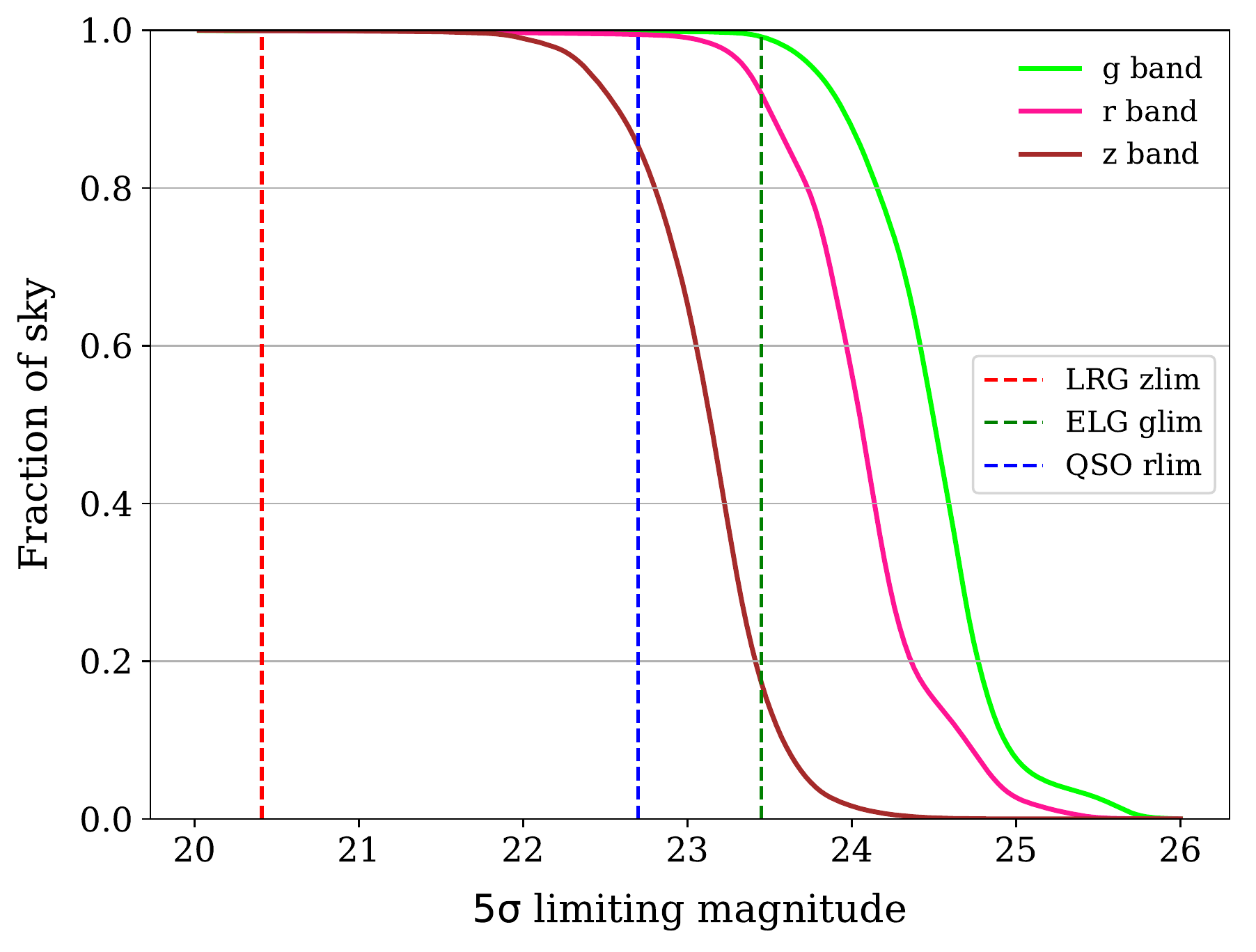}
\caption{Cumulative sky fraction vs. $5\sigma$ limiting magnitudes, with target selection cuts shown as vertical lines to demonstrate the effect that shifting a magnitude limit up or down would have on the completeness of the corresponding target.}
\label{fig:depths_colorcuts}
\end{figure}

To select a uniform and complete sample, we implement a ``completeness'' mask. Since two exposures, at minimum, are needed to meet the nominal depth requirements of DESI over most of the footprint \citep{Dey18}, a reasonable approach might be to mask out areas with fewer than two exposures in each band. However, not every pixel will exceed target depth for a given number of overlapping exposures (see Figure~\ref{fig:depths_nexp}), and thus the result will still be biased towards regions with more passes. Even with perfectly uniform coverage, variations in observing conditions affect depths and therefore the homogeneity of the resulting catalog.


Instead, using the $5\sigma$ point-source depths as limiting magnitudes, combined with DESI target definitions, we construct a ``binary completeness mask,'' in which a particular observed pixel is in the ``complete'' area for a target type if and only if it meets the following conditions: 
\begin{enumerate}
\item the limiting magnitudes in the bands used for magnitude cuts are sufficient to observe even the faintest targets with 5$\sigma$ confidence;
\item imaging exists for all bands used in the target definitions.
\end{enumerate}
This ensures that only the ``deep enough'' regions of the sky are used to generate DESI catalogs and randoms for analysis. Figure~\ref{fig:depths_colorcuts} shows the sky fraction as a function of depth for each band, with the magnitude cuts for the three targets plotted as vertical lines, to visualize how shifting a magnitude cut up or down would affect usable sky area. As Figure~\ref{fig:depths_colorcuts} shows, all depths are sufficient to detect the full target samples.

By only requiring the existence of imaging in the bands used for color cuts, we are implicitly assuming that uncertainties in the colors do not affect the reliability of the target selection. Alternatively, we could require a 5$\sigma$ detection in all bands, not just the primary bands used to apply magnitude cuts. However, it is not clear how to define a detection limit in a band that is only used for color cuts, as the targets are not necessarily bounded in these bands. The other advantage of defining the completeness mask on only the bands used for magnitude cuts is that it makes it easier to apply the definition to target selections of finer granularity in the color space, for example, represented by a forest of decision trees.

\subsection{Bright foregrounds}
\label{sec:masks/foregrounds}

As discussed in Section~\ref{sec:catalogs}, we correct for galactic extinction by adjusting the color of objects using the SFD98 dust map. In addition to extinction, bright objects in the foreground (point sources such as stars, and extended sources such as nearby galaxies) can also affect the detection of targets in their angular vicinity, due to CCD saturation and diffraction spikes that contaminate the surrounding pixels. The systematics due to these bright objects in the target catalogs are spatially localized and uncorrelated with the underlying true density of the targets. Thus, we can quantify these systematic effects by measuring the densities of targets as a function of their proximity to the bright foreground objects, and mitigate them via masks.

\subsubsection{Bright stars}
\label{sec:masks/foregrounds/stars}

We use two bright star catalogs, Tycho-2 and WISE.  Tycho-2 is a reference catalog of the 2.5 million brightest stars in the sky, with photometry in two optical bands, $\lambda_{BT} = 435$ nm and $\lambda_{VT} = 505$ nm. It has highly accurate astrometric positions and is 99\% complete out to $VT \sim 11$ magnitude, making it well suited to our analysis. However, some stars not included in the Tycho-2 catalog may still be bright enough in the near-infrared bands to affect the detection of LRGs and QSOs,  both of which use the WISE band $W1$ in their target selection (QSOs also use $W2$). To be safe, we create separate veto masks for each star catalog and apply both to our data. Table~\ref{table:star_masks} compares the effective areas and sky fractions of the two star masks. 

\begin{table}
{\centering
\noindent\begin{tabular}{p{1.0cm}p{1.9cm}p{1.3cm}p{1.5cm}p{0.5cm}}
\toprule
Target & Mask & Number & Area (deg$^2$) & $f_{\rm sky}$ \\
\midrule
\multirow{4}{*}{LRG} & no mask & 4882206 & 9298.91 & 1.0 \\
                     & Tycho mask only      & 4385463 & 8752.25 & 0.94 \\
                     & WISE mask only   & 4529935 & 8970.73 & 0.96 \\
                     \cmidrule{2-5}
                     & \textbf{both star masks} & 4304375 & 8584.68 & 0.92 \\
\hline
\multirow{4}{*}{ELG} & no mask & 23224353 & 9298.91 & 1.0 \\
                     & Tycho mask only      & 21378695 & 8811.44 & 0.95 \\
                     & WISE mask only   & 21065984 & 8970.73 & 0.96 \\
                     \cmidrule{2-5}
                     & \textbf{both star masks} & 20329143 & 8638.84 & 0.93 \\
\hline
\multirow{4}{*}{QSO} & no mask & 3125148 & 9298.91 & 1.0 \\
                     & Tycho mask only      & 2192932 & 7541.98 & 0.81 \\
                     & WISE mask only   & 2030414 & 7269.13 & 0.78 \\
                     \cmidrule{2-5}
                     & \textbf{both star masks} & 1814359 & 6655.11 & 0.72 \\
\bottomrule
\end{tabular}}
\caption{Summary of star masks and how each affects the number of targets and the effective area and sky fraction. Here, sky fractions are quoted relative to the ``no mask'' case, which is simply the joint regions of the footprint where imaging is available in all three optical bands.}
\label{table:star_masks}
\end{table}

\begin{figure*}
\centering
\includegraphics[width=0.45\linewidth]{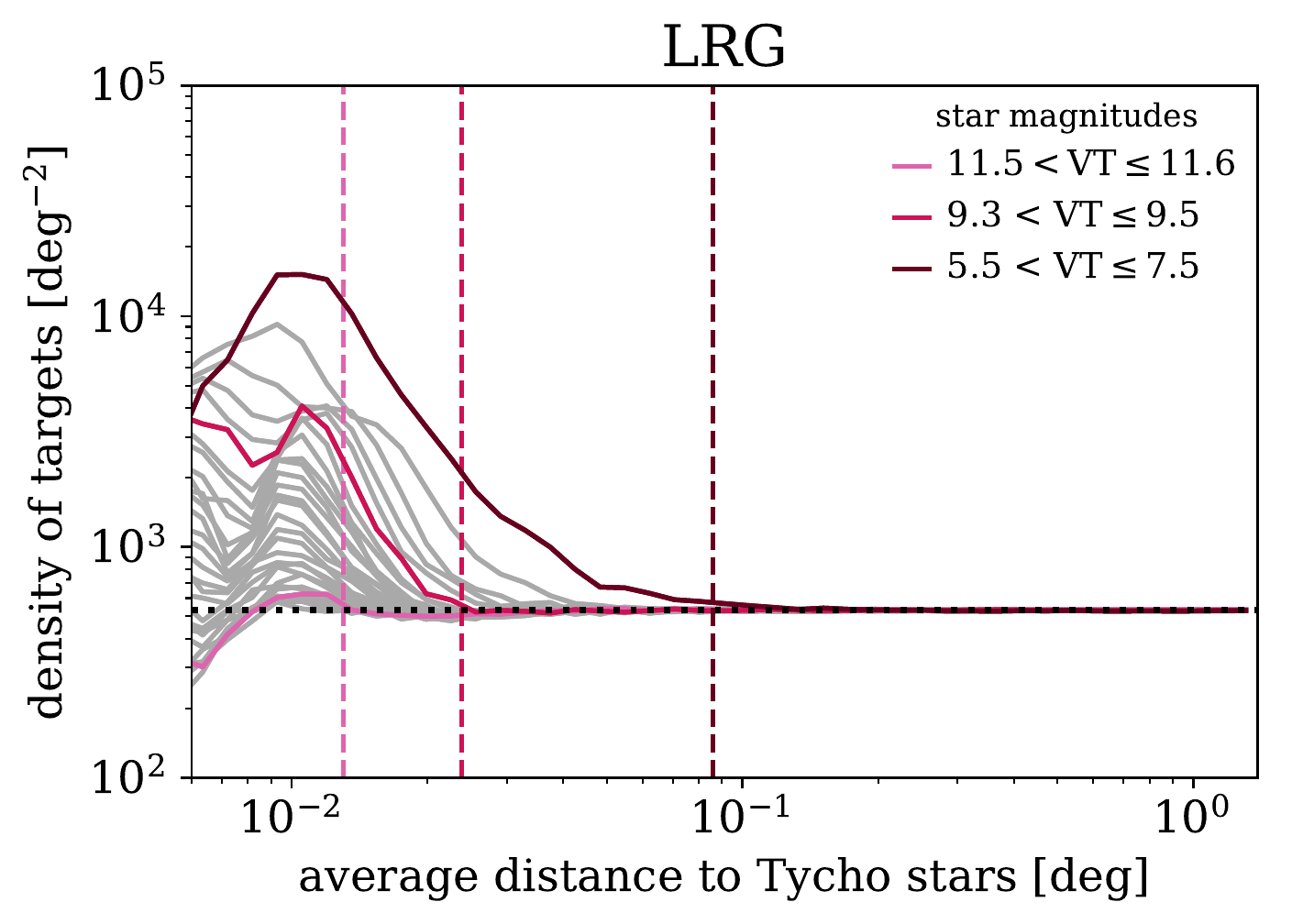}
\includegraphics[width=0.45\linewidth]{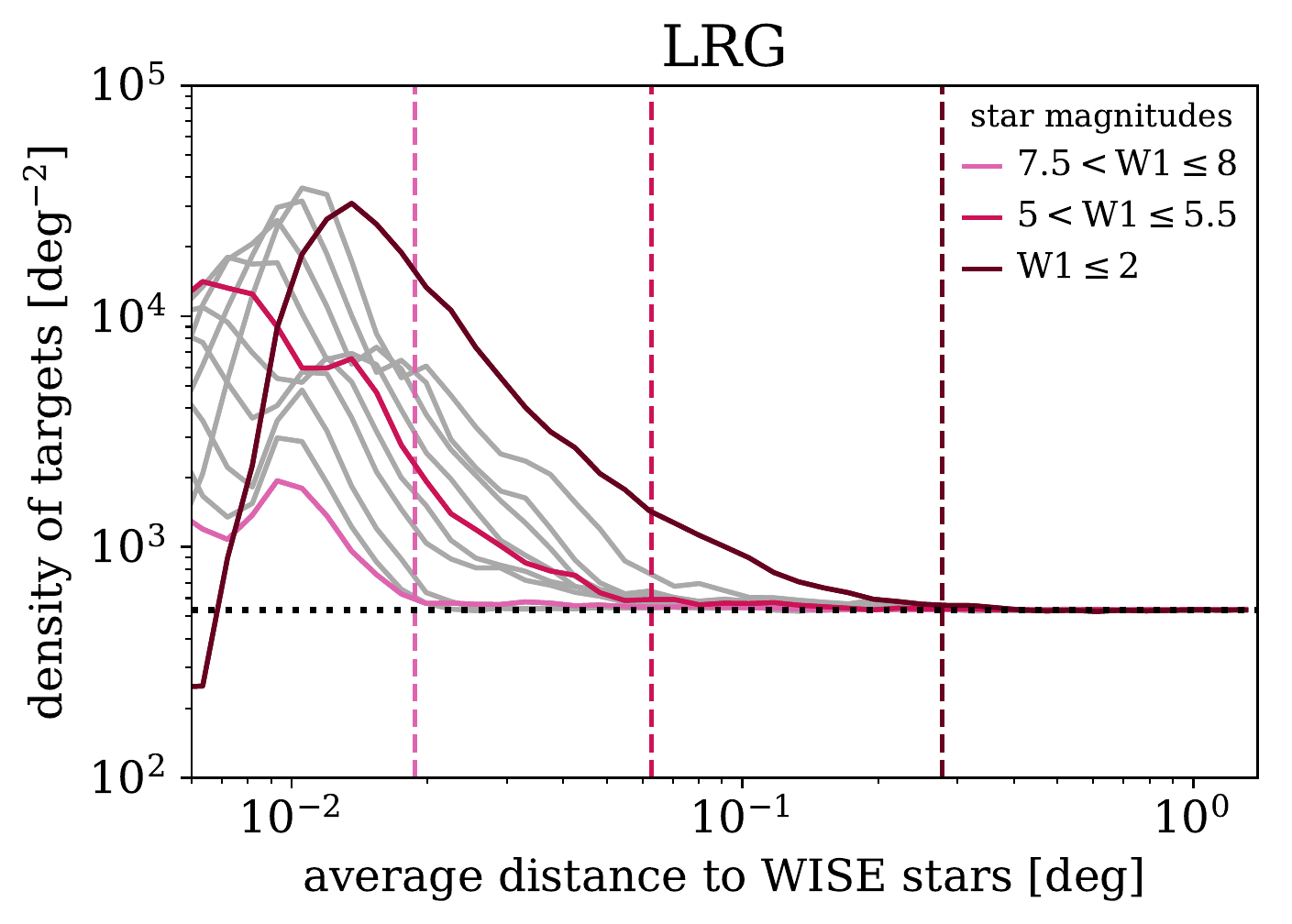}
\includegraphics[width=0.45\linewidth]{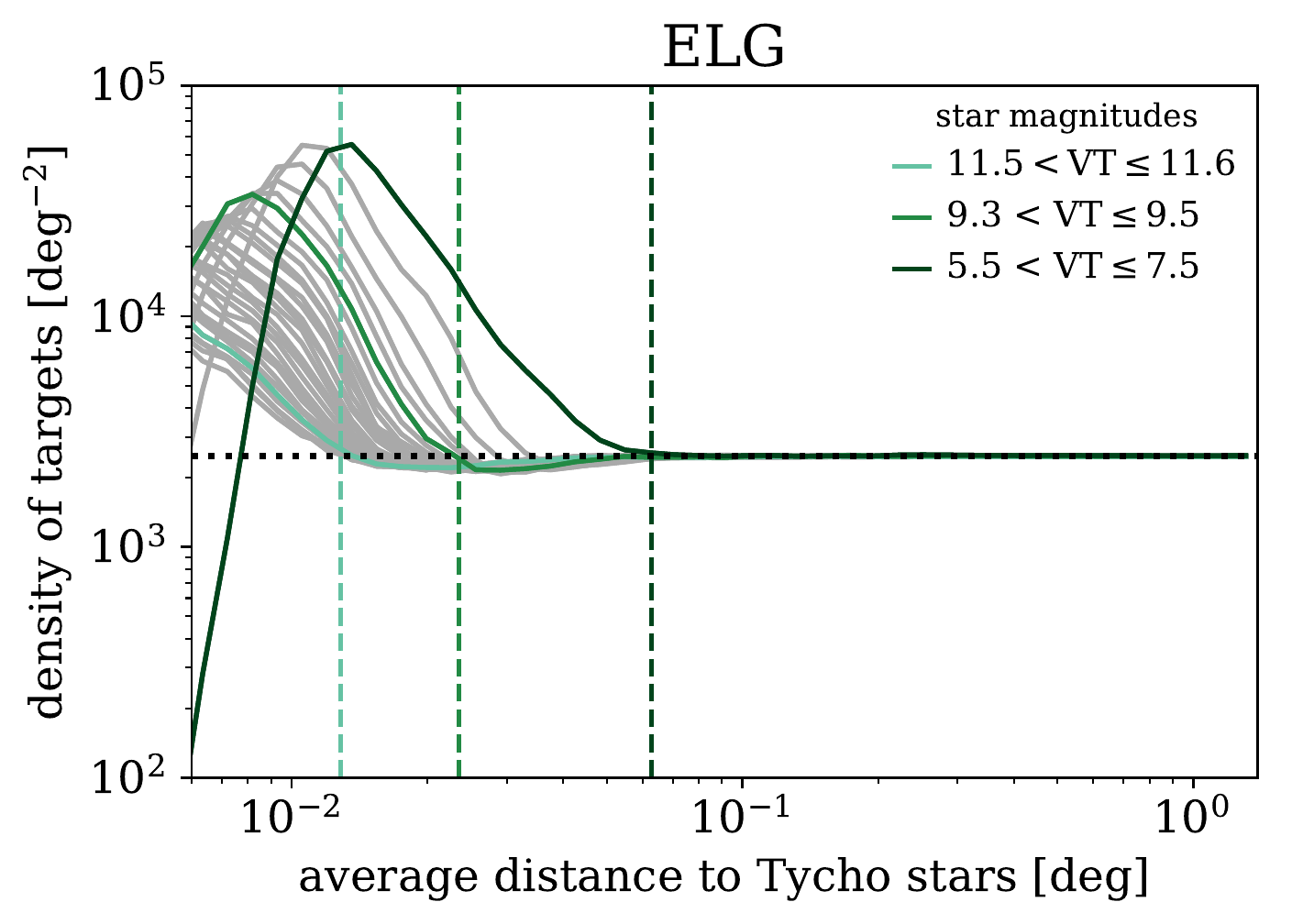}
\includegraphics[width=0.45\linewidth]{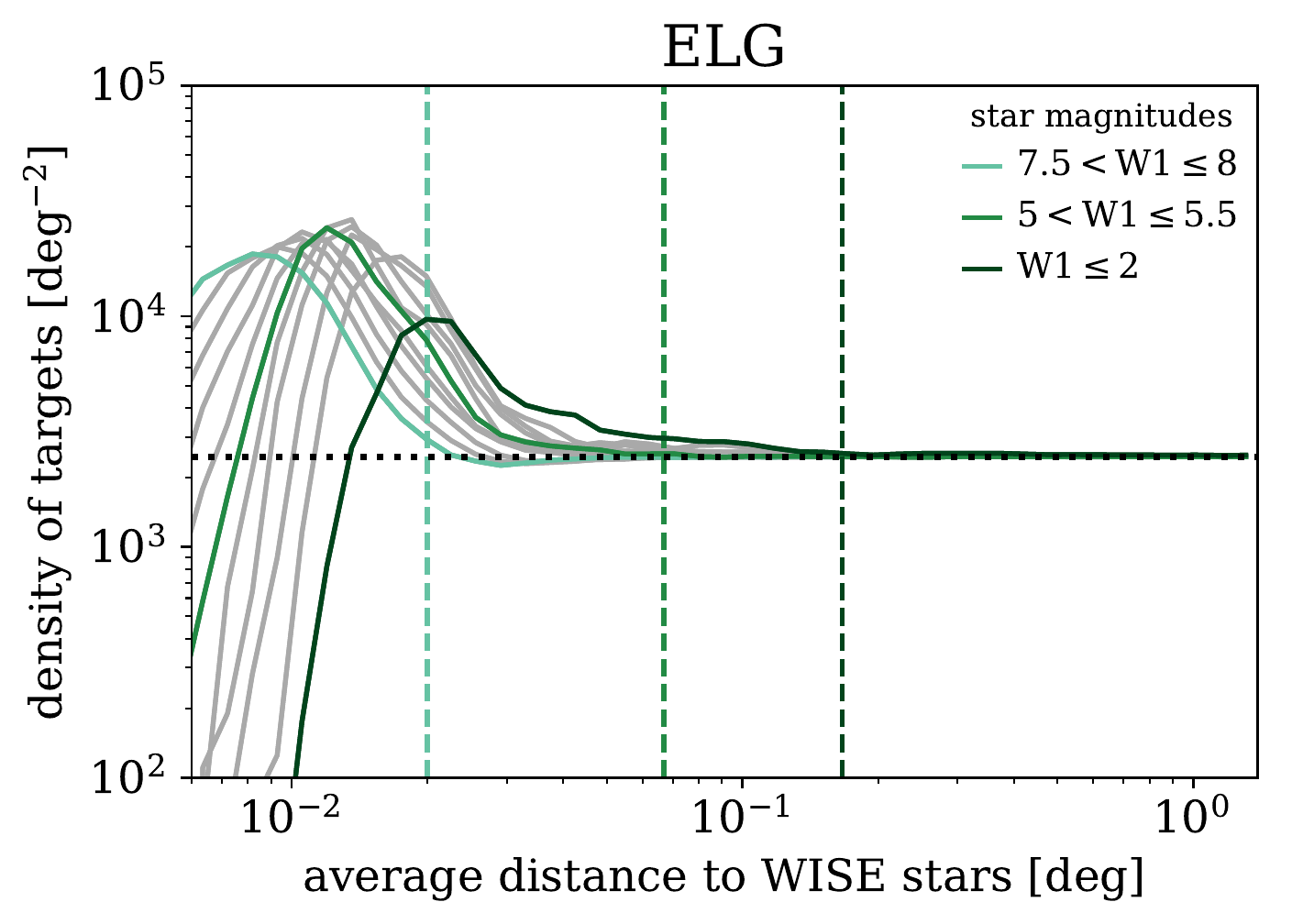}
\includegraphics[width=0.45\linewidth]{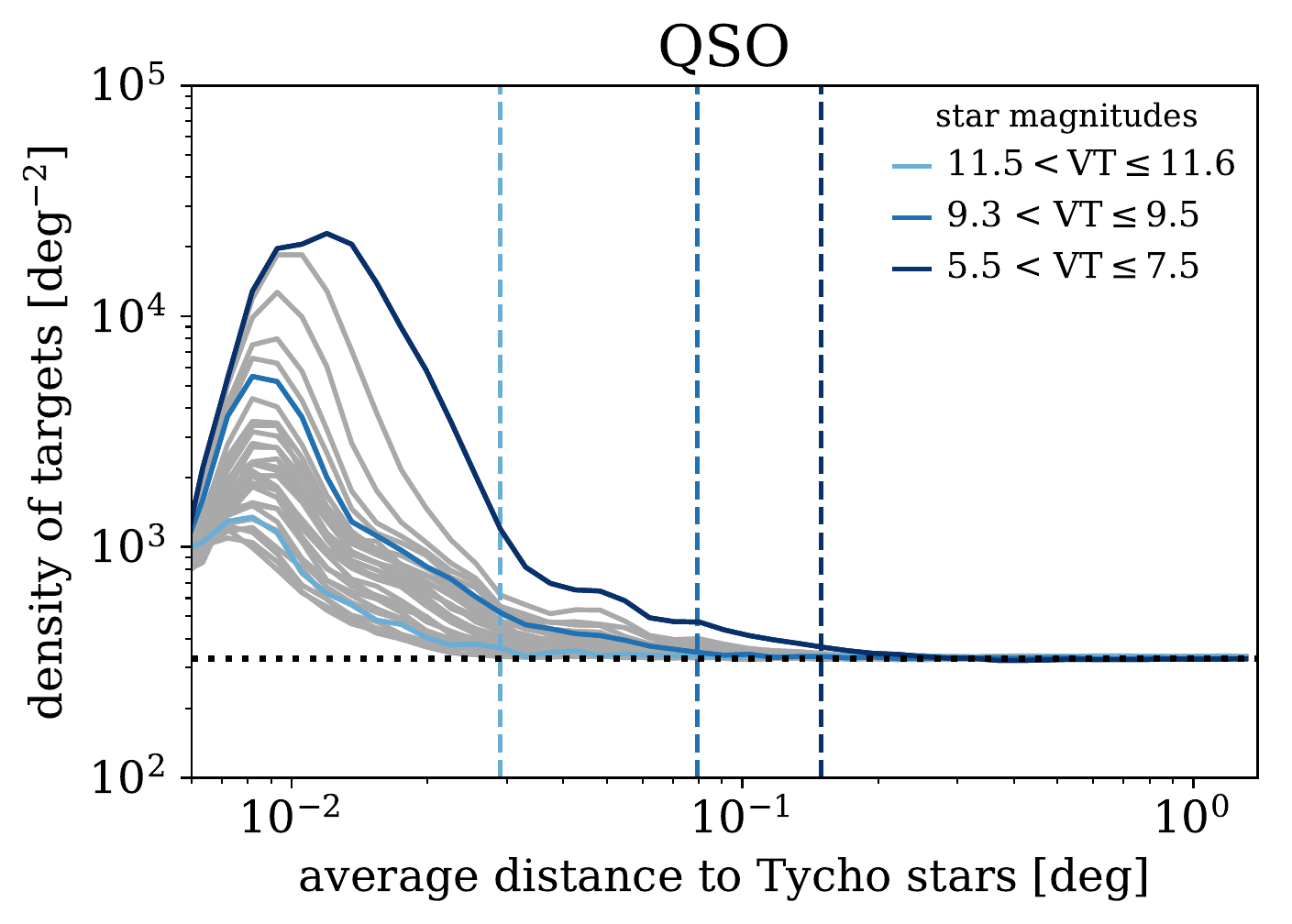}	
\includegraphics[width=0.45\linewidth]{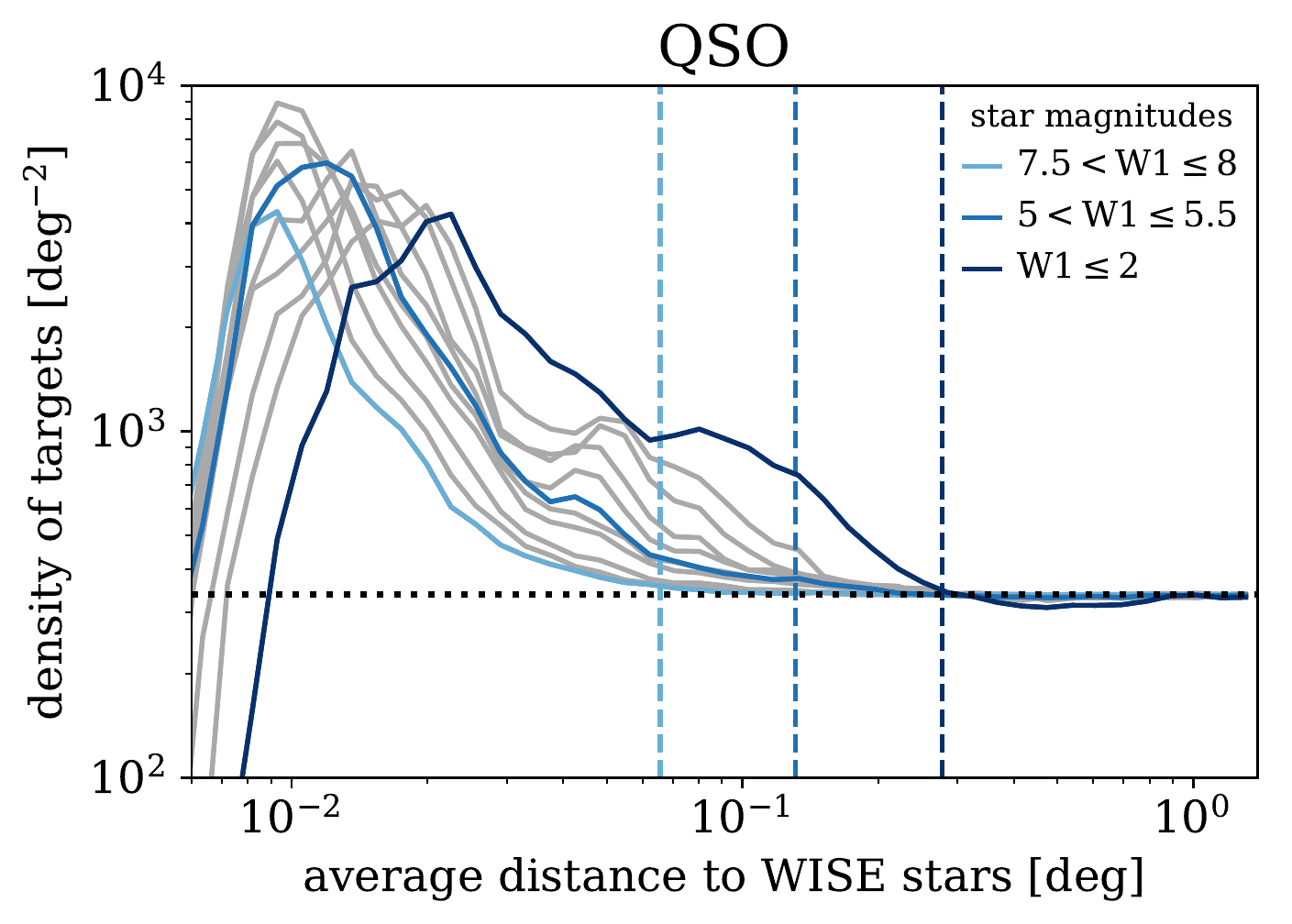}	
\caption{Average density of DESI targets as a function of distance to bright stars from Tycho-2 (left) and WISE (right). Both star catalogs are divided into magnitude bins, which are spaced such that each contains a roughly comparable number of stars. In each plot, three bins are highlighted for illustration, with the dashed vertical lines representing the corresponding masking radius (Equations \ref{eq:tycho-mr} and \ref{eq:wise-mr}) calculated using the average magnitude of that bin.}
\label{fig:density_star}
\end{figure*}

\begin{figure*}
\centering
\includegraphics[width=0.29\linewidth]{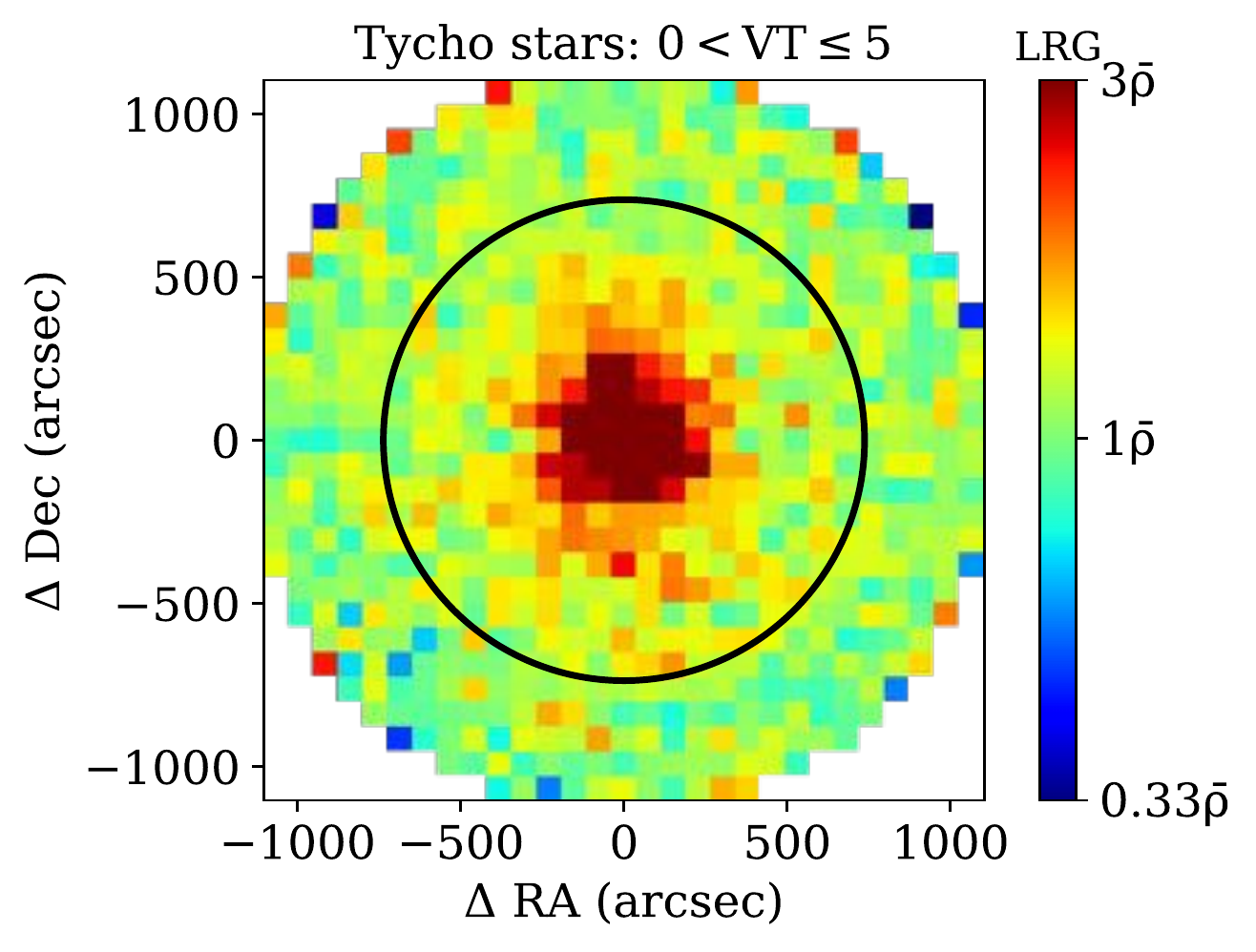}
\includegraphics[width=0.29\linewidth]{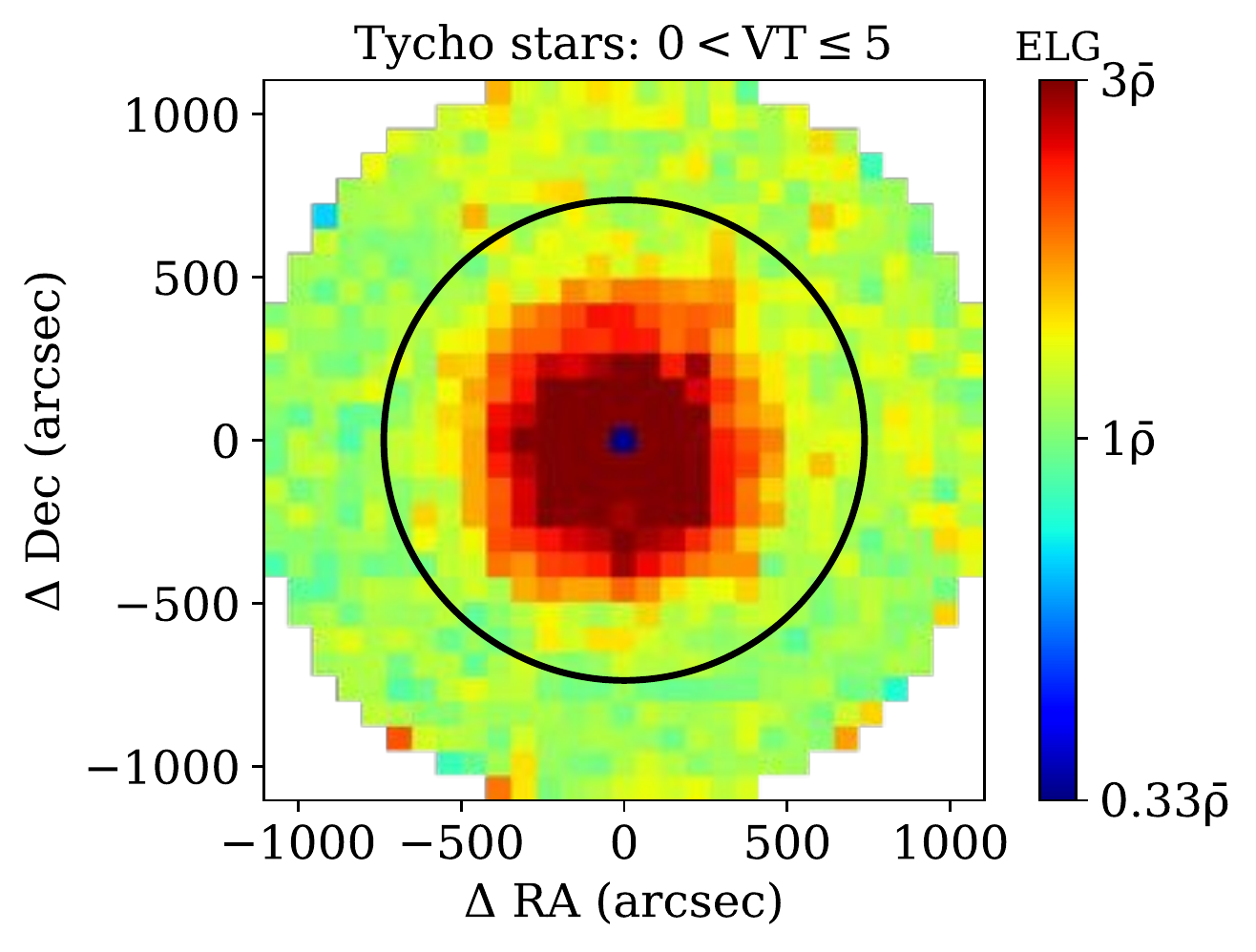}
\includegraphics[width=0.29\linewidth]{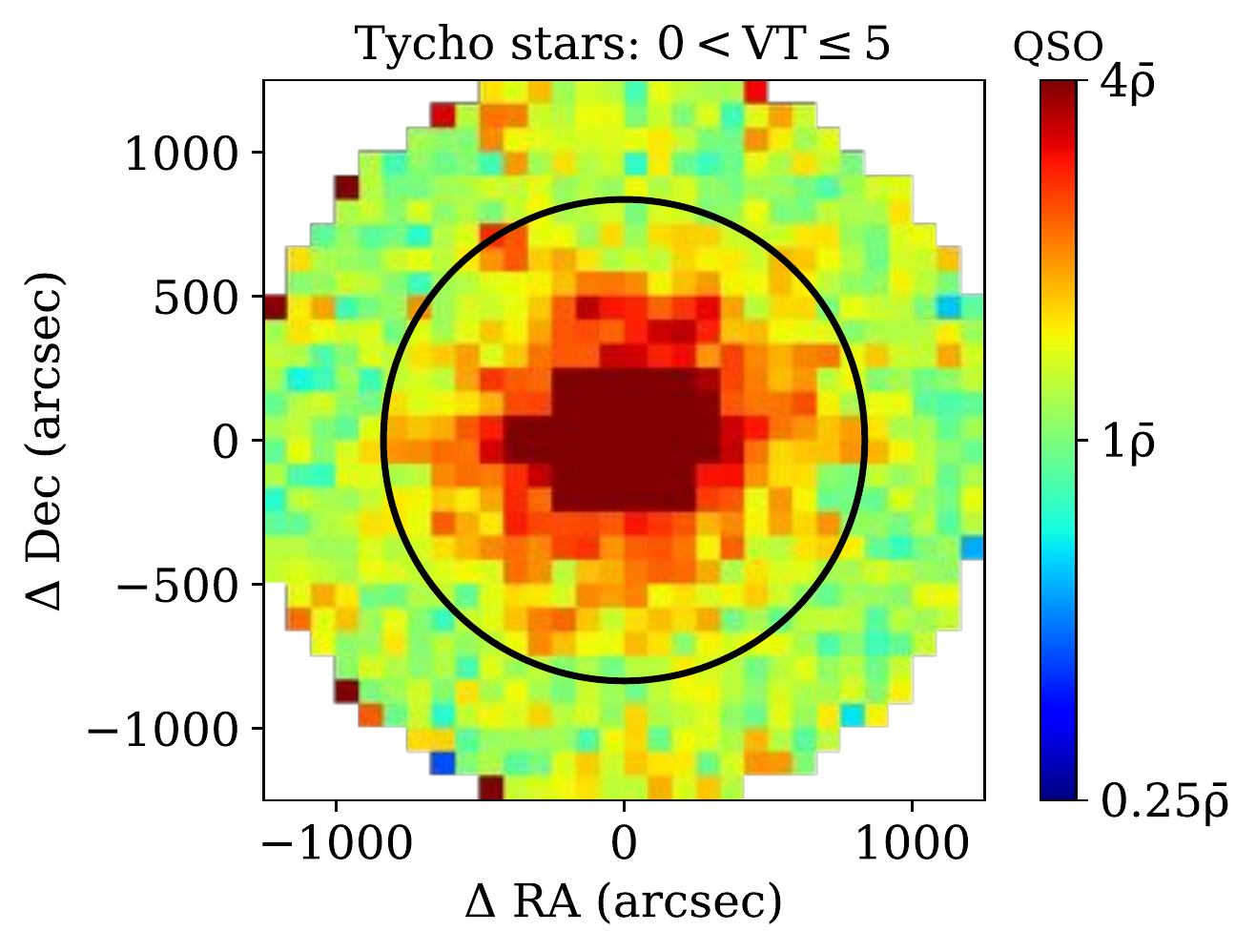}
\includegraphics[width=0.29\linewidth]{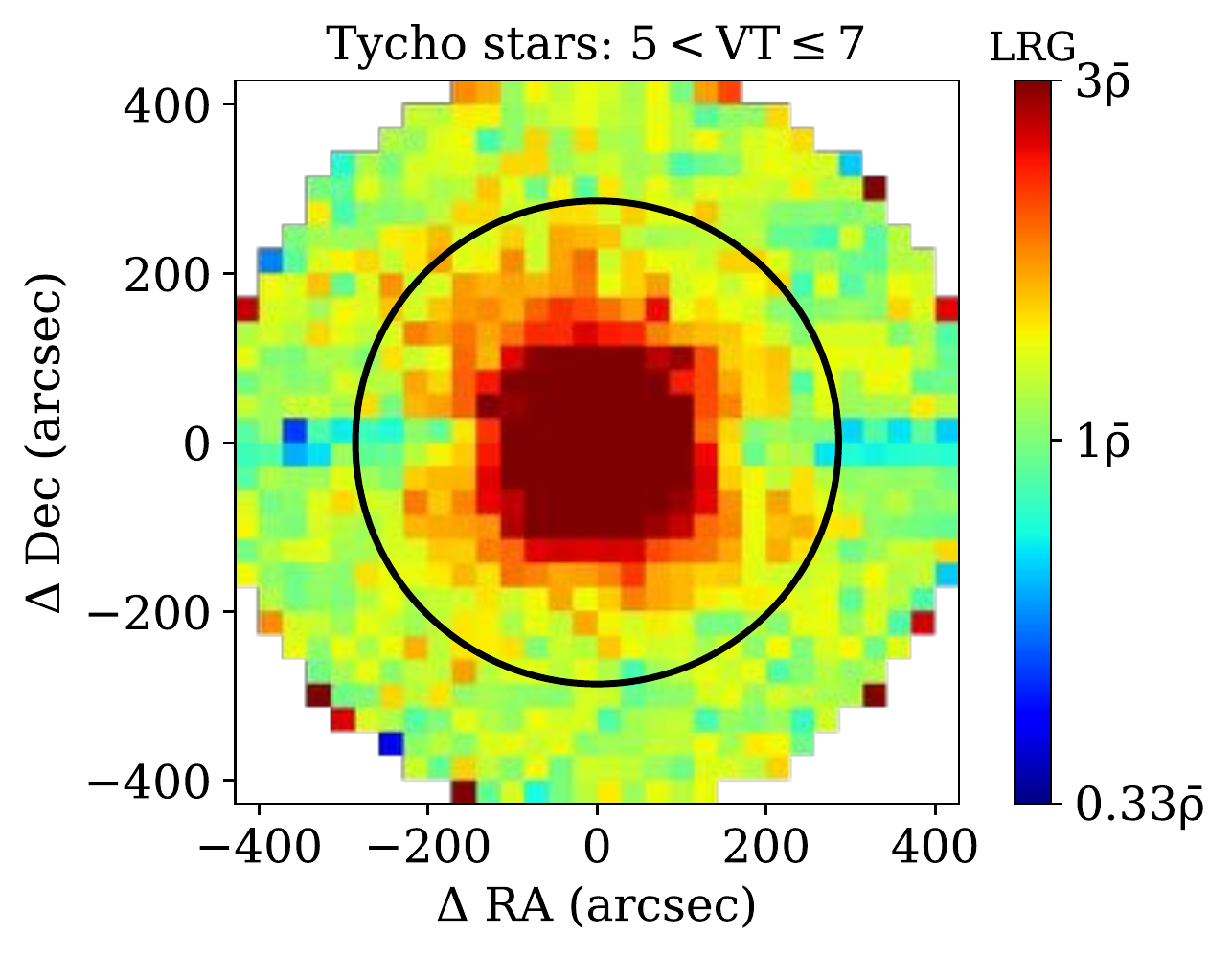}
\includegraphics[width=0.29\linewidth]{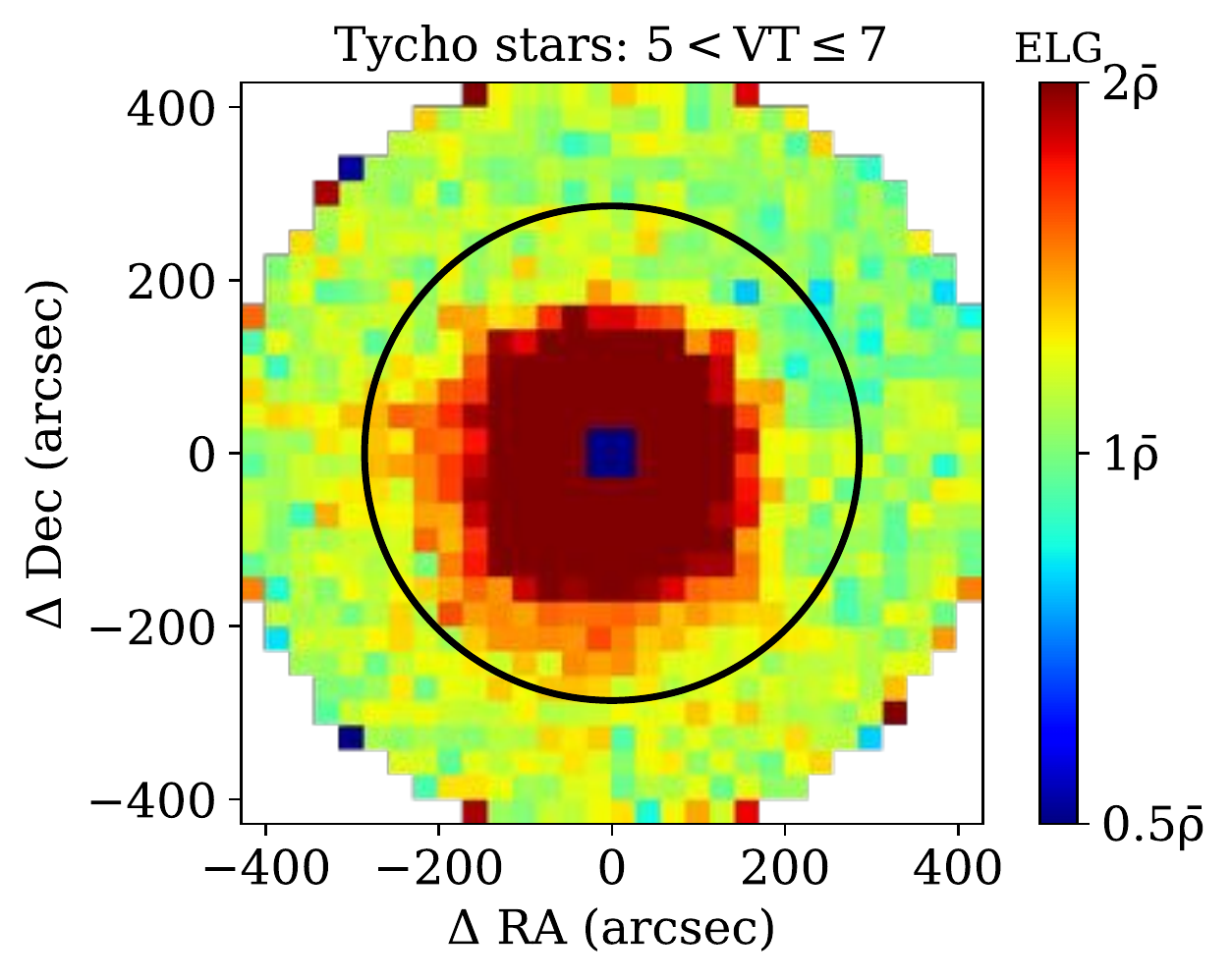}
\includegraphics[width=0.29\linewidth]{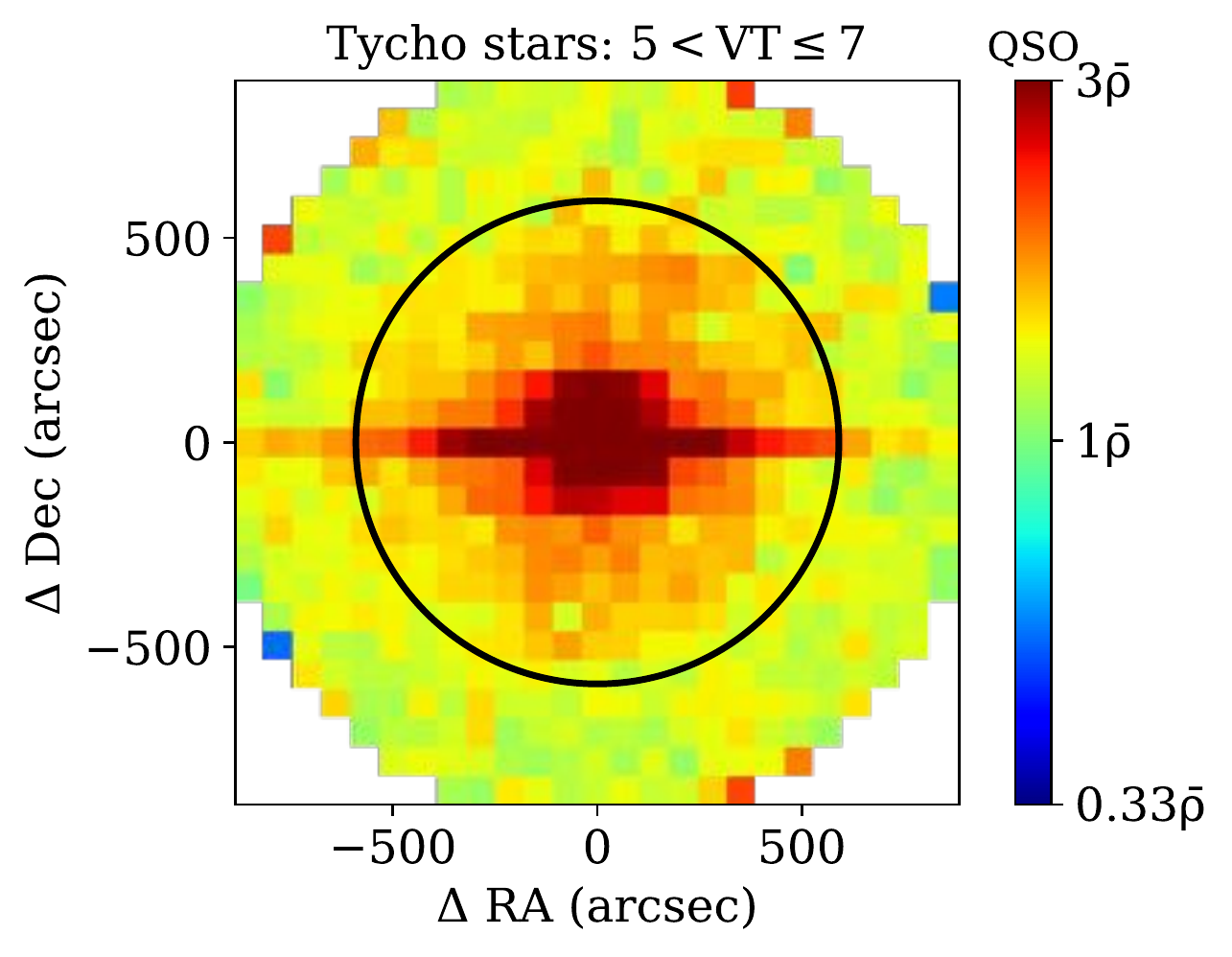}
\includegraphics[width=0.29\linewidth]{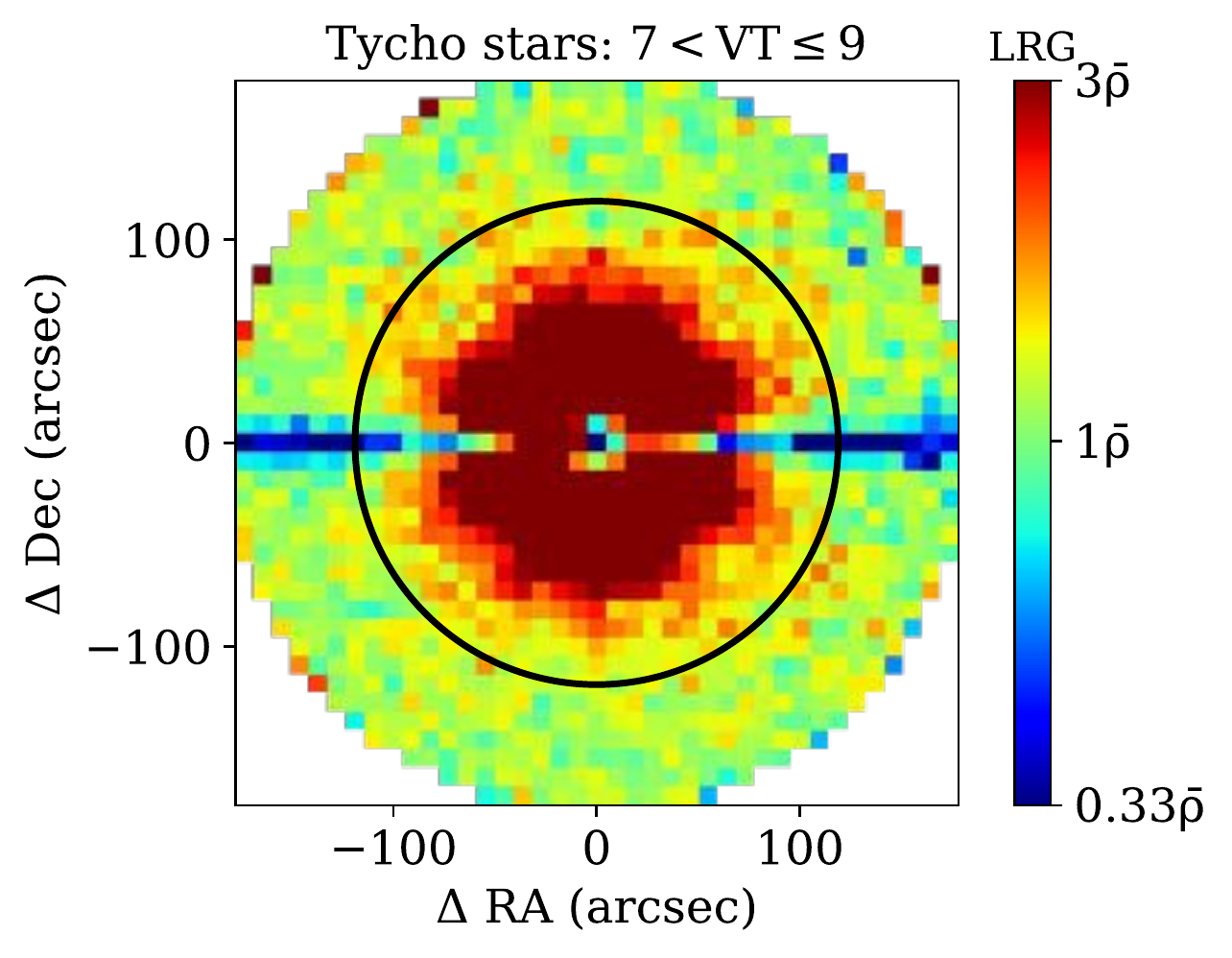}
\includegraphics[width=0.29\linewidth]{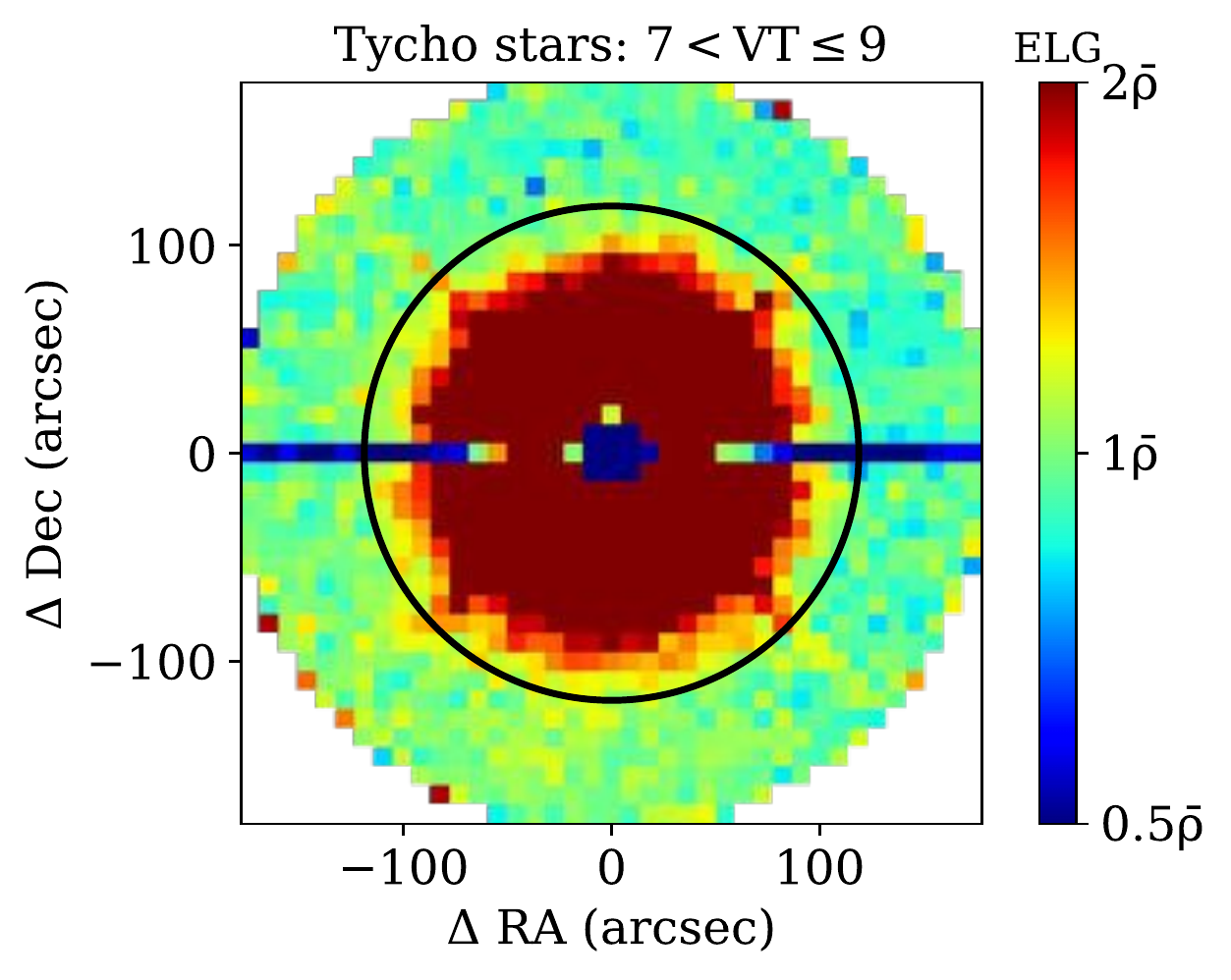}
\includegraphics[width=0.29\linewidth]{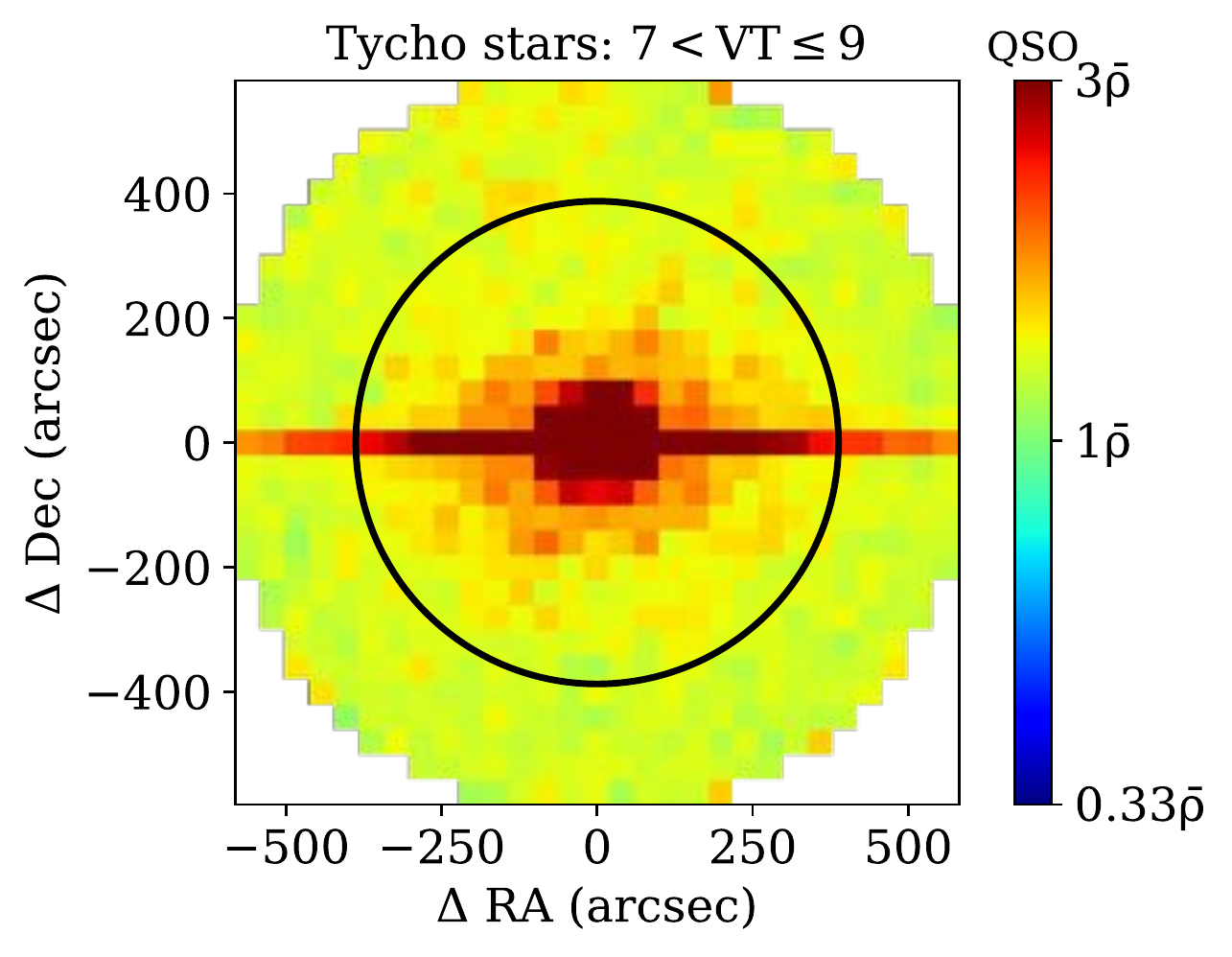}
\includegraphics[width=0.29\linewidth]{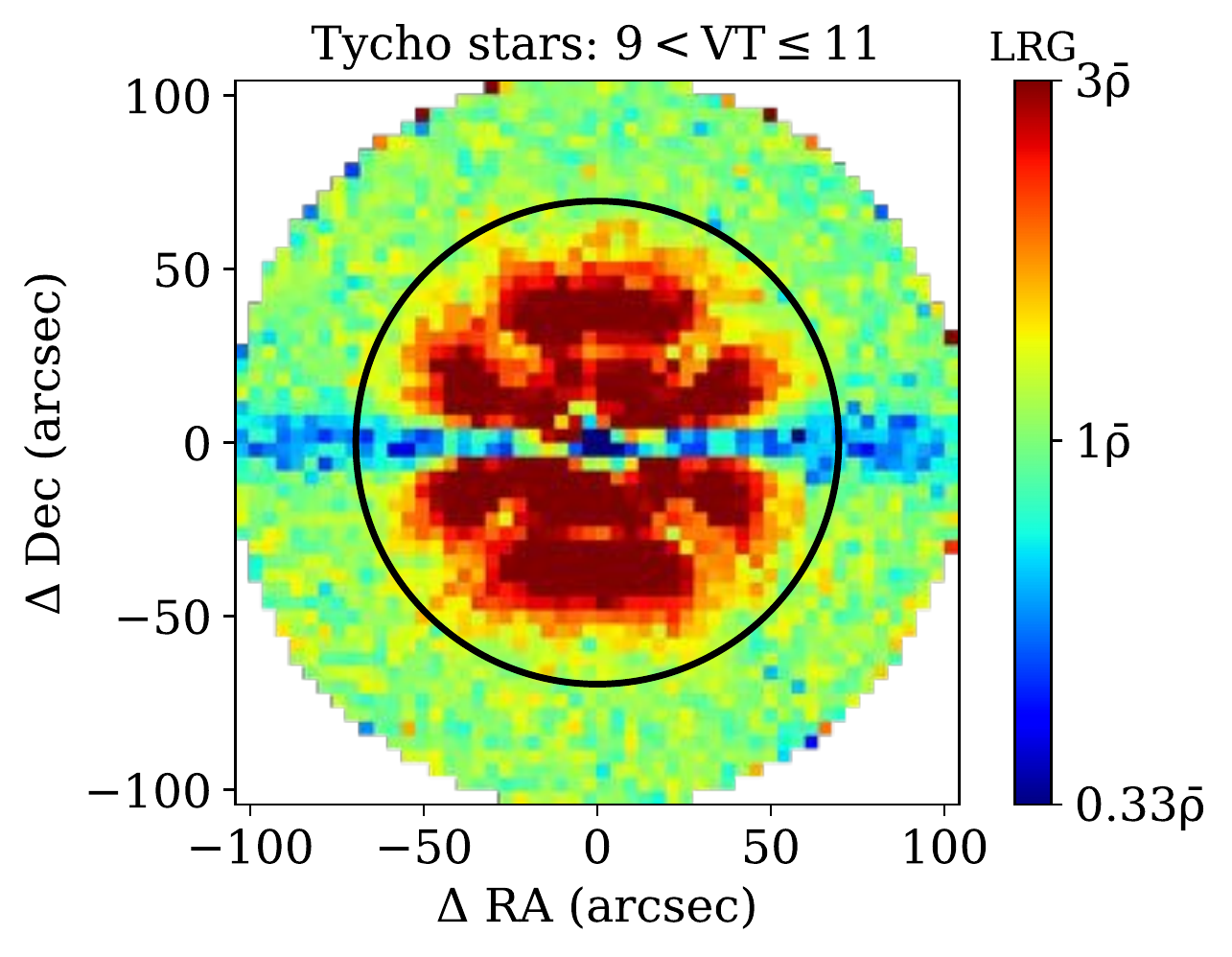}
\includegraphics[width=0.29\linewidth]{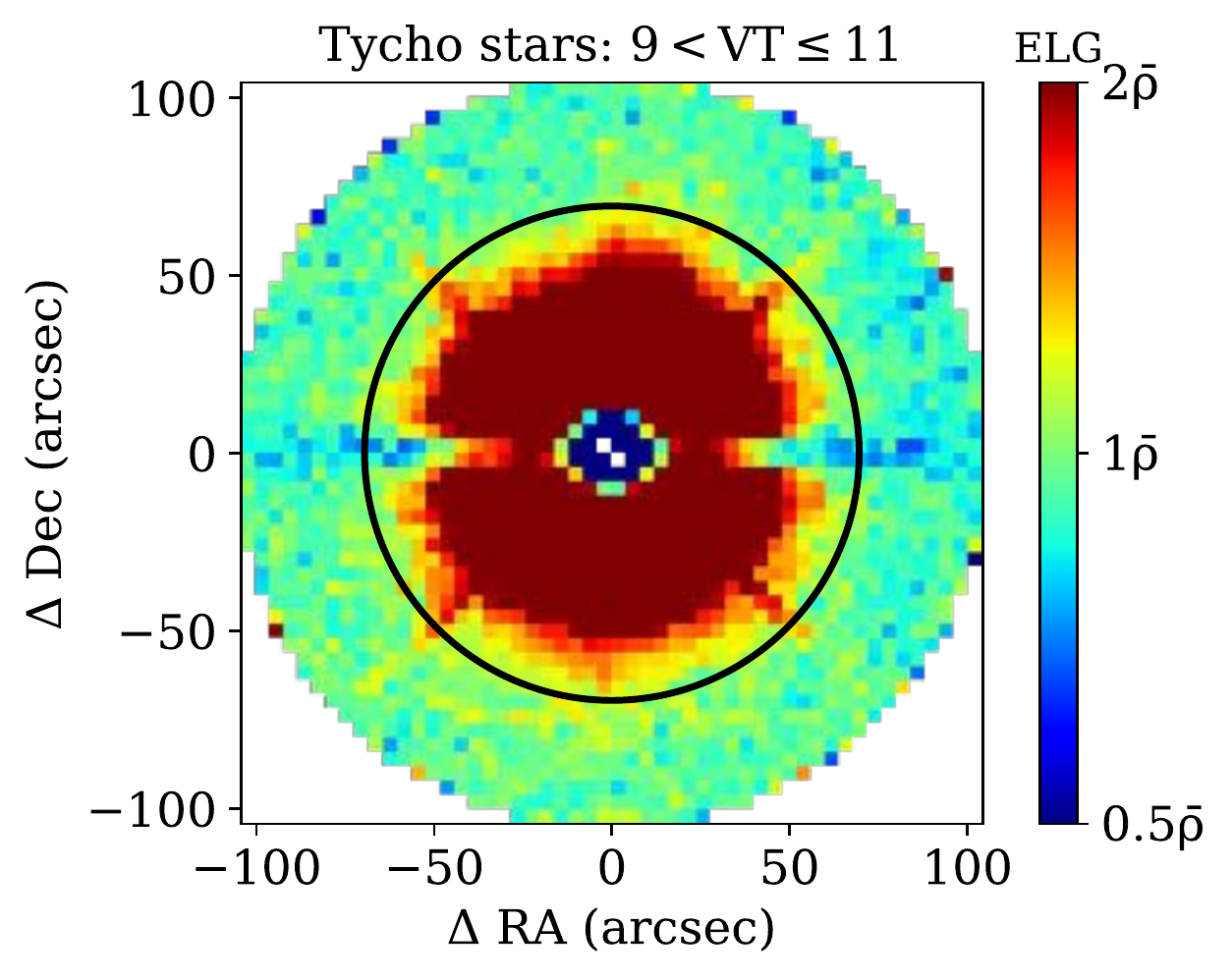}
\includegraphics[width=0.29\linewidth]{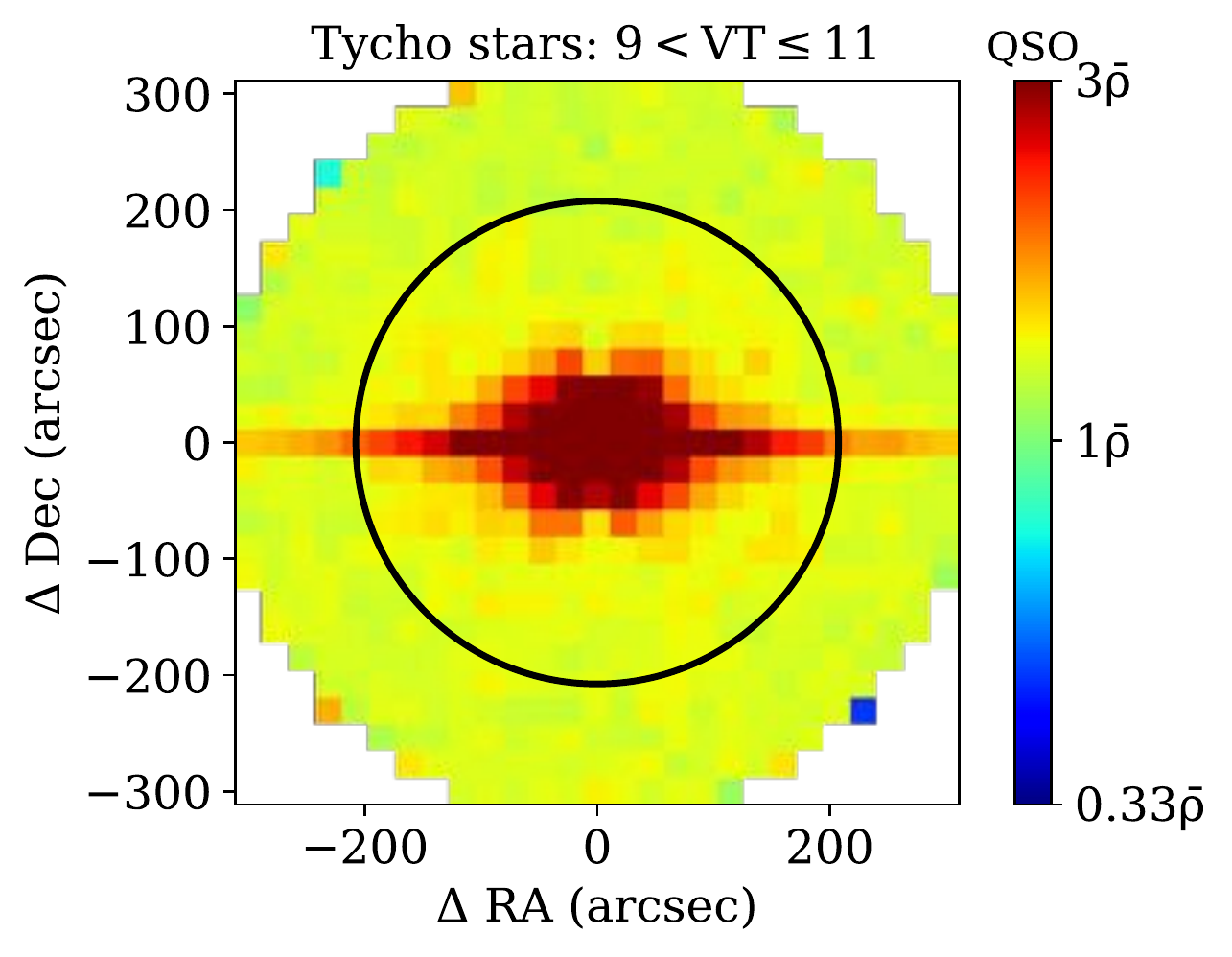}
\includegraphics[width=0.29\linewidth]{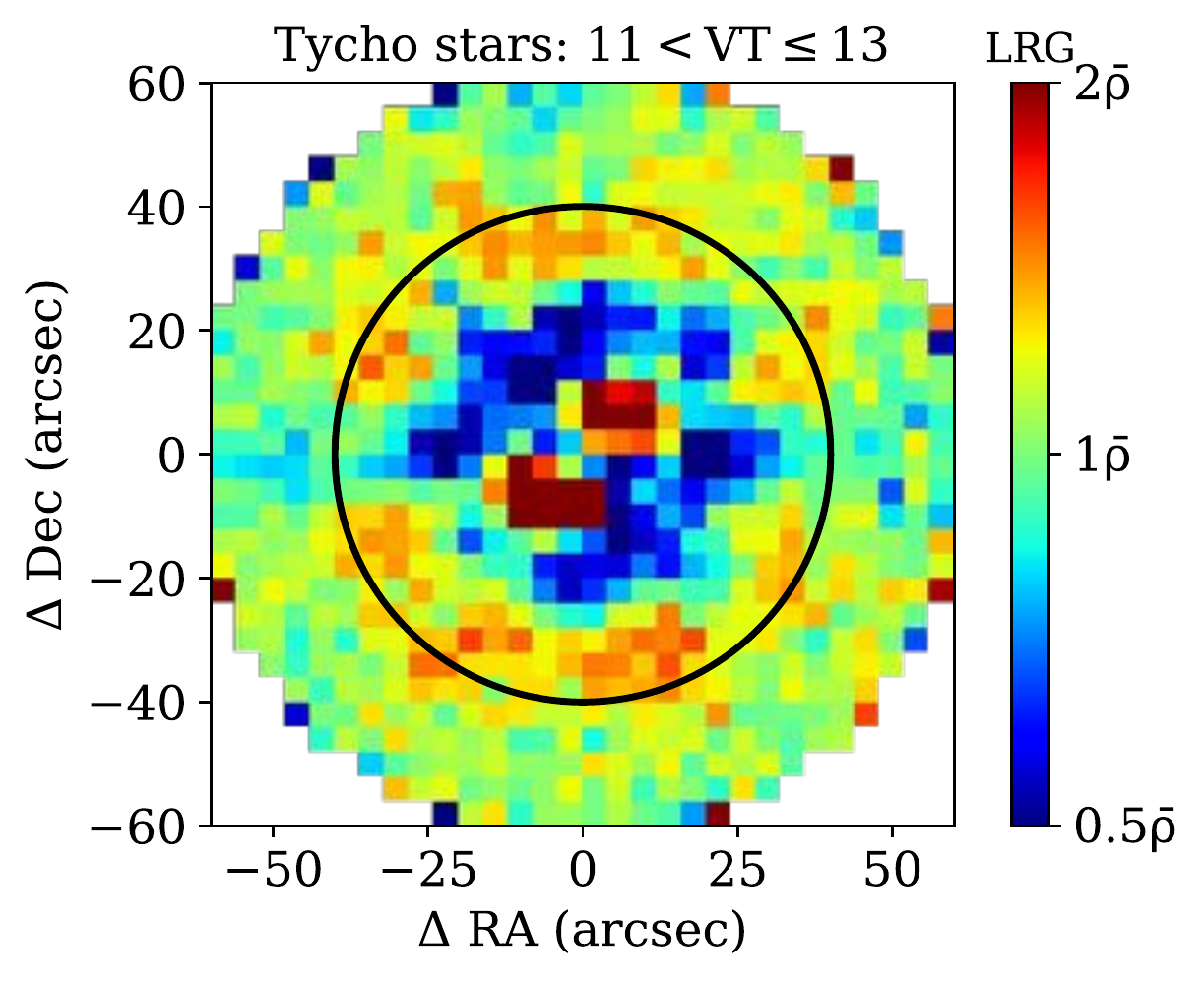}
\includegraphics[width=0.29\linewidth]{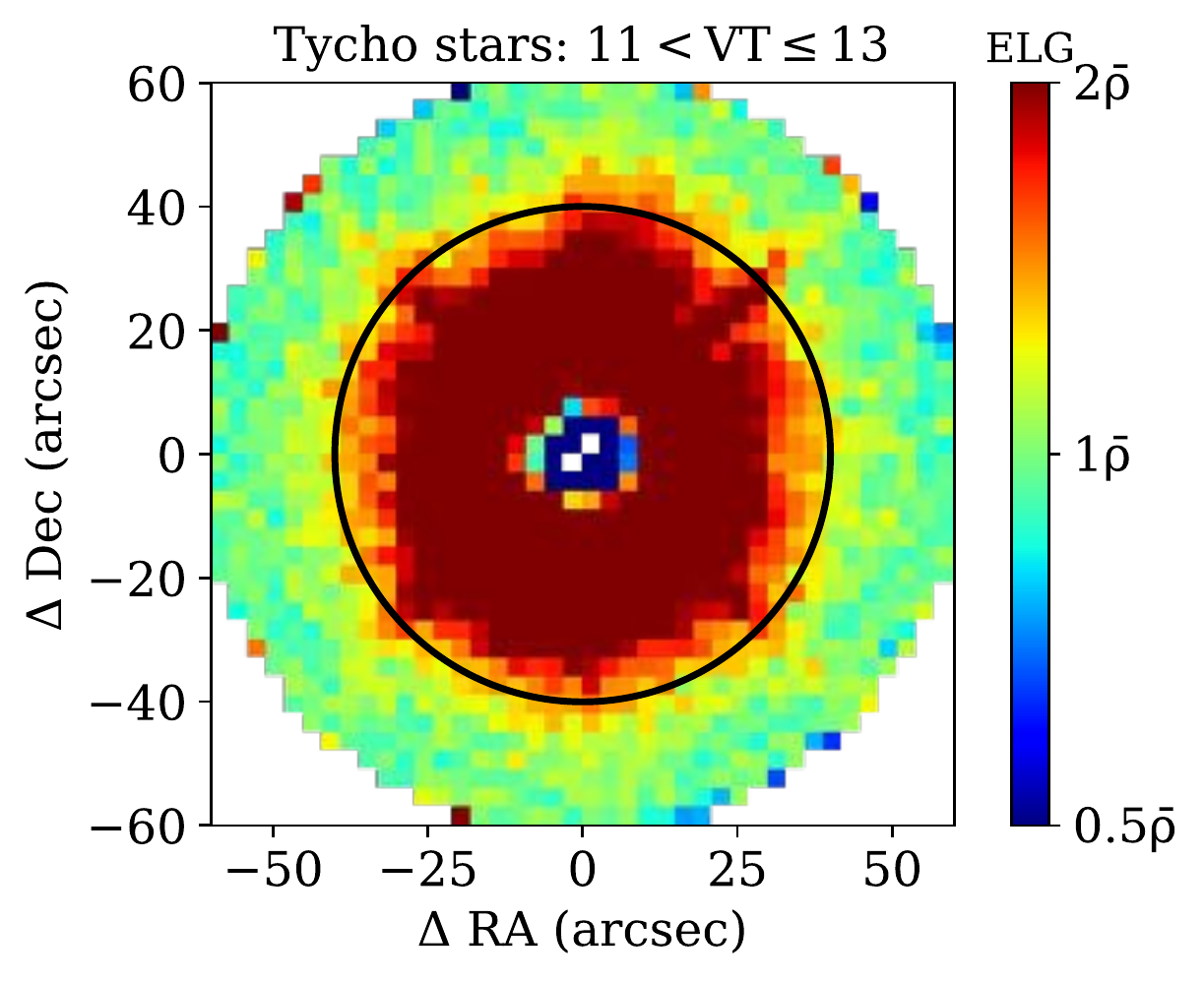}
\includegraphics[width=0.29\linewidth]{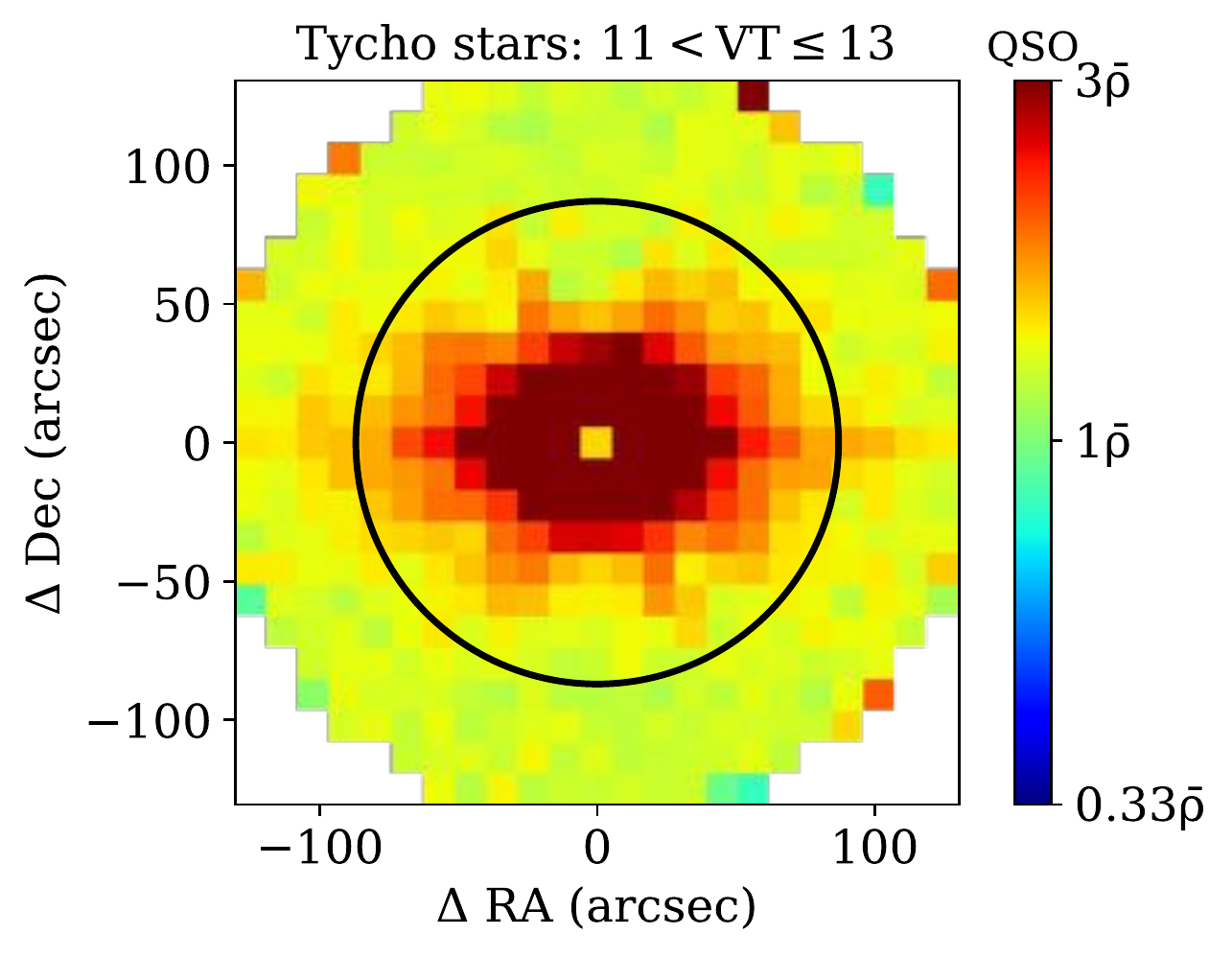}
\caption[]{2D histograms of the density of DESI targets around stacks of bright stars from Tycho-2. The solid black circles represent our star masks (Equation~\ref{eq:tycho-mr}). The horizontal features appearing in some maps, which are due to insufficient masking of charge bleed trails in the CCDs, are only a few arcsec in width, and we find that removing them with a separate rectangular mask has no perceptible impact on the densities around stars.}
\label{fig:star2d-tycho}
\end{figure*}

\begin{figure*}
\centering
\includegraphics[width=0.29\linewidth]{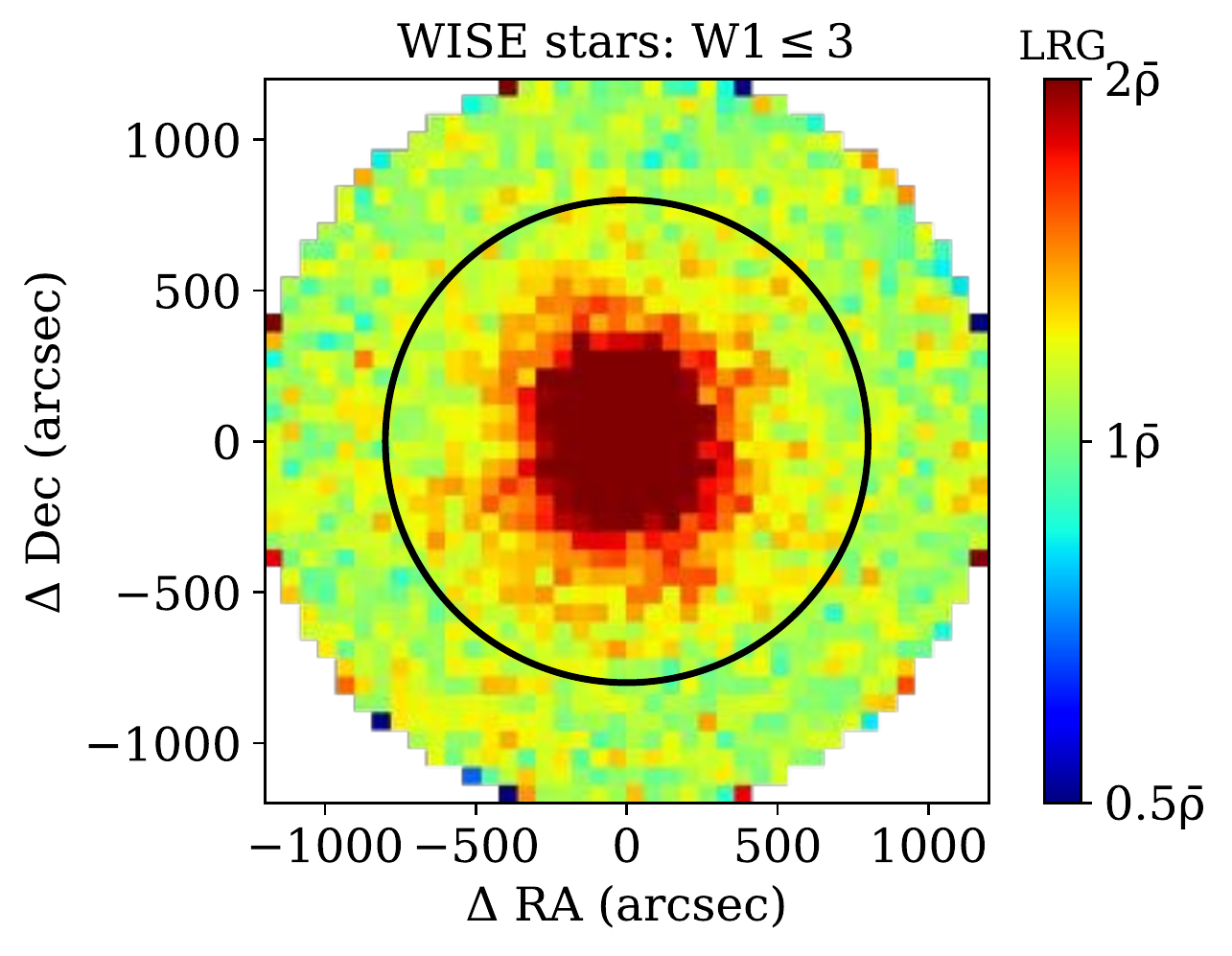}
\includegraphics[width=0.29\linewidth]{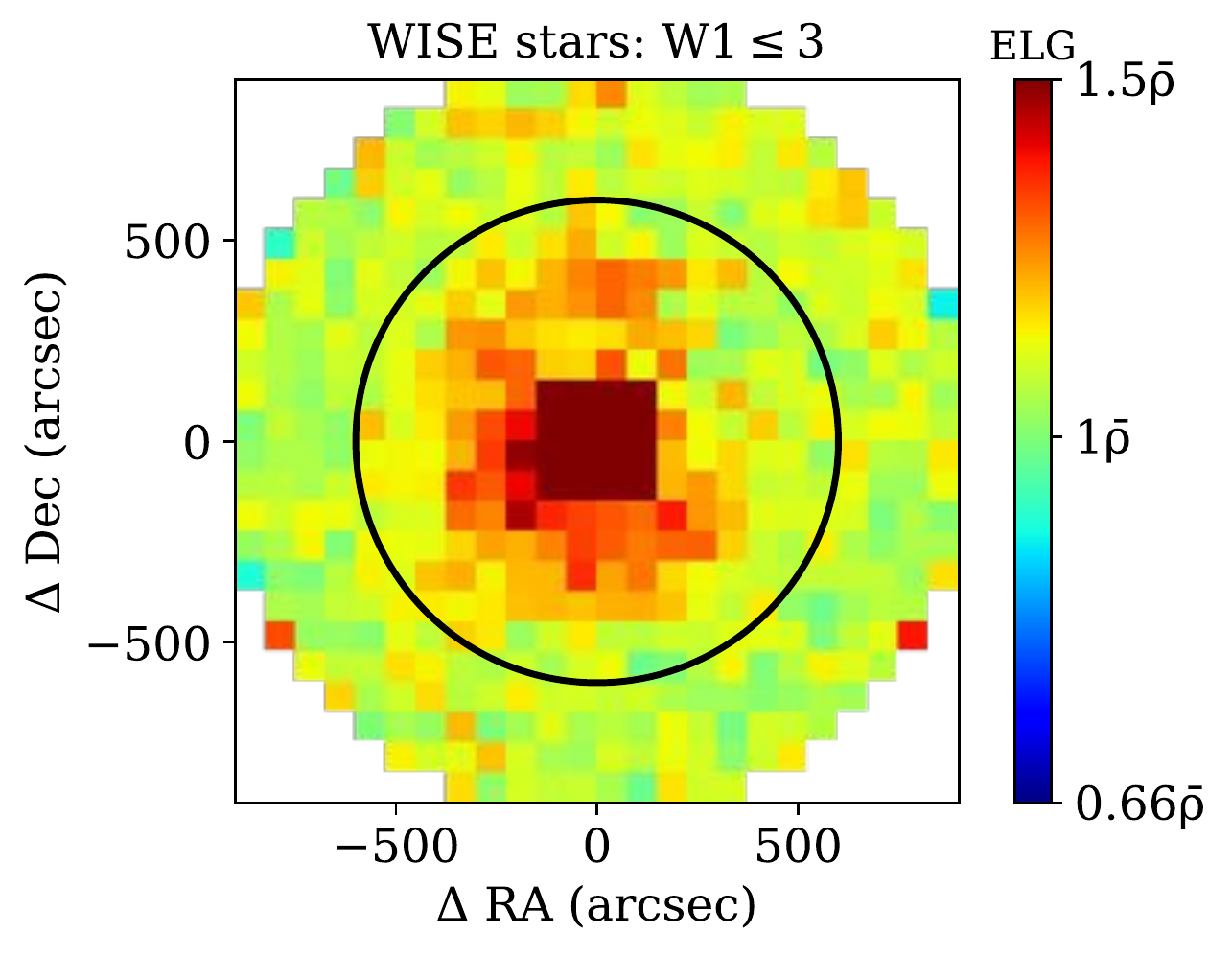}
\includegraphics[width=0.29\linewidth]{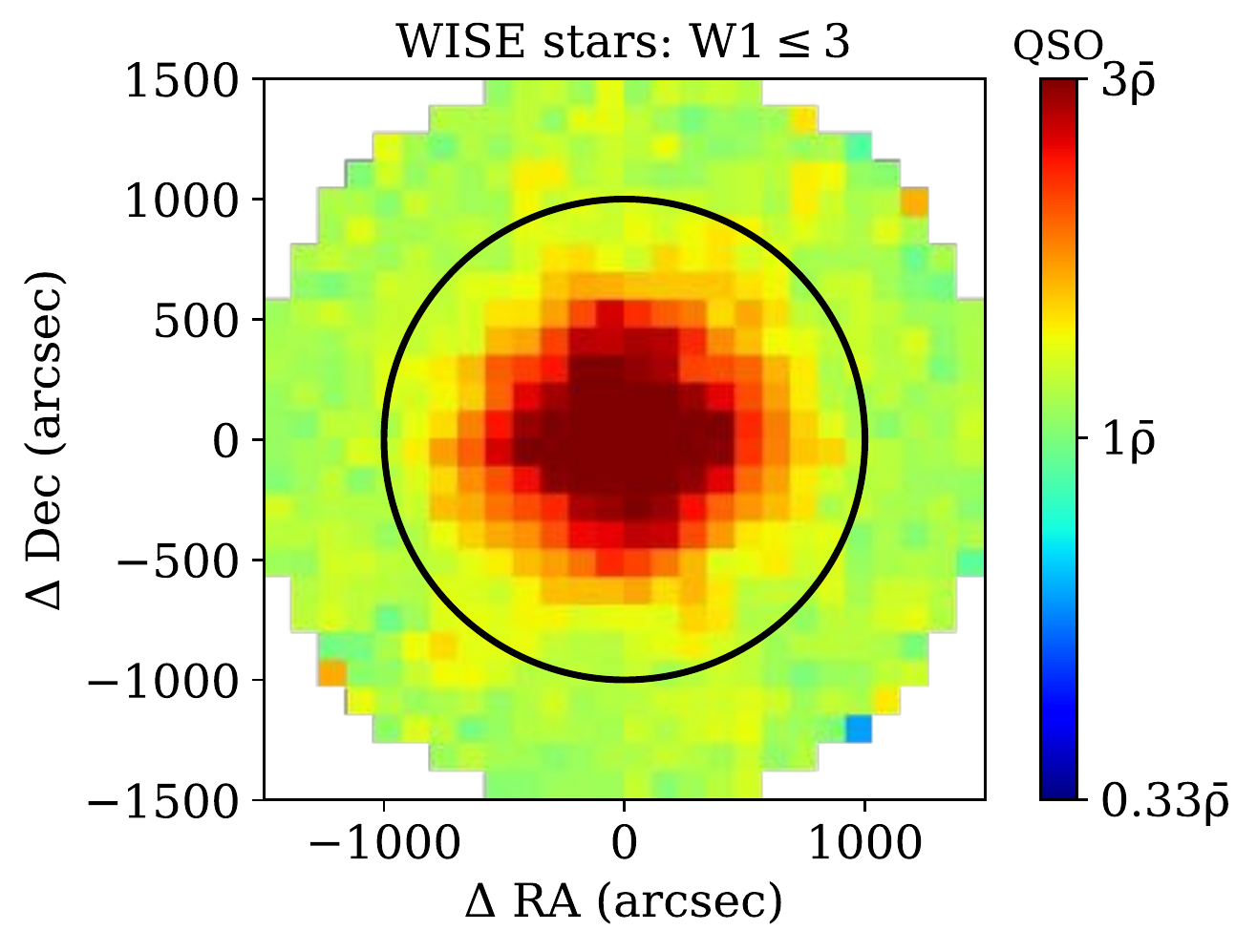}
\includegraphics[width=0.29\linewidth]{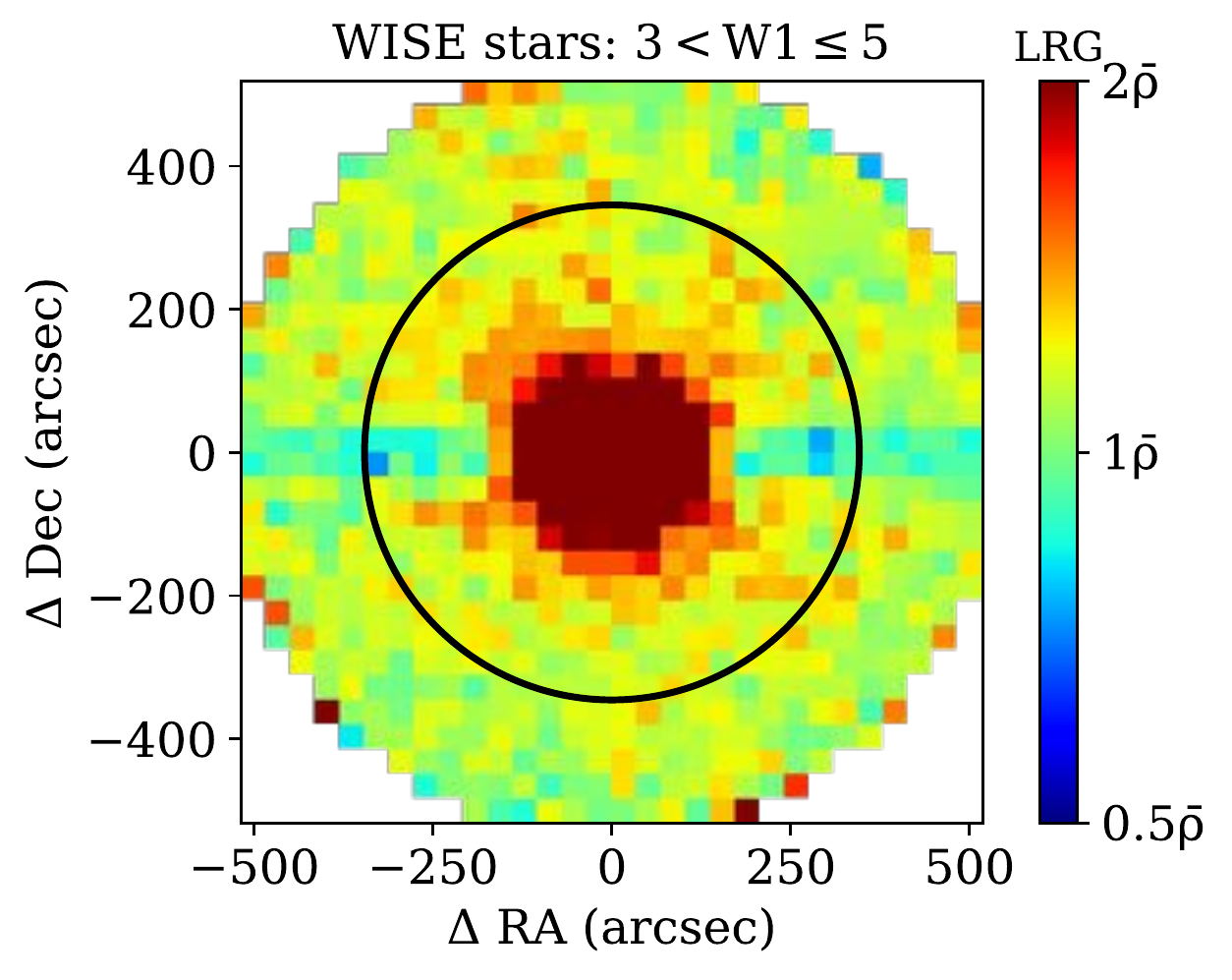}
\includegraphics[width=0.29\linewidth]{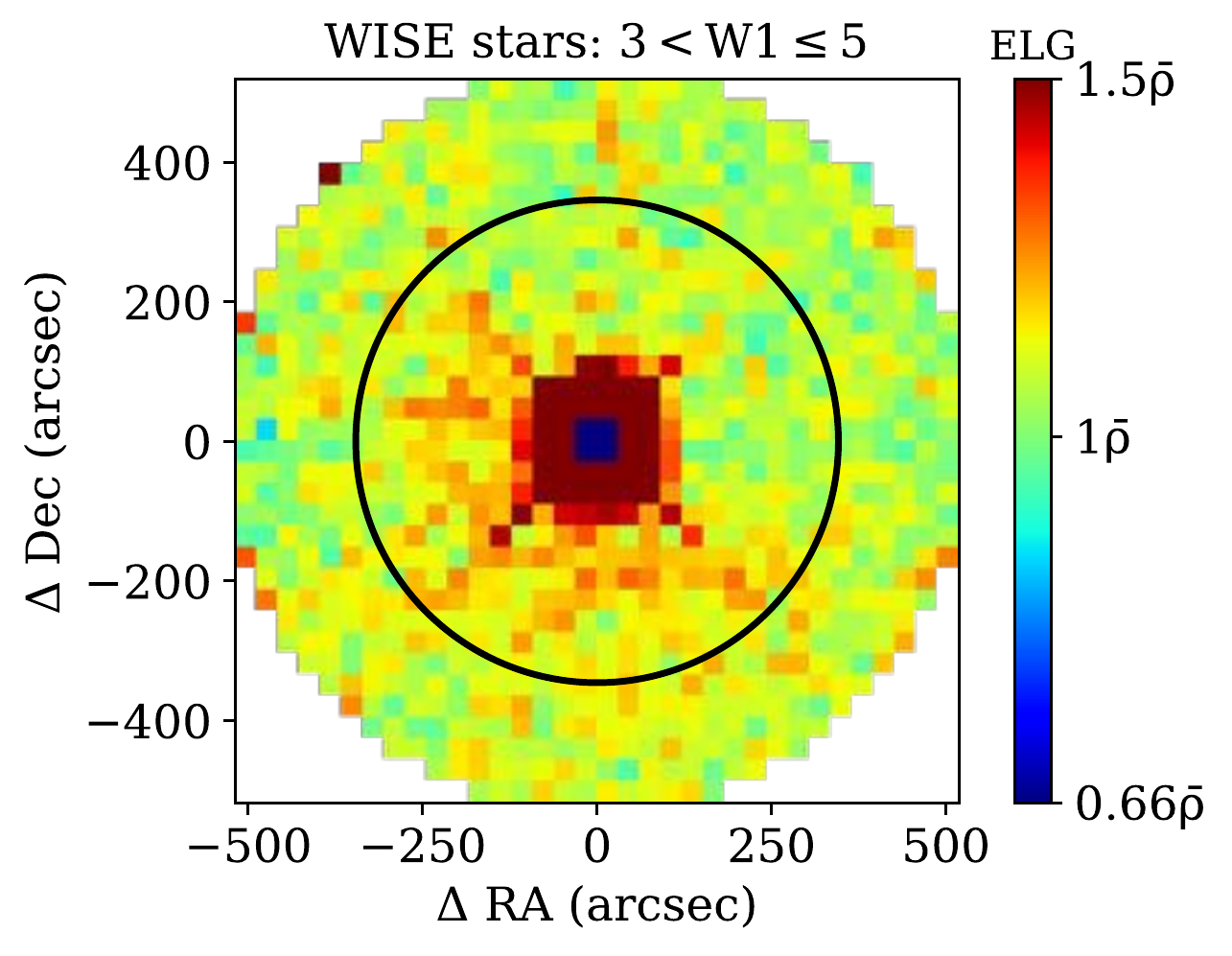}
\includegraphics[width=0.29\linewidth]{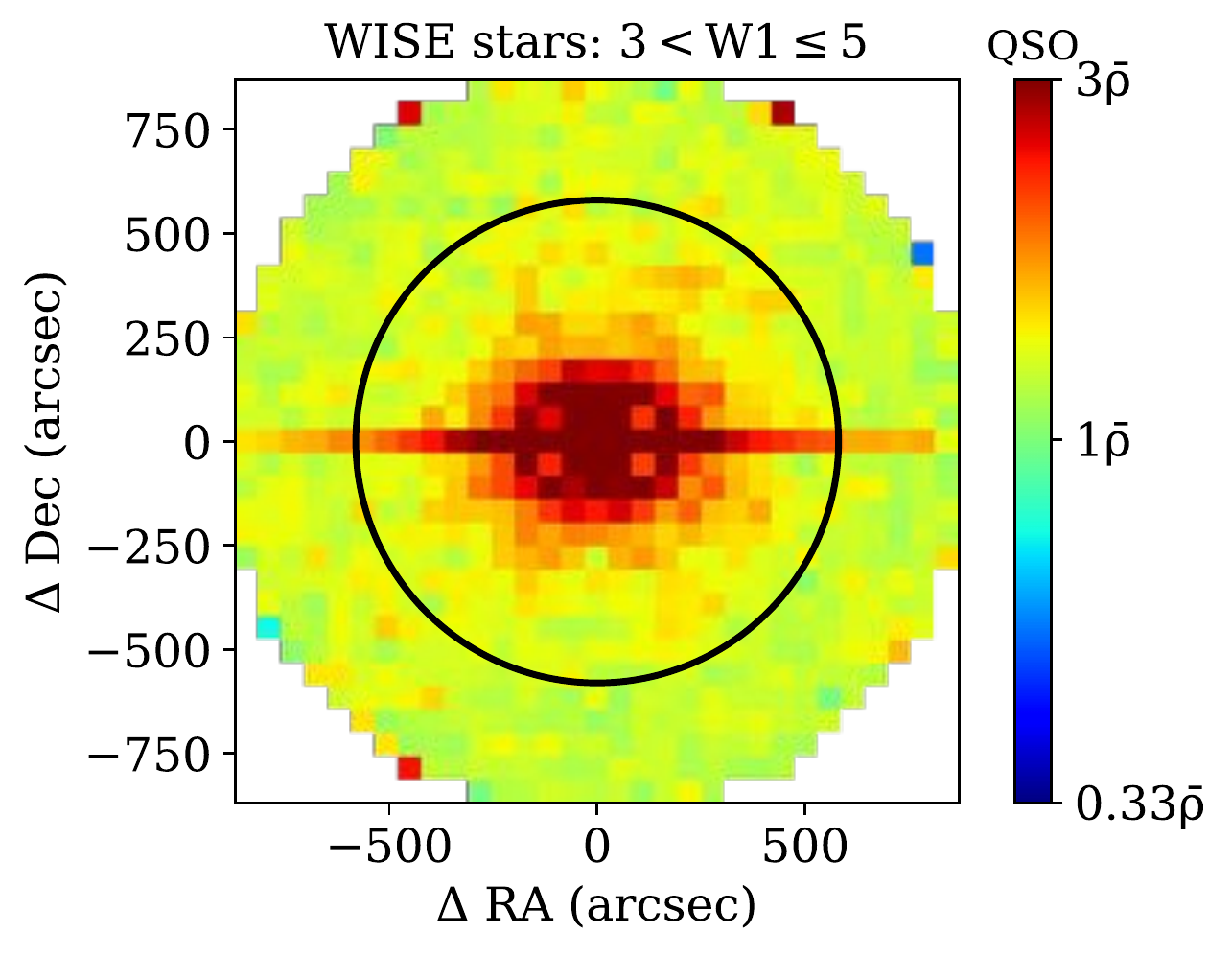}
\includegraphics[width=0.29\linewidth]{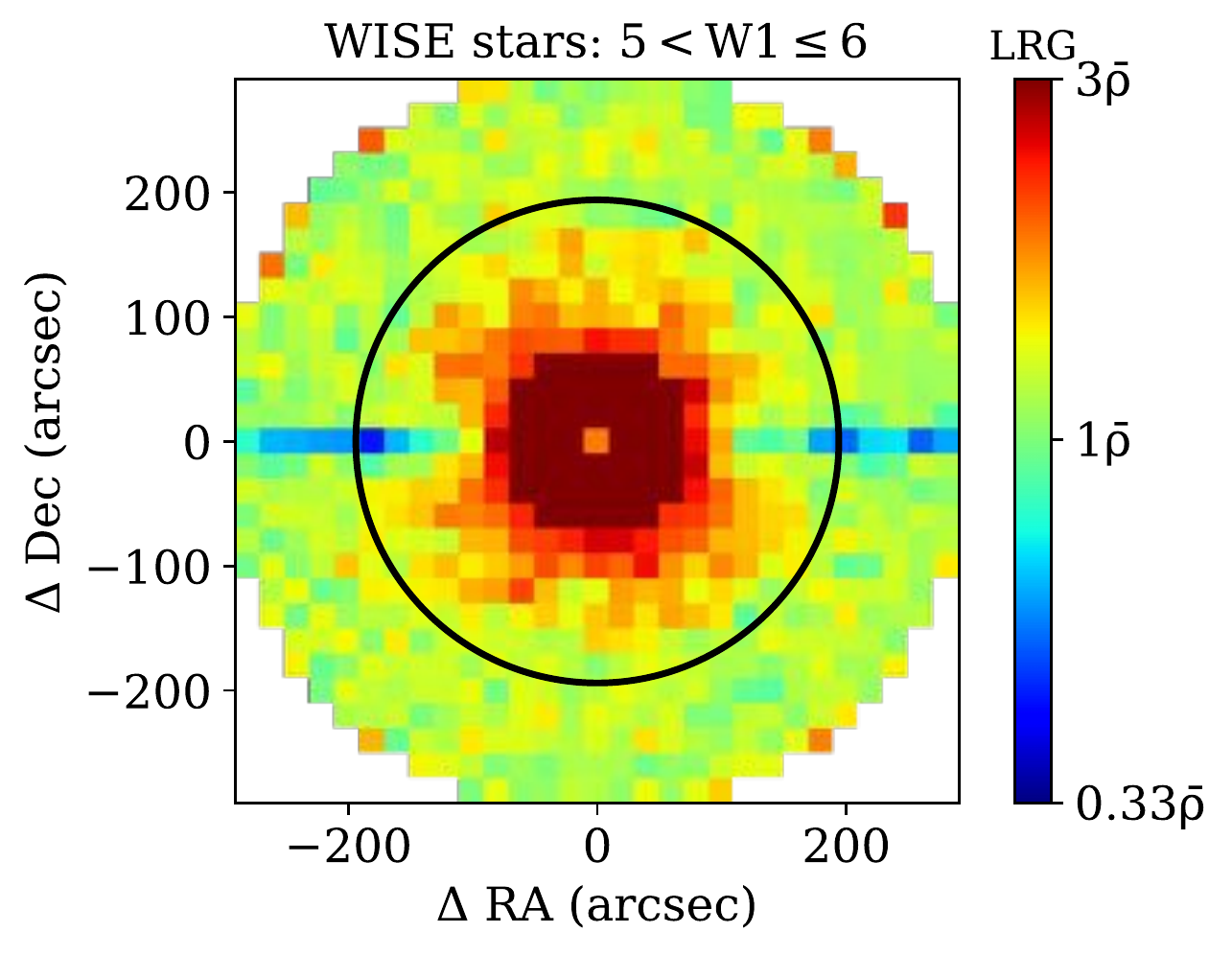}
\includegraphics[width=0.29\linewidth]{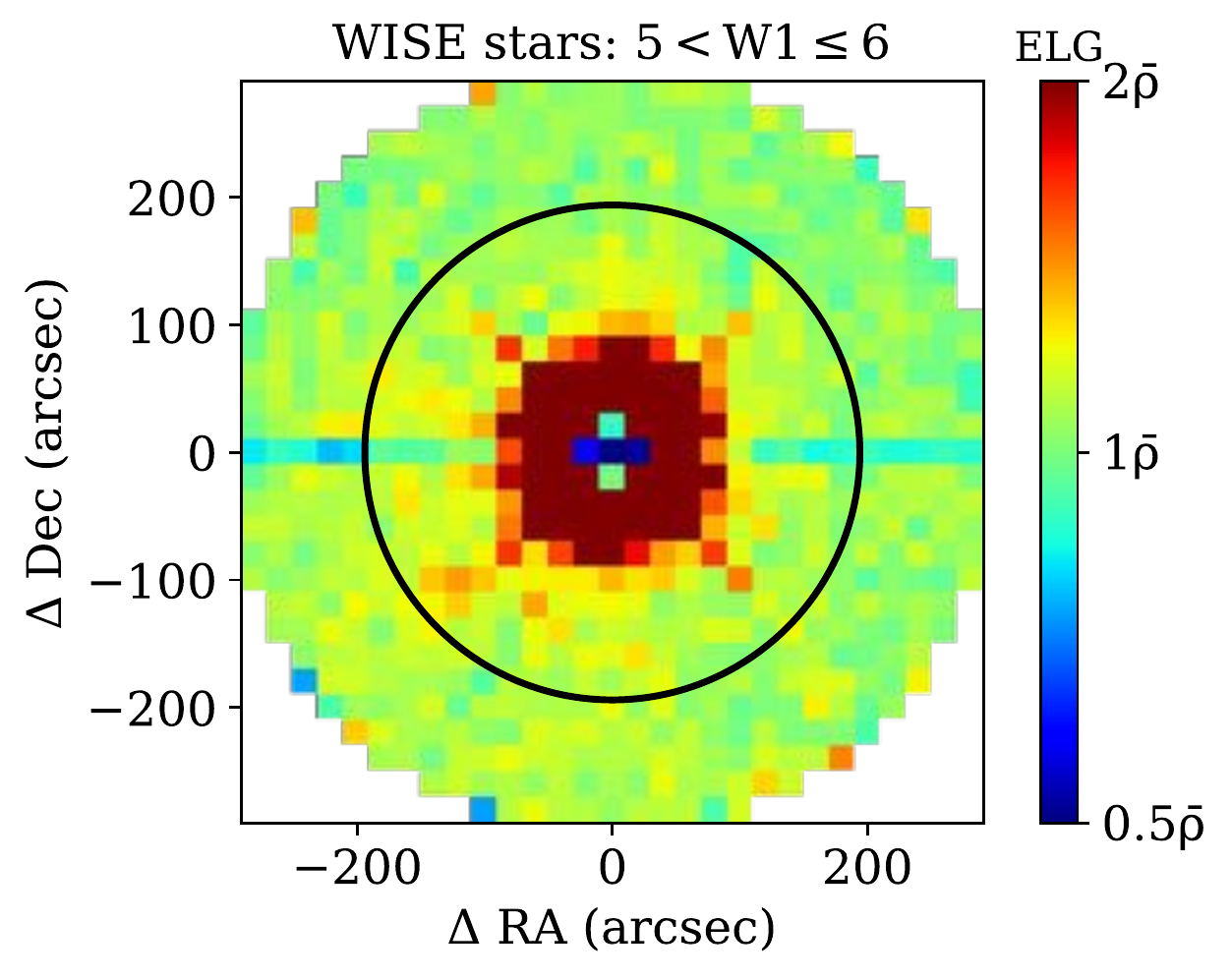}
\includegraphics[width=0.29\linewidth]{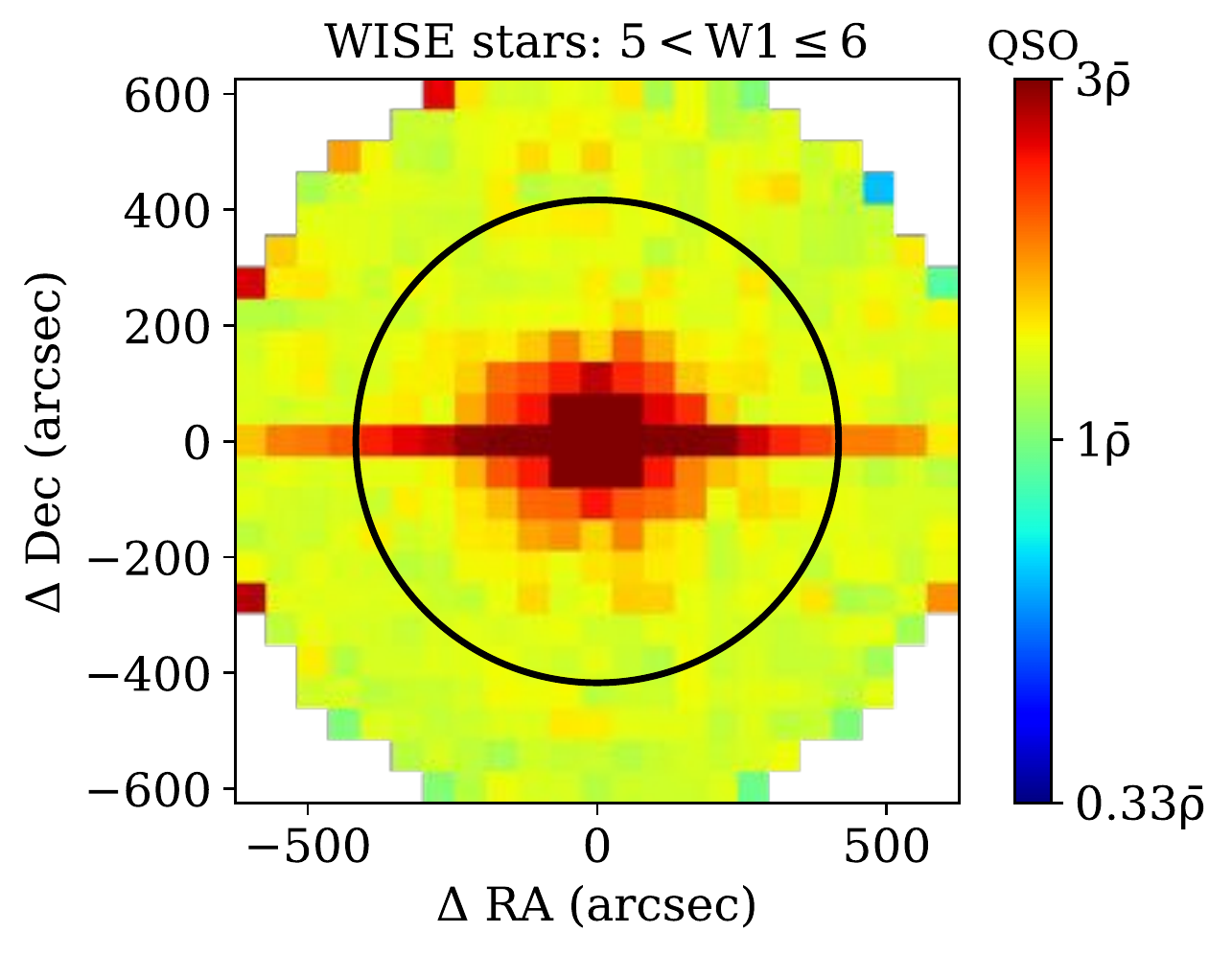}
\includegraphics[width=0.29\linewidth]{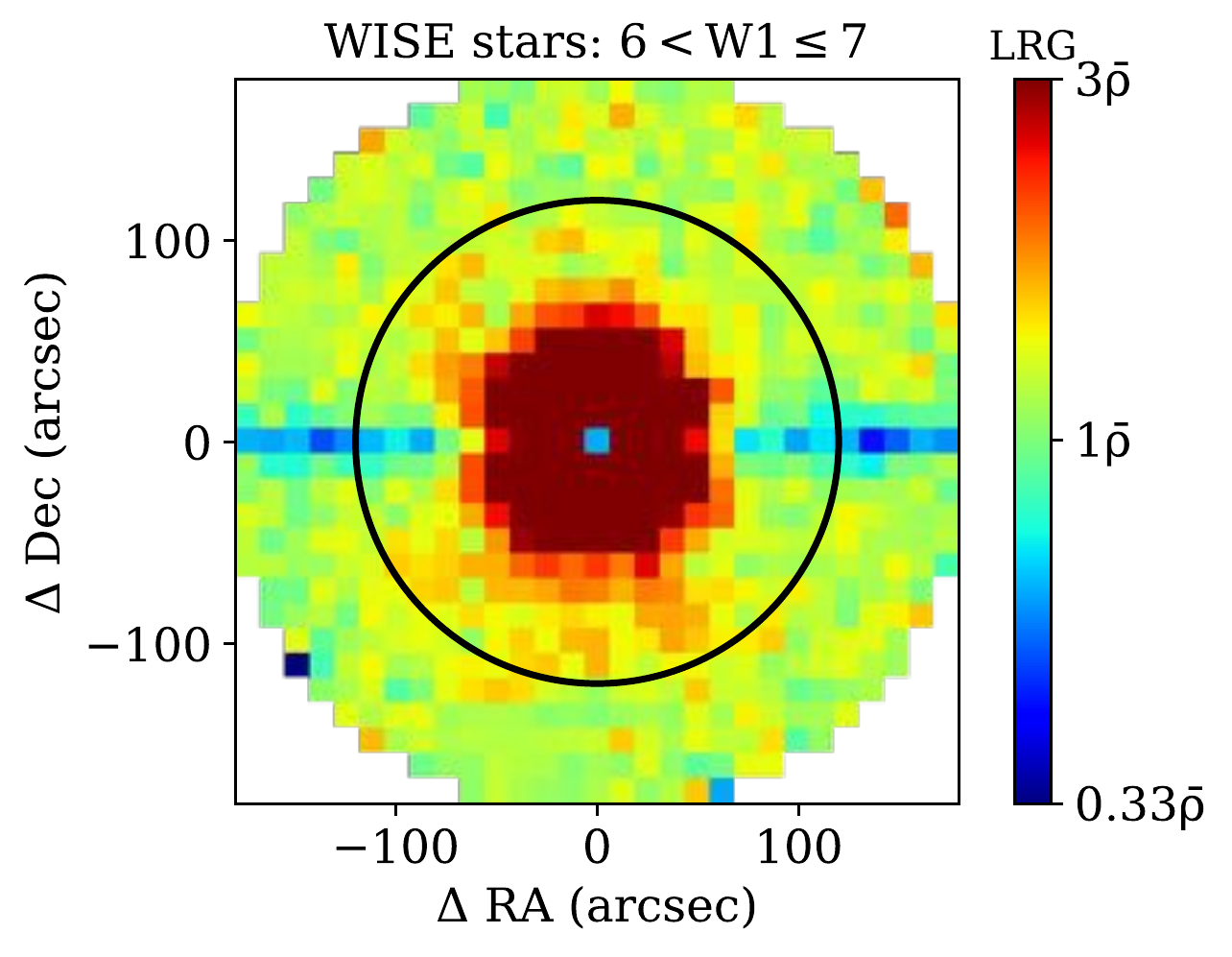}
\includegraphics[width=0.29\linewidth]{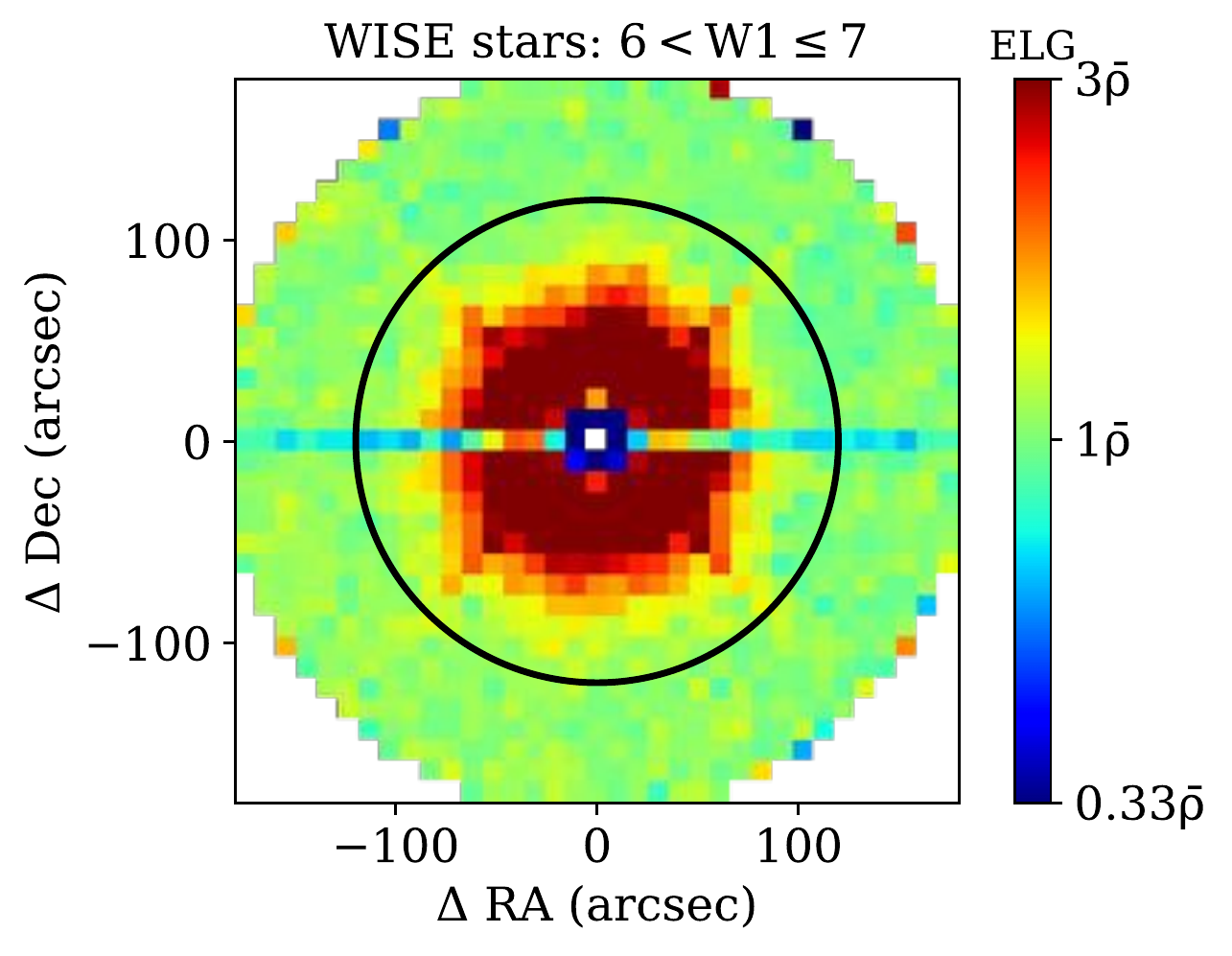}
\includegraphics[width=0.29\linewidth]{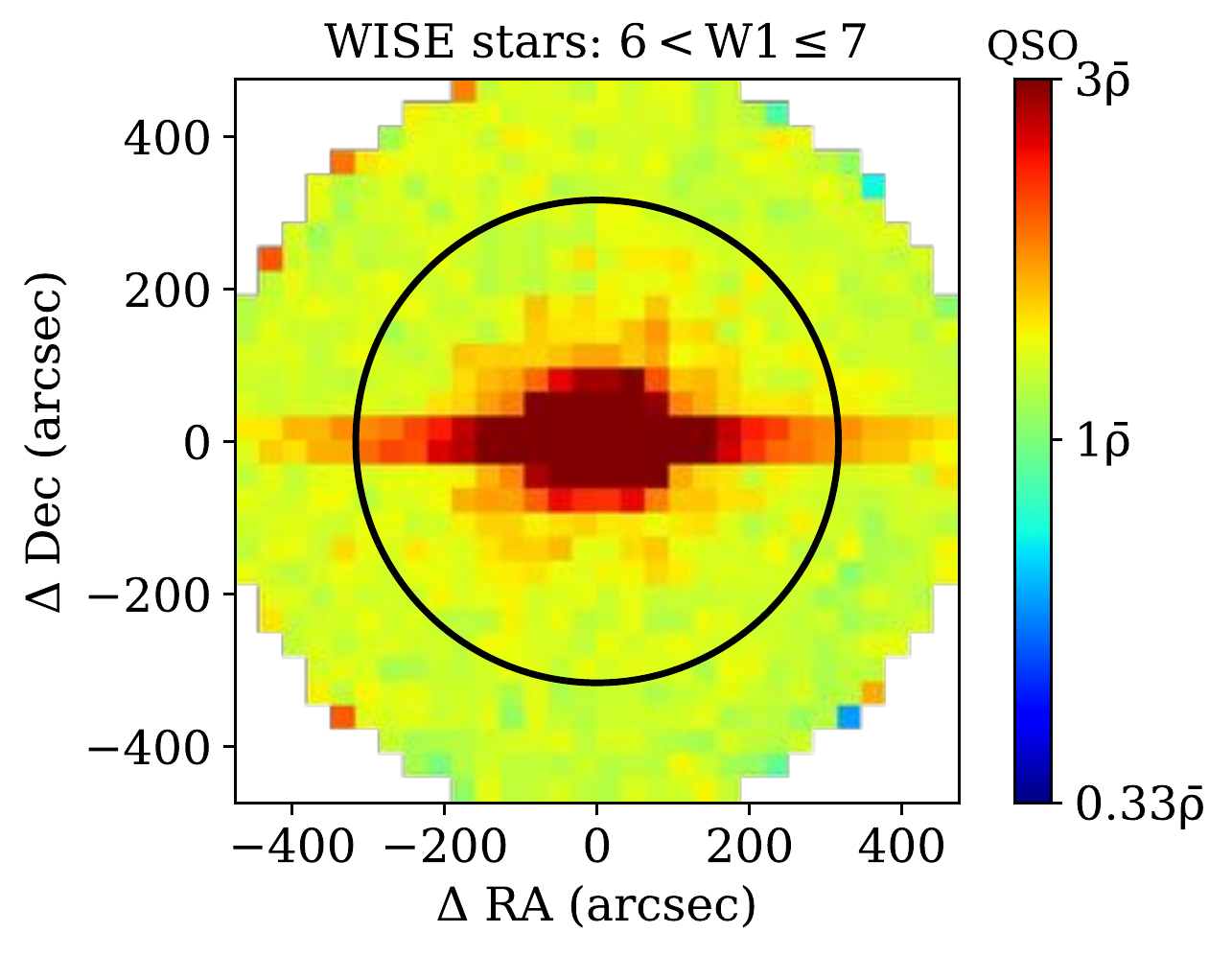}
\includegraphics[width=0.29\linewidth]{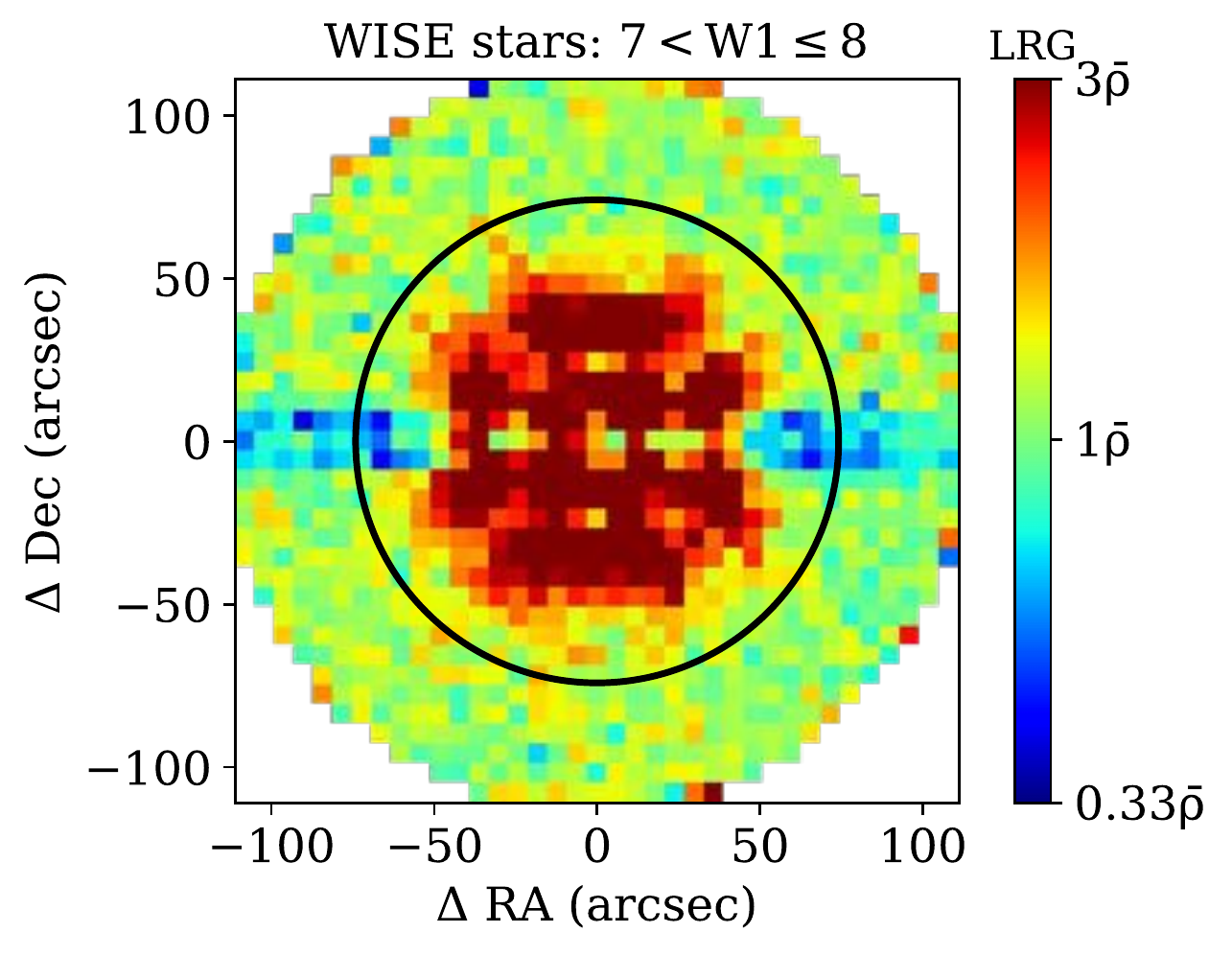}
\includegraphics[width=0.29\linewidth]{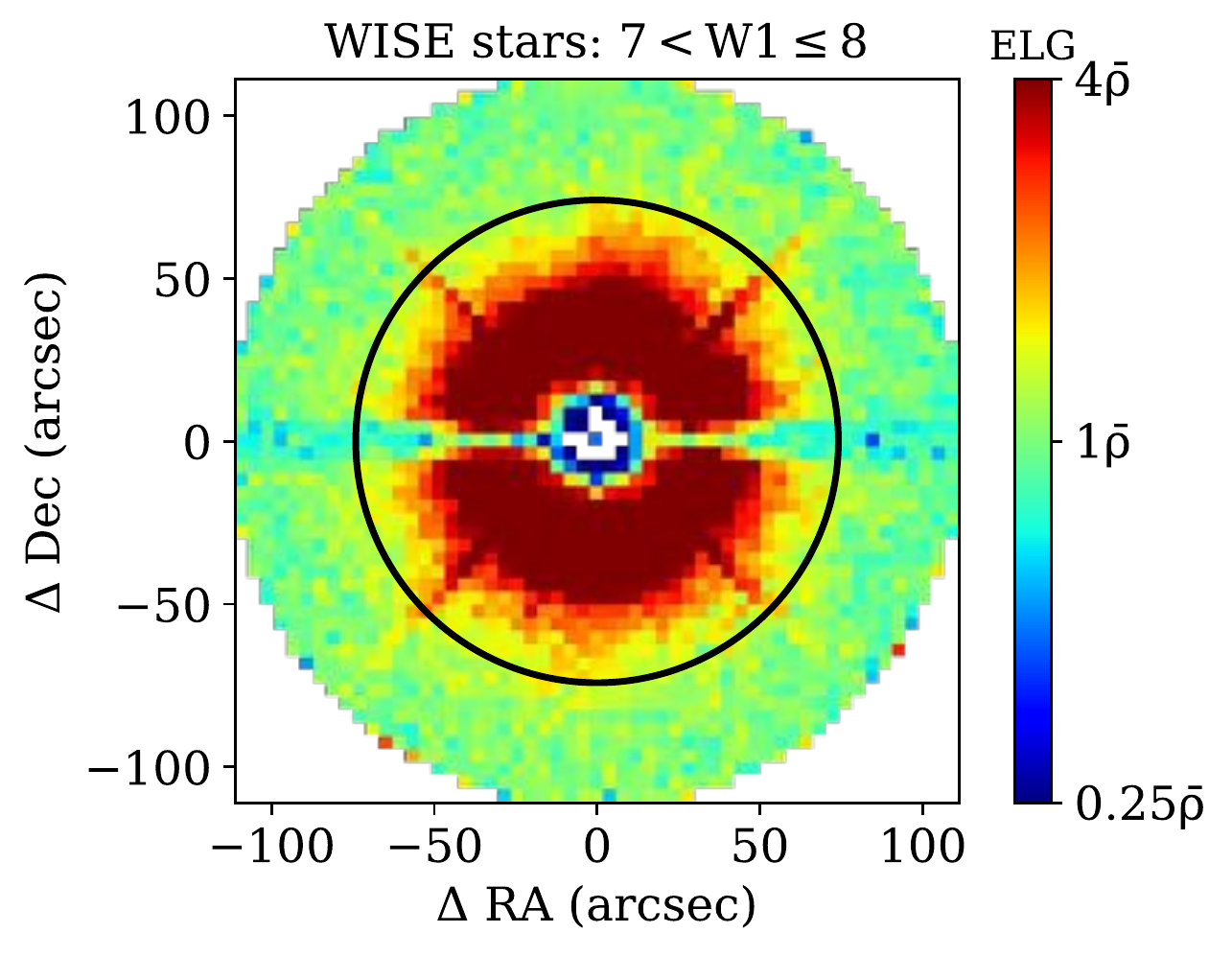}
\includegraphics[width=0.29\linewidth]{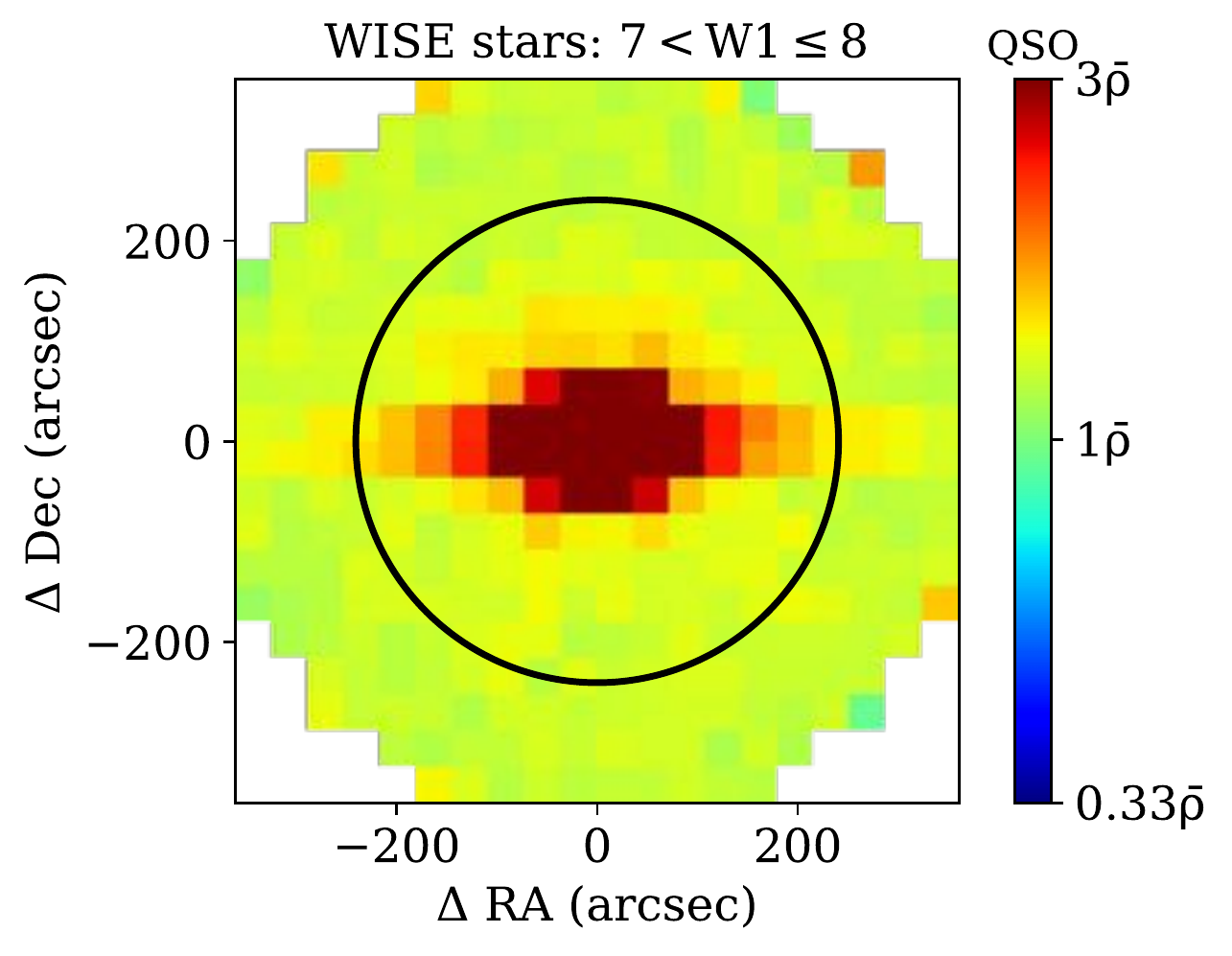}
\caption[]{2D histograms of the density of DESI targets around stacks of bright stars from WISE. The solid black circles represent our star masks (Equation \ref{eq:wise-mr}). The horizontal features appearing in some maps, which are due to insufficient masking of charge bleed trails in the CCDs, are only a few arcsec in width, and we find that removing them with a separate rectangular mask has no perceptible impact on the densities around stars.}
\label{fig:star2d-wise}
\end{figure*}

For both catalogs, we follow a similar procedure to construct and test a radial mask for each target class. We begin by splitting the sample of stars into magnitude bins, since the masking radius will be magnitude-dependent. The bin widths are chosen such that each bin contains a similar number of stars and therefore has comparable Poisson errors. For a given magnitude bin, we use the pair-enumeration algorithm in \texttt{KDcount}\footnote{\url{https://github.com/rainwoodman/kdcount}} to efficiently locate all star-galaxy pairs within a distance of $\theta = 0.05$ rad $\approx 10^4$ arcsec of one another. We calculate the density of targets around each star in logarithmically-spaced annular bins up to this maximum separation, then average across all stars to determine the mean density of targets in each annular bin. By assigning a conservative cutoff radius to each magnitude bin by eye and fitting the results, we obtain the following magnitude-radius relations: 
\begin{align}\label{eq:tycho-mr}
& \text{For Tycho-2:} \nonumber \\
& R = 
\begin{cases}
   - 2.5 \times VT^3 + 77.4 \times VT^2 \\ \ \ \ \ \ \ - 813.6 \times VT + 2969 \ \text{arcsec}, & \text{LRG, ELG} \\
   2.8 VT^2 - 143.4 VT + 1387.1 \ \text{arcsec},   & \text{QSO}\\
\end{cases} \\
& \text{For WISE:} \nonumber \\
& R = 
\begin{cases}\label{eq:wise-mr}
    10^{\ 3.29 \ - \ 0.18 \ \times \ W1} \ \text{arcsec}, \hspace{1.3cm} & \text{LRG, ELG} \\
    10^{\ 3.29 \ - \ 0.12 \ \times \ W1} \ \text{arcsec}, & \text{QSO} \\
\end{cases}
\end{align}
For both types of star mask, the LRGs and ELGs can be fit to the same magnitude-radius relation, while the QSOs require their own, more conservative mask. We found that relaxing the QSO stellar masks (by, for instance, applying the LRG/ELG masks instead) led to a measurable increase in stellar contamination, evidenced by inflated QSO autocorrelation and QSO-star cross-correlation measurements. 

In our analysis, we implement Equations \ref{eq:tycho-mr} and \ref{eq:wise-mr} instead of using the \texttt{MASKBITS} column provided by DECaLS to mask out stars. We have found that our geometric masks are more aggressive than the combination of available stellar bitmasks; in particular, they are necessary for removing contaminated areas at larger radii.

Figure~\ref{fig:density_star} shows the resulting target densities vs. distance to bright stars, sectioned by target type and star catalog, with each magnitude bin plotted separately. Three bins are highlighted for illustration, with the corresponding dashed vertical lines representing the relevant masking radii calculated with the average magnitude of that bin. These plots show how the masks eliminate spurious clustering due to bright stars. In Figures~\ref{fig:star2d-tycho} and \ref{fig:star2d-wise}, we also present 2D histograms of target densities around stacks of bright stars, plotted in equatorial coordinates\footnote{We also performed this analysis in ecliptic coordinates, to see if additional structure could be identified. In some of the fainter magnitude bins, resolution and contrast could be manipulated to resolve an x-shaped feature in the 2D stacked plots, but this feature was very fine, and we found that removing it beforehand had no perceptible impact on the 1D density plots.} with star masks drawn on as circles, again showing well-fiting mask radii.

\subsubsection{Bright extended sources}
\label{sec:masks/foregrounds/xs}

\begin{figure}
\centering
\includegraphics[width=0.97\columnwidth]{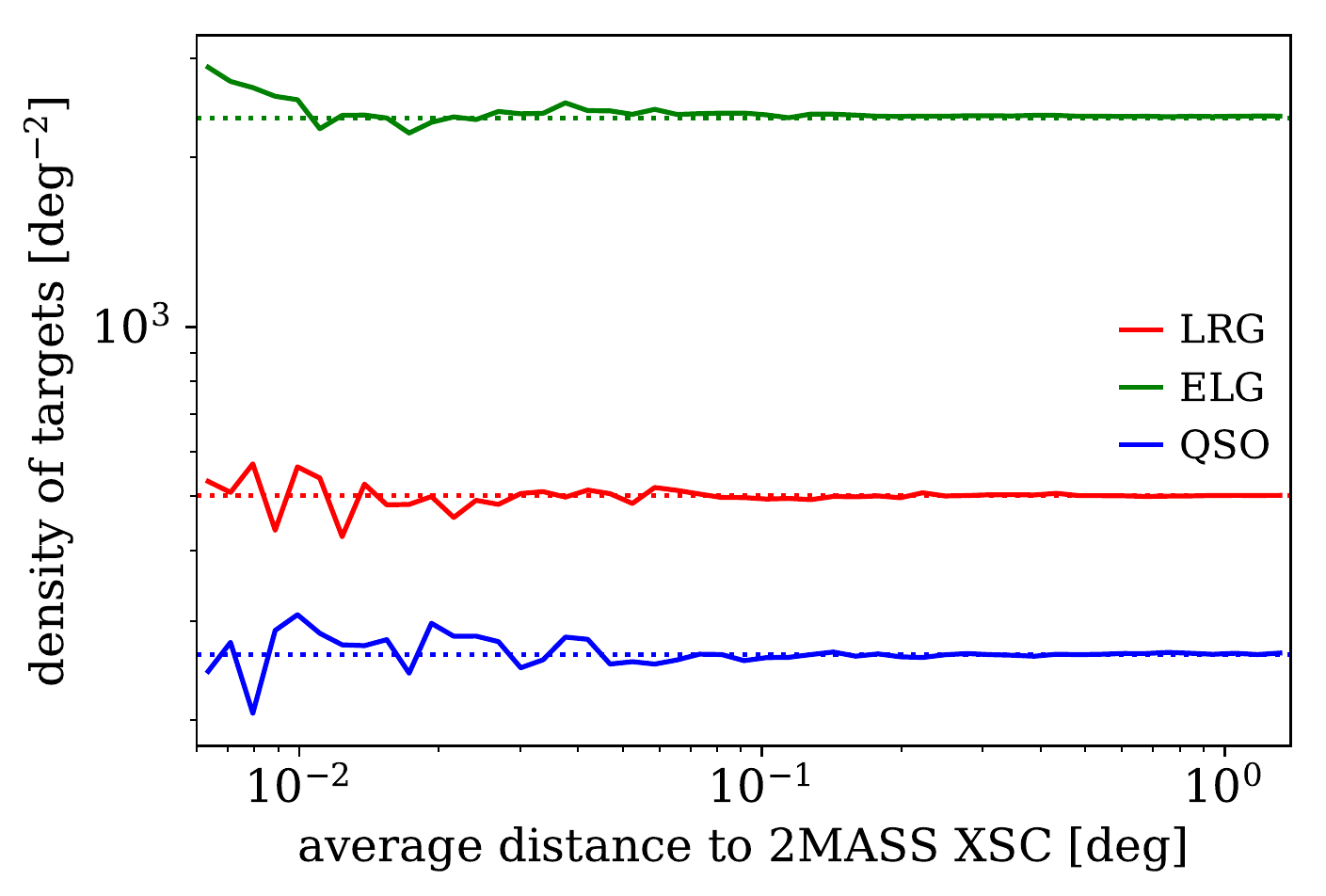}
\caption{Average density of DESI dark time targets as a function of distance to extended sources from 2MASS-XSC.}
\label{fig:density_xs}
\end{figure}

Similarly to bright stars, we examine the density of targets near bright extended sources such as nearby galaxies. We use the 2MASS Extended Source Catalog \citep{Jarrett00}, a catalog of near-IR extended sources complete for angular sizes greater than $\sim$10 arcsec. Restricting to $10 < J < 15$ total $J$-band magnitude, we find no appreciable impact on the density of our DESI dark time targets (Figure~\ref{fig:density_xs}), and thus we do not apply a mask.

\section{SPATIAL VARIATIONS}
\label{sec:spatial}
\subsection{NGC vs. SGC}\label{sec:NvS}

\begin{figure}
    \centering
    \includegraphics[width=\columnwidth]{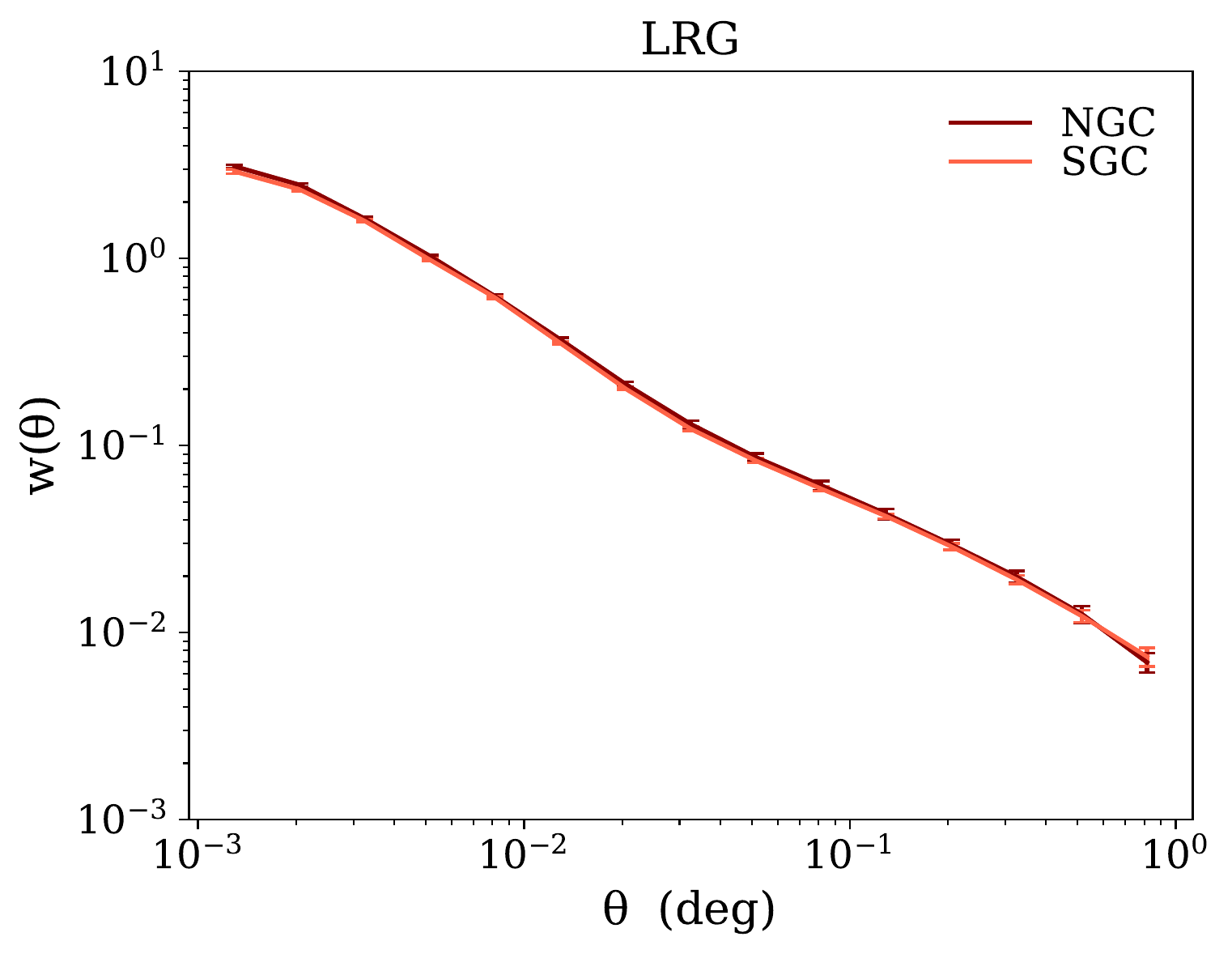}
    \includegraphics[width=\columnwidth]{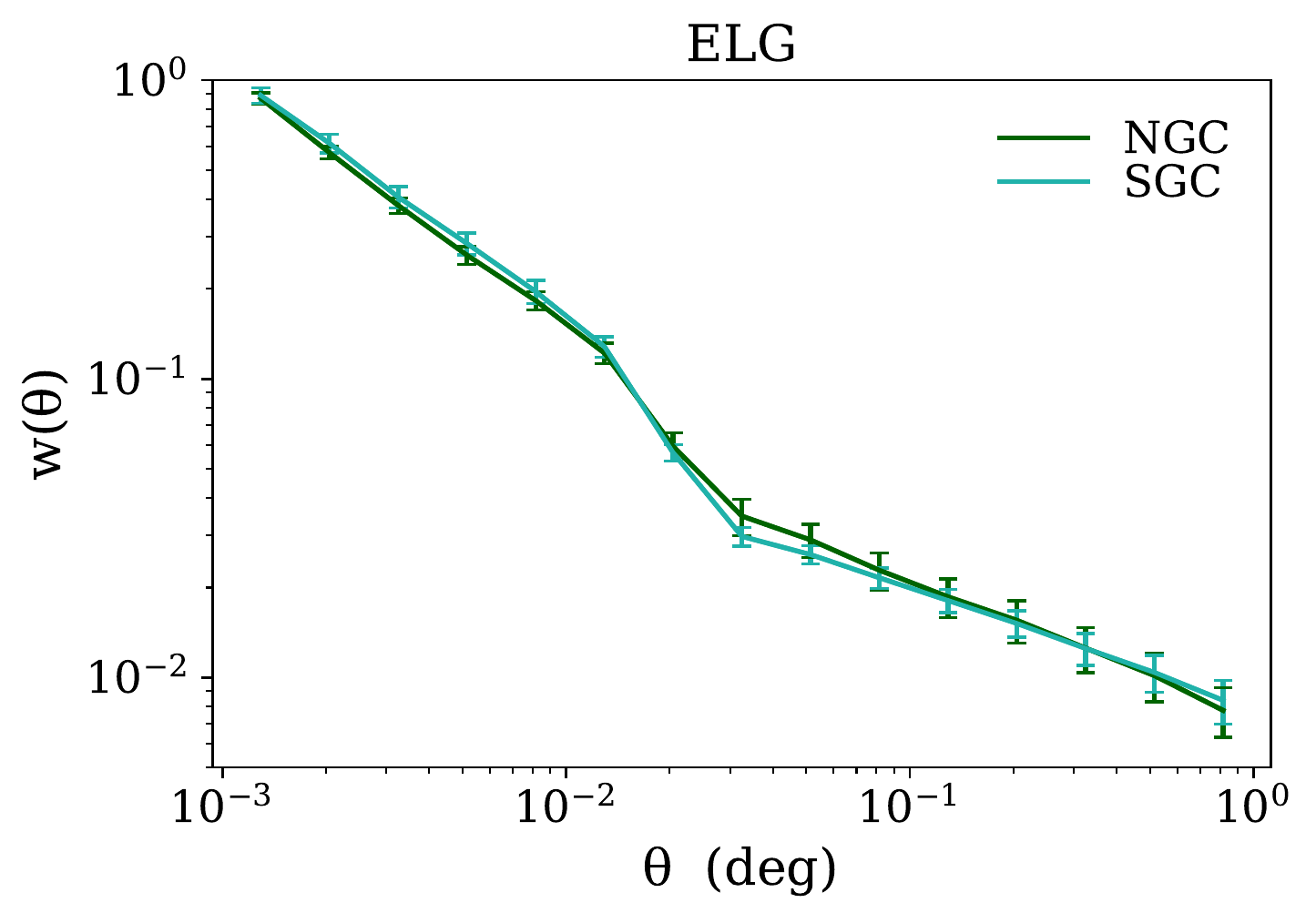}
    \includegraphics[width=\columnwidth]{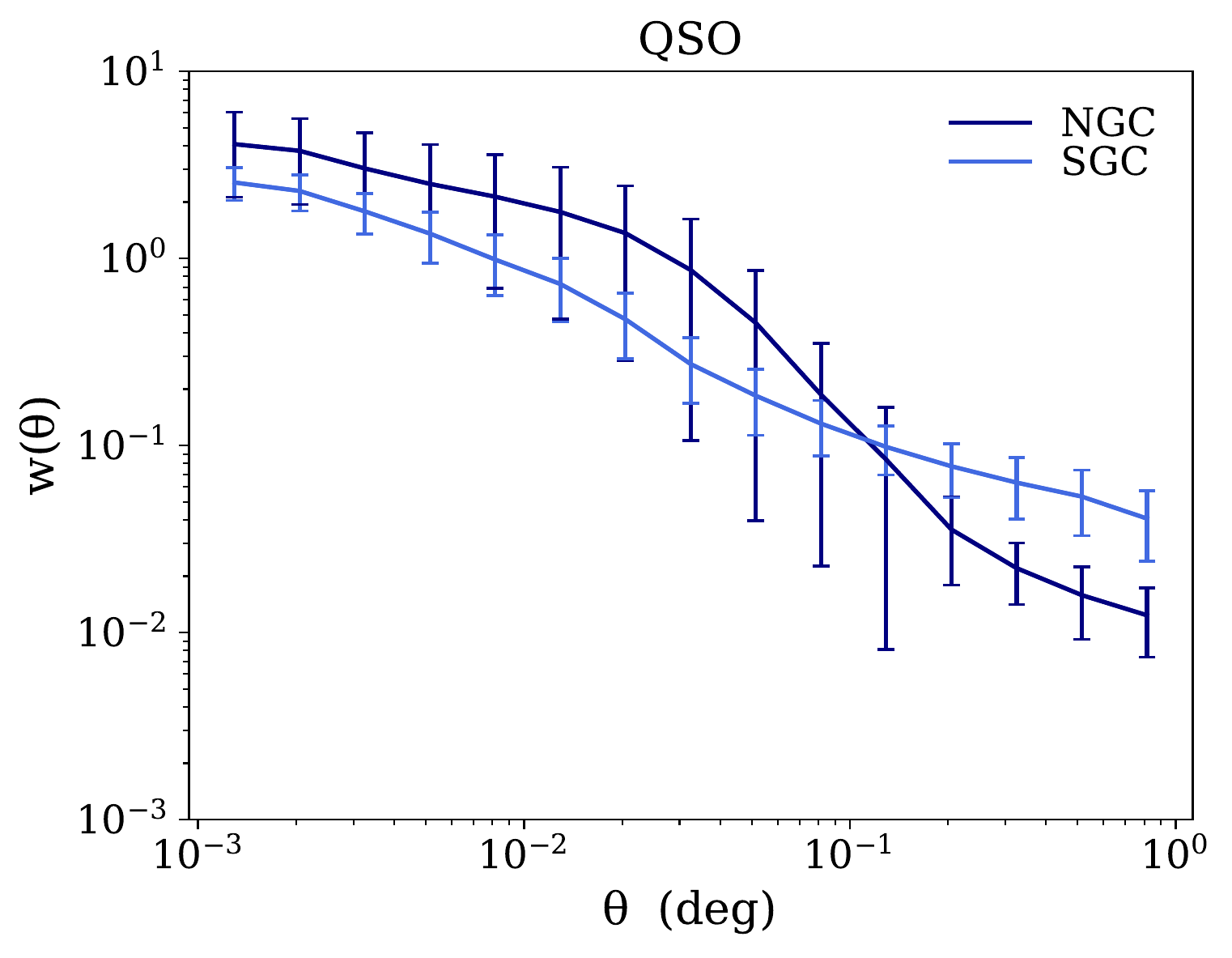}
    \caption{$w(\theta)$ for LRGs, ELGs, and QSOs calculated in NGC and SGC separately. Error bars are from bootstrapping.}
    \label{fig:NvS}
\end{figure}

We calculate the angular correlation functions in the north galactic cap (NGC) and the south galactic cap (SGC) individually, as the two hemispheres may suffer from different systematics, and earlier analyses have found NGC/SGC variations in BOSS data (see \citealt{Ross12} Section 4.1 for an explanation of the origin of this difference in number density between the NGC and SGC in BOSS). The results are shown in Figure~\ref{fig:NvS}. For LRGs, the autocorrelations are virtually identical. For ELGs, there is a slight divergence, most noticeably in the $0.02^{\circ} < \theta < 0.09^{\circ}$ range. For QSOs, the difference between NGC and SGC is significantly more pronounced. The NGC results appear more impacted by systematics, as indicated by a bulge in the correlation function with extremely large bootstrap errors. This is likely due to the fact that parts of the SGC, where there is DES imaging, are very deep (see also Figure~\ref{fig:visual_inspec} in the following section to visually observe how the NGC appears more impacted by systematics for QSO).

\subsection{Visual inspection}\label{sec:visinspec}

\begin{figure*}
    \centering
    \begin{subfigure}[b]{0.5\textwidth}
        \includegraphics[width=\textwidth]{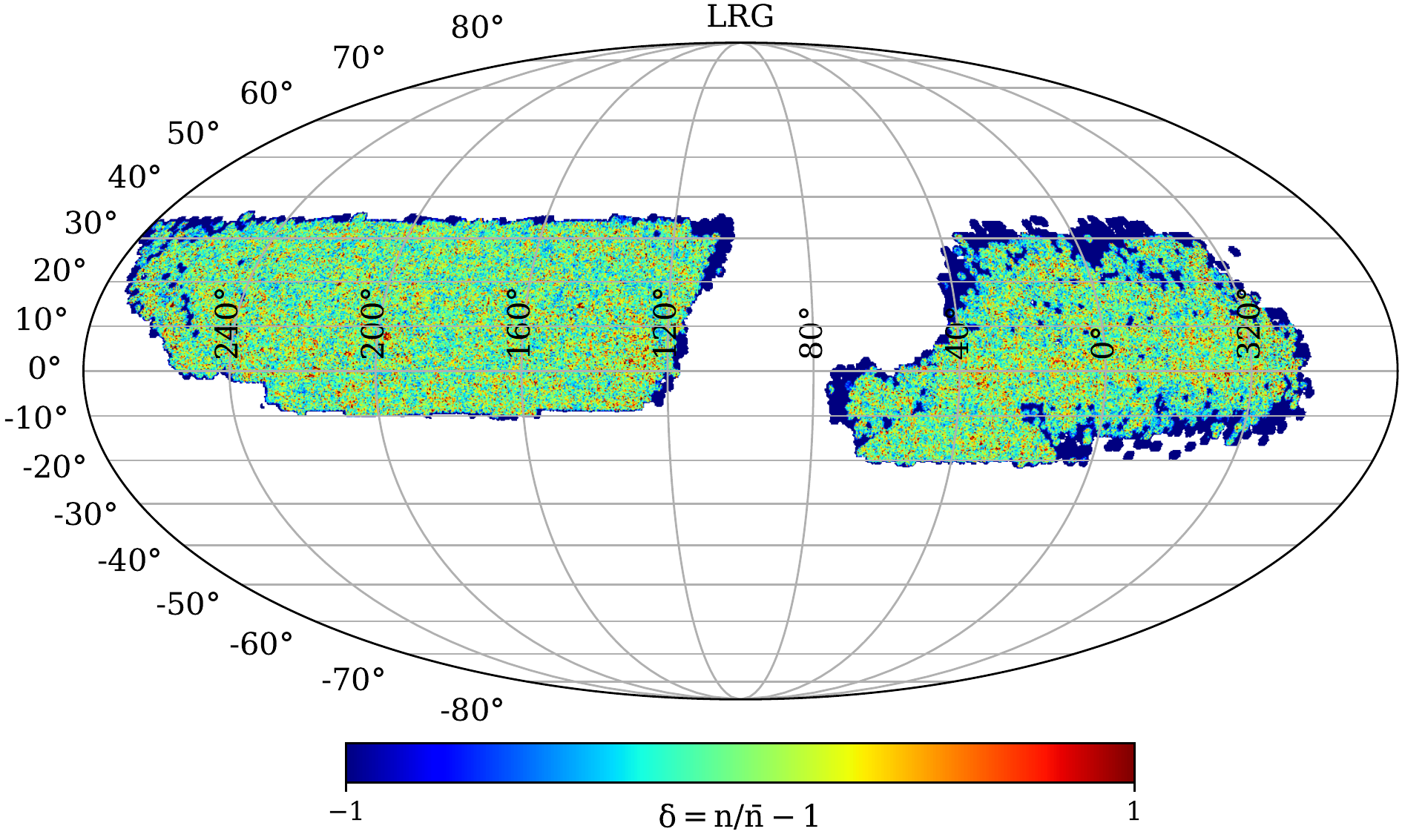}
        \includegraphics[width=\textwidth]{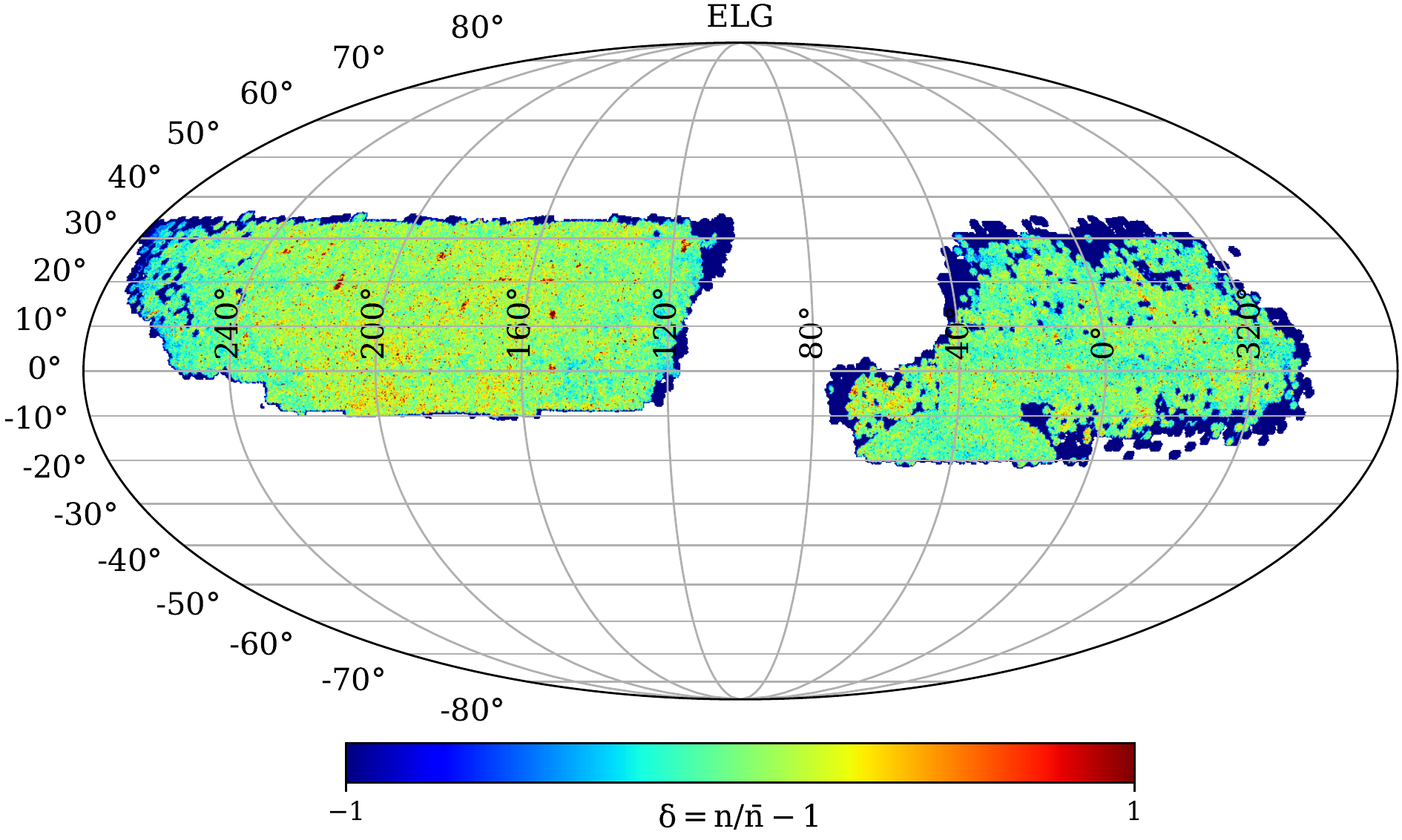}        \includegraphics[width=\textwidth]{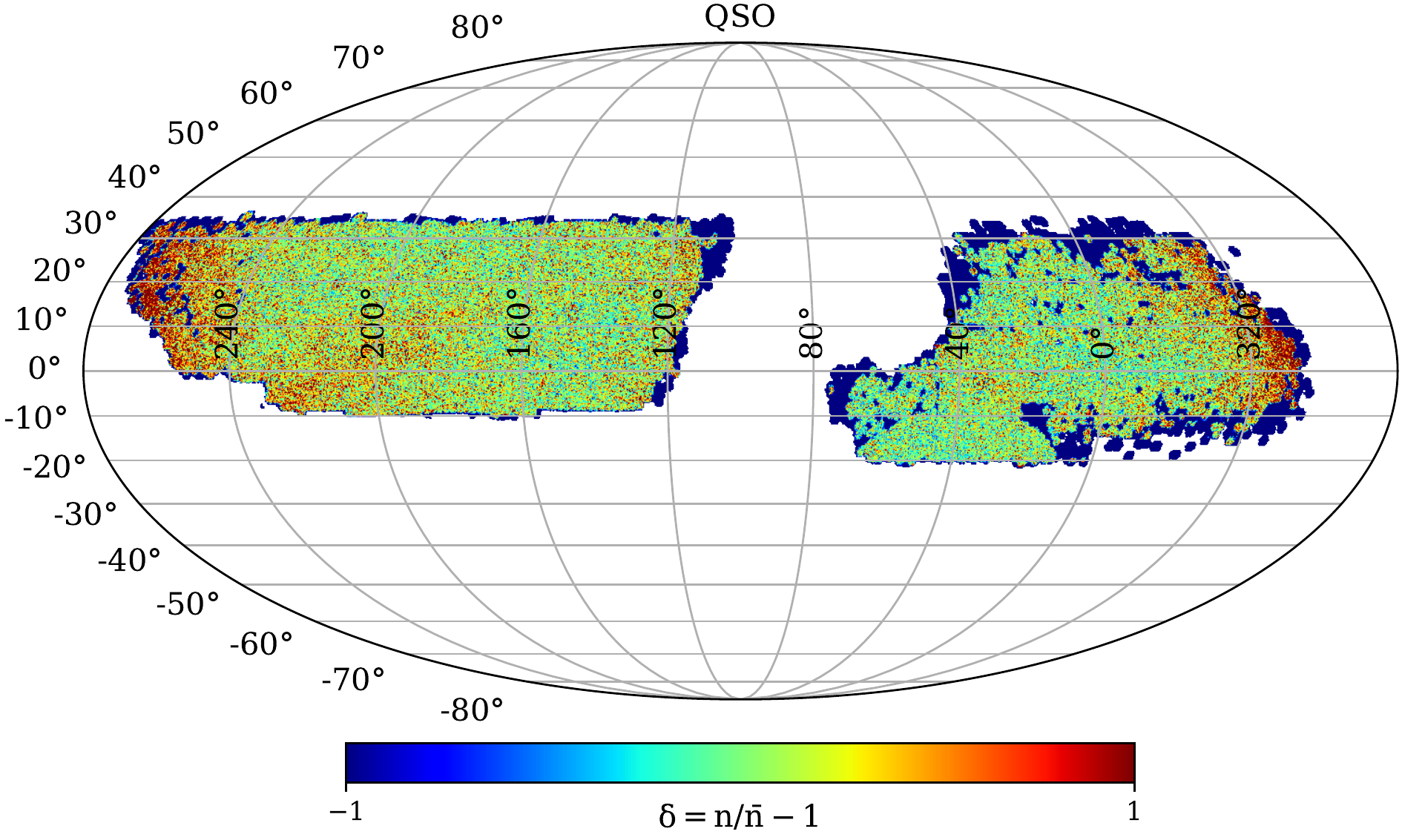}
        \caption{Unmasked.}
    \end{subfigure}%
    \begin{subfigure}[b]{0.5\textwidth}
        \includegraphics[width=\textwidth]{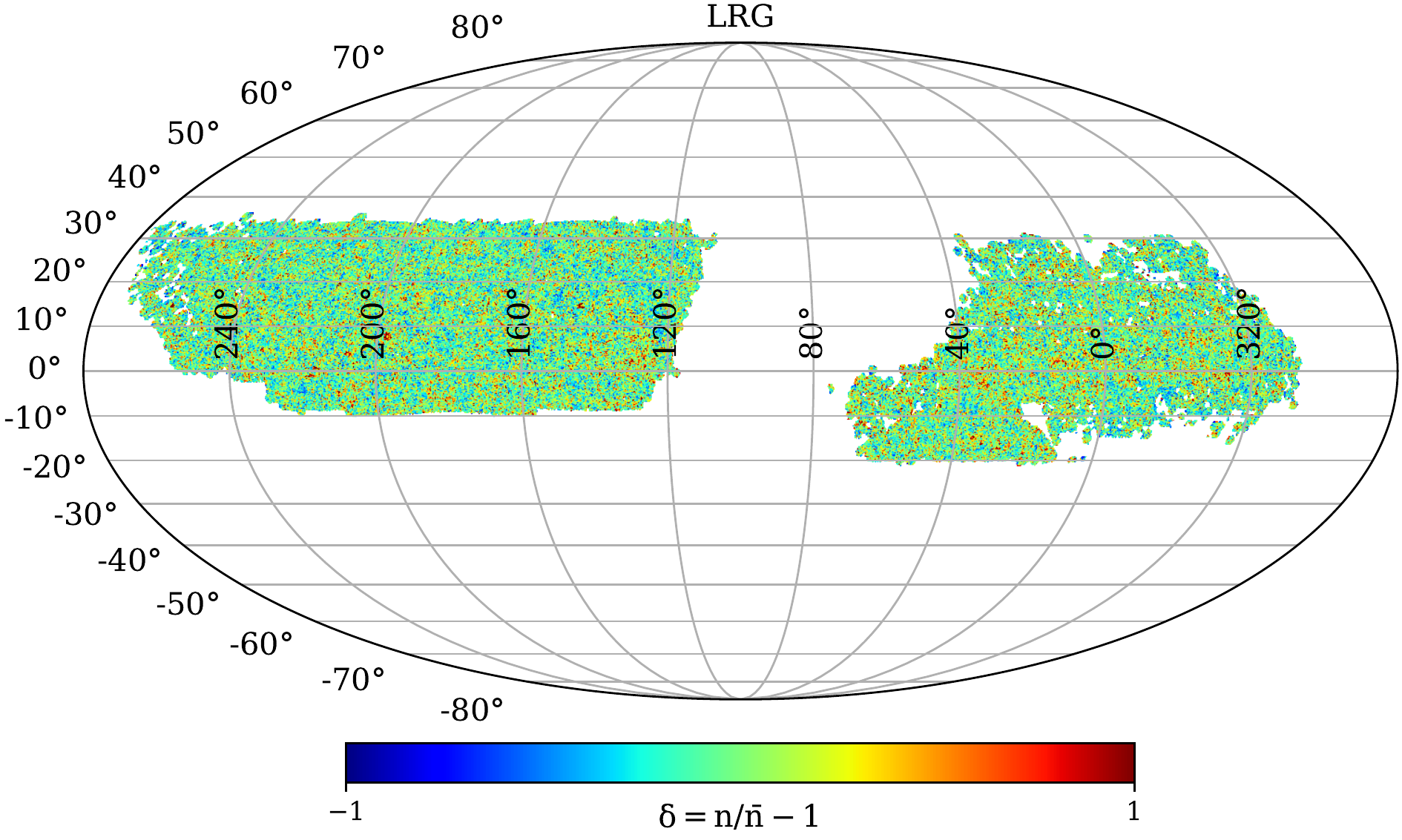}
        \includegraphics[width=\textwidth]{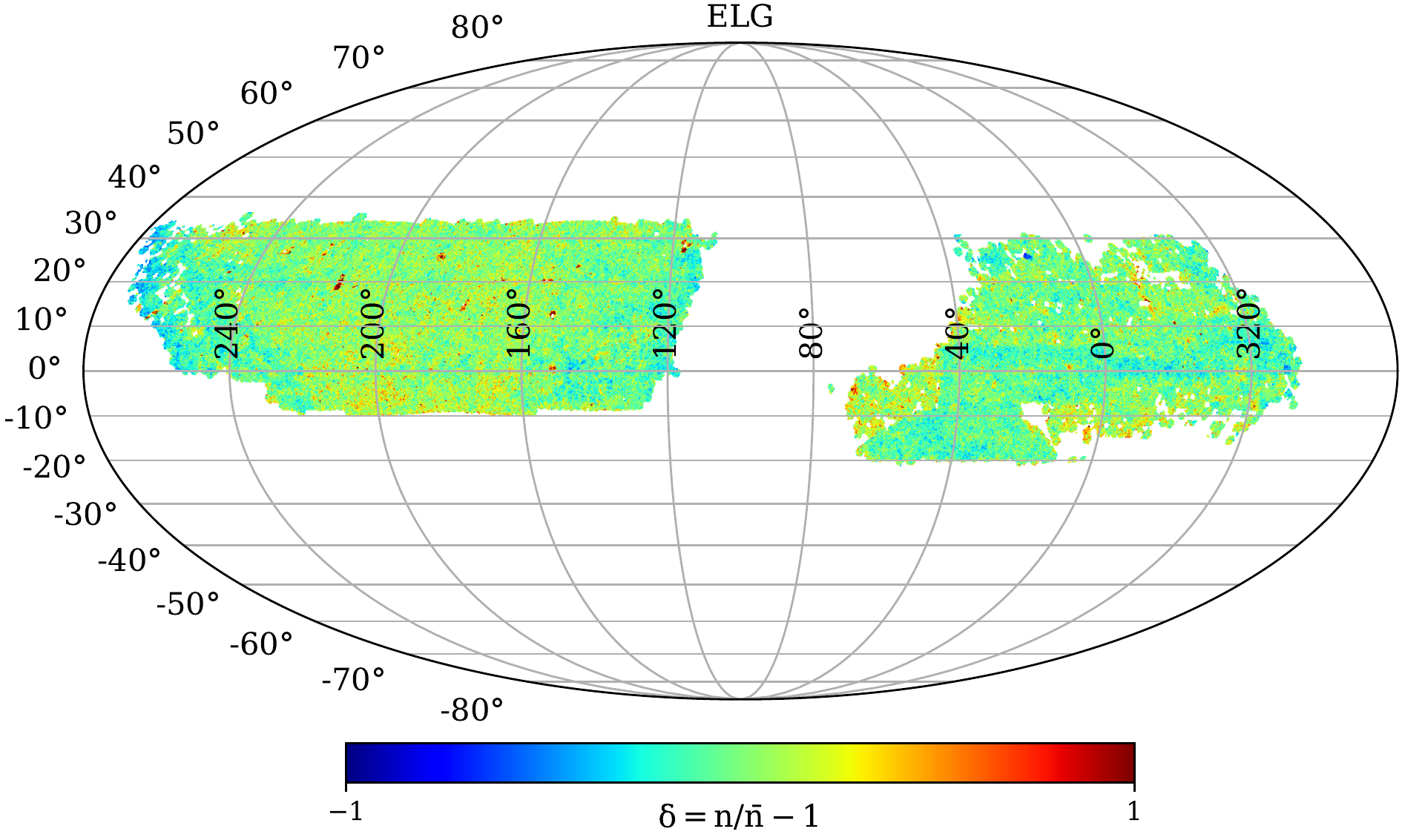} \includegraphics[width=\textwidth]{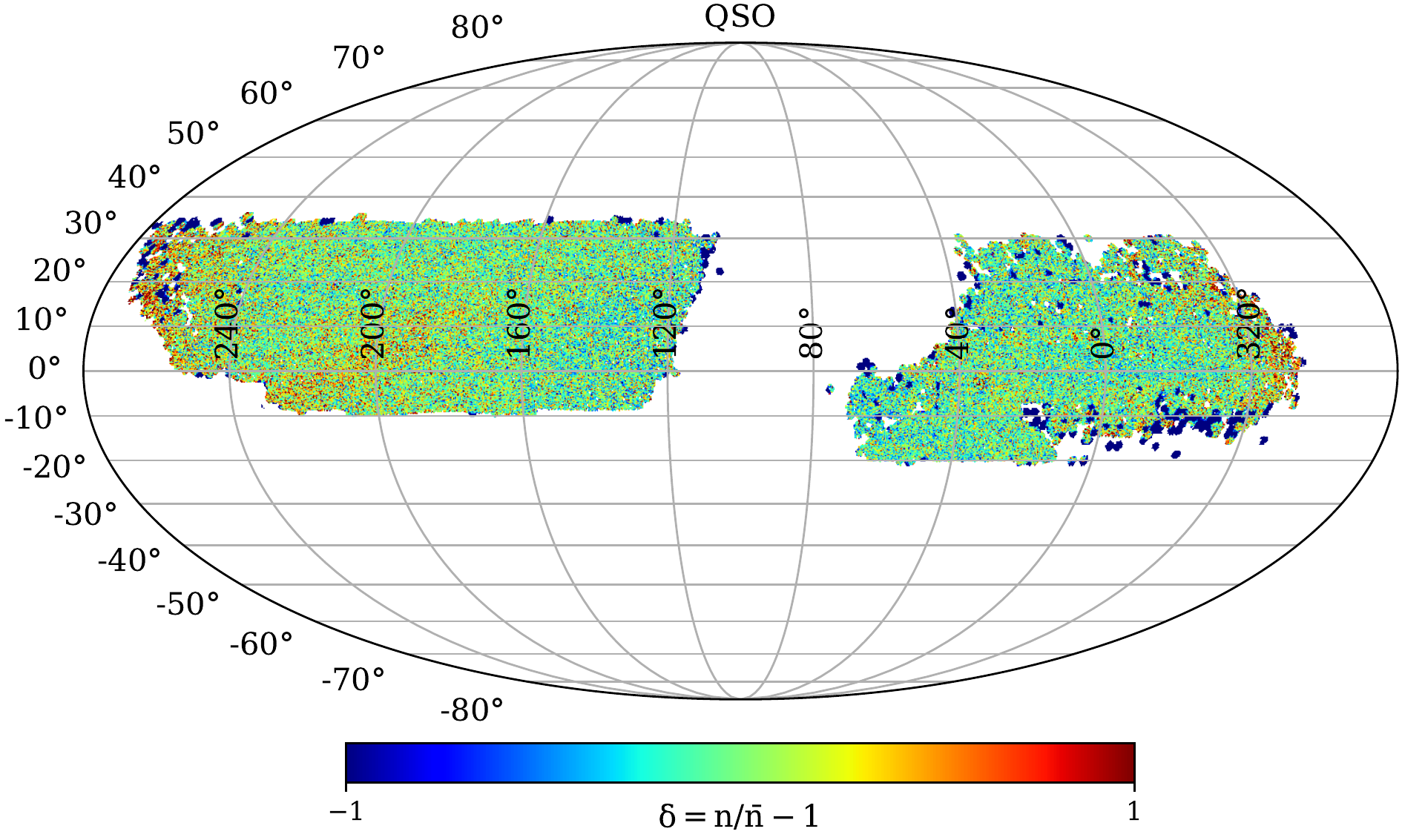}
        \caption{Masked.}
    \end{subfigure}    
    \caption{Maps of the density contrast $\delta = n/\bar{n} - 1$ calculated with \texttt{HEALPix} resolution $N_{\text{SIDE}}=256$. Mollweide projection in equatorial coordinates with right ascension centered at RA $= 100^{\circ}$. Masked data (right) is less impacted by stellar contamination and variations in imaging depth than raw data (left).}
    \label{fig:visual_inspec}
\end{figure*}

As an initial sanity check, we create maps of the density contrast $\delta = n/\bar{n} - 1$ averaged over \texttt{HEALPix} pixels of $N_{\text{SIDE}}=256$. The results are shown in Figure~\ref{fig:visual_inspec}, with the unmasked catalogs mapped on the left and their masked counterparts on the right. The masked catalogs are visually cleaner, with the star masks reducing stellar contamination and the completeness masks cancelling the imprint of imaging depth on target density. Several features remaining in the masked maps are highlighted and discussed below.

While LRG clustering appears relatively uniform, ELG clustering shows some troubling large-scale trends. For example, the shape of the DES region in the south is detectable, appearing under-dense despite its superior depth. This suggests contamination in non-DES regions, likely due to the effect of low redshift ($z < 0.8$) objects preferentially scattering into the ELG target selection across the low-z color cut $(g \ - \ r) < 1.15 \ (r \ - \ z) - 0.15$ in regions of worse depth. This was tested (Ashley Ross, private communication) by injecting artificial noise into regions of very deep imaging and examining the photometric redshifts of the resulting scattered objects. Additionally, a few suspicious ``hot spots'' appear in the ELG density. When examined closely, most of these occur around a small set of very bright stars (such as Arcturus) which have been successfully masked, but which cause dramatic and complex artifacts in the image beyond the expected masking radius due to reflections of pupil ghosts. Finally, looking at the density of QSOs, there remains noticeable stellar contamination along the galactic plane even after applying conservative masks, as well as the Sagittarius Stream in the north, and there are also some effects at the edges of the footprint.

\subsection{Outlier analysis}\label{sec:jackknife}

\begin{figure}
    \centering
    \includegraphics[width=\columnwidth]{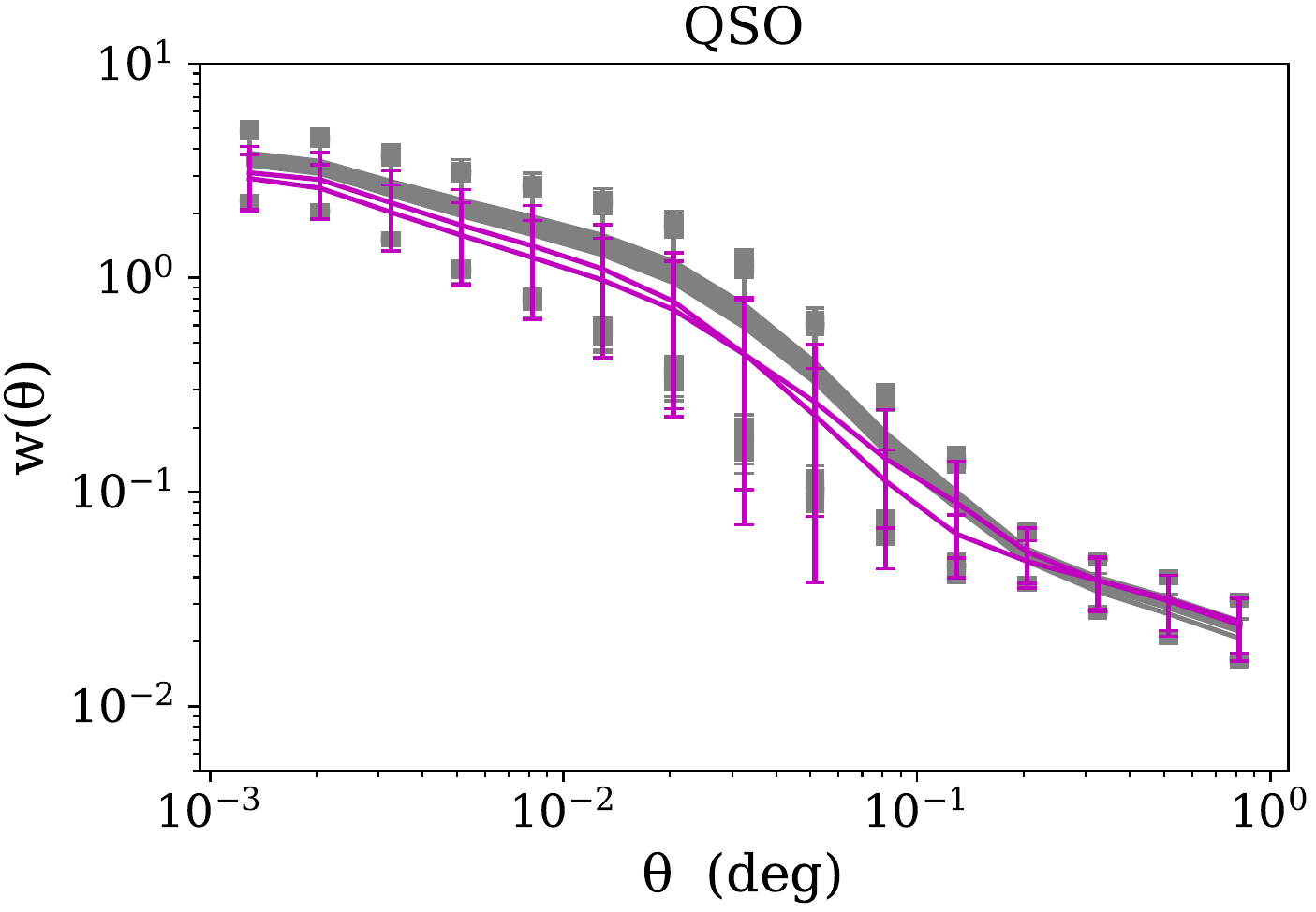}
    \includegraphics[width=\columnwidth]{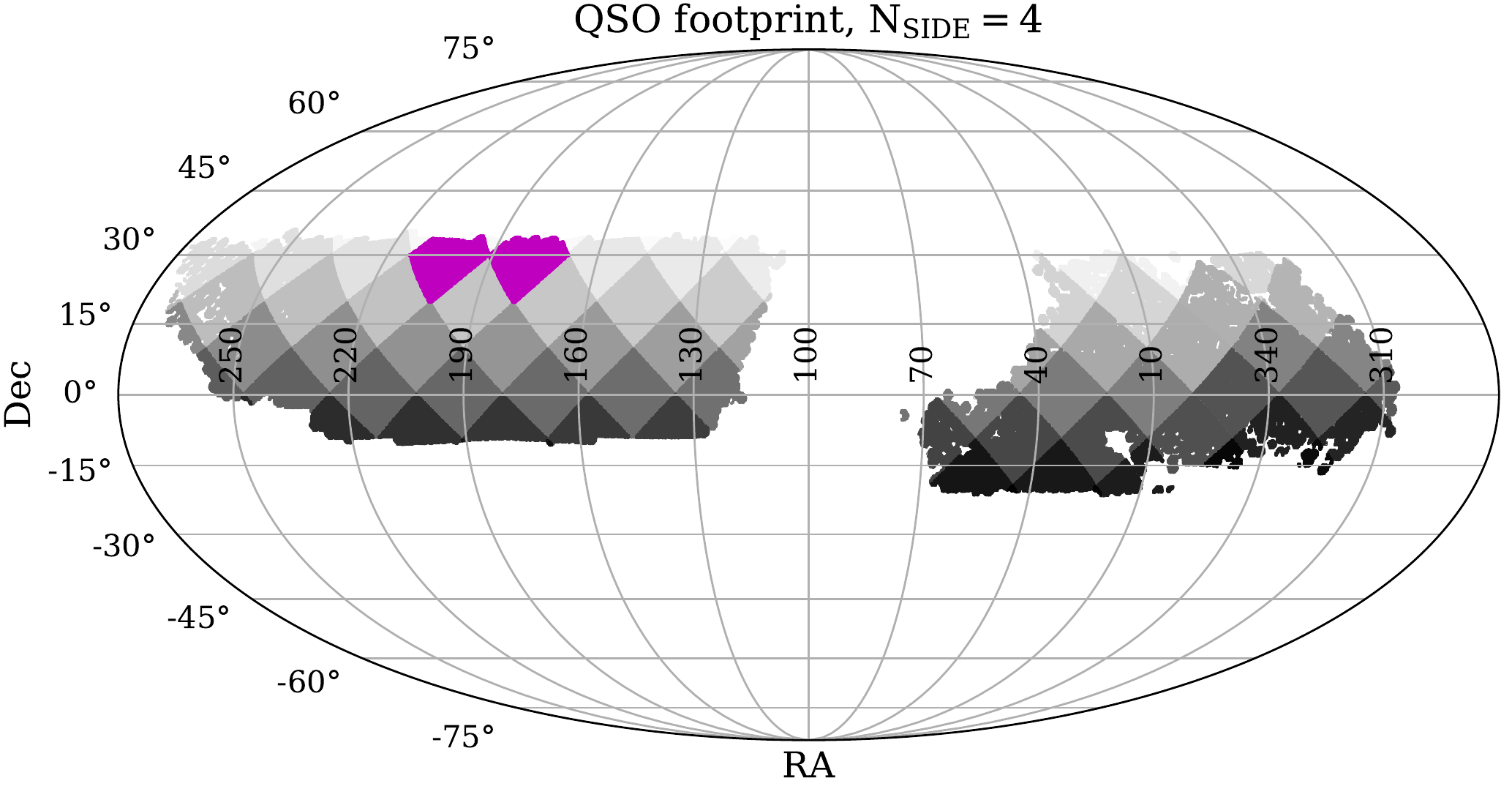}
    \includegraphics[width=\columnwidth]{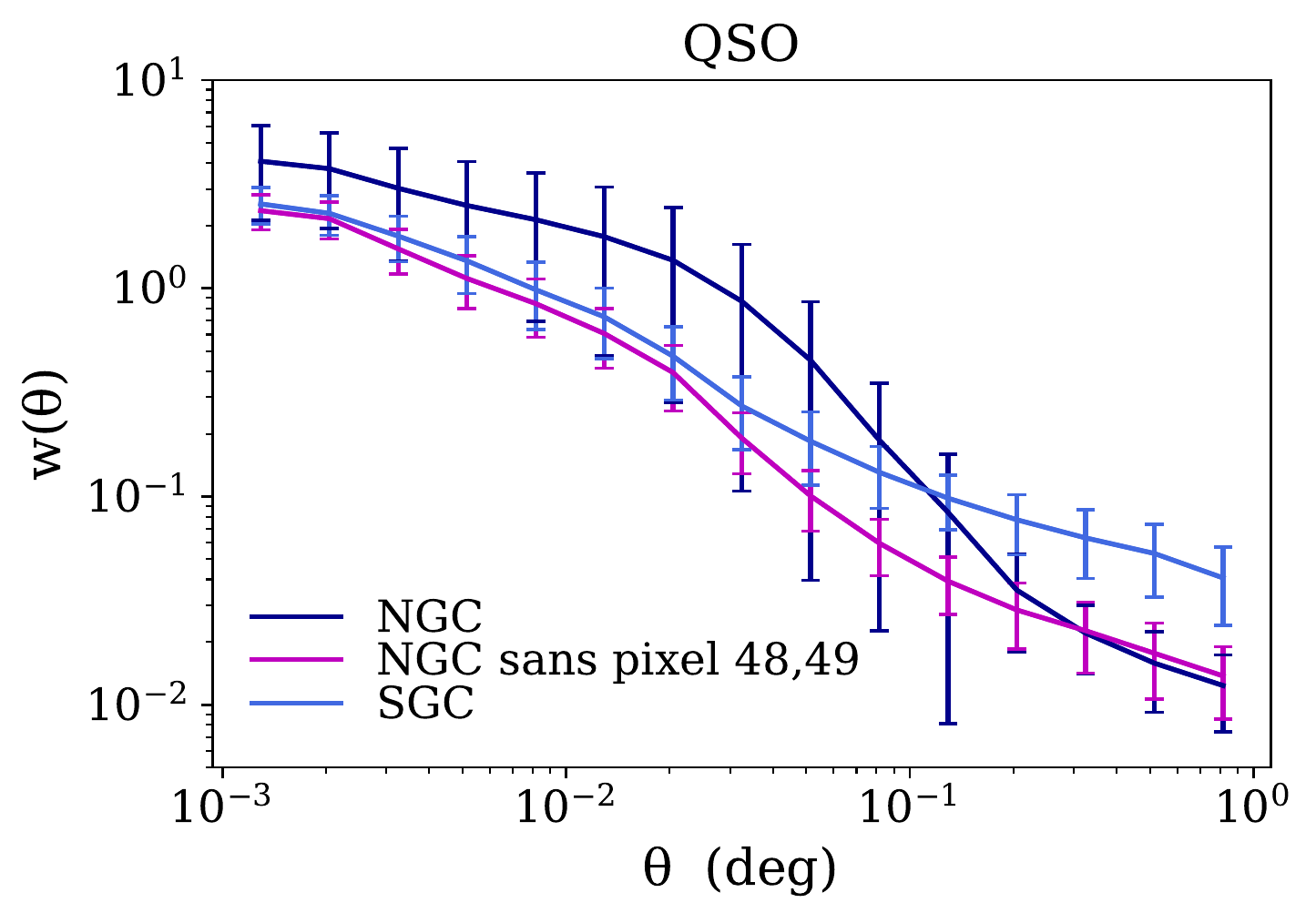}
    \caption{Upper: The angular correlation function is re-calculated with each pixel excluded in turn, still performing a complete bootstrap analysis on the remaining pixels. Two neighboring pixels in the NGC affect the result when either is excluded (violet lines). Middle: Map highlighting the location of the two pixels which were found to affect clustering. Lower: When both these pixels are excluded, the disparity between the NGC and SGC is dramatically reduced on small scales.}
    \label{fig:QSO_jackknife}
\end{figure}

As another test of spatial variation, we perform a jackknife-inspired outlier analysis on the data. Again using the \texttt{HEALPix} scheme, we divide the footprint into large pixels and re-calculate $w(\theta)$ with each non-empty pixel excluded in turn, still performing the full bootstrap error analysis on the remaining pixels for each iteration. We begin with the coarsest  pixels, corresponding to $N_{\text{SIDE}}=1$, and increase the resolution as needed to resolve any anomalies that are detected. For LRGs and ELGs, the results are indistinguishable even at this minimum resolution. However, for QSOs, we find two pixels at resolution $N_{\text{SIDE}}=4$ which, when either is excluded, lead to a significant change in the correlation function (see Figure~\ref{fig:QSO_jackknife}). Likely culprits are the Coma Cluster (Abell 1656), which contains over 1,000 galaxies, and M3/NGC 5272, one of the largest and brightest globular clusters in the sky.

\section{POTENTIAL SYSTEMATICS}
\label{sec:systematics}
Potential systematics include astrophysical foregrounds, variations in observing conditions, and uncertainties in data calibration, processing, and reduction (for similar studies in the context of SDSS, see e.g. \citealt{Myers06}, \citealt{Crocce11}, \citealt{Ross11}; for similar analyses using DES Verification Data, see e.g. \citealt{Suchyta++16}, \citealt{Crocce++16}, \citealt{Leistedt++16}, \citealt{ElvinPoole18}). We introduce maps of spatially-varying potential systematics in Section~\ref{sec:systematics_maps}, and examine their impact on the densities of DESI targets, before and after applying photometric weights, in Section~\ref{sec:systematics_densities}. The purpose of these weights is to mitigate the density trends by up-weighting (or down-weighting) regions where target density is diminished (or enhanced) due to systematics. Finally, in Section~\ref{sec:stellar_contam}, we cross-correlate the targets with stars and attempt to quantify stellar contamination in the QSO sample.

\subsection{Maps of potential systematics}
\label{sec:systematics_maps}

We begin by using the \texttt{HEALPix} scheme with $N_{\text{SIDE}} = 256$ to divide the data into equal-area pixels of approximately $0.05$ square degrees each. This resolution was chosen to avoid the shot noise limit in which most pixels contain zero or one targets. For our LRG, ELG, QSO samples with approximate mean densities (per square degree) 500, 2400, and 260, respectively, it produces an average of 10-20 LRGs/QSOs and $\sim$100 ELGs per pixel.\footnote{Note that most of the systematics studied here (with the exception of $E_{B-V}$ and stellar density) are not available at higher resolution than this pixelisation scheme in any case, as they are measured per CCD.} In every pixel, an average value for the potential systematic is calculated. For survey properties measured per CCD, unless otherwise noted, we first average over overlapping exposures to obtain a mean value for each random, then pixelise using the randoms. The resultant maps are shown below, along with brief descriptions of how they are determined and why they are included in the analysis.

\begin{figure*}
\includegraphics[width=0.49\linewidth, trim={1.5cm 1.2cm 1.5cm 1.6cm},clip]{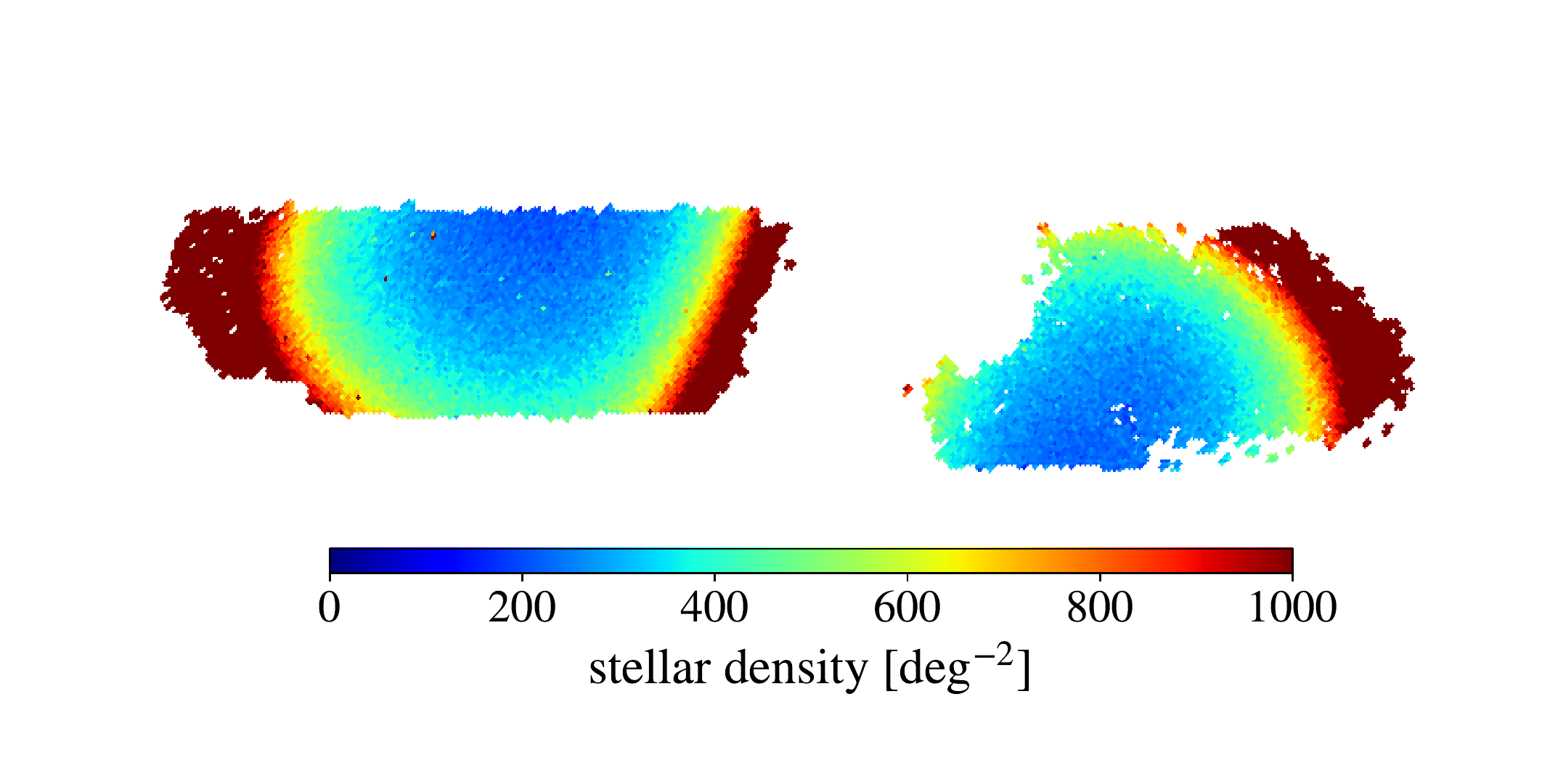}
\includegraphics[width=0.49\linewidth, trim={1.5cm 1.2cm 1.5cm 1.6cm},clip]{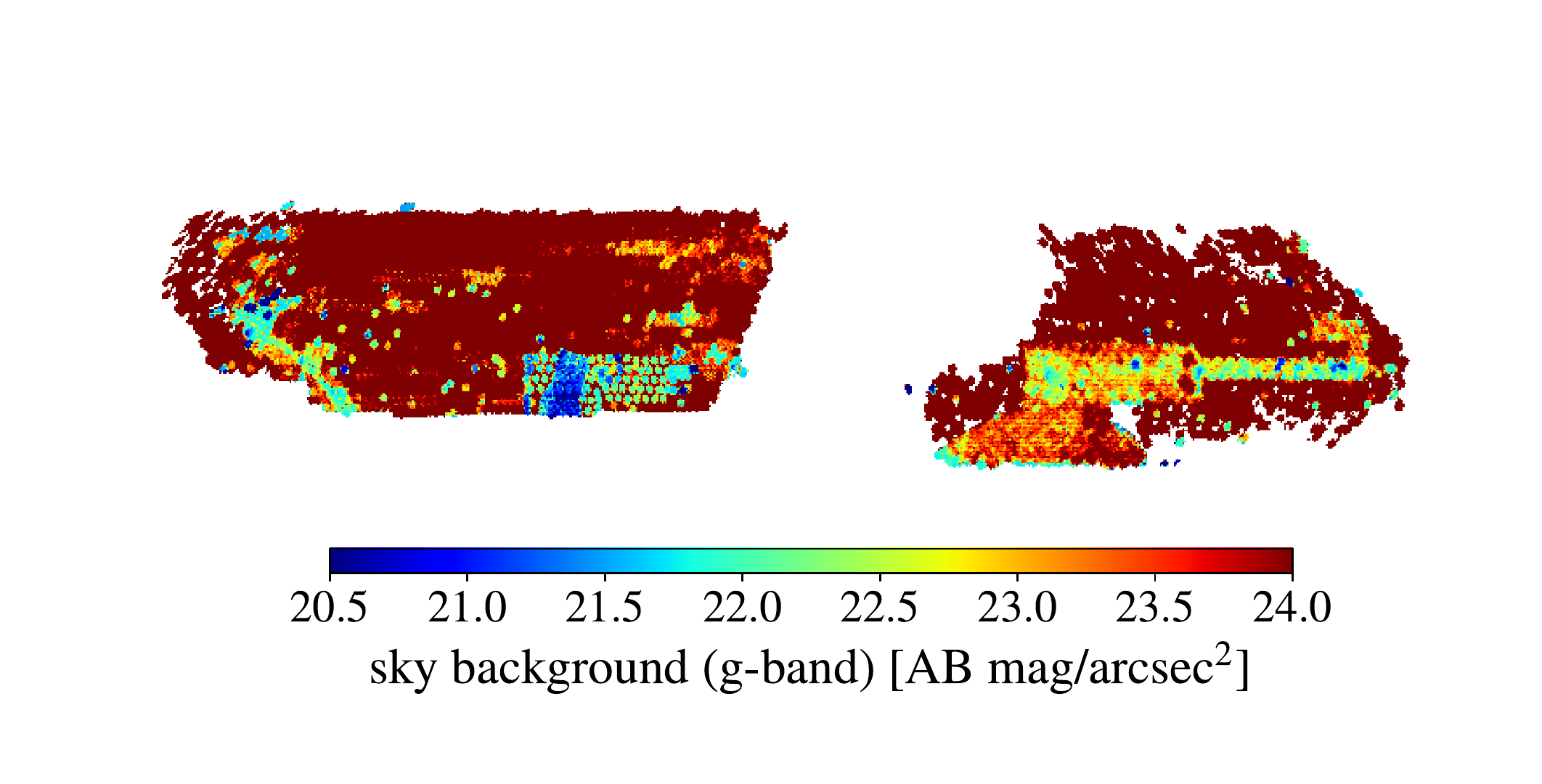}
\includegraphics[width=0.49\linewidth, trim={1.5cm 1.2cm 1.5cm 1.6cm},clip]{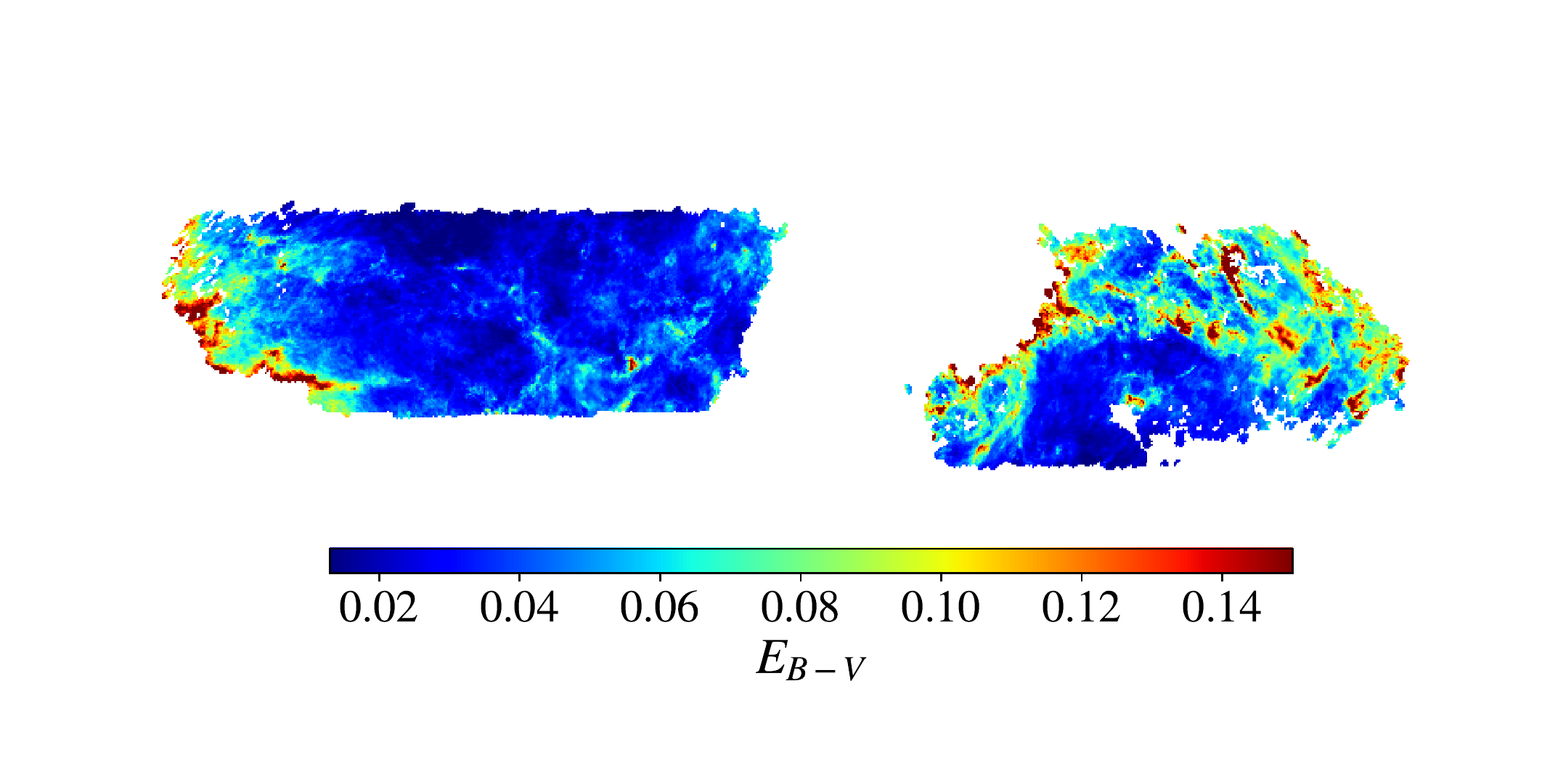}
\includegraphics[width=0.49\linewidth, trim={1.5cm 1.2cm 1.5cm 1.6cm},clip]{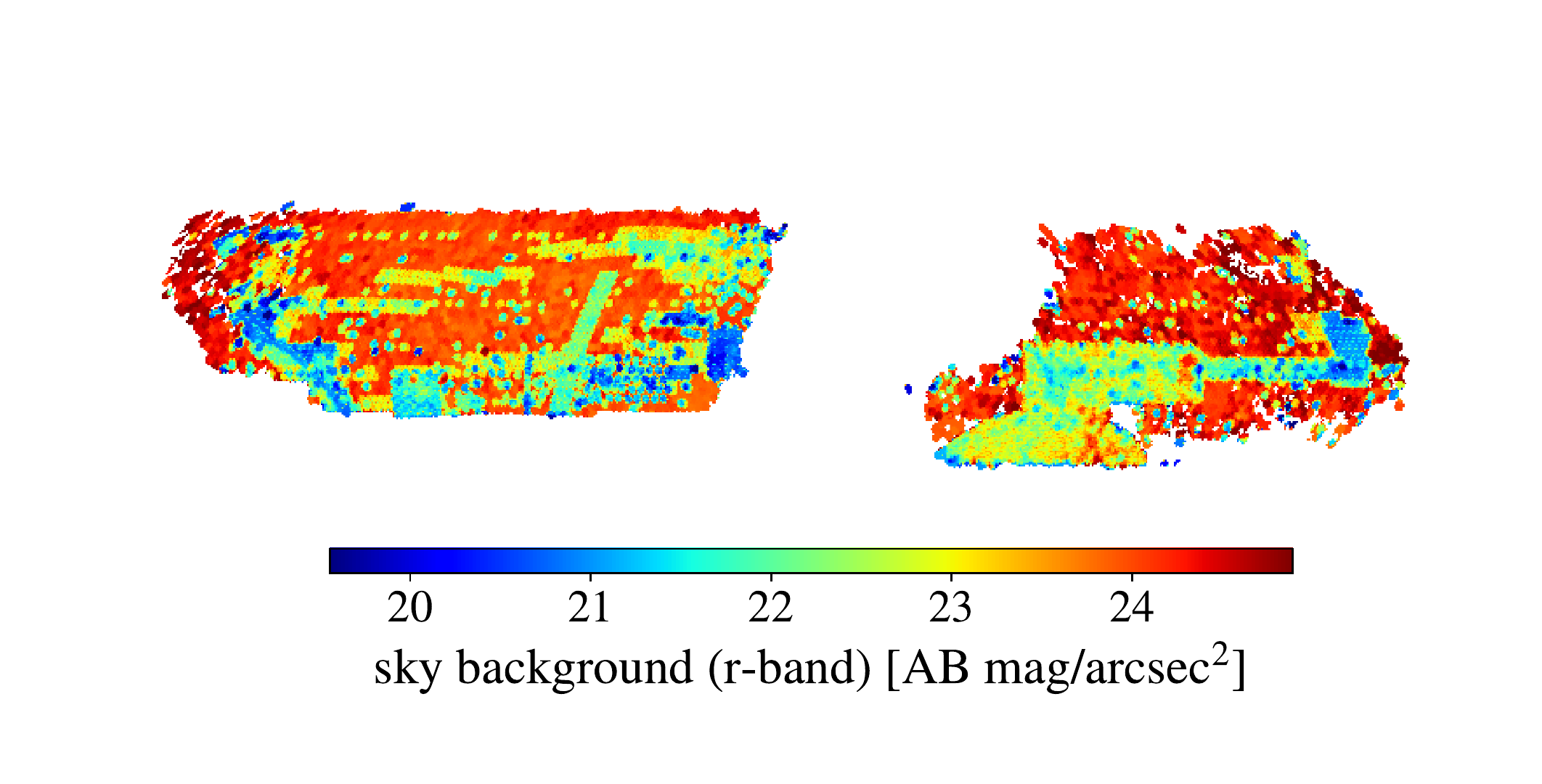}
\includegraphics[width=0.49\linewidth, trim={1.5cm 1.2cm 1.5cm 1.6cm},clip]{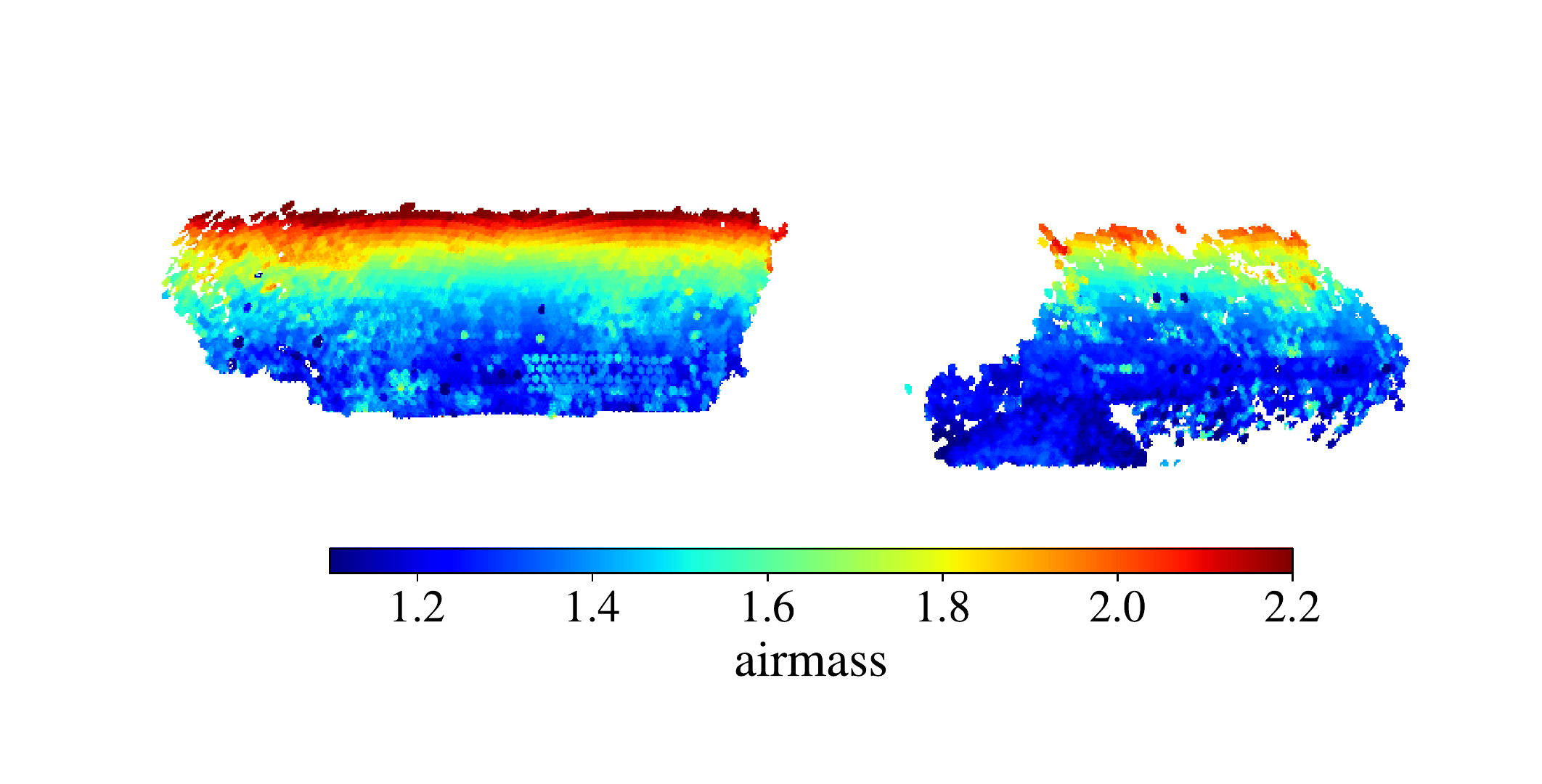}
\includegraphics[width=0.49\linewidth, trim={1.5cm 1.2cm 1.5cm 1.6cm},clip]{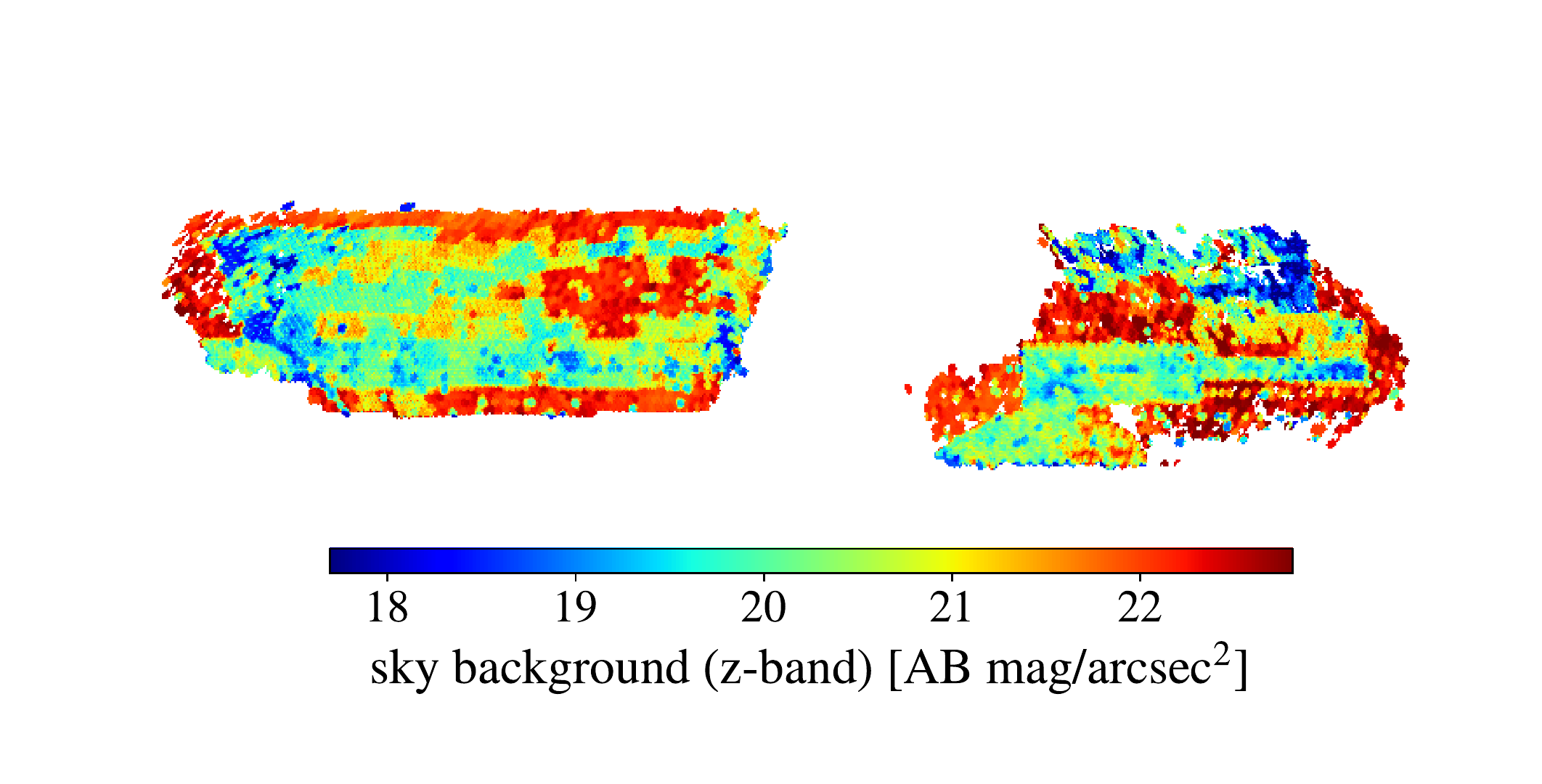}
\includegraphics[width=0.49\linewidth, trim={1.5cm 1.2cm 1.5cm 1.6cm},clip]{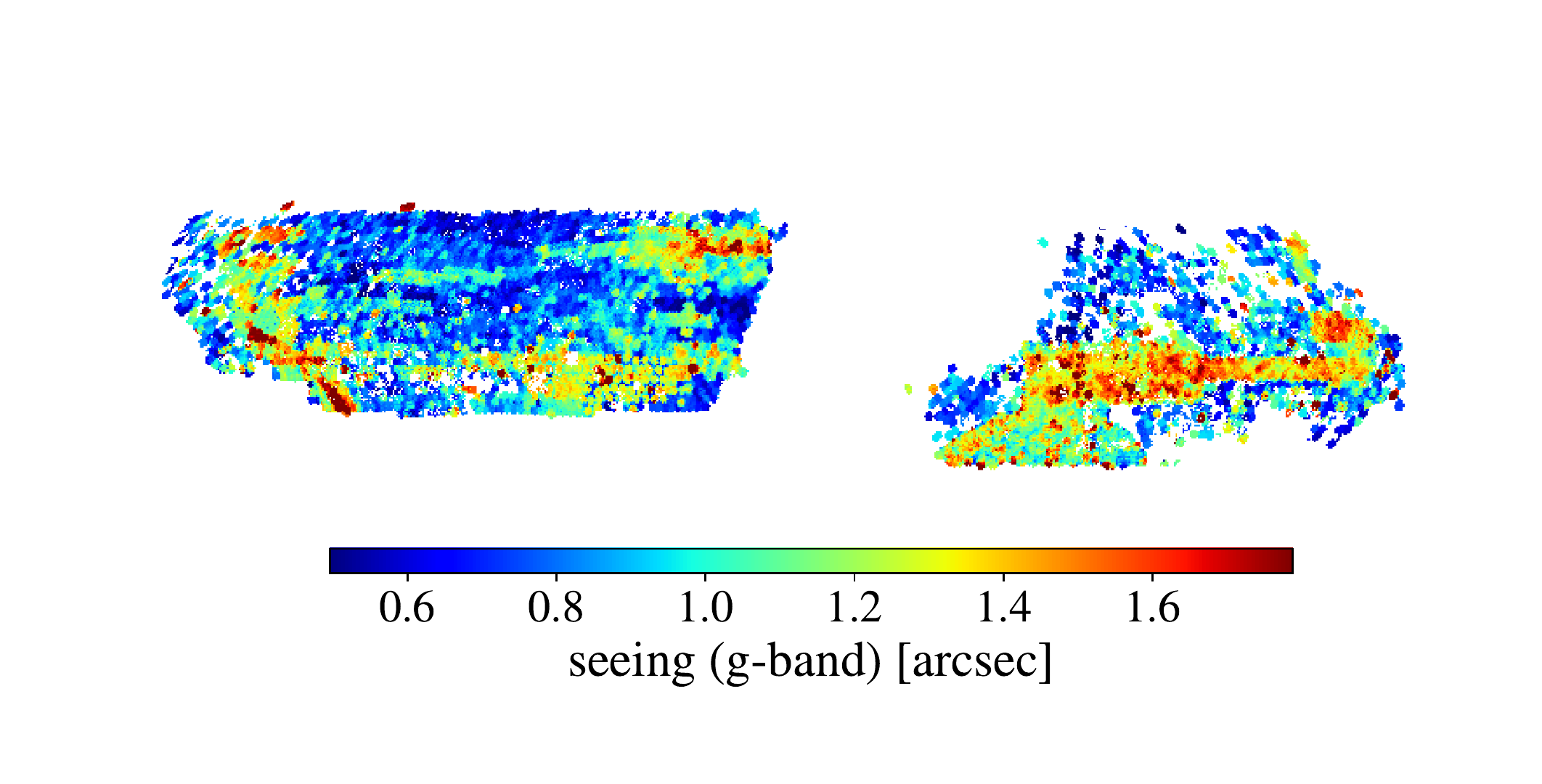}
\includegraphics[width=0.49\linewidth, trim={1.5cm 1.2cm 1.5cm 1.6cm},clip]{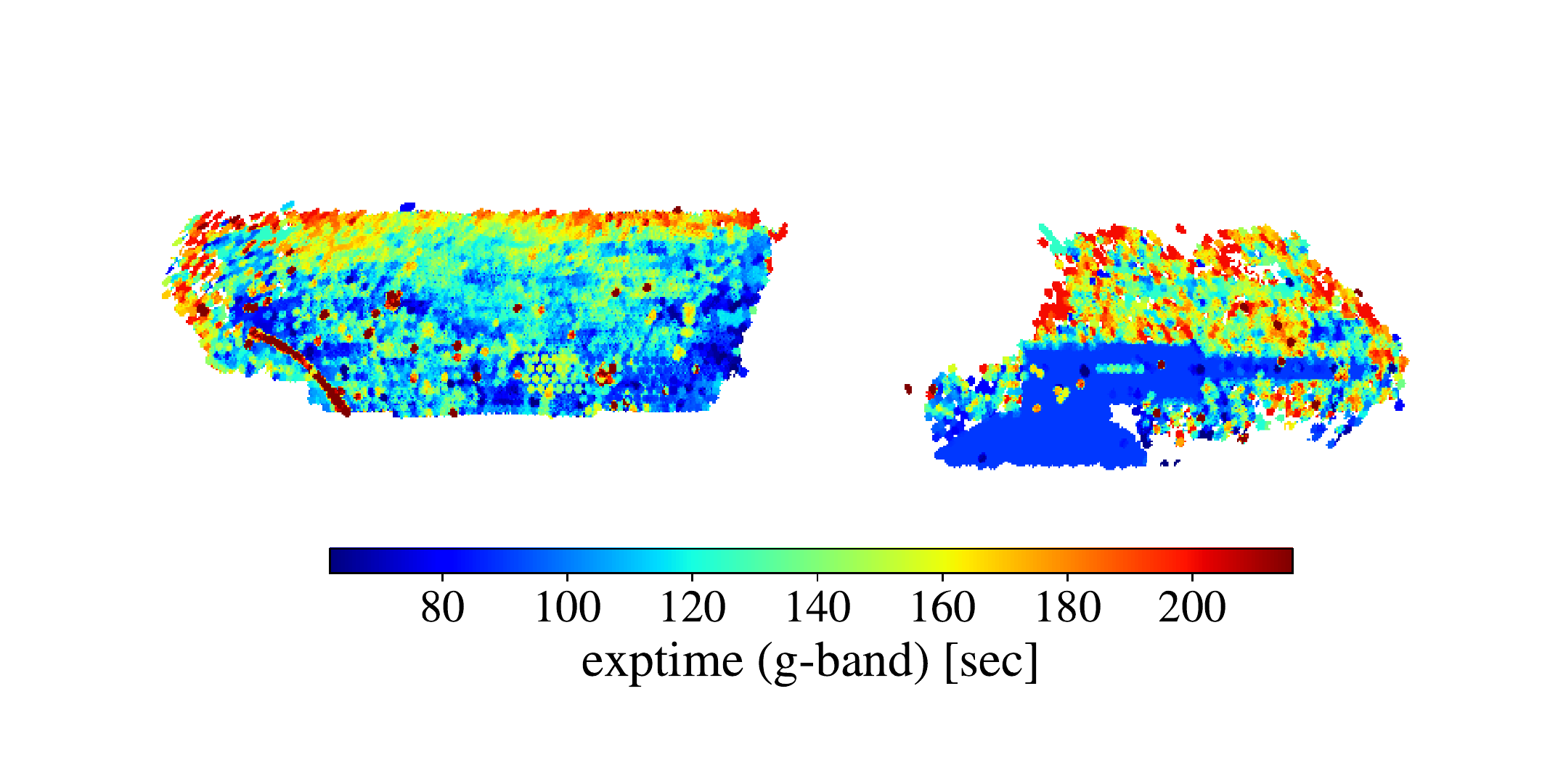}
\includegraphics[width=0.49\linewidth, trim={1.5cm 1.2cm 1.5cm 1.6cm},clip]{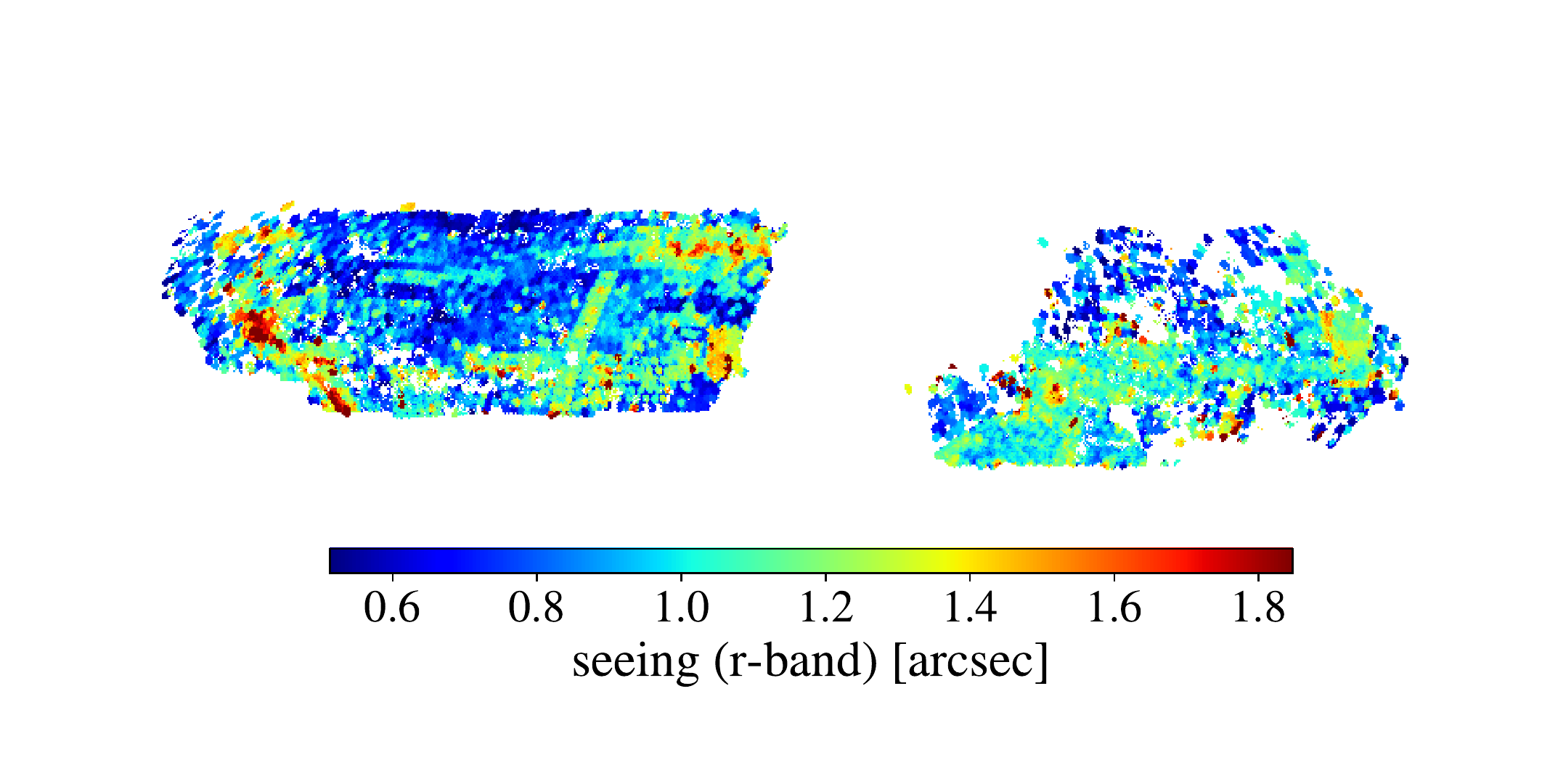}
\includegraphics[width=0.49\linewidth, trim={1.5cm 1.2cm 1.5cm 1.6cm},clip]{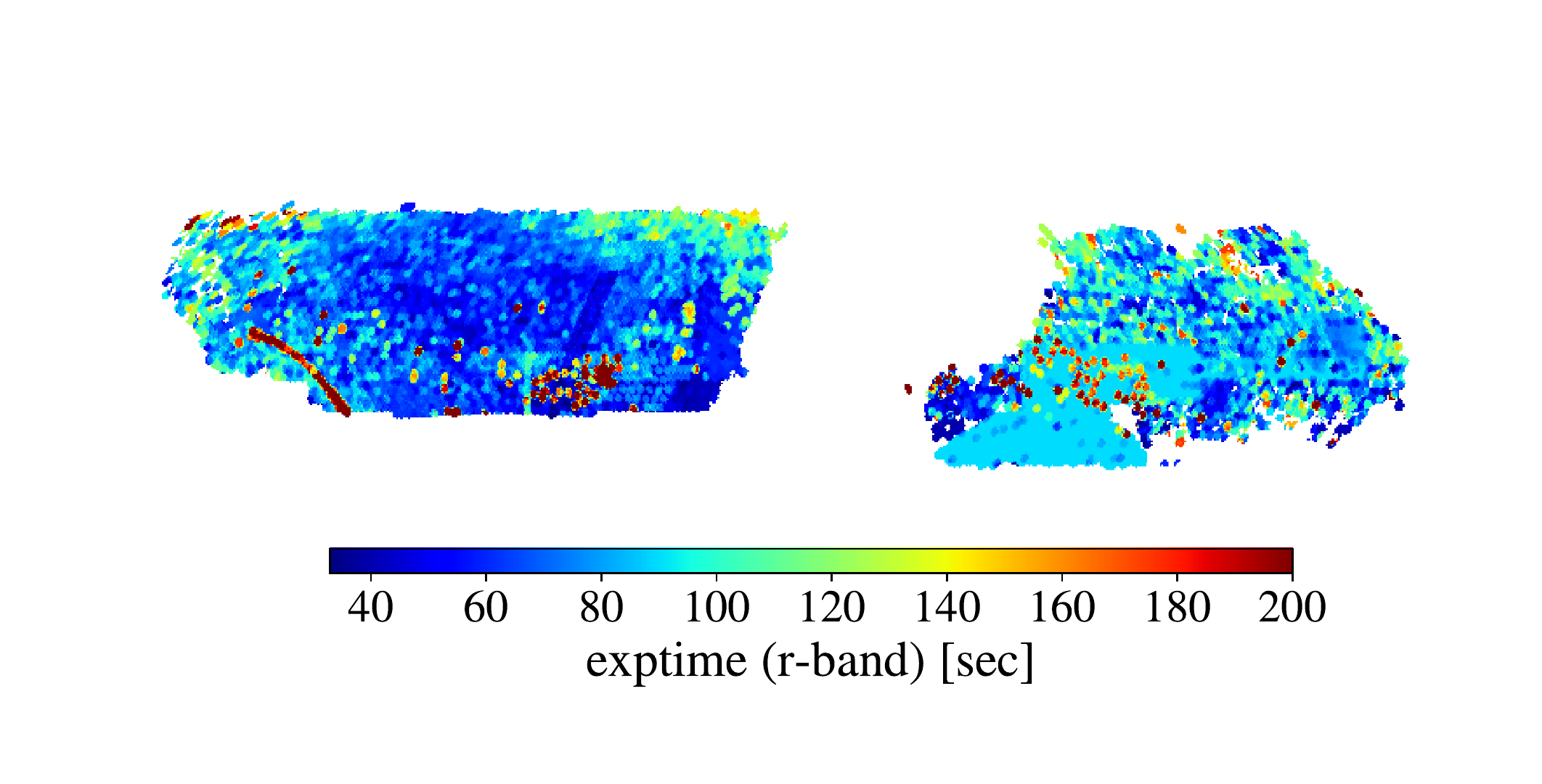}
\includegraphics[width=0.49\linewidth, trim={1.5cm 1.2cm 1.5cm 1.6cm},clip]{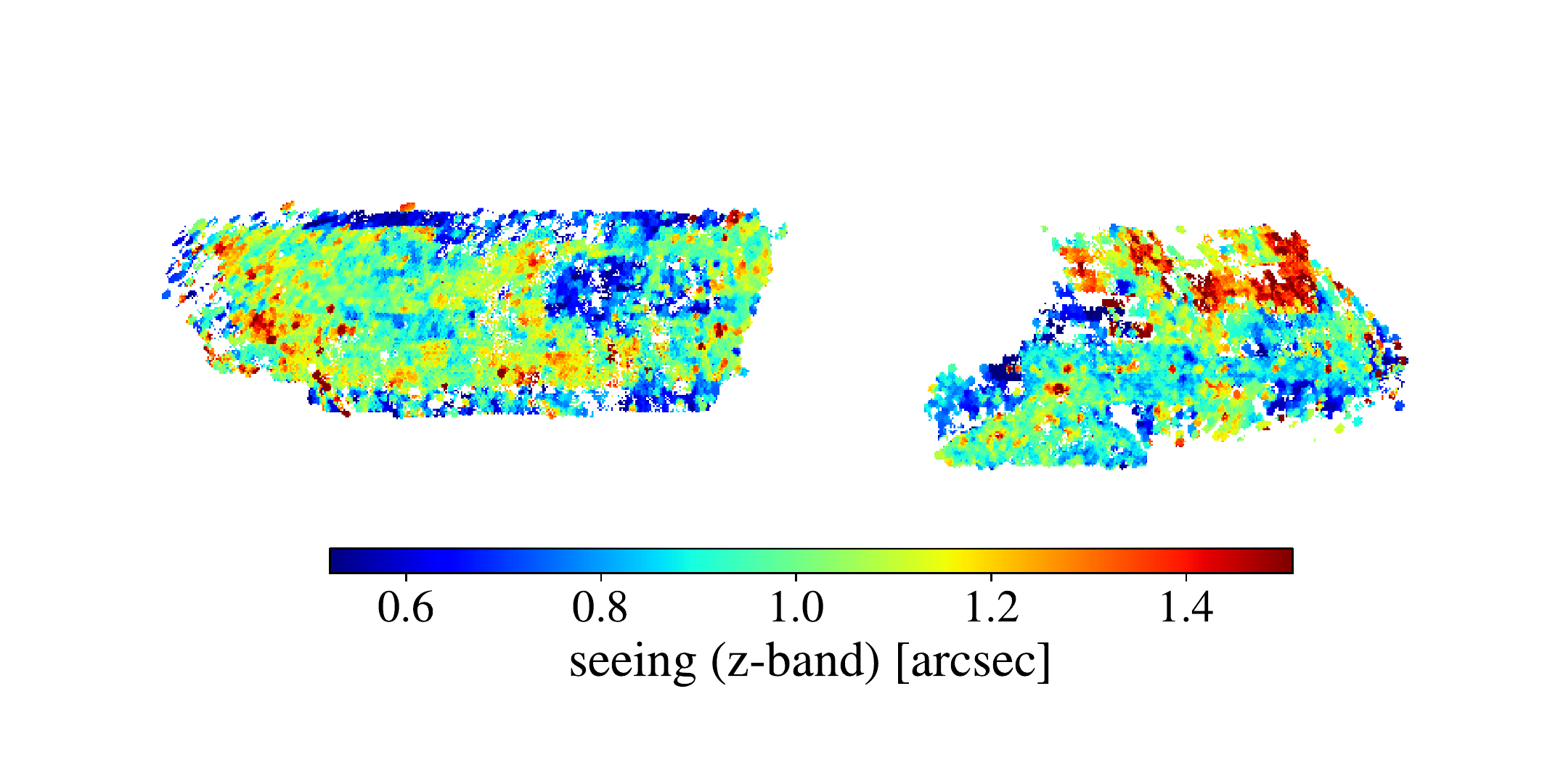}
\includegraphics[width=0.49\linewidth, trim={1.5cm 1.2cm 1.5cm 1.6cm},clip]{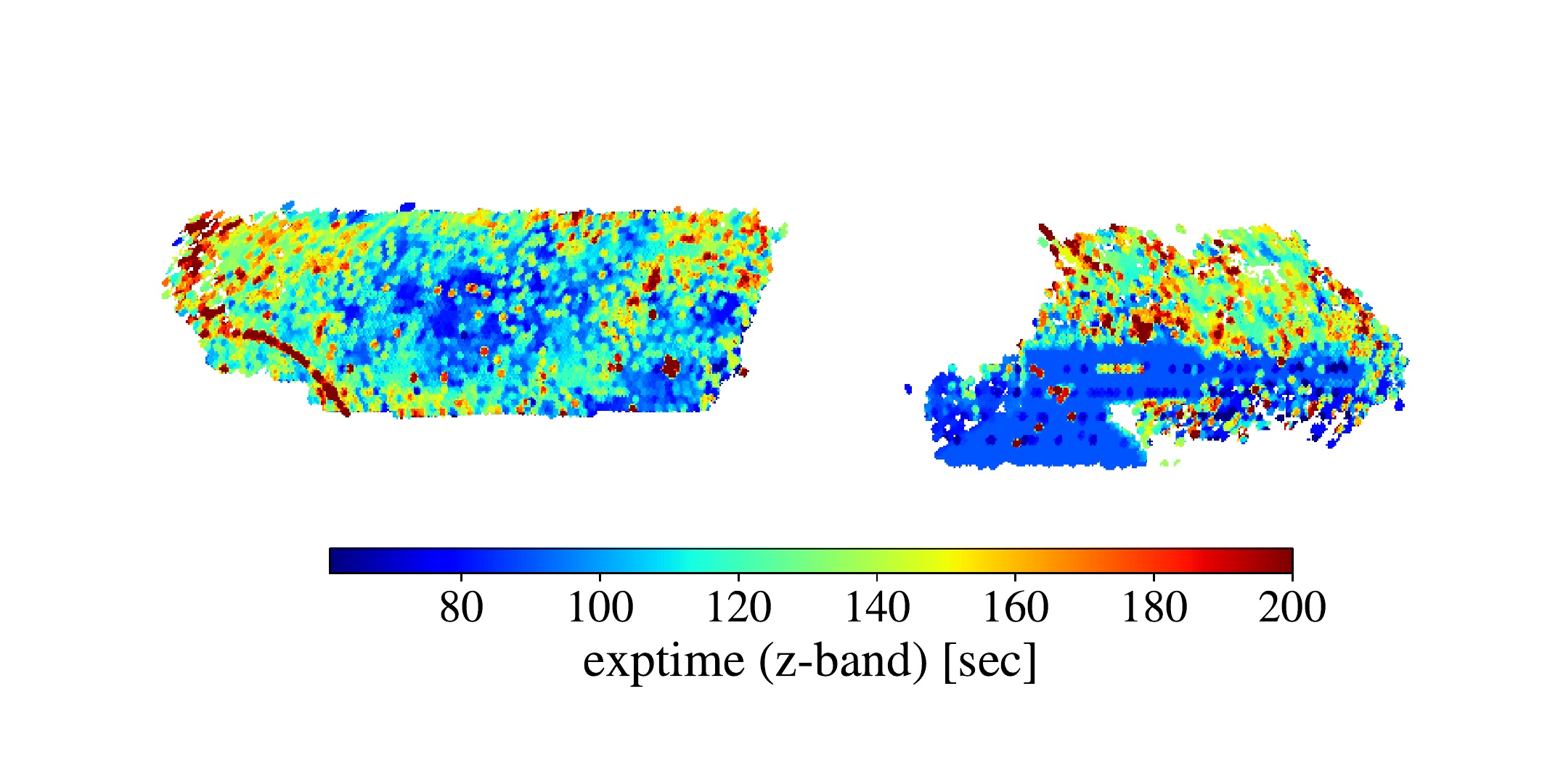}
\caption{Maps of spatially-varying potential systematics in equatorial coordinates with Mollweide projection and  astronomy convention (east towards left).}
\label{fig:systematic_maps}
\end{figure*}

\begin{enumerate}

    \item Stellar density
    \subitem In addition to the detection issues near bright foreground stars discussed in Section~\ref{sec:masks/foregrounds/stars}, the presence of stars impacts the measured density and clustering of galaxies in other ways (e.g. \citealt{Crocce11}, \citealt{Ross11}). Stars with similar colors (see Figure~\ref{fig:stellar_contams}) can contaminate the samples, and the inclusion of a separate population manifests as an enhanced clustering signal. The observed clustering may also be imprinted with the density gradient of stars, which increases towards the galactic plane. Furthermore, residual PSF tails of some fainter stars may pollute the pixels used to calculate sky background and therefore affect target photometry.
    
    \hspace{\parindent} We create a catalog of stars from DECaLS by selecting objects lying in the stellar locus (using the color cut $17 < r < 18$) with PSF morphology. The density of this class of objects, shown in the top left panel of Figure~\ref{fig:systematic_maps}, indicates that it is a reliable stellar template. 

    \begin{figure}
    \centering
    \includegraphics[width=\linewidth]{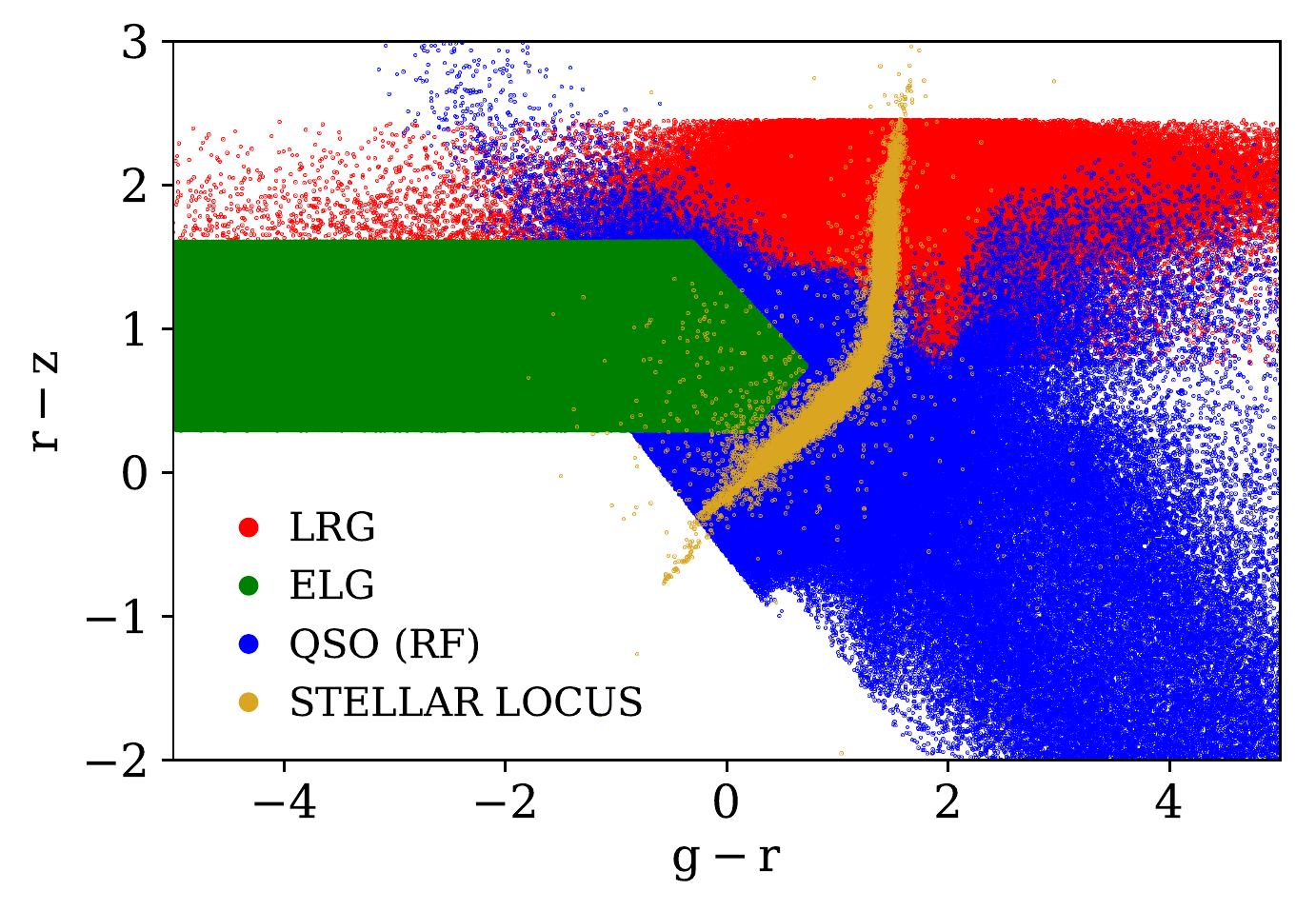}
    \caption{Color-color plot of LRG, ELG, QSO target selection, with the shape of the stellar locus selected from DECaLS as PSF-type objects with $17 < r < 18$.}
    \label{fig:stellar_contams}
    \end{figure}

	\item Galactic extinction
    \subitem Galactic extinction is the wavelength-dependent absorption and scattering of light by interstellar dust in the Milky Way, causing sources to appear redder. DESI and its imaging surveys deliberately avoid regions of high extinction along the galactic plane. In addition, we use extinction-corrected fluxes in our analysis. The total extinction in each band is provided by the DECaLS pipeline, calculated from the SFD98 dust map combined with a set of extinction coefficients $A_\lambda / E_{B-V}$ for each DECam and WISE filter. The extinction coefficients were determined at airmass $= 1.3$ from the values recommended by \cite{SchlaflyFinkbeiner11} using the \cite{Fitzpatrick99} extinction curve at $R_V = 3.1$. These values are 3.214, 2.165, 1.211 for $g$, $r$, $z$. However, it has been shown in other surveys that residual errors in this correction may cause spurious clustering (see e.g. \citealt{Scranton02}, \citealt{Myers06}, \citealt{Ross06}). In addition, erroneously correcting the photometry of stars in the foreground of the dust could potentially bias their color and cause some of them to scatter into target selection. Hence, we treat $E_{B-V}$ as a potential systematic and test its effect on the density field. \\
    
    \item Seeing
    \subitem Astronomical seeing is the blurring of an image due to turbulence in the Earth's atmosphere. The distortions fill out a point spread function (PSF) whose full width at half maximum quantifies the quality of the seeing conditions. In DECaLS, seeing is determined by fitting the median PSF of stars on the CCD to a 2D Gaussian. Since seeing varies between nights and even exposures, a mean value is reported, calculated by averaging the inverse of the effective number of pixels in the PSF (such that images with better seeing dominate the mean, as they contain more information). We use this mean PSF size to determine the impact of seeing conditions on the density field, since bad seeing causes larger magnitude errors as well as more cross-contamination with stars due to poor morphological fits.\\
    
    \item Airmass
    \subitem Airmass is the optical path length of light through the Earth's atmosphere. When photons from a celestial source travel to a terrestrial observer, they are absorbed and scattered by the atmosphere. Light that must traverse more atmosphere will be attenuated more, so sources appear dimmer at the horizon than at the zenith. For zenith angles $ \lessapprox 60^{\circ}$, we can approximate the atmosphere as plane-parallel, and also assume its density is more or less constant. In this limit, the airmass is simply the secant of the angle from the zenith to the source location on the sky. We treat the mean airmass as a potential systematic, as it contributes to magnitude errors (for instance, airmass induces atmospheric differential refraction as a function of color, which can affect the photometry) and is entangled with the extinction and seeing corrections. \\

	\item Sky brightness
    \subitem Variations in the background brightness of the sky (due to various sources such as airglow, scattered starlight, Moon phase, light pollution) can affect the measured flux errors and therefore the density of targets by scattering objects in or out of color cuts. Sky brightness also has a strong dependence on airmass, which increases the brightness of airglow. We include the mean sky background in each band measured on the individual CCDs as a potential systematic. \\

	\item Exposure time
    \subitem DECam can attain the depths required for DESI targeting in total exposure times of 166, 134, and 200 sec for $g$, $r$, $z$ bands, given median observing conditions. As part of the imaging strategy, dynamic exposure times are increased to compensate for poor observing conditions in order to obtain a more uniform sample. We look at variations with mean exposure time in each band, which affects depth and is correlated with other potential systematics, to see how it modulates the observed density.

\end{enumerate}

\subsection{Target densities vs.\ potential systematics}
\label{sec:systematics_densities}

We can determine the post-masking target density per pixel using the random catalogs. Since the randoms are uniformly distributed, counting the number of post-masking randoms in a pixel is equivalent to measuring its effective area, up to a proportionality factor: 
\begin{equation}\label{delta_i}
    \delta_i = n^{\rm gal}_i / \bar{n}^{\rm gal} - 1 = N^{\rm gal}_i / N^{\rm ran}_i \times N^{\rm ran}_{\rm masked} / N^{\rm gal}_{\rm masked} - 1
\end{equation}

\begin{figure*}
\centering
\includegraphics[width=0.3\textwidth, trim=0.2cm 0 0 0, clip]{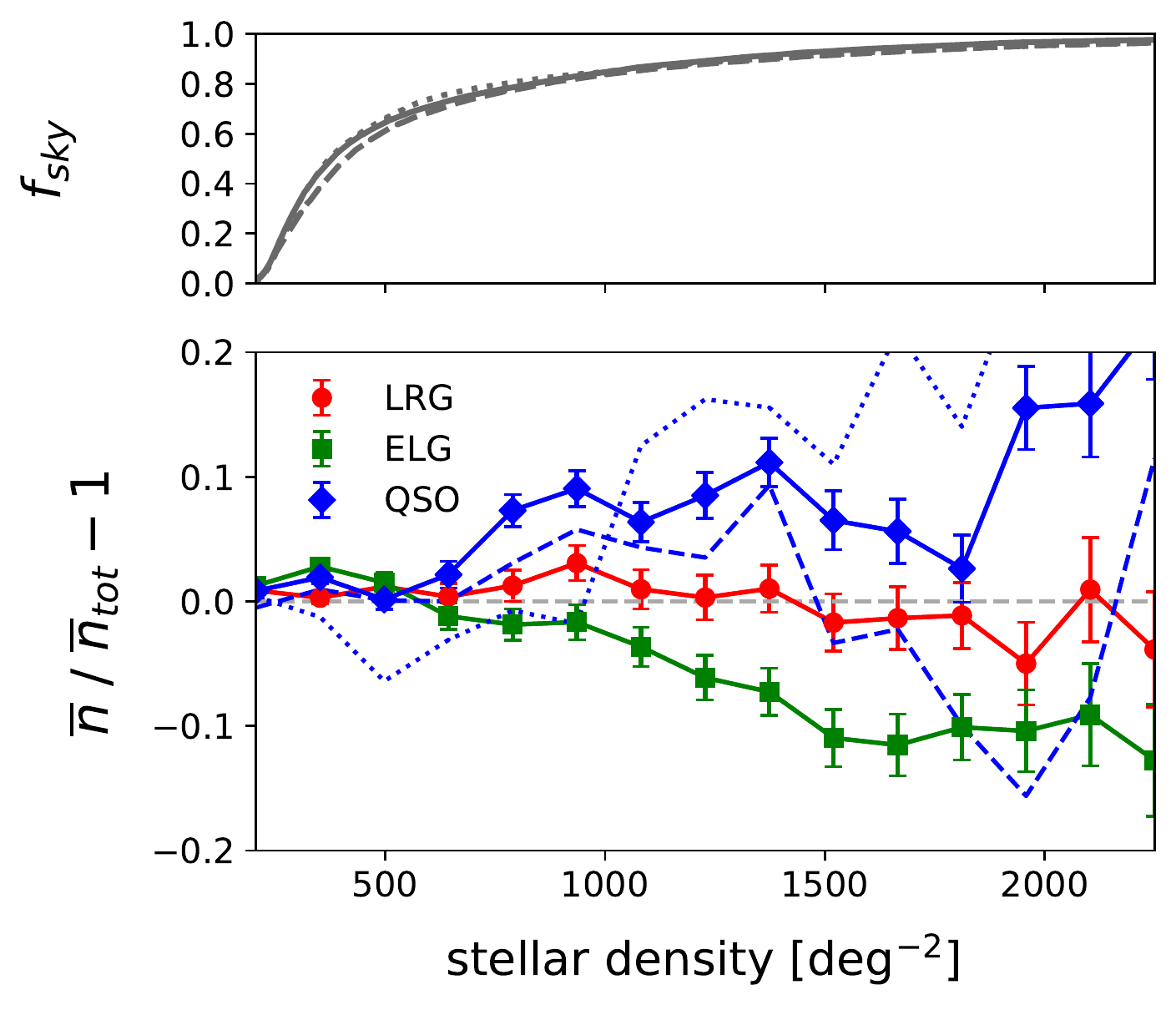}
\includegraphics[width=0.3\textwidth, trim=0.2cm 0 0 0, clip]{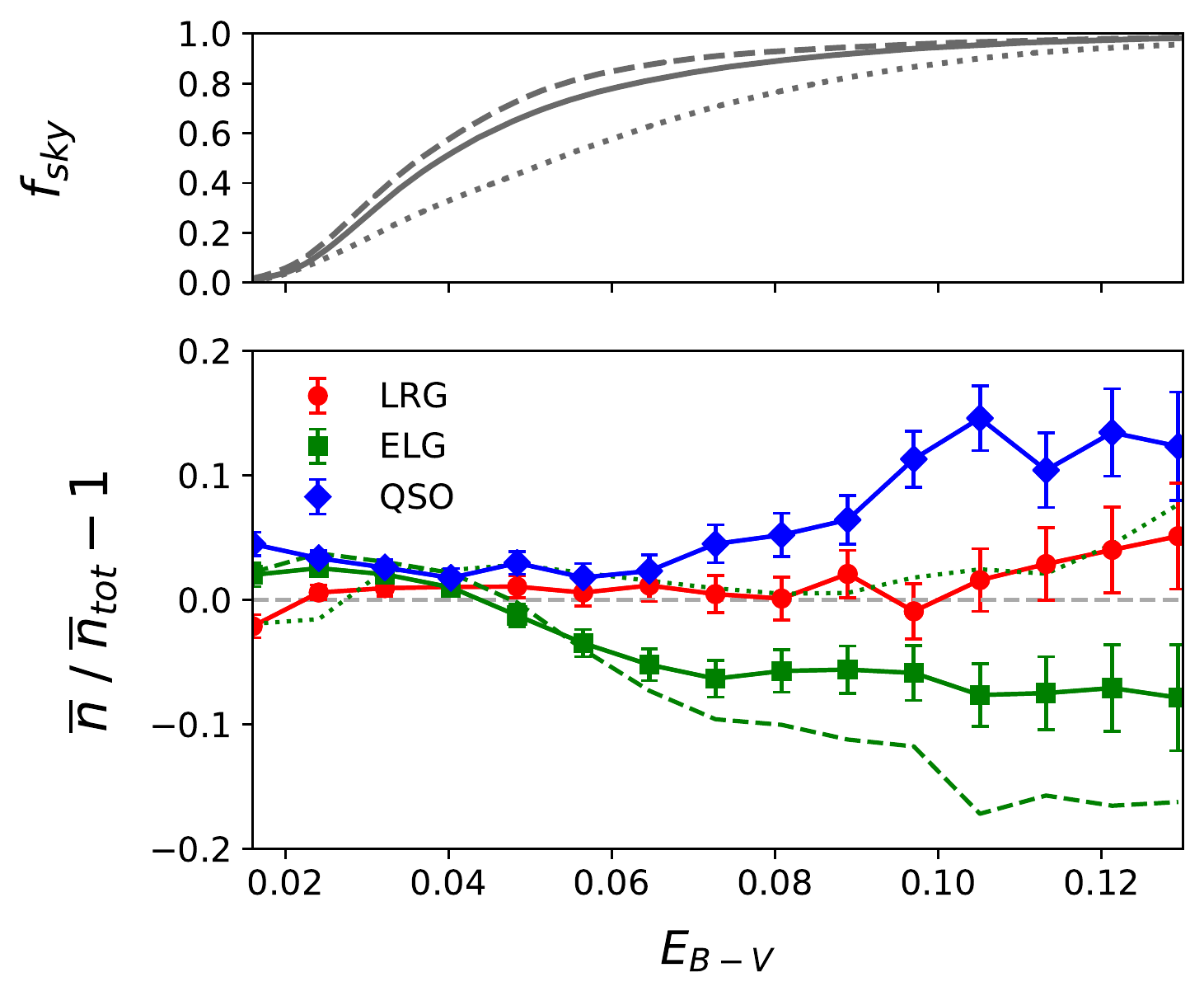}
\includegraphics[width=0.3\textwidth, trim=0.2cm 0 0 0, clip]{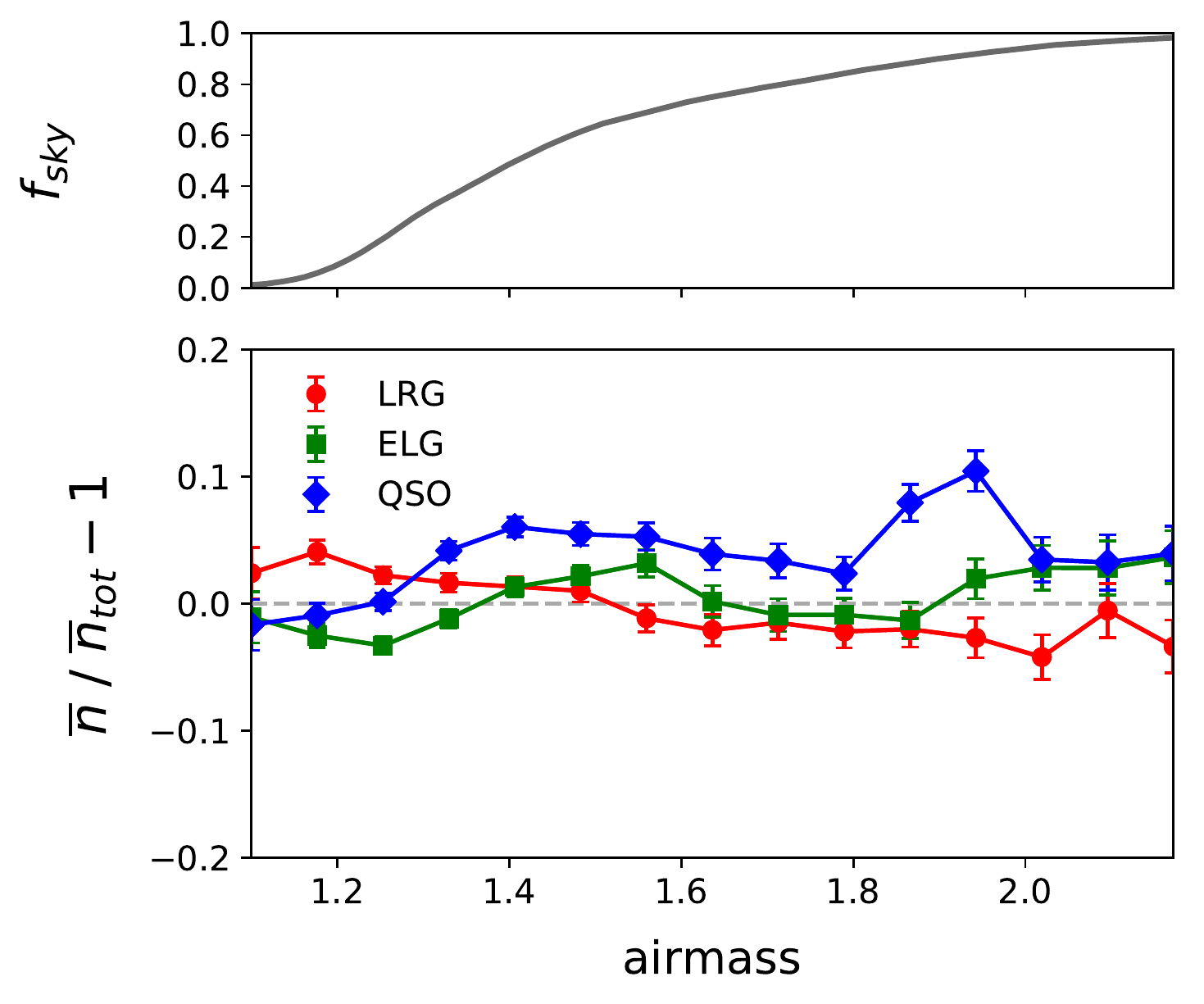}
\includegraphics[width=0.3\textwidth, trim=0.2cm 0 0 0, clip]{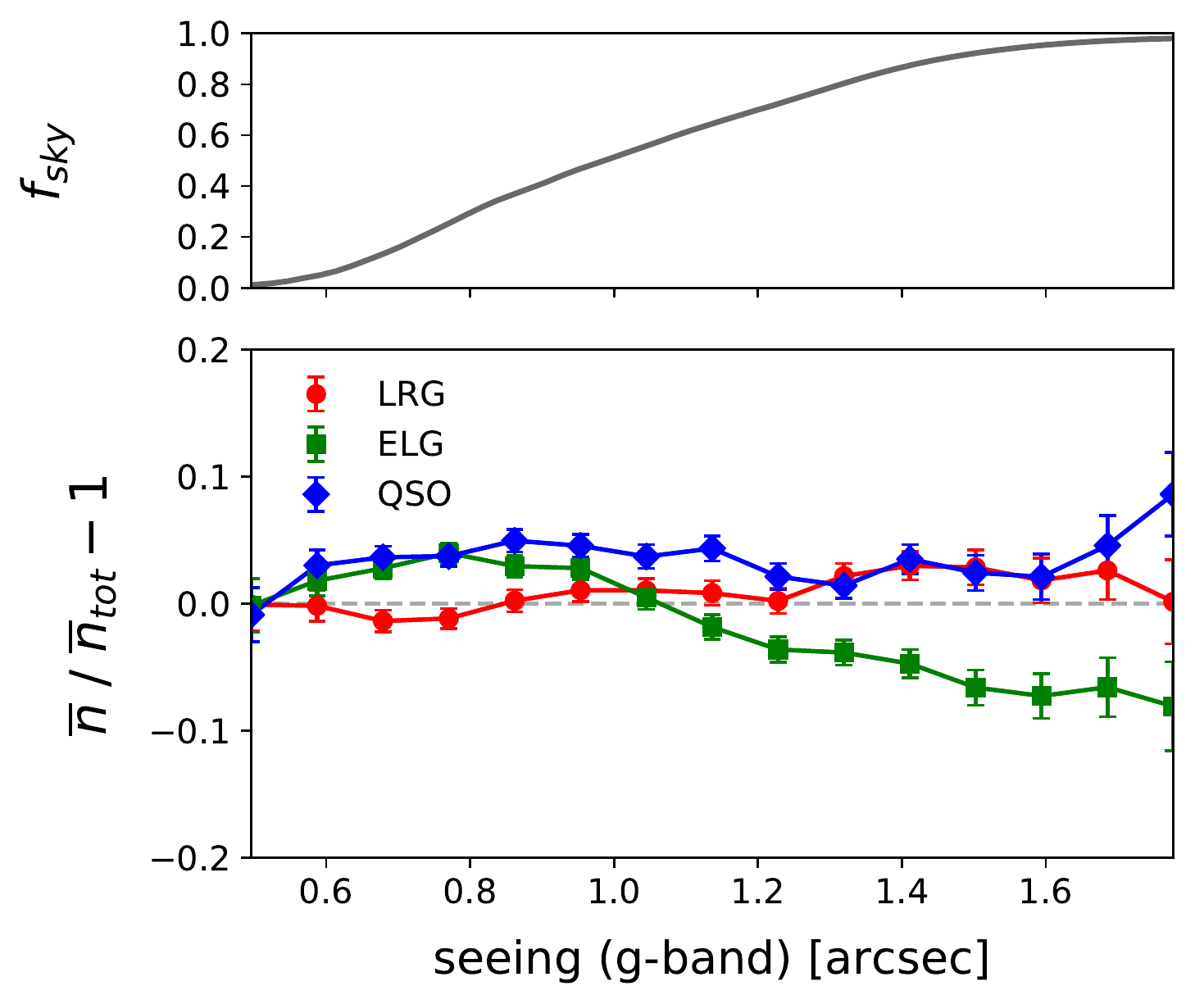}
\includegraphics[width=0.3\textwidth, trim=0.2cm 0 0 0, clip]{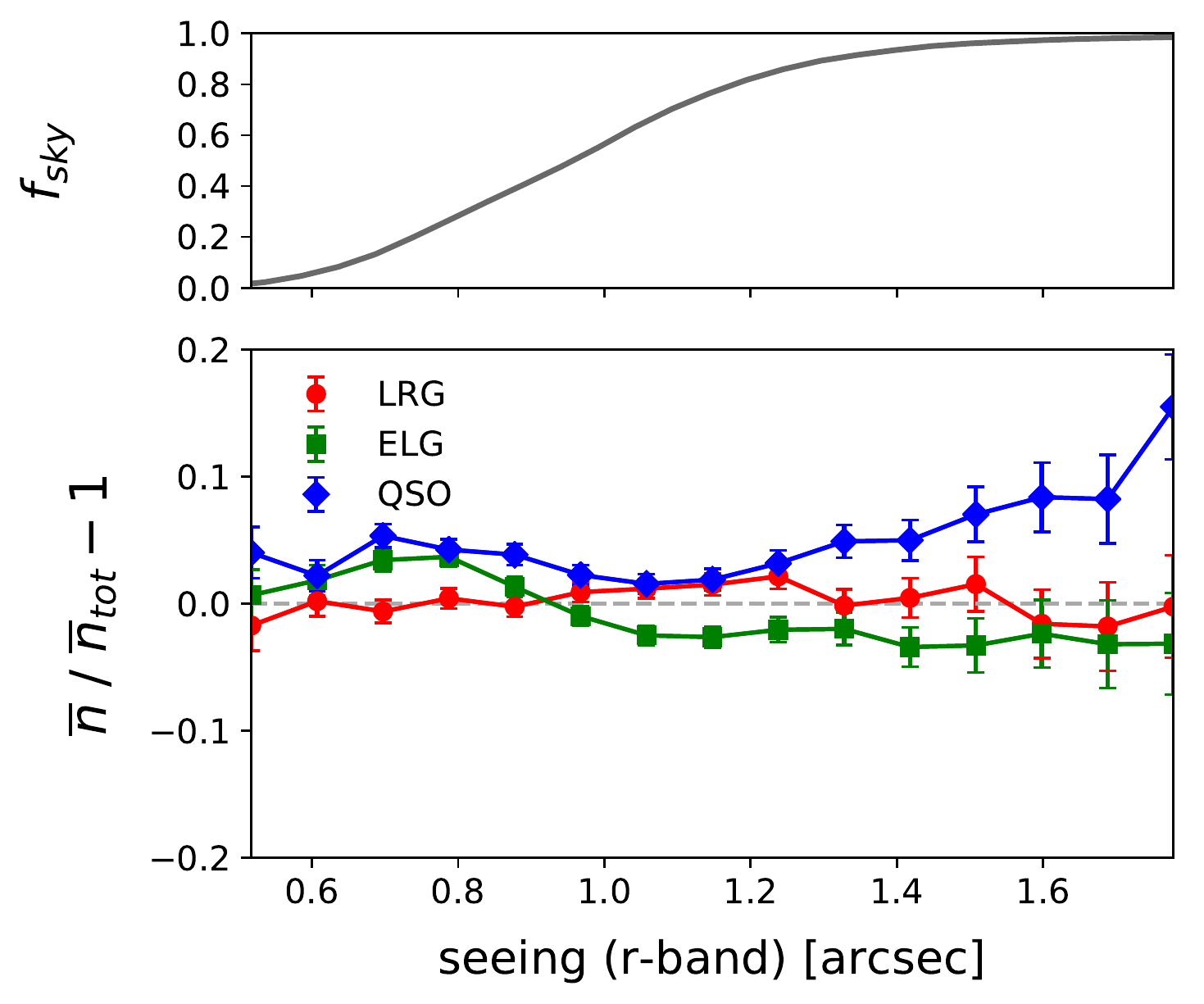}
\includegraphics[width=0.3\textwidth, trim=0.2cm 0 0 0, clip]{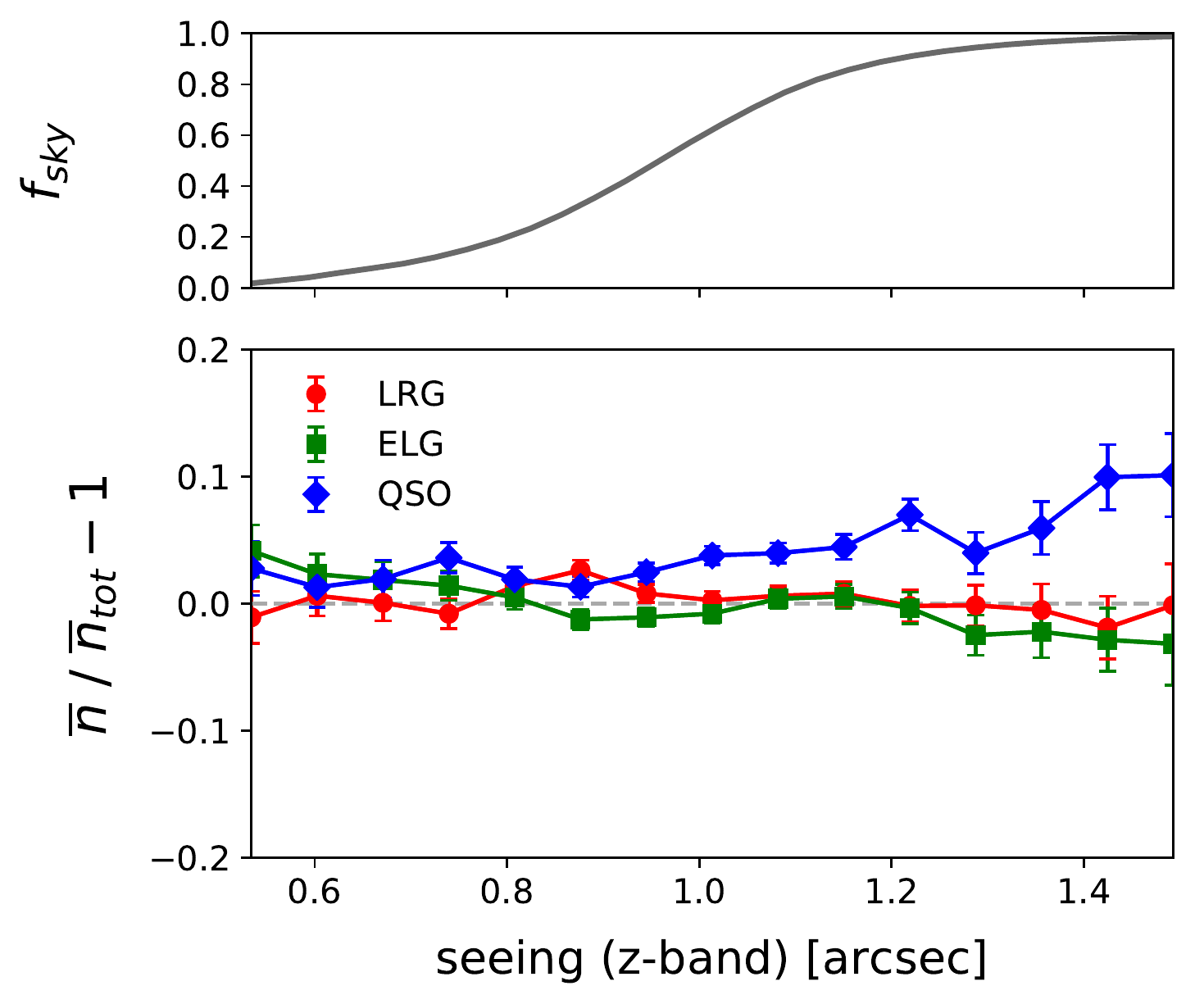}
\includegraphics[width=0.3\textwidth, trim=0.2cm 0 0 0, clip]{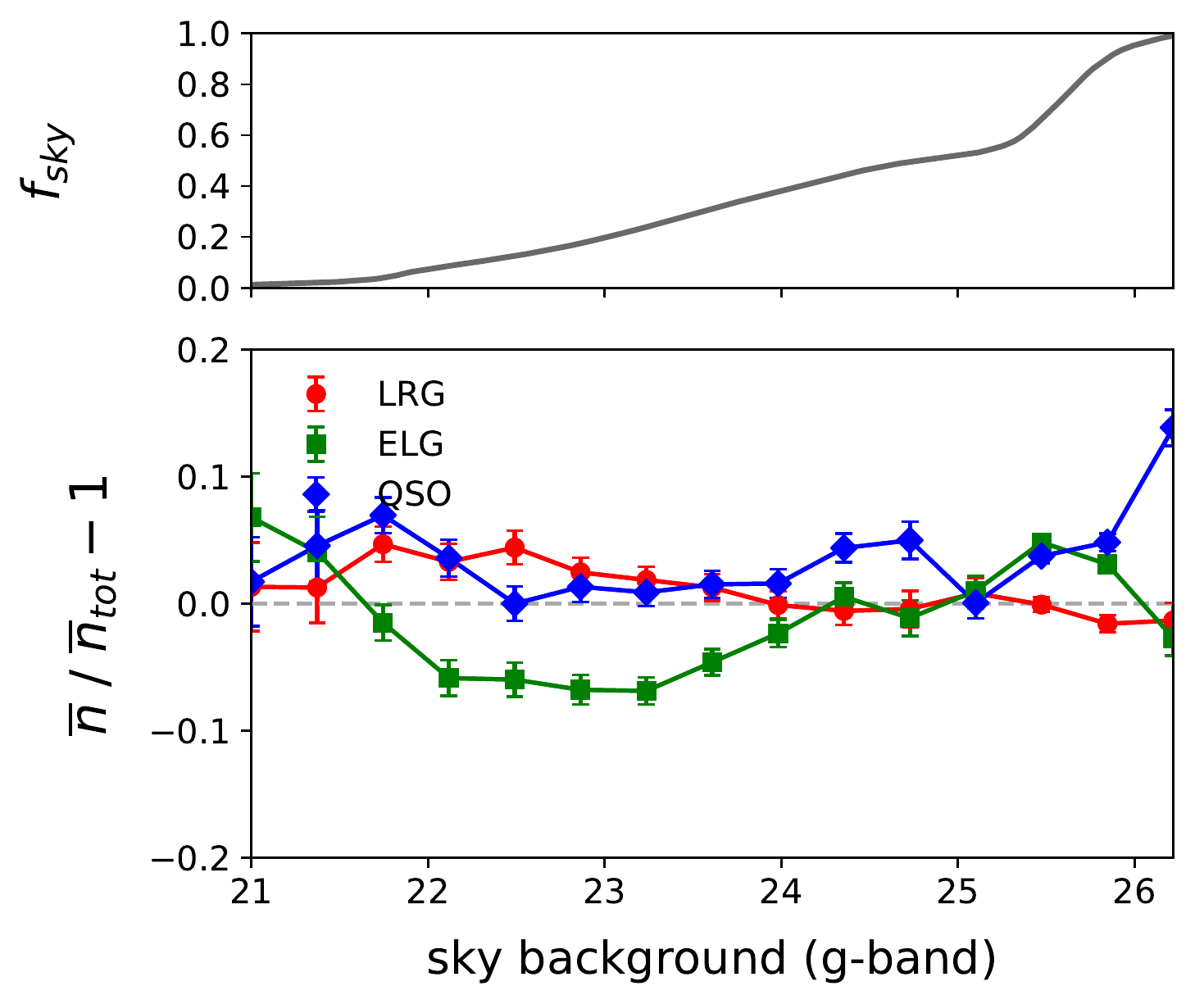}
\includegraphics[width=0.3\textwidth, trim=0.2cm 0 0 0, clip]{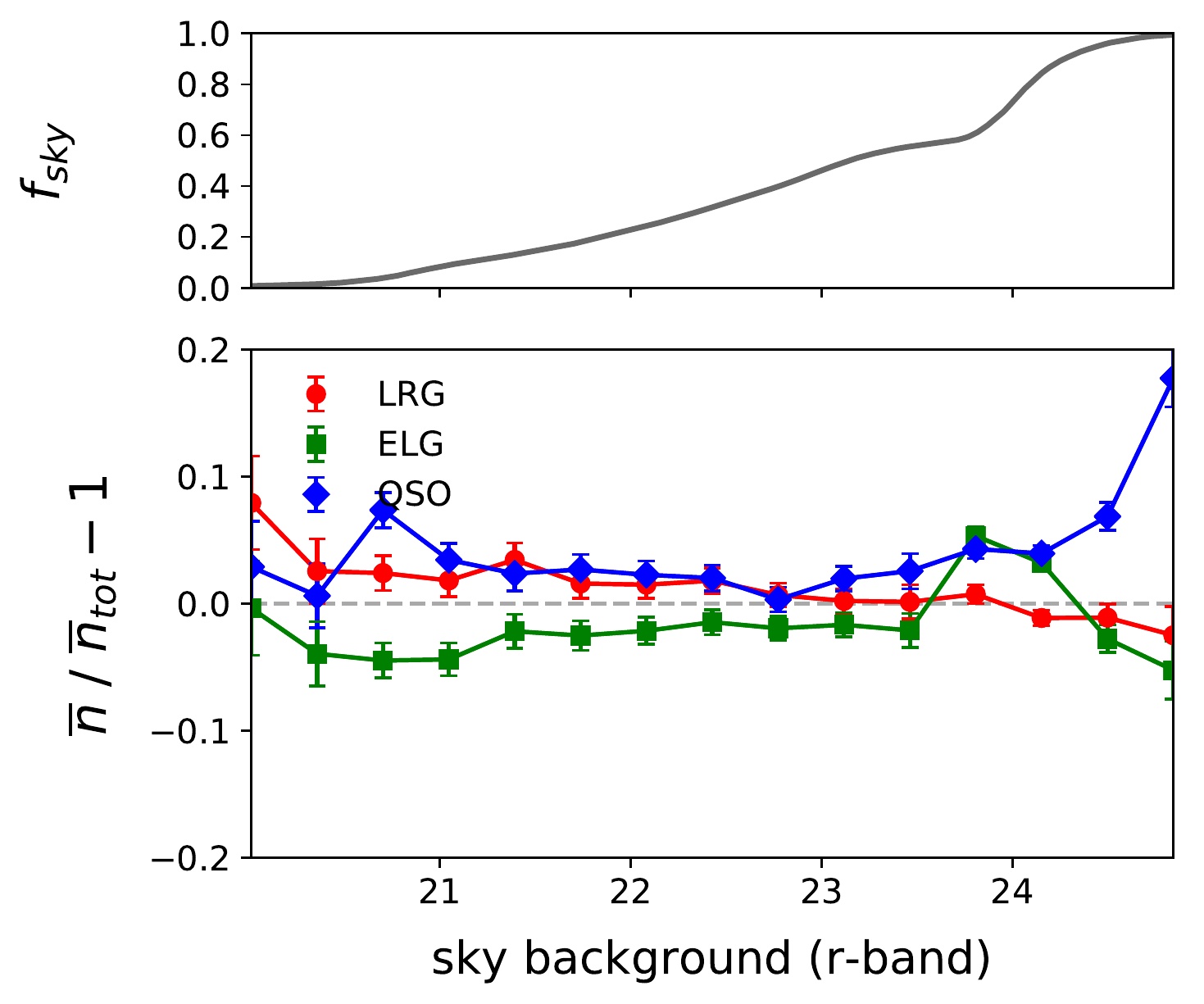}
\includegraphics[width=0.3\textwidth, trim=0.2cm 0 0 0, clip]{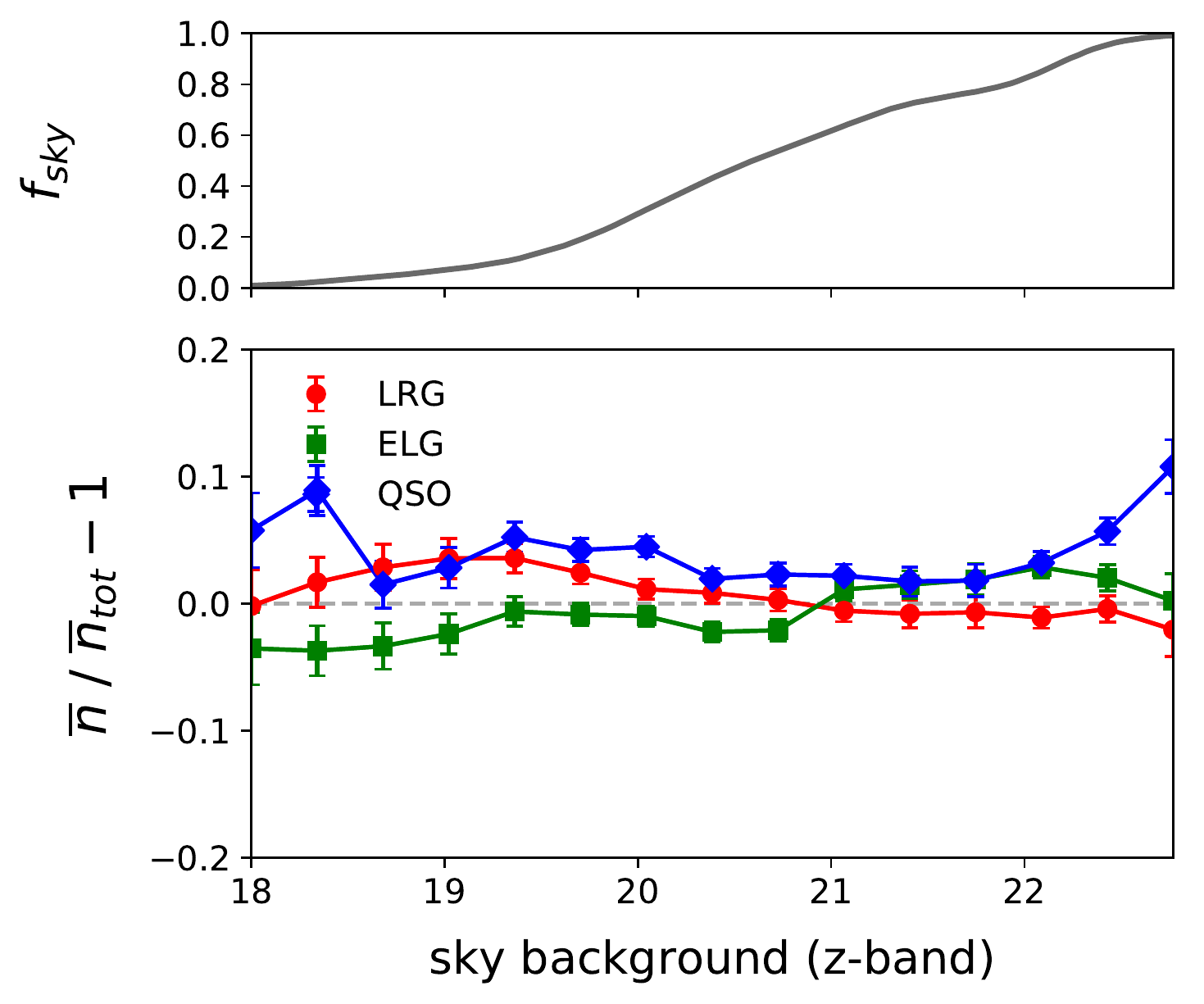}
\includegraphics[width=0.3\textwidth, trim=0.2cm 0 0 0, clip]{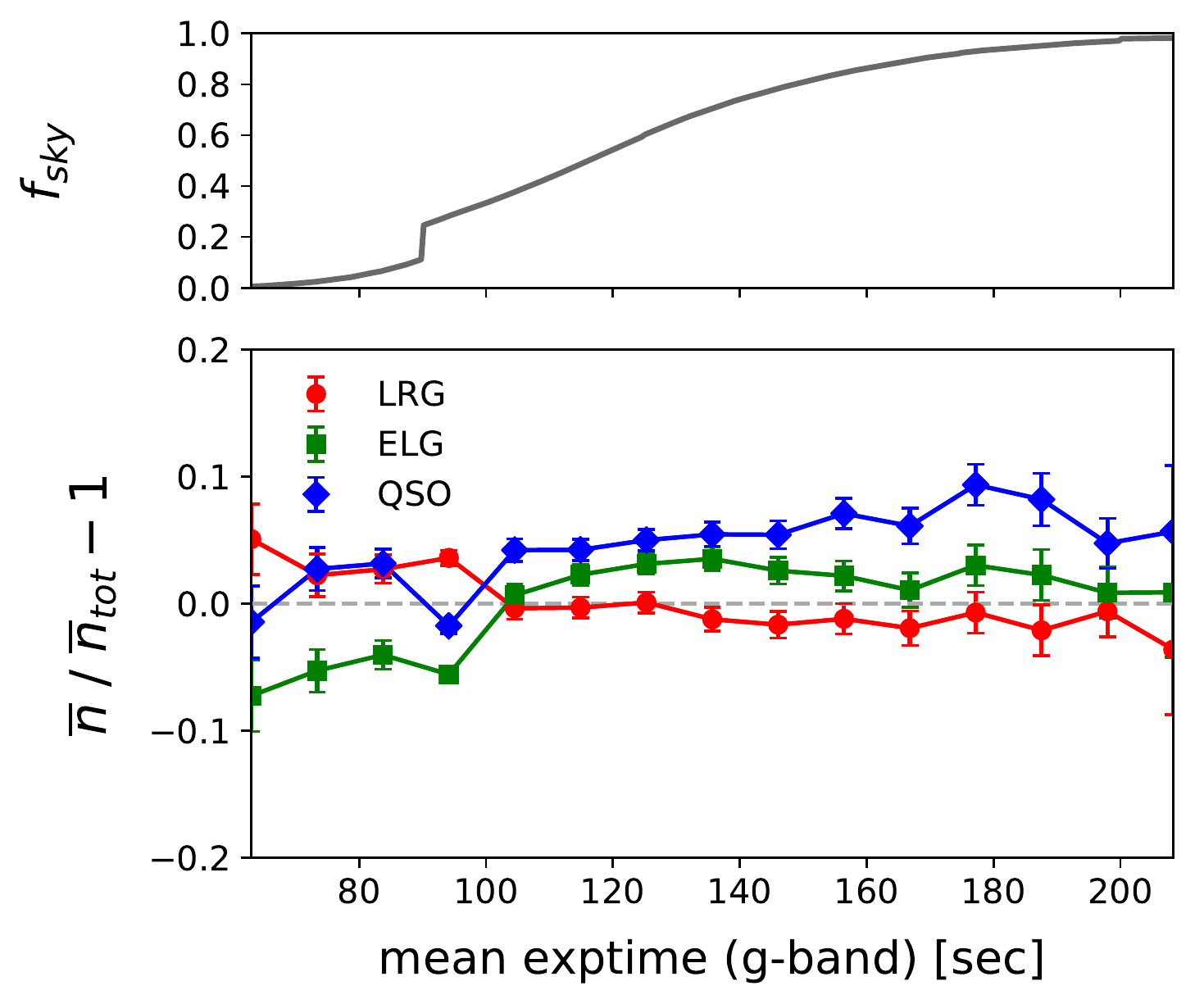}
\includegraphics[width=0.3\textwidth, trim=0.2cm 0 0 0, clip]{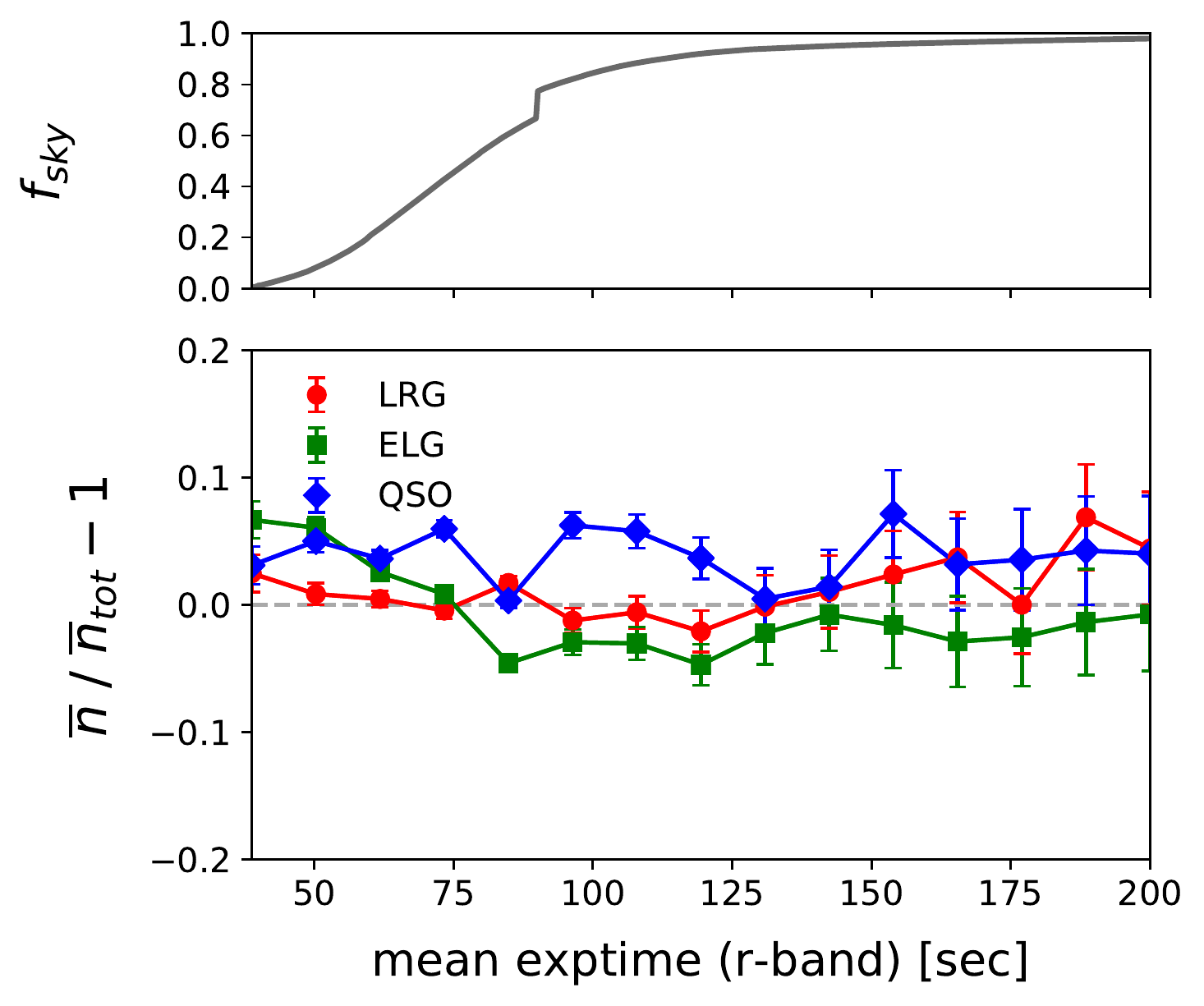}
\includegraphics[width=0.3\textwidth, trim=0.2cm 0 0 0, clip]{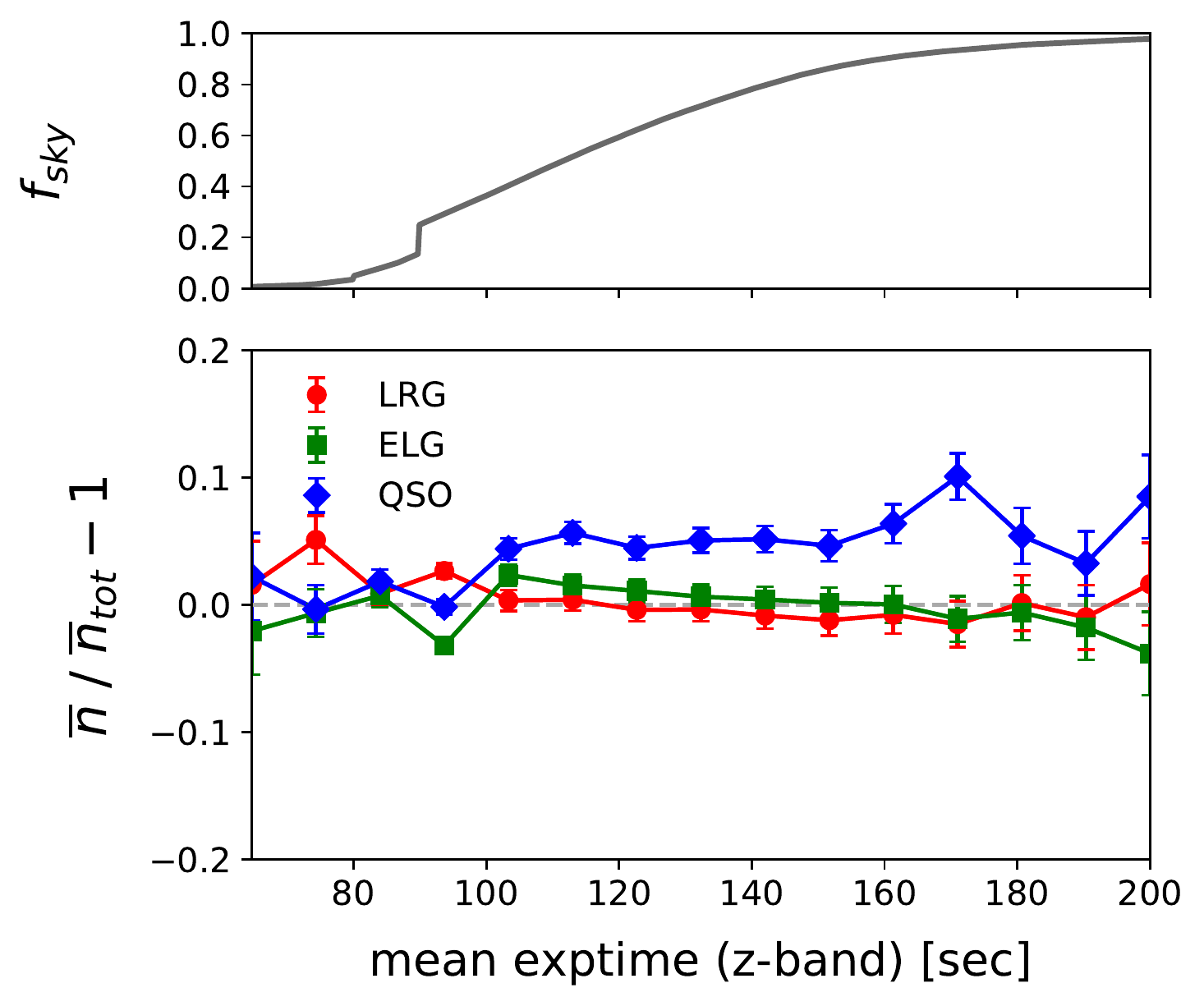}
\caption{Systematic dependences before applying photometric weights: Mean density fluctuations for LRGs (red circles), ELGs (green squares), and QSOs (blue diamonds), as a function of (from top-left to bottom-right) mean stellar density, color excess $E_{B-V}$, airmass, seeing, sky background, and exposure time in each band, with Poisson errors. The top panel in each figure is the cumulative sky fraction for each systematic. We generally find that splitting between NGC and SGC has negligible effects on the 1D density trends, with the exception of $E_{B-V}$ for ELGs and stellar density for QSOs; for these two cases, we have added the NGC-only (dashed) and SGC-only (dotted) trend lines.}
\label{fig:density_systematics}
\end{figure*}

For each potential systematic, we bin the pixels by systematic value and then plot the average density versus the average systematic value of the bins. The results are shown in Figure~\ref{fig:density_systematics}, with LRGs, ELGs, and QSOs plotted together in each subplot and cumulative sky fraction displayed in the upper panels. The errors bars represent the Poisson noise in each bin; using standard error of the mean gives minuscule error bars, as the variance within each bin is very small, regardless of the exact bin size or pixel resolution used. In general, LRGs show very little dependence on survey properties, while ELGs and QSOs appear more impacted, with QSOs often displaying a nonlinear dependence likely due to the more complex selection function. We find that the NGC and SGC density trends are similar and thus do not need to be plotted separately, with the exception of $E_{B-V}$ for ELGs and stellar density for QSOs. For these two special cases, we include the NGC-only (dashed) and SGC-only (dotted) trend lines in Figure~\ref{fig:density_systematics}.

The most significant systematic effects indicated are from stellar density and extinction. ELG density decreases significantly with increasing stellar density and extinction, with $10$\% level effects in some areas, while QSO density presents the opposite trend. The observed correlation between stellar density and QSO density, and anti-correlation between stellar density and ELG density, is also present in the angular cross-correlation results (Section~\ref{sec:stellar_contam}). One possible explanation for the extinction dependence is the issue of infrared emission from background galaxies contaminating the SFD dust maps used to correct DECaLS magnitudes. This has been shown to lead to underestimation of the reddening in low extinction ($E_{B-V} < 0.15$) regions (\citealt{Yahata07}, \citealt{Kashiwagi12}, \citealt{Kashiwagi15}), attributed to the fact that the extragalactic contamination dominates the dust signal in such regions. Looking at the shape of the targets selection functions in colorspace (Figure~\ref{fig:stellar_contams}), underestimation of reddening could preferentially scatter objects out of the ELG selection or into the QSO selection. While the resultant underestimation of extinction in these regions is small, it may be highly correlated with the targets. 

Some minor dependence on seeing and sky background are also observable, particularly for QSOs, as it is more difficult to distinguish between QSOs and stars in regions with bad seeing and bright sky backgrounds. The relationships between target densities and mean exposure times is more complex. Due to the use of dynamic exposure times, there is entanglement with other systematics; for example, exposures are scaled longer for higher airmass or regions of higher galactic extinction. CCDs with exposure times less than 30 sec are automatically removed in the image reduction pipeline, and the ``jumps'' or discontinuities in sky fraction at various other times are artifacts of the observing strategy. The fact that we are averaging the exposure times over multiple overlapping CCDs slightly muddies the interpretation as well. Nevertheless, it is clear that the attempt to obtain more uniform depths through dynamic exposure times is not perfectly successful for QSOs.

Based on these findings, we create photometric weights to reduce the variance in target densities due to systematics. Working directly with the HEALPix pixels defined in the previous section, we use principal component analysis (PCA) to transform the list of potential systematics into a minimum set of linearly uncorrelated variables. PCA using a full SVD solver reduces the dimensionality from 12 scaled features to 11 components. The first component explains $\sim$22\% of the variance and last component explains $\sim$3\% of the variance, with $\sim$50\% of the variance explained by the first 3 components and $\sim$75\% of the variance explained by the first 6 components

Unsurprisingly, we find that exposure times contain a great deal of information, as they are correlated with all of the other systematics by design. However, since exposure times are difficult to interpret, we also perform a version of the component analysis with the exposure times removed from consideration, in order to show more clearly how the other features contribute and in what combinations. We find that nearly equal contributions from stellar density and galactic extinction tend to strongly dominate a few components, while more complex mixtures of sky background, seeing, and airmass features dominate the others, as physical intuition might lead us to expect. 

\begin{figure*}
\centering
\includegraphics[width=0.3\textwidth, trim=0.2cm 0 0 0, clip]{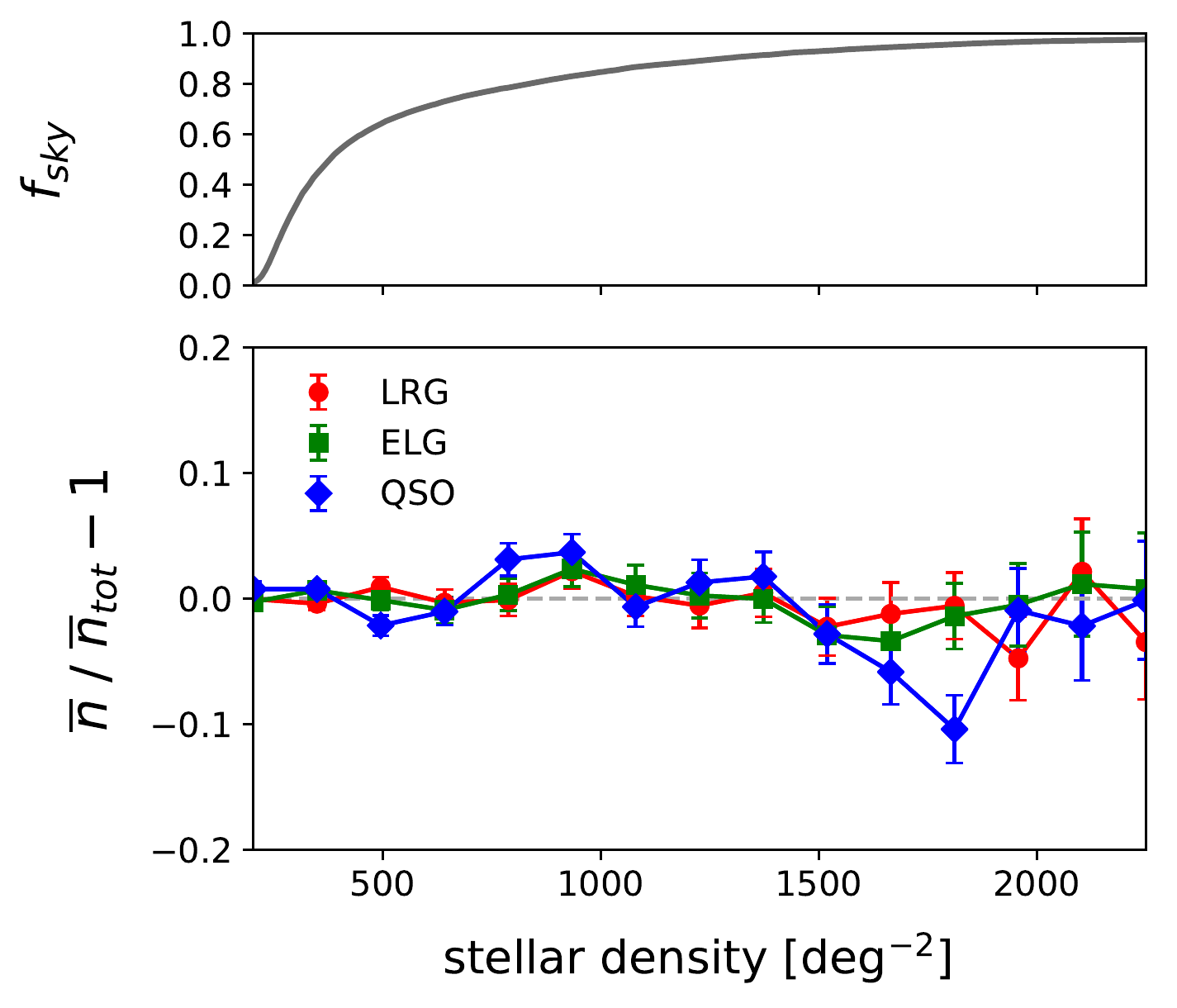}
\includegraphics[width=0.3\textwidth, trim=0.2cm 0 0 0, clip]{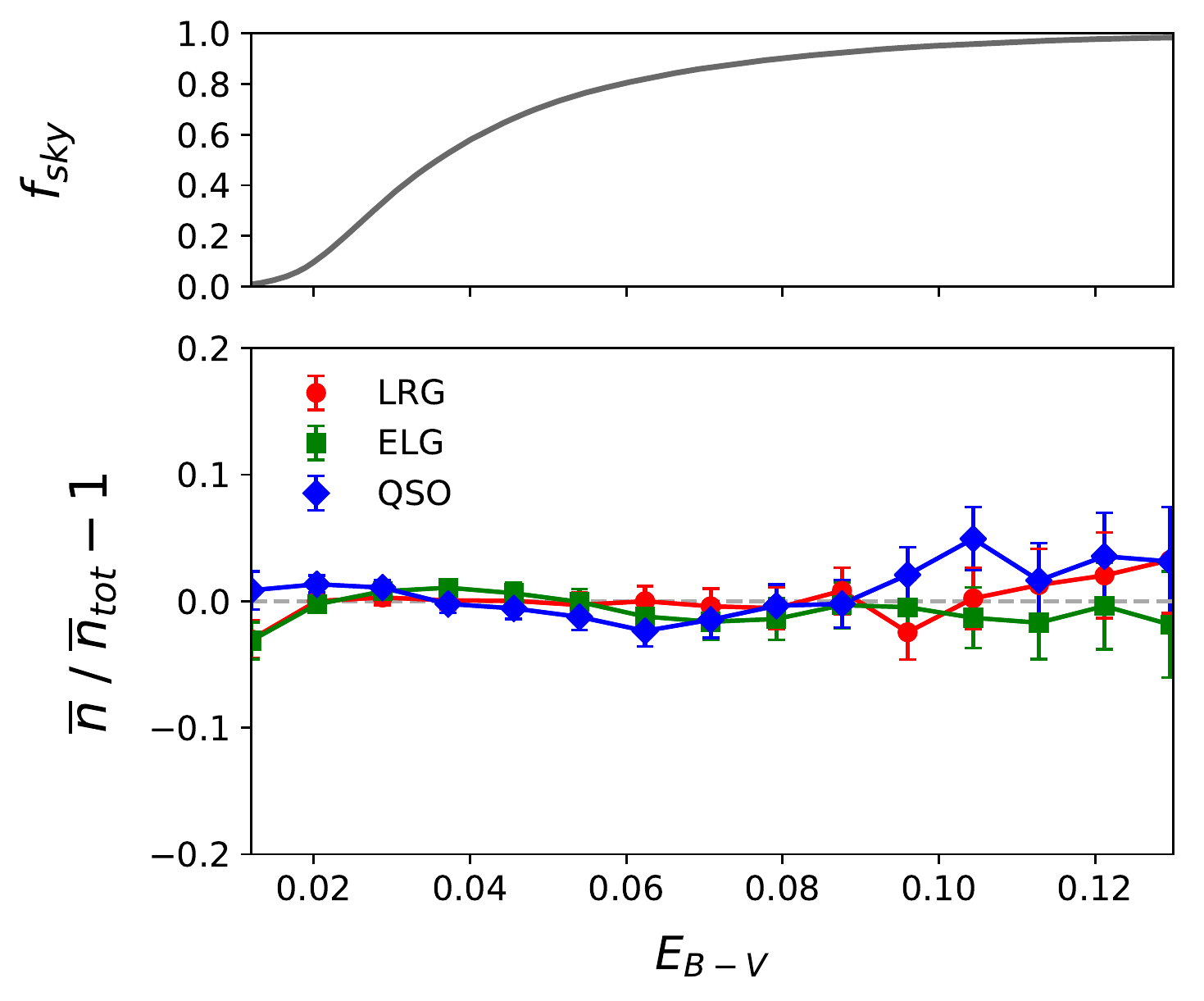}
\includegraphics[width=0.3\textwidth, trim=0.2cm 0 0 0, clip]{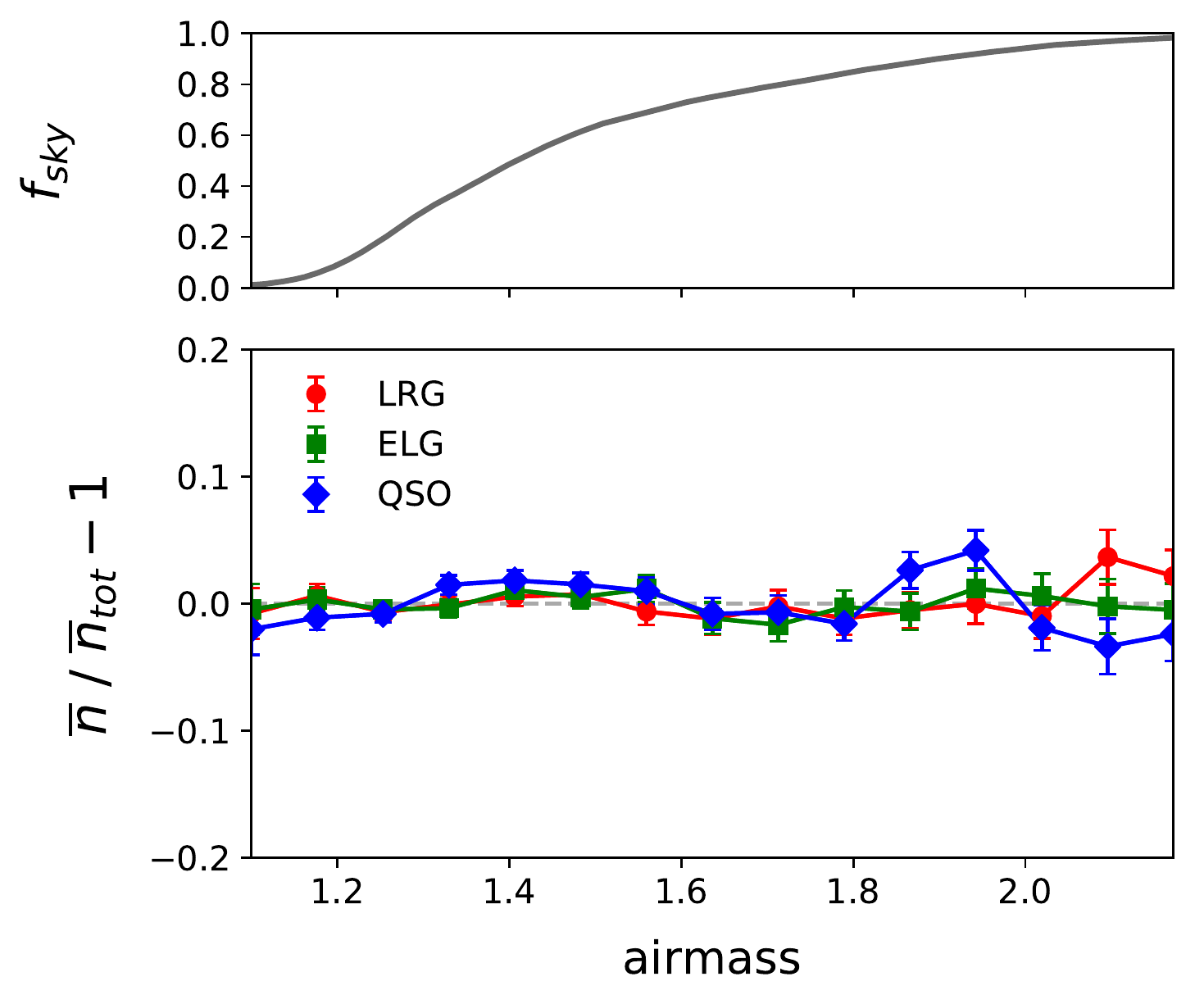}
\includegraphics[width=0.3\textwidth, trim=0.2cm 0 0 0, clip]{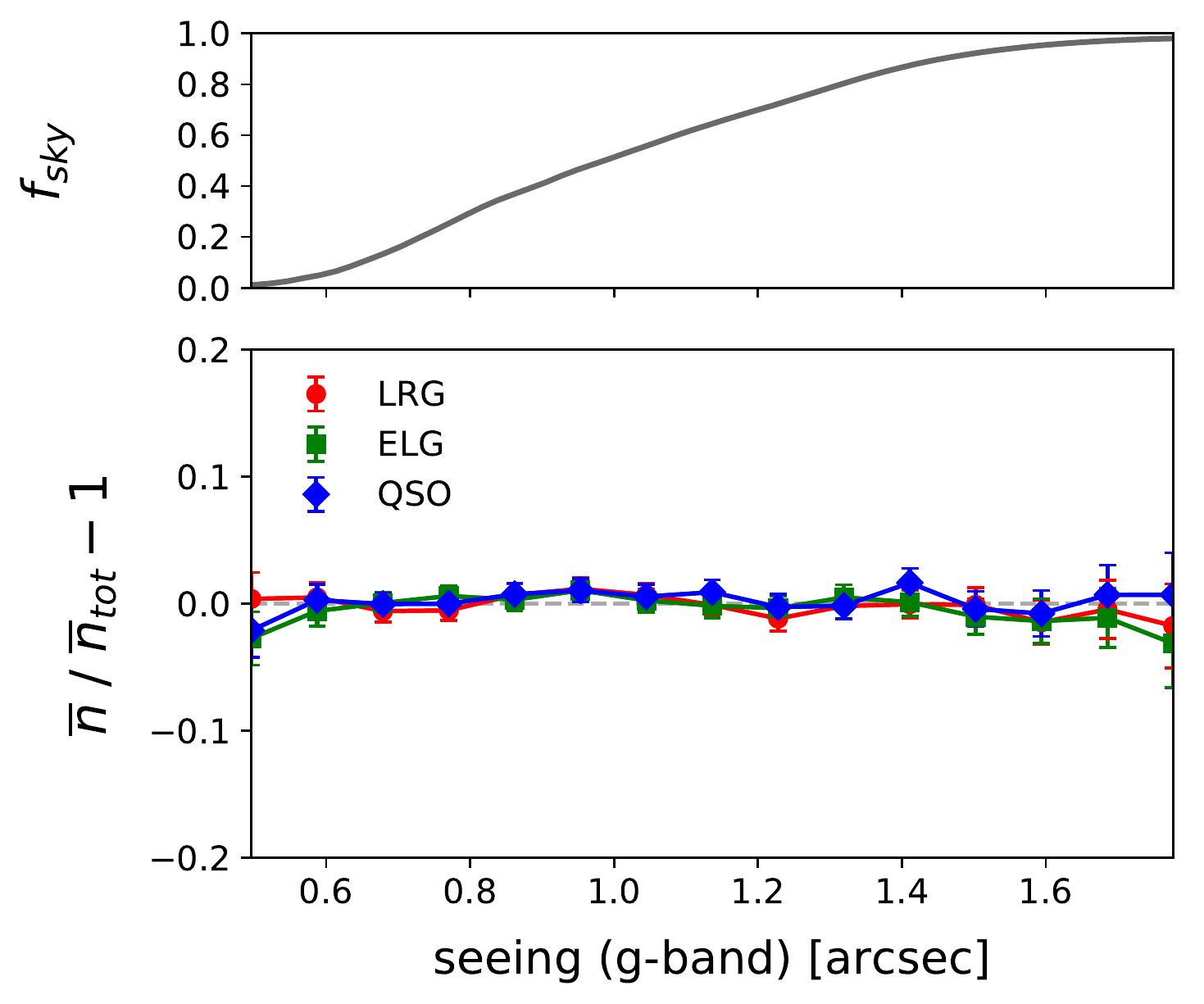}
\includegraphics[width=0.3\textwidth, trim=0.2cm 0 0 0, clip]{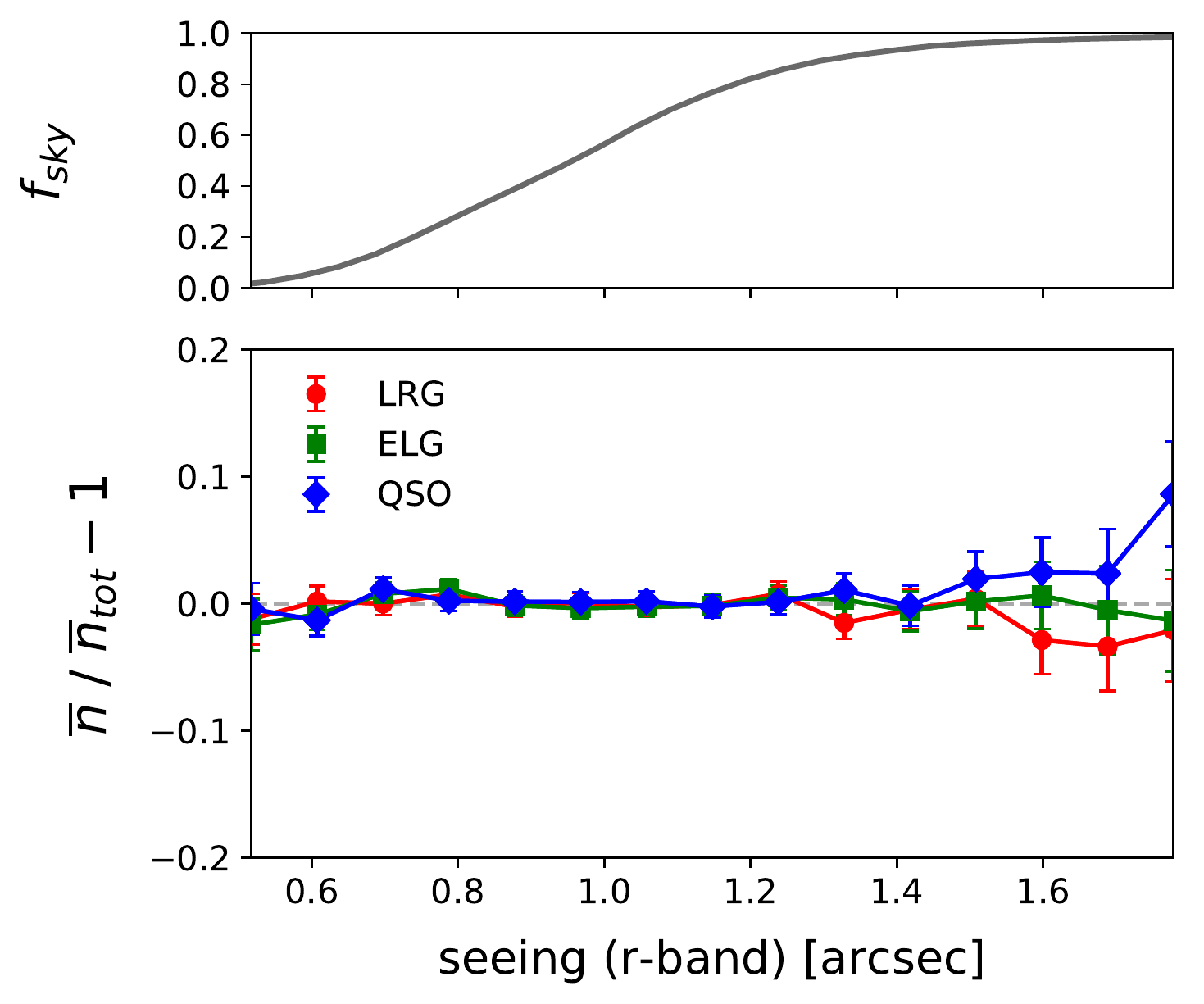}
\includegraphics[width=0.3\textwidth, trim=0.2cm 0 0 0, clip]{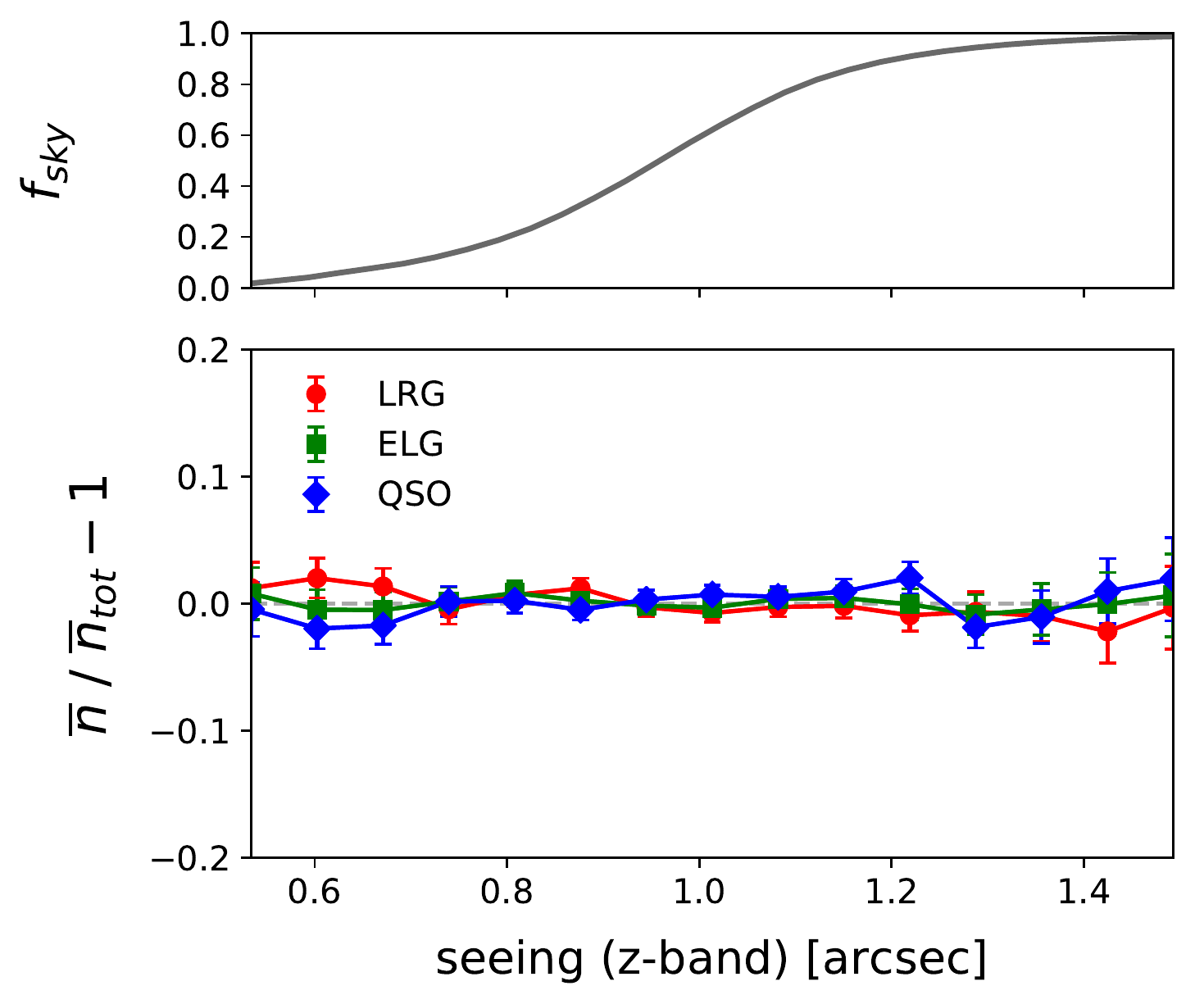}
\includegraphics[width=0.3\textwidth, trim=0.2cm 0 0 0, clip]{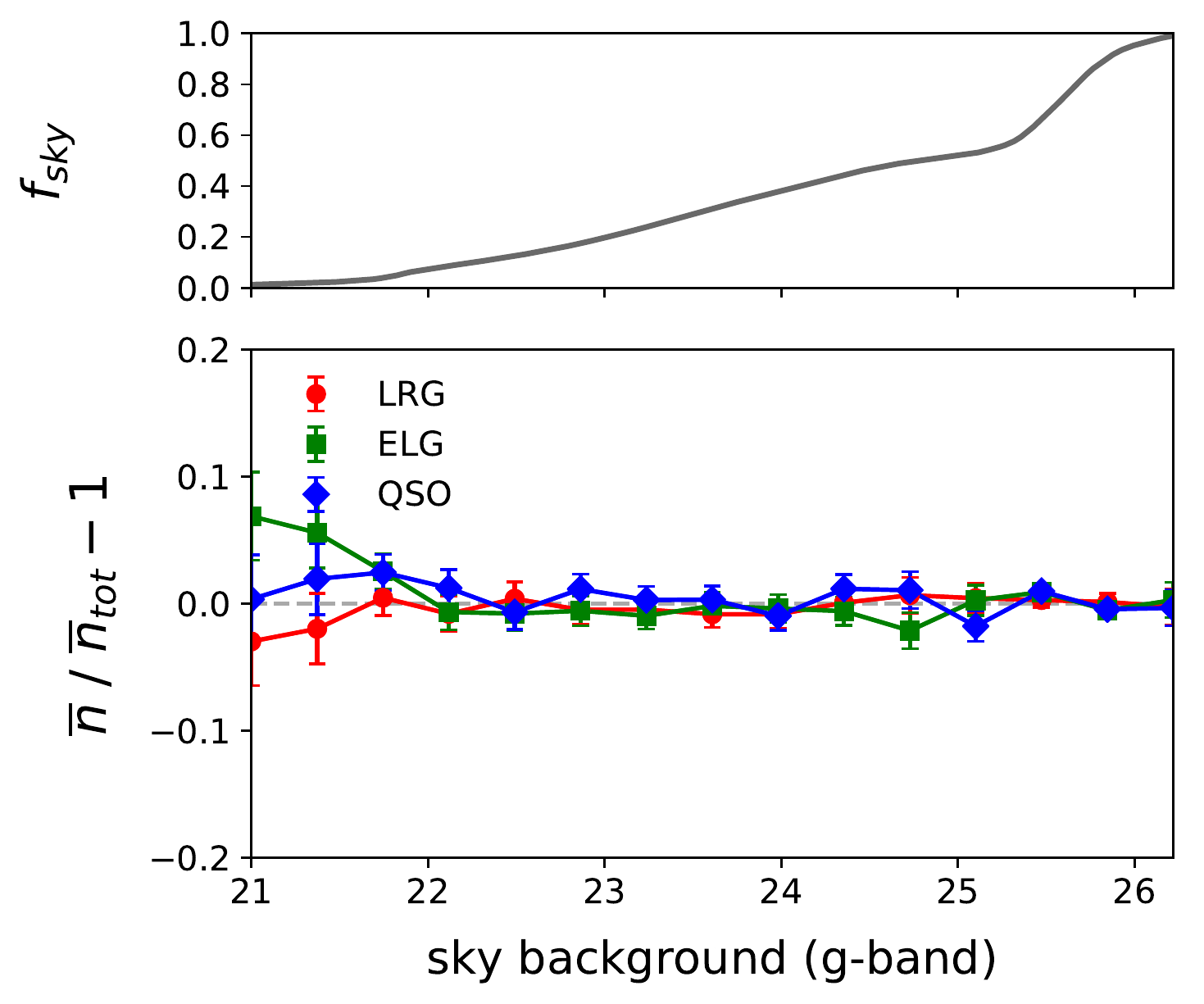}
\includegraphics[width=0.3\textwidth, trim=0.2cm 0 0 0, clip]{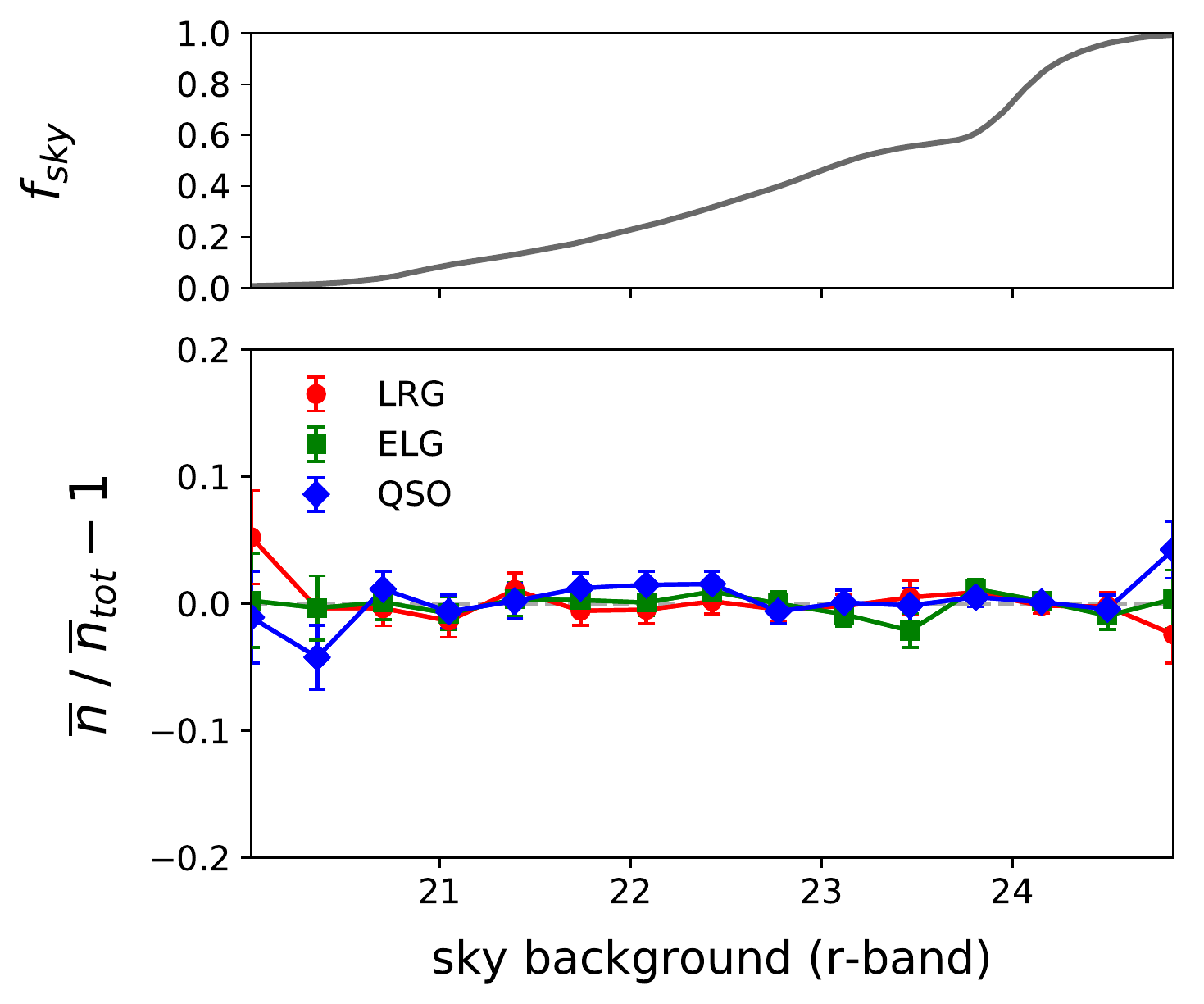}
\includegraphics[width=0.3\textwidth, trim=0.2cm 0 0 0, clip]{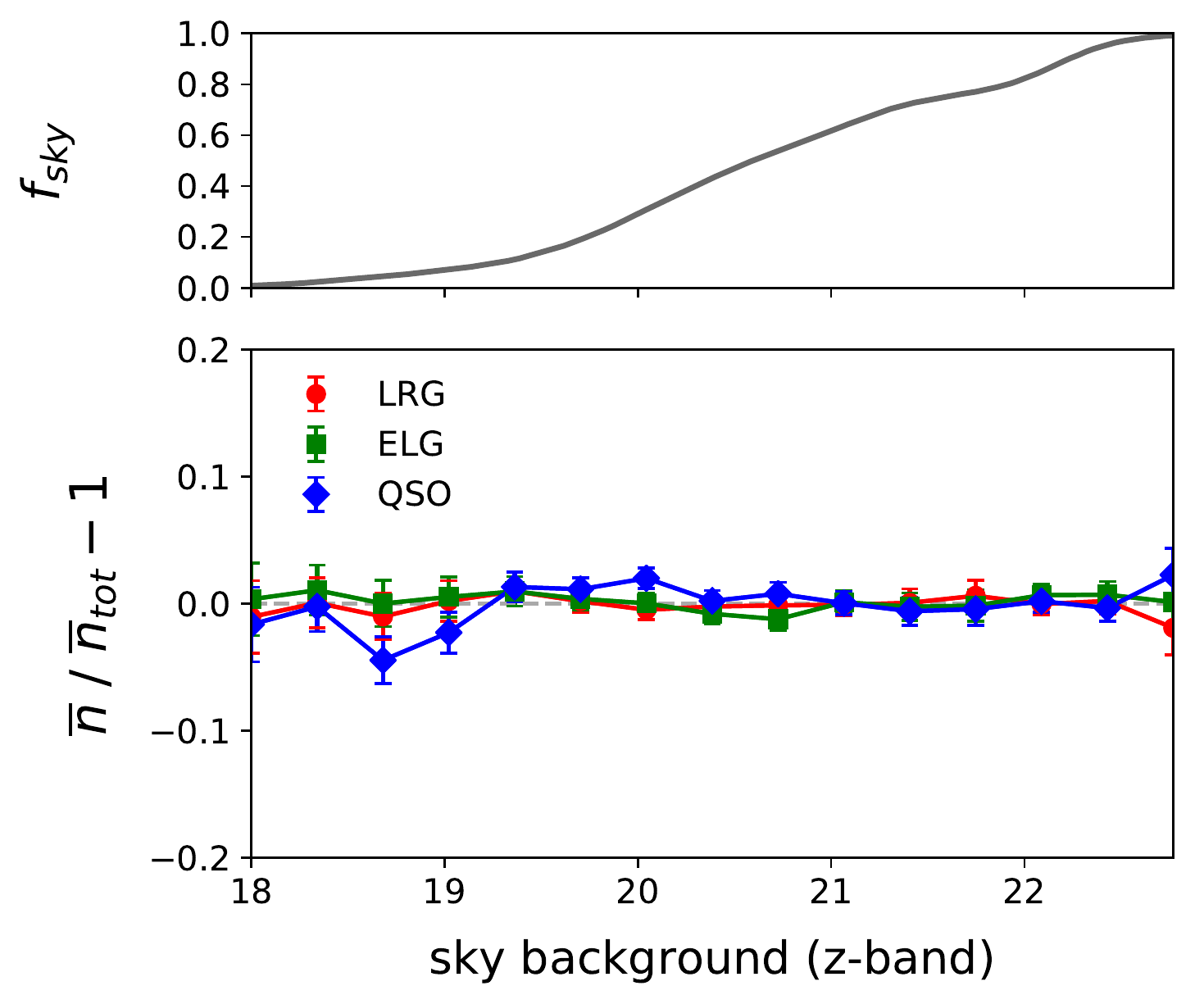}
\includegraphics[width=0.3\textwidth, trim=0.2cm 0 0 0, clip]{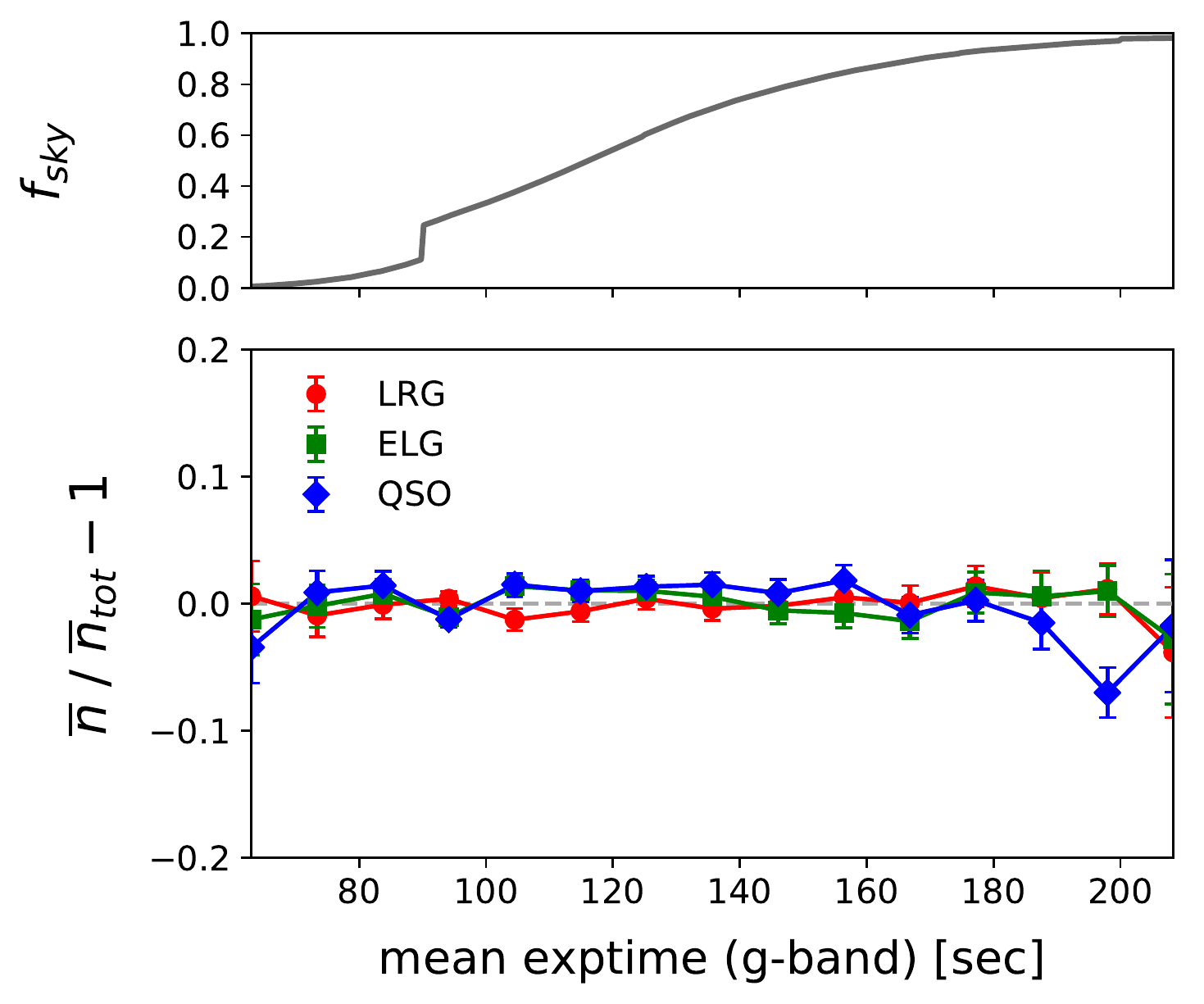}
\includegraphics[width=0.3\textwidth, trim=0.2cm 0 0 0, clip]{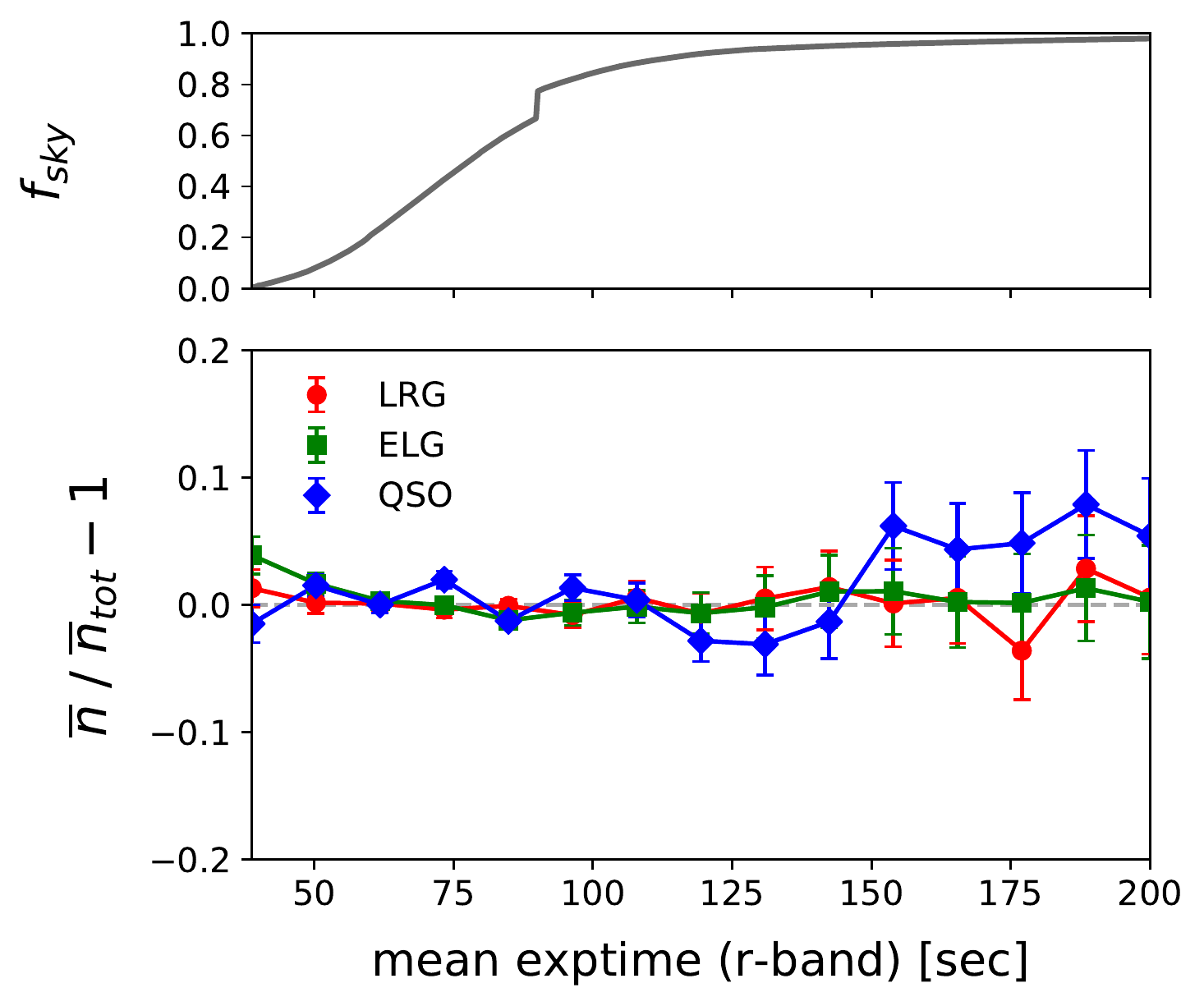}
\includegraphics[width=0.3\textwidth, trim=0.2cm 0 0 0, clip]{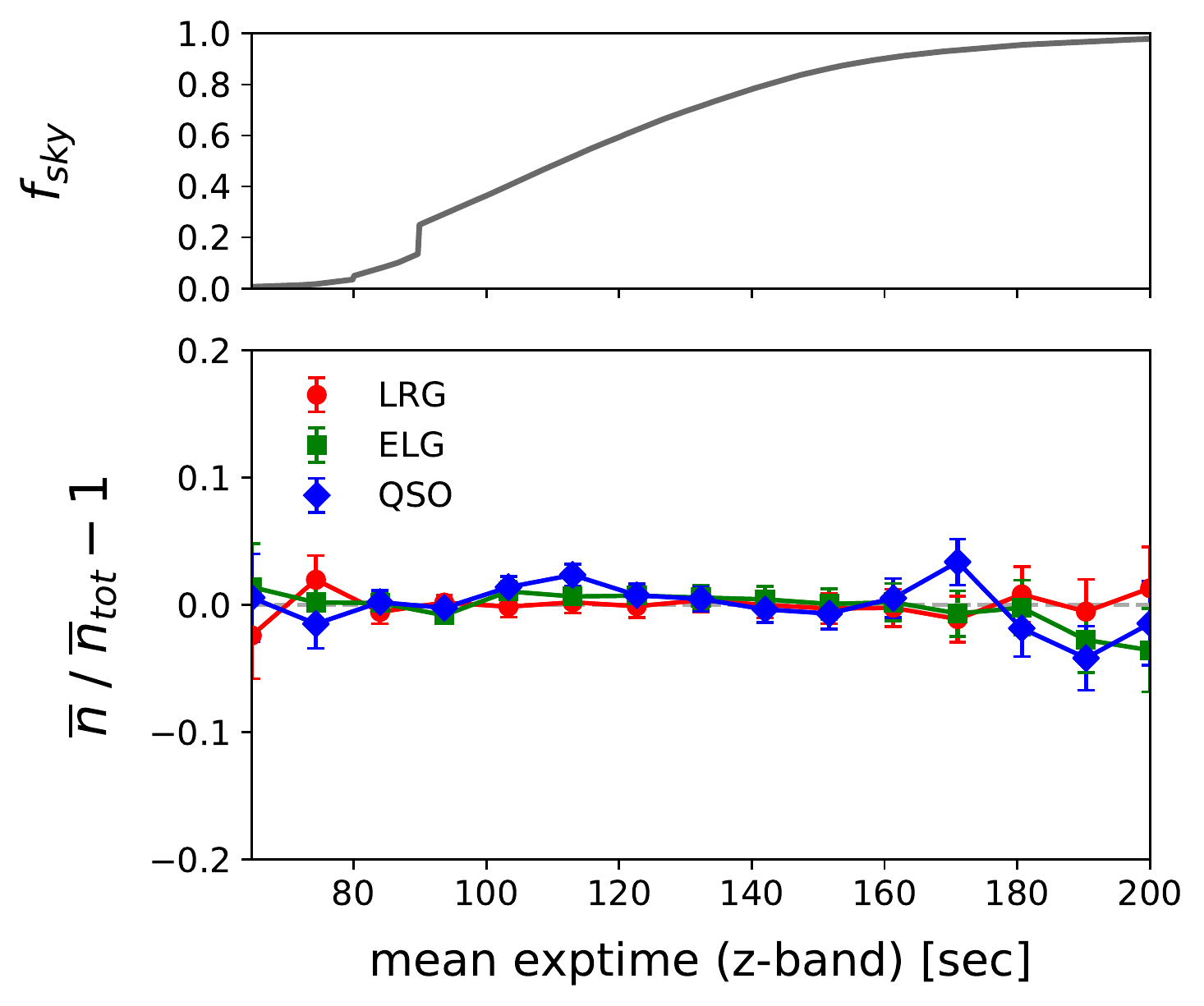}
\caption{Systematic dependences after applying photometric weights: Mean density fluctuations for LRGs (red circles), ELGs (green squares), and QSOs (blue diamonds), as a function of (from top-left to bottom-right) mean stellar density, color excess $E_{B-V}$, airmass, seeing, sky background, and exposure time in each band, with Poisson errors. The top panel in each figure is the cumulative sky fraction for each systematic.}
\label{fig:density_systematics_wt}
\end{figure*}

We discretize the feature-space to reduce the impact of noisy pixels and outliers, then apply multilinear regression. The resulting model of density as a function of potential systematics is used to generate weights.\footnote{For randoms where any of the potential systematics were undefined due to lack of exposures in one or more bands, the weights were manually set to one. The randoms used in our clustering analysis have had the completeness mask of Section~\ref{sec:masks/complete} applied, and thus are not affected by this.} We apply our photometric weights to the randoms, modulating the effective area (and therefore target density) of the survey based on the local values of systematics. Weights are normalized in the sense that the proportionality factor in Equation~{\ref{delta_i}} changes from the number of randoms in the masked footprint to the sum of their weights. Plots of the density vs. systematics after applying weights is  shown in Figure~\ref{fig:density_systematics_wt}, with the linear parts of the trends improved. Figure~\ref{fig:final-wtheta} demonstrates the effect of applying these weights on the angular correlation functions. All clustering results presented in Section~\ref{sec:characterization_angular} and Section~\ref{sec:characterization_cross} are computed using the weighted values.

\subsection{Correlation with stars and stellar contamination fraction}
\label{sec:stellar_contam}

\begin{figure}
\includegraphics[width=\columnwidth]{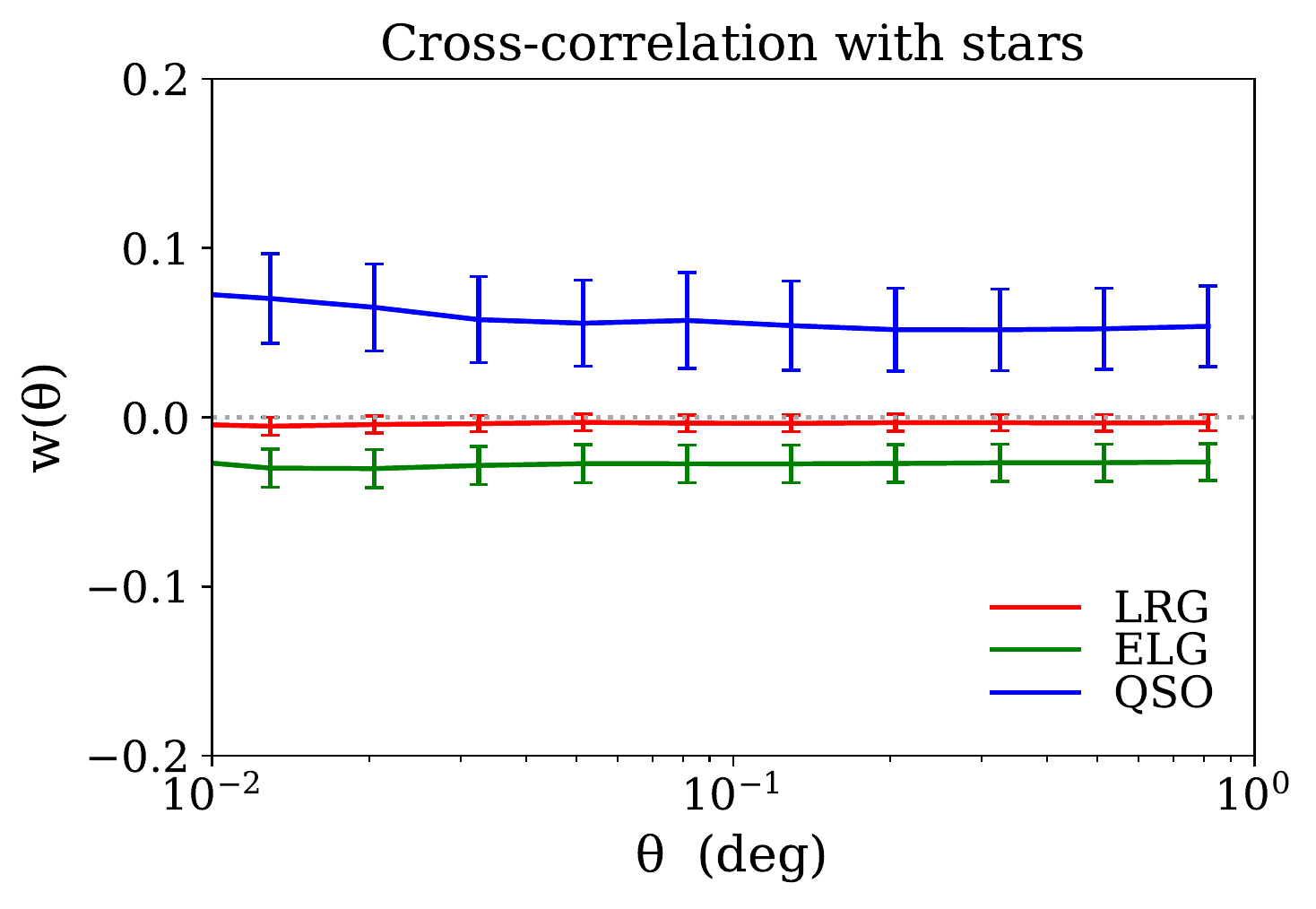}
\caption{Angular cross-correlation between DESI targets and stars, with errors from bootstrapping on the area.}
\label{fig:stellarxcorr}
\end{figure}

We also measure the angular cross-correlation between the targets (after masking but before weighting) and our stellar catalog, with the results shown in Figure~\ref{fig:stellarxcorr}. Consistent with the density trends observed in Figure~\ref{fig:density_systematics}, LRGs appear uncorrelated with stars, while ELGs demonstrate a small constant anticorrelation, and QSOs show a more significant constant correlation. 

Using the angular cross-correlation, we can estimate the fraction of stellar contamination in the QSO sample. Let us assume that the observed number of QSOs at any given location includes some non-trivial number of contaminants, as seems strongly indicated. Let $N_{\text{star}}$ be the total number of stars that modulate the QSO density in some way:
\begin{equation}
N_{\text{obs}} = N_{\text{true}} + \bar{\epsilon}N_{\text{star}}
\end{equation}
where $\bar{\epsilon}$ is the average number of impacted sources associated with each star. For stars which are simply misclassified as QSOs, $\epsilon = 1$. For spurious QSO detections in the immediate vicinity of stars, $\epsilon > 0$. For occulted QSOs in the immediate vicinity of stars, $\epsilon < 0$. Note that for the latter two effects (spurious or occulted sources near stars) we are considering fainter stars that were not masked out in Section~\ref{sec:masks/foregrounds/stars}. Thus, any cross-correlations between these sources and their own associated star are negligible beyond very small scales, so the cross-correlation between contaminants and stars is dominated by the autocorrelation of stars times the multiplicative factor $\epsilon$.

The fraction of the total sample which is made up of these problematic stars is approximately given by
\begin{equation}
f_{\text{star}} = \frac{\langle N_{\text{star}} \rangle}{\langle N_{\text{obs}}\rangle}
\end{equation}
where the brackets signify a spatial average. Similarly, the fraction of true objects is 
\begin{align}
f_{\text{true}} = \frac{\langle N_{\text{true}} \rangle}{\langle N_{\text{obs}} \rangle}
\end{align}
We can rewrite this in terms of the density contrasts $ \delta = N/\langle N \rangle - 1 $, by exploiting the fact that $\langle N_{\text{obs}}\rangle = \langle N_{\text{star}}\rangle / f_{\text{star}} = \langle N_{\text{true}}\rangle / f_{\text{true}}$, such that the observed density of objects is
\begin{align}
\delta_{\text{obs}} &= \frac{N_{\text{obs}}}{\langle N_{\text{obs}} \rangle} - 1 = \frac{N_{\text{true}}}{\langle N_{\text{obs}} \rangle} + \bar{\epsilon}\frac{N_{\text{star}}}{\langle N_{\text{obs}} \rangle} - 1 \nonumber \\  
&= f_{\text{true}}\frac{N_{\text{true}}}{\langle N_{\text{true}} \rangle} + \bar{\epsilon}f_{\text{star}}\frac{N_{\text{star}}}{\langle N_{\text{star}} \rangle} - 1 \nonumber \\
&= f_{\text{true}}(\delta_{\text{true}} + 1) + \bar{\epsilon}f_{\text{star}}(\delta_{\text{star}} + 1) - 1 \nonumber \\ 
&= f_{\text{true}}\delta_{\text{true}} + \bar{\epsilon}f_{\text{star}}\delta_{\text{star}}
\end{align}
Thus we can extract the contamination fraction by dividing the QSO-star cross-correlation by the stellar autocorrelation function, $w_{\text{cross}}(\theta)/w_{\text{star}}(\theta) = \langle \delta_{\text{obs}}, \delta_{\text{star}} \rangle / \langle \delta_{\text{star}}, \delta_{\text{star}} \rangle = \bar{\epsilon}f_{\text{star}} \equiv f_{\text{contam}}$, since the cross-correlation between true QSOs and stars vanishes.\footnote{We have assumed an ideal stellar template; in reality, there may be a small fraction of true QSOs in the star sample, or a fraction of galaxies in both the QSO and star samples which correlate with each other, but these fractions should be much smaller than the fraction of stellar contaminants in the QSO sample, and hence we can ignore them to first order.}

As the stellar density varies significantly across the sky (see Figure~\ref{fig:systematic_maps}), with a strong gradient towards the galactic plane, we first divide the sky into three bins: $|b| < 40$, $40 < |b| < 60$, and $|b| > 60$. For each galactic latitude bin, we calculate $w_{\text{cross}}(\theta)$ and $w_{\text{star}}(\theta)$, averaged across all angular scales (as both correlation functions are flat, within error bars, for all bins), and then bootstrap upon these averaged values to obtain error bars. The resulting stellar contamination fractions are $f_{\text{contam}}$ = 7\% $\pm$ 4.9\% for $|b| < 40$, $f_{\text{contam}}$ = 4.9\% $\pm$ 2.7\% for $40 < |b| < 60$, and $f_{\text{contam}}$ = 4.1\% $\pm$ 2.3\% for $|b| > 60$.

\label{sec:characterization}
\section{ANGULAR CLUSTERING MEASUREMENTS}
\label{sec:characterization_angular}

\subsection{Mean densities}
\label{sec:characterization/densities}
The average target densities for DR7 are given in Table~\ref{table:densities}. We present the raw densities as well as the densities after observational effects have been accounted for using the masks and weights described in this paper. Here, densities are calculated by taking the ratio of the total number of objects and the total area, with the latter being estimated using counts of uniform randoms with the  masks and weights applied to them. For an independent calculation of raw target densities, column 2 of the counts-in-cells tables in Section~\ref{sec:characterization/cic} gives the average number of objects $\bar{N}$ within a square cell of some width, and thus can be divided by the corresponding cell area to give the mean (raw) density smoothed over that scale.

\begin{table}
\centering{
\begin{tabular}{*{5}{c}}
\toprule
 & \multicolumn{4}{c}{Avg. density (deg$^{-2}$)} \\
\cmidrule{2-5}
Target & Raw & Masked & Masked \& Weighted & FDR \\
\midrule
LRG & 525.0 & 501.4 & 498.9 & 480 \\
ELG & 2497.5 & 2351.8 & 2352.7 & 2400 \\
QSO & 336.1 & 261.2 & 256.2 & 260 \\
\bottomrule
\end{tabular}}

\caption{Average densities for each target type, calculated over the available footprint. The first column is the uncorrected densities, the second column has only had the masks of Section~\ref{sec:masks} applied, and the third column has additionally had the photometric weights of Section~\ref{sec:systematics_densities} applied. The projected target densities from the FDR are also included for reference.}
\label{table:densities}
\end{table}

\subsection{Angular correlation functions with $r_0,\gamma$ fits}
\label{sec:characterization/corrfuns}
\begin{figure}
\centering
\includegraphics[width=0.99\linewidth]{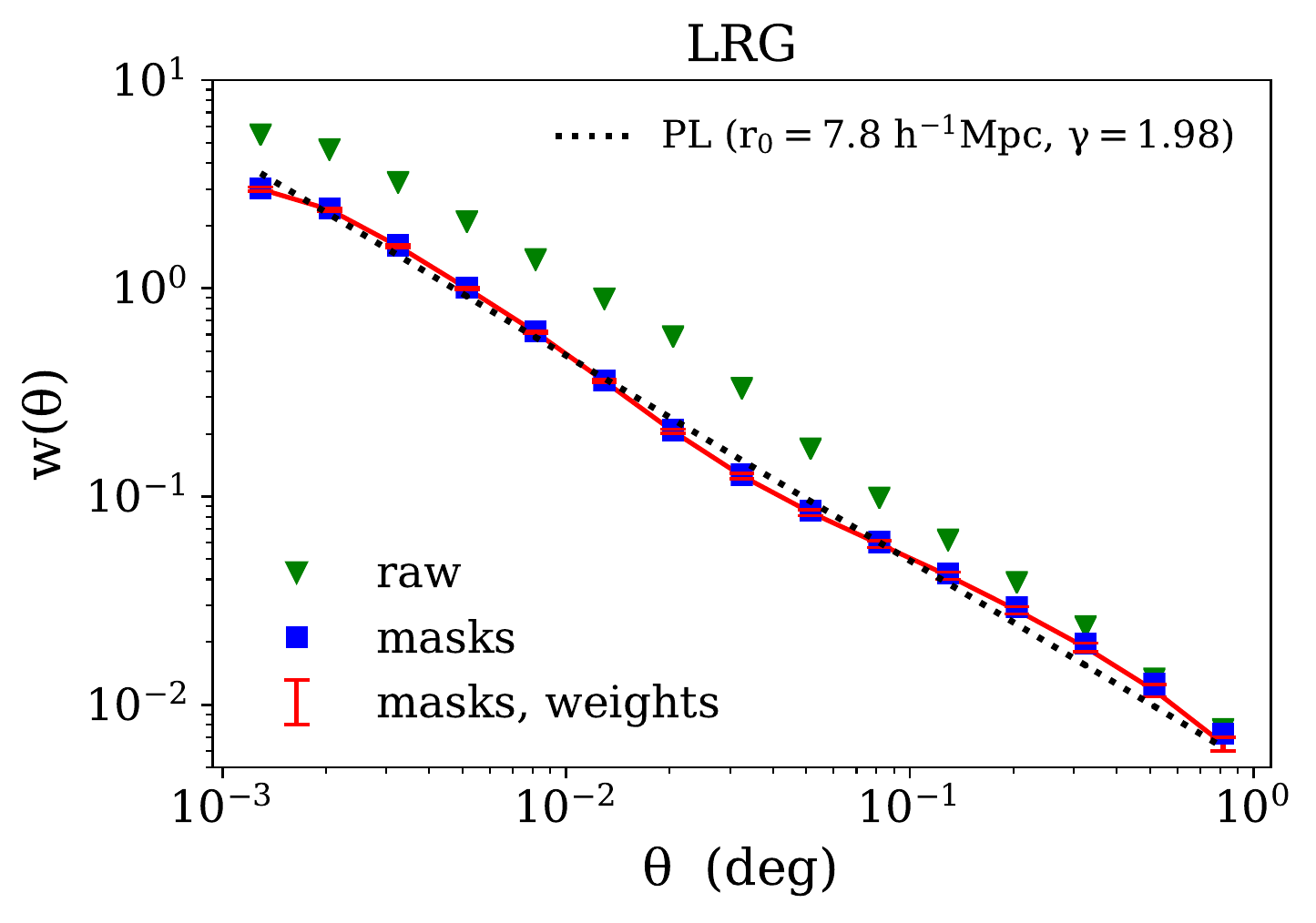}
\includegraphics[width=0.99\linewidth]{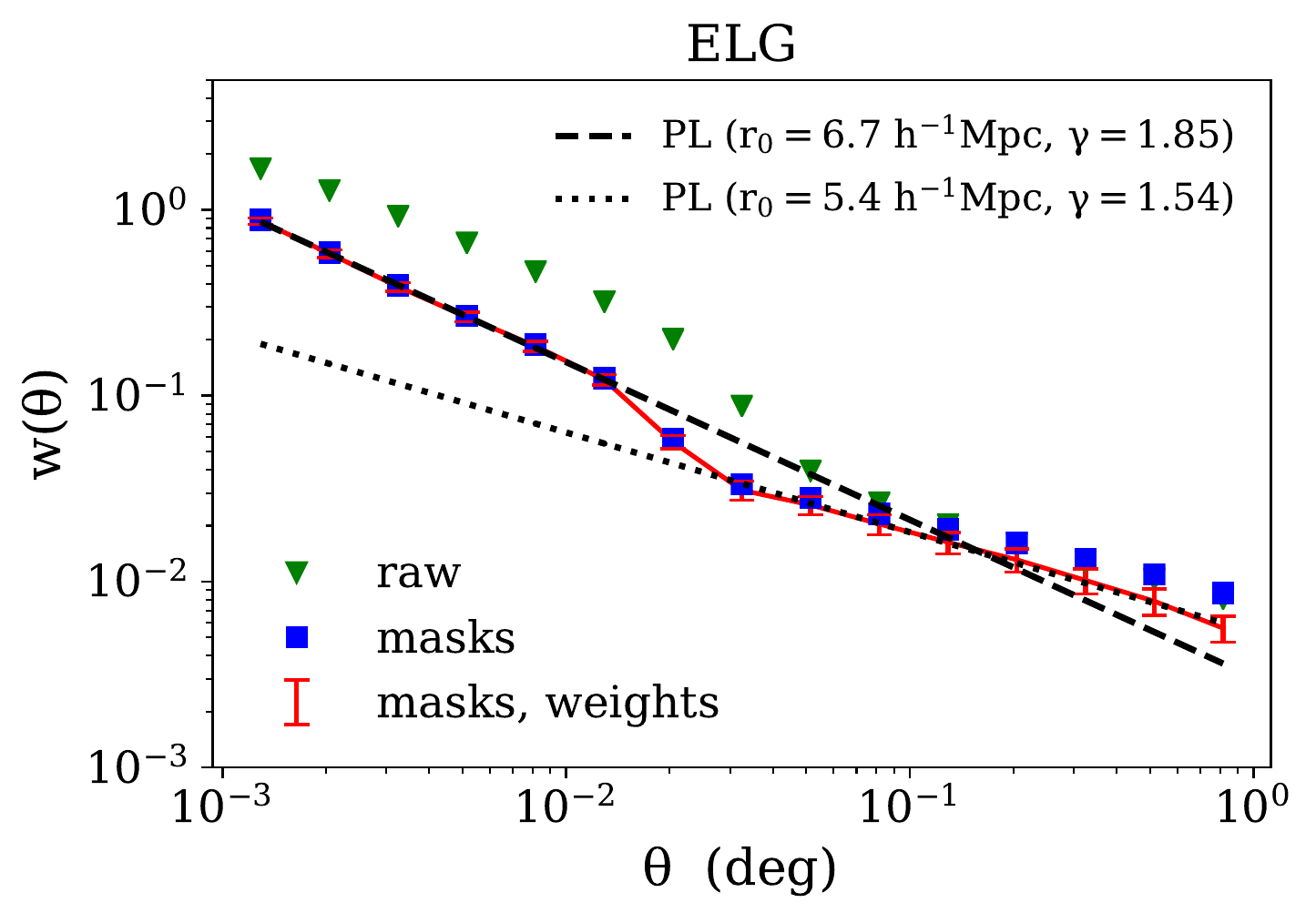}
\includegraphics[width=0.99\linewidth]{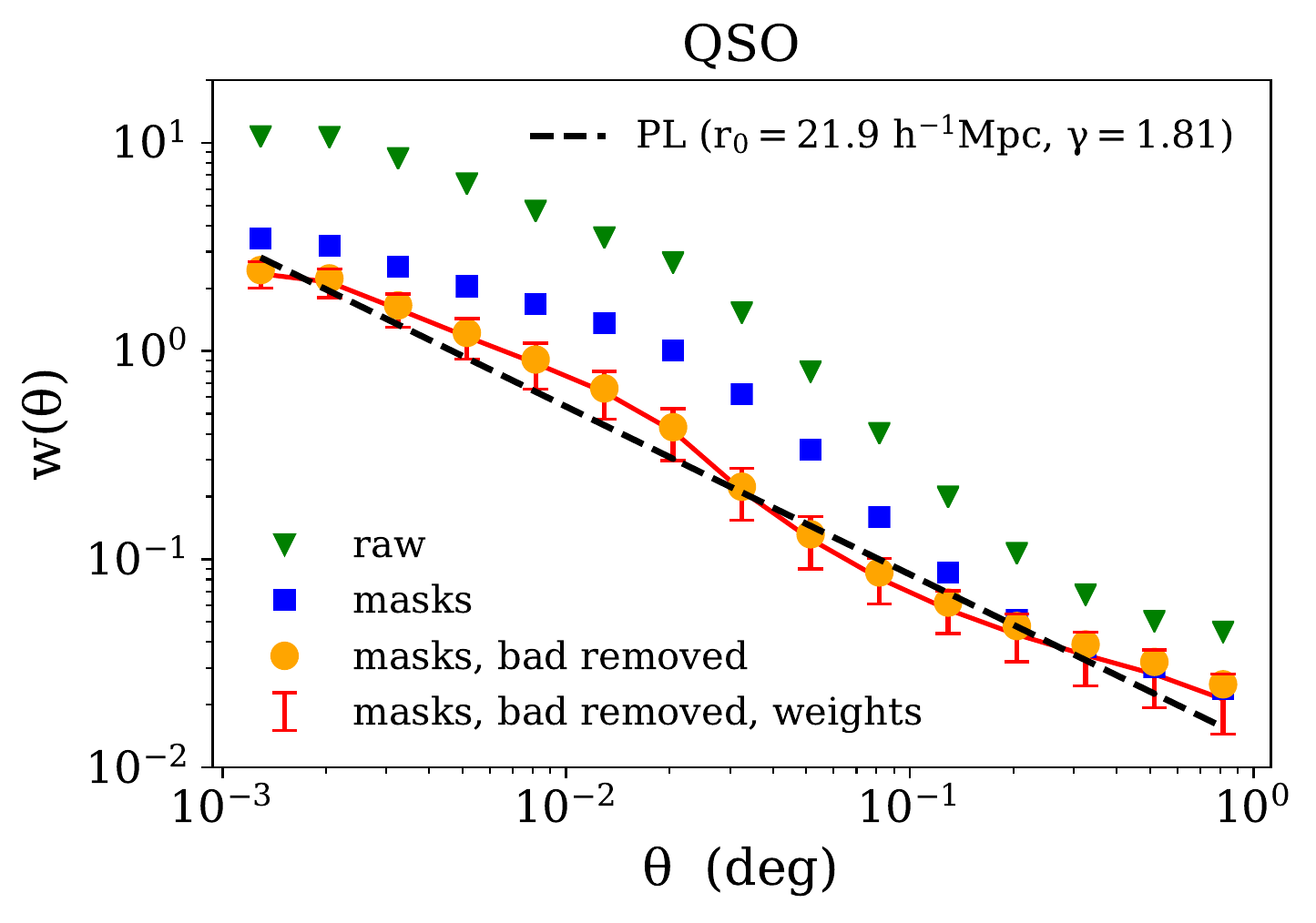}
\caption{Two-point angular correlation functions for LRGs, ELGs, and QSOs at several levels of systematics analysis: without any corrections (green triangles), after applying masks (blue squares), and after applying both masks and photometric weights (red error bars). For QSOs, we perform the additional intermediate step of removing the anomalous regions discussed in Section~\ref{sec:spatial} (orange circles). The red lines are fit to a power-law model for the three-dimensional clustering $\xi(r) = (r/r_0)^{-\gamma}$. The values of these fits are listed in more detail in Table~\ref{table:fiducial_powerlaw}.}
\label{fig:final-wtheta}
\end{figure}

The measured angular correlation functions for the three target classes are shown in Figure~\ref{fig:final-wtheta}. We present the correlation functions at various stages of analysis, to demonstrate the effects of applying masks, weights, and so on. The final, ``cleanest'' version is fit to theory, as described below.

According to the current paradigm of galaxy formation, galaxies form within collapsed overdensities of dark matter called ``halos'' (for a recent review on the galaxy-halo connection, see \citealt{WechslerTinker18}). Under this model, the correlation function of galaxies is the sum of two contributions: a 2-halo term corresponding to pairs of galaxies within different halos, and a 1-halo term corresponding to pairs of galaxies within the same halo. On small scales, where the 1-halo term dominates, the correlation function depends on the complex baryonic physics of galaxy formation and evolution, while on larger scales, where the 2-halo term dominates, the correlation function is characterized by the halo bias describing how dark matter halos trace the dark matter distribution. When combined, these two terms result in an approximate power law, with a feature corresponding to the 1-halo to 2-halo transition occurring around 1-2 $h^{-1}$ Mpc, the typical virial radius of a halo. Motivated by this, we assume the real-space correlation function is a simple power law of the form $\xi(r) = (r/r_0)^{-\gamma}$. Using tabulated $dN/dz$ for each target and applying the Limber approximation (Equation~\ref{eqn:limber-pl}), we obtain constraints on $r_0$ and $\gamma$, listed in Table~\ref{table:fiducial_powerlaw}.

For LRGs, we determine $r_0 = 7.78 \pm 0.26$ $h^{-1}$Mpc and $\gamma = 1.98 \pm 0.02$, which agrees well with previous results for similar samples from the literature; for example, \cite{Sawangwit++11} (Table 2, row 4) finds $r_0 = 7.56 \pm 0.03$ $h^{-1}$Mpc and $\gamma = 1.96 \pm 0.01$ for a photometric subsample of LRGs from SDSS imaging with $\bar{z}=0.68$ and a similar redshift distribution, over approximately the same range of angular scales. While the LRG correlation function shows some additional structure that is not fit perfectly by a power law model, no strong features are observed on these scales, which is generally consistent with earlier findings from eBOSS and SDSS LRG studies (see e.g. \citealt{Zehavi++05}). \\ 

We find that the ELG correlation function has a broken form. When fitting from $\theta = 0.001^\circ$ to $\theta = 0.01^\circ$, the correlation function is well fit by $r_0 = 6.70 \pm 0.10$ $h^{-1}$Mpc and $\gamma = 1.85 \pm 0.01$. However, for scales $\theta > 0.05^\circ$, the slope becomes shallower, and the correlation function is better fit by $r_0 = 5.45 \pm 0.10$ $h^{-1}$Mpc and $\gamma = 1.54 \pm 0.01$. At the mean effective redshift of the DESI ELG sample, $z \sim 0.85$, the co-moving scale of this break is approximately 1 $h^{-1}$Mpc, consistent with a 1-halo to 2-halo transition. The second slope matches with the findings of \cite{Favole++16}, who modeled a sample of ELGs selected from the Canada-France Hawaii Telescope Legacy Survey (CFHTLS), cross-matched with BOSS ELG and VIPERS redshifts, at mean redshift $\bar{z} \approx 0.8$, to obtain $s_0 = 5.3 \pm 0.2$ $h^{-1}$Mpc and $\gamma = 1.6 \pm 0.1$. Furthermore, when calculating the angular correlation function over the full CFHTLS footprint, they also observed a change of slope occurring at $\theta \approx 0.01^{\circ} - 0.05^{\circ}$, and found that this clustering was consistent with an HOD model having halo masses on the order of $10^{12} M_{\odot}$ and satellite fraction $f_{\rm sat} \sim 22\%$. Similarly, \cite{Jouvel++15} found $s_0 = 4.2 \pm 0.26$ $h^{-1}$Mpc  and $\gamma = 1.48 \pm 0.04$ for a bright sample of eBOSS ELGs selected from DES photometry at $\bar{z} =  0.86$. The real-space clustering amplitudes and slopes for both LRGs and ELGs are also consistent with the results from \cite{Mostek++13} for red and blue galaxy populations in DEEP2.

The QSO correlation function still contains a significantly enhanced clustering signal due to systematics and contamination, with $r_0 = 21.9 \pm 0.10$ $h^{-1}$Mpc and $\gamma = 1.81 \pm 0.02$. By comparison, some fiducial values of QSO clustering amplitude at $z \sim 2$ from the literature are: $r_0 = 5.84 \pm 0.33$ and $\gamma = 1.65 \pm 0.05$ \citep{Croom++05}, or $r_0 = 4.56 \pm 0.48$ at fixed $\gamma = 1.5$ \citep{Myers++09}. 

\begin{table}
\centering{
\begin{tabular*}{\linewidth}{*{5}c}
\toprule
Target & $\theta_{\rm min}, \theta_{\rm max}$ & $r_0$ ($h^{-1}$Mpc)  & $\gamma$ & $\log_{10}A_w$ \\
\midrule
LRG  &  $0.001^\circ$, $1^\circ$ & $7.78 \pm 0.26$ & $1.98 \pm 0.02$  & $-4.01 \pm 0.05$ \\ 
ELG  &  $0.001^\circ$, $1^\circ$ & $ 5.98 \pm 0.30$ & $1.90 \pm 0.03$ & $-4.25 \pm 0.06$ \\ 
     &  $0.001^\circ$, $0.01^\circ$ & $6.70 \pm 0.10$ & $1.85 \pm 0.01$ & $-4.01 \pm 0.02$ \\ 
     &  $0.05^\circ$, $1^\circ$ & $5.45 \pm 0.10$ & $1.54 \pm 0.01$ & $-3.22 \pm 0.02$ \\ 
QSO  &  $0.001^\circ$, $1^\circ$ & $21.9 \pm 1.01$ & $1.81 \pm 0.03$ & $-3.31 \pm 0.07$ \\ 
\bottomrule
\end{tabular*}}
\caption{Results from fitting the correlation functions to a power-law $\xi(r) = (r/r_0)^{-\gamma}$, where we have used the expected redshift distributions of DESI targets in the Limber approximation to convert between angular clustering and real-space clustering. We list the angular scales fitted over, the expected $\bar{z}$ of each sample, and the best fit parameter values and errors.}
\label{table:fiducial_powerlaw} 
\end{table}

\subsection{Angular power spectra with $b_0, b(k)$ fits}
\label{sec:characterization/cls}
We measure the angular power spectra of the three main DESI target samples using the methods described in Section~\ref{sec:methods/cls}. We reiterate that the results presented here have already had the masks and weights of the previous sections applied. Similar to their effect on the angular correlation functions, the impact of the photometric weights derived in Section~\ref{sec:systematics_densities} on the angular power spectra is to reduce power on large scales. For LRGs, not using the weights would increase the amplitude by $\sim$ 15\% at $\ell \sim 20$ down to $\sim$ 1\% at $\ell \sim 75$. For ELGs, it would increase by $\sim$ 43\% at $\ell \sim 20$ down to $\sim$ 1\% at $\ell \sim 150$. For QSOs, it would increase by $\sim$ 12\% at $\ell \sim 20$ down to $<$ 1\% at $\ell \sim 75$. 

From the angular power spectra, we fit the linear bias. First, we restrict to very large scales where the bias is approximately constant, then relax this restriction as we probe the scale dependence of the bias using the ``$P$ model'' (\citealt{Smith++07}, \citealt{Hamann++08}, \citealt{CresswellPercival09}), which treats the nonlinear correction to the bias as an extra non-Poissonian shot noise term arising from the assumption that galaxies populate halos (\citealt{Seljak01}, \citealt{SchulzWhite06}, \citealt{Guzik++07}):
%
%
%
%
\begin{align}
    P_{g} &\xrightarrow{} P_{g} + P \implies \\ 
    b(z)^2 = \frac{b_0^2}{D(z)^2}  &\xrightarrow{} b(k,z)^2 = \frac{b_0^2}{D(z)^2} \Bigg( 1 + \frac{P}{b_0^2 P_{m}(k,z)} \Bigg)  \nonumber
\end{align}
In terms of the angular power spectra, which involve convolution with the radial distributions (see Equation~\ref{eqn:Cls}), we have
\begin{align}
    &C_{\ell} \xrightarrow{} C_{\ell} + C \nonumber \\ 
    C &= P\int d\chi \ f(\chi)^2 \frac{1}{\chi^2}\frac{1}{D(z)^2}
\end{align}
Using \texttt{CLASS}, we compare $C_{\ell}$ derived from linear and HALOFIT \citep{Smith++03} predictions of the matter power spectrum to estimate $\ell_{\rm max}$ where they begin to diverge, taking the common assumption that the nonlinear correction to the matter power spectrum becomes significant at approximately the same scale as nonlinear effects in the galaxy bias (see e.g. \citealt{FryGaztanaga93}, \citealt{Modi++17}, \citealt{Desjacques++18}, \citealt{Wilson++19}). We find that $\ell_{\rm max} \approx 200$ is appropriate for LRGs and ELGs, and slightly conservative for QSOs, which are at higher mean redshift. 

\subsubsection{LRGs}

\begin{figure}
    \centering
    \includegraphics[width=0.97\columnwidth]{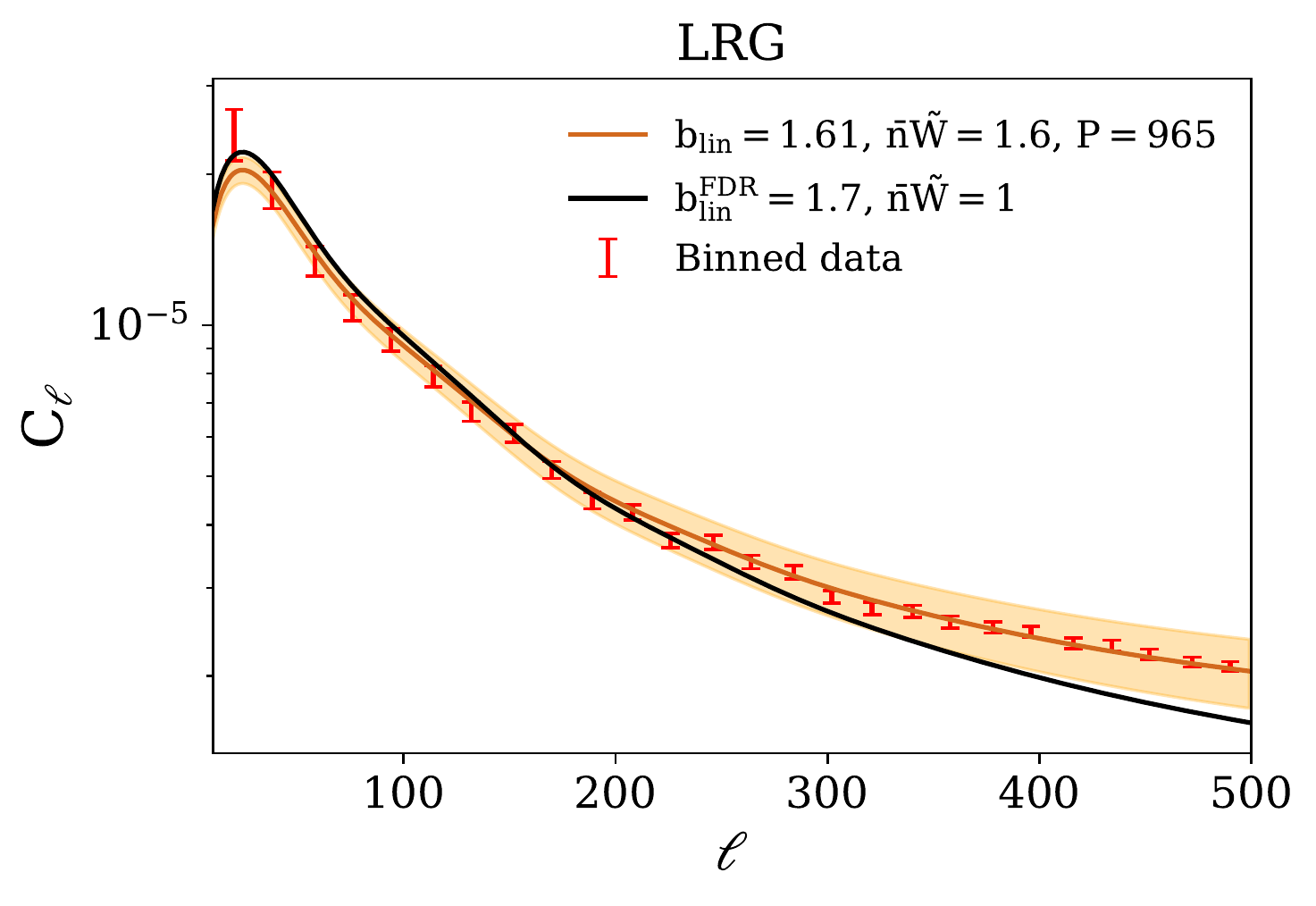}
    \caption{Angular power spectrum $C_{\ell}$ for LRGs. The red error bars are the binned errors from Equation~\ref{eqn:binnedCls}. The solid black line is the theoretical curve using the FDR value for the linear bias. The orange envelope is our best fitting P-model result, with the model errors dominated by uncertainty in the shot noise terms.}
    \label{fig:lrg_cls}
\end{figure}

With $\ell_{\rm max} = 200$, we use the linear theory matter power spectrum and fit the scale-independent bias in two ways: first, by fixing the Poisson shot noise term $\tilde{W} = 1/\bar{n}$ and only fitting $b_0$, and second, by simultaneously fitting $b_0$ and $\tilde{W}$. Then, we extend out to $\ell_{\rm max} = 500$ and add the additional $P$ parameter to our model. We fit P in several ways: both using the previously found values of $b_0$ and $\tilde{W}$ from the $\ell_{\rm max} = 200$ case, and also doing a simultaneous fit to $b_0$ and $P$, with $\tilde{W}$ absorbed into $P$. The results of all fits are shown in Table~\ref{table:lrg_bias} for LRGs, with all models giving similar results and showing agreement with expectation. In Figure~\ref{fig:lrg_cls}, we plot the binned data, the best fit model, and the FDR expectation curve.

\begin{table*}
{\centering
\noindent\begin{tabular}{ccccccccc}
\toprule
 & & \multicolumn{3}{c}{$\ell_{\rm max} = 200$} & \multicolumn{4}{c}{$\ell_{\rm max} = 500$} \\
 & & \multicolumn{3}{c}{scale-independent bias} & \multicolumn{4}{c}{$P$ model} \\
\cmidrule(lr){3-5} \cmidrule(lr){6-9}
Target & $b_0^{\text{FDR}}$ & $b_0$ & $\bar{n}\tilde{W}$ & $\frac{\chi^2}{\text{d.o.f}}$ & $b_0$ & $\bar{n}\tilde{W}$ & P & $\frac{\chi^2}{\text{d.o.f}}$ \\
\midrule
\multirow{3}{*}{LRG} & \multirow{3}{*}{1.7} & $1.570 \pm 0.014$ & 1 (fixed) & 5.0 / 9 & - & - & $3539 \pm 99$ & 12.6 / 25 \\
                     &                      & $1.607 \pm 0.040$ & $1.61 \pm 0.43$ & 3.7 / 8 & - & - & $965 \pm 99$ & 18 / 25 \\
                     &                      & & & & $1.569 \pm 0.017$ & 0 (fixed) & $7120 \pm 198$ & 12.6 / 24 \\
\bottomrule
\end{tabular}} 
\caption{Fitting the LRG large-scale bias from the angular power spectra. We initially limit to $\ell \le 200$ and fit the linear bias to a constant; first, by fixing the Poisson shot noise term as $1/\bar{n}$, and second, by fitting the bias and noise simultaneously. Then we extend to $\ell = 500$ and fit an additional parameter $P$ for the scale-dependent bias model; first, by holding the previously found values fixed and fitting only the non-Poisson shot noise term P at larger $\ell$, and second, by fitting the linear bias and $P$ simultaneously, with $P$ now absorbing both shot noise terms.}
\label{table:lrg_bias}
\end{table*}

\subsubsection{ELGs}

For ELGs, we find that attempting to co-fit the bias and shot noise terms simultaneously returns unphysical negative values for the latter, due to enhanced power at scales $\ell < 150$ even after applying masks and weights (fortunately, these scales should not directly impact BAO and RSD measurements). However, when fixing the shot noise as $\tilde{W} = 1/\bar{n}$, we obtain $b_0 = 1.273 \pm 0.005$, which agrees well with e.g. \cite{Comparat++13}, \cite{Delubac++17}. 

\begin{figure}
    \centering
    \includegraphics[width=0.97\columnwidth]{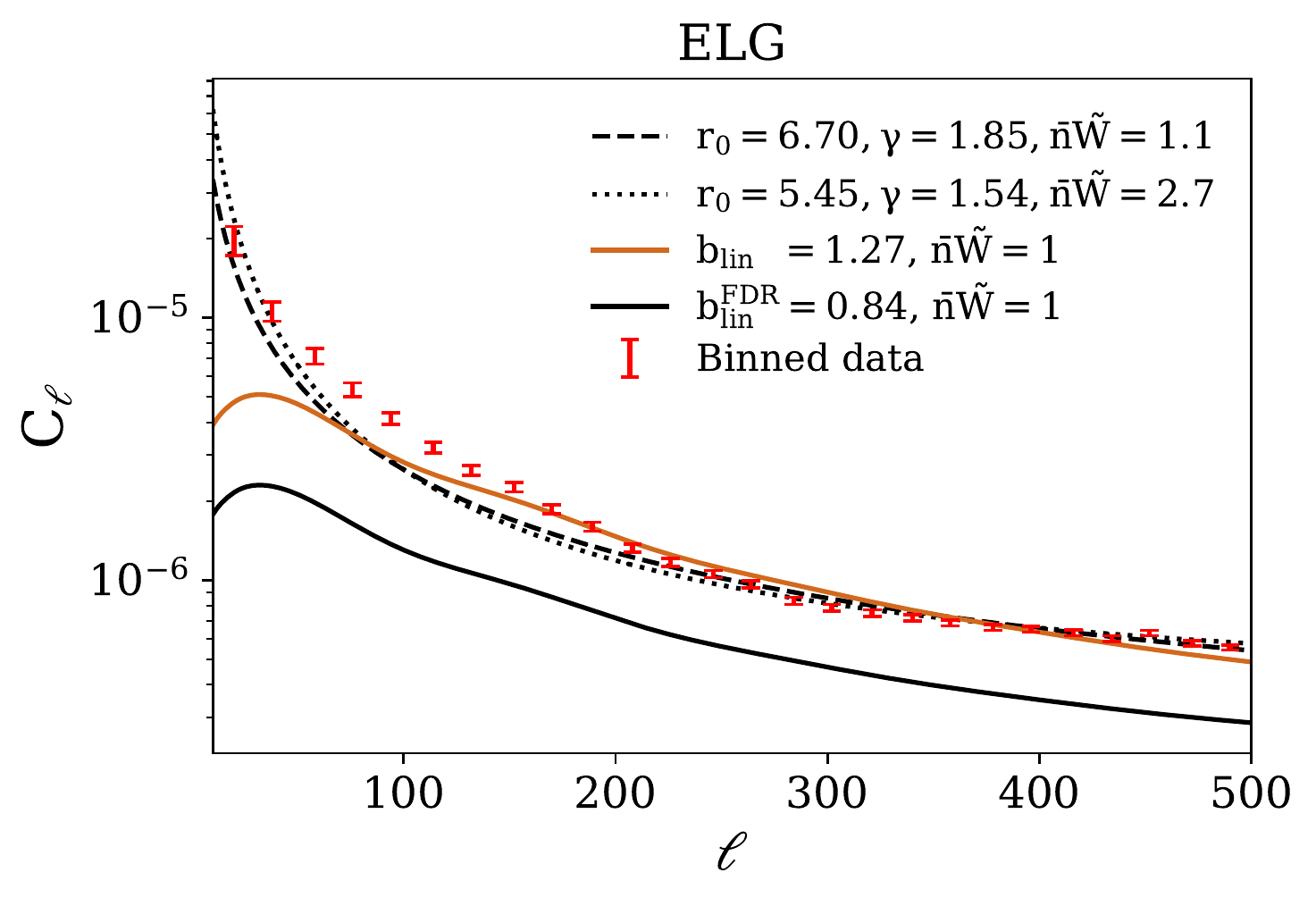}
    \caption[]{Angular power spectrum $C_{\ell}$ for ELGs. The red error bars are the binned errors from Equation~\ref{eqn:binnedCls}. The solid black line is the theoretical curve using the FDR value for the linear bias. The orange envelope\footnotemark{} is our best fit for the linear bias, using fixed shot noise $\tilde{W} = 1/\bar{n}$. The dashed and dotted lines are the angular power spectra corresponding to the power-law fits to the angular clustering determined in Section~\ref{sec:characterization/corrfuns}, with the shot noise terms fit as extra free parameters.}
    \label{fig:elg_cls}
\end{figure}

We also calculate the corresponding $C_{\ell}$ for each of the two power-law model $w(\theta)$ fits in Figure~\ref{fig:final-wtheta} and plot these as well in Figure~\ref{fig:elg_cls}, with shot noise contributions fitted as additional free parameters. The results show consistency between our $w(\theta)$ and $C_{\ell}$ results, both of which give clustering parameters falling within the range of fiducial values found in previous studies. 

The DESI FDR assumes a conservative lower limit of $b_0 = 0.84$, also plotted in Figure~\ref{fig:elg_cls}, and we confirm that the ELG clustering bias is higher than this value. This is significant as it has the effect of improving the statistical errors on BAO, while also somewhat degrading the RSD forecasts, since more strongly biased tracers exhibit weaker anisotropy. We note that allowing the shot noise term to float in the FDR curve in order to raise its amplitude still results in a very poor fit, as it flattens the curve such that it can only achieve artificial agreement with observation at very large $\ell$.

\footnotetext{Since uncertainty in the shot noise terms would normally dominate the model errors (see Figure~\ref{fig:lrg_cls}), the ``envelope'' with fixed shot noise appears very thin.}

\subsubsection{QSOs}

\begin{figure}
    \centering
    \includegraphics[width=0.97\columnwidth]{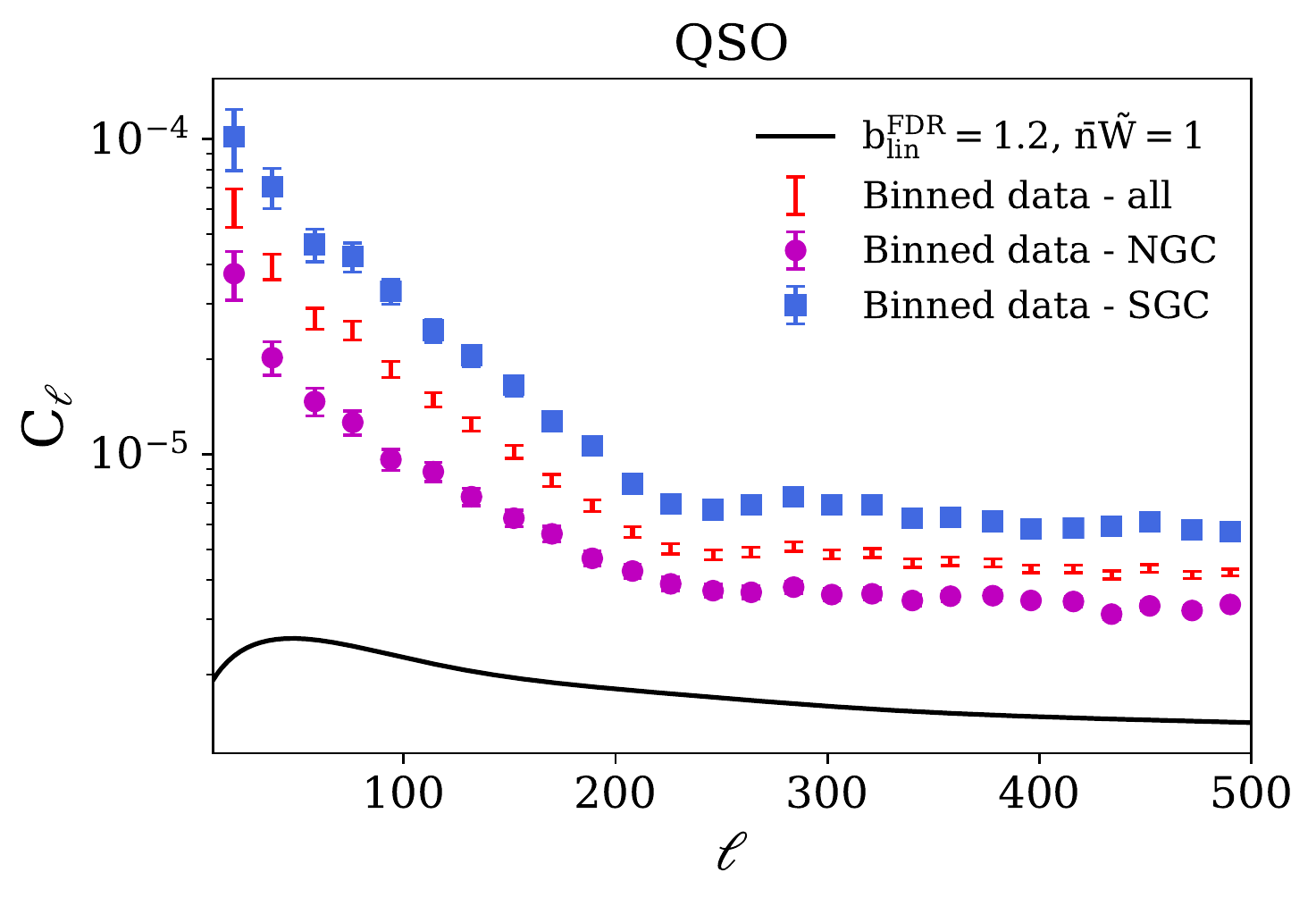}
    \caption{Angular power spectrum $C_{\ell}$ for QSOs. The red error bars are the binned errors from Equation~\ref{eqn:binnedCls}. The magenta circles and blue squares are for measurements restricted to the NGC and SGC, respectively. The solid black line is the theoretical curve using the FDR value for the linear bias and fixed shot noise $\tilde{W} = 1/\bar{n}$.}
    \label{fig:qso_cls}
\end{figure}

For QSOs, the angular power spectrum, like the angular correlation function, is significantly inflated with non-cosmological signals. As such, we do not report fitted values, but merely plot the results in Figure~\ref{fig:qso_cls} alongside FDR expectation to demonstrate the discrepancy. Furthermore, whereas the LRGs and ELGs show no difference when comparing NGC and SGC measurements, QSOs once again show a mismatch between galactic hemispheres; after removing the problematic pixels found in Section~\ref{sec:spatial}, which caused disproportionately strong small-scale clustering in the NGC, the shapes of the angular power spectra in the NGC and SGC become identical, but the SGC is enhanced on all scales compared to the NGC. This is consistent with our other findings, namely that the SGC has more stellar contamination than the NGC and that the clustering of these stellar contaminants is relatively flat across scales (Section~\ref{sec:stellar_contam}). 

\subsection{Counts-in-cells moments}
\label{sec:characterization/cic}
As discussed in Section~\ref{sec:methods/cic}, detailed small-scale clustering information is invaluable for accurate modelling and mock calibration.

\begin{figure}
	\centering
	\includegraphics[width=\columnwidth]{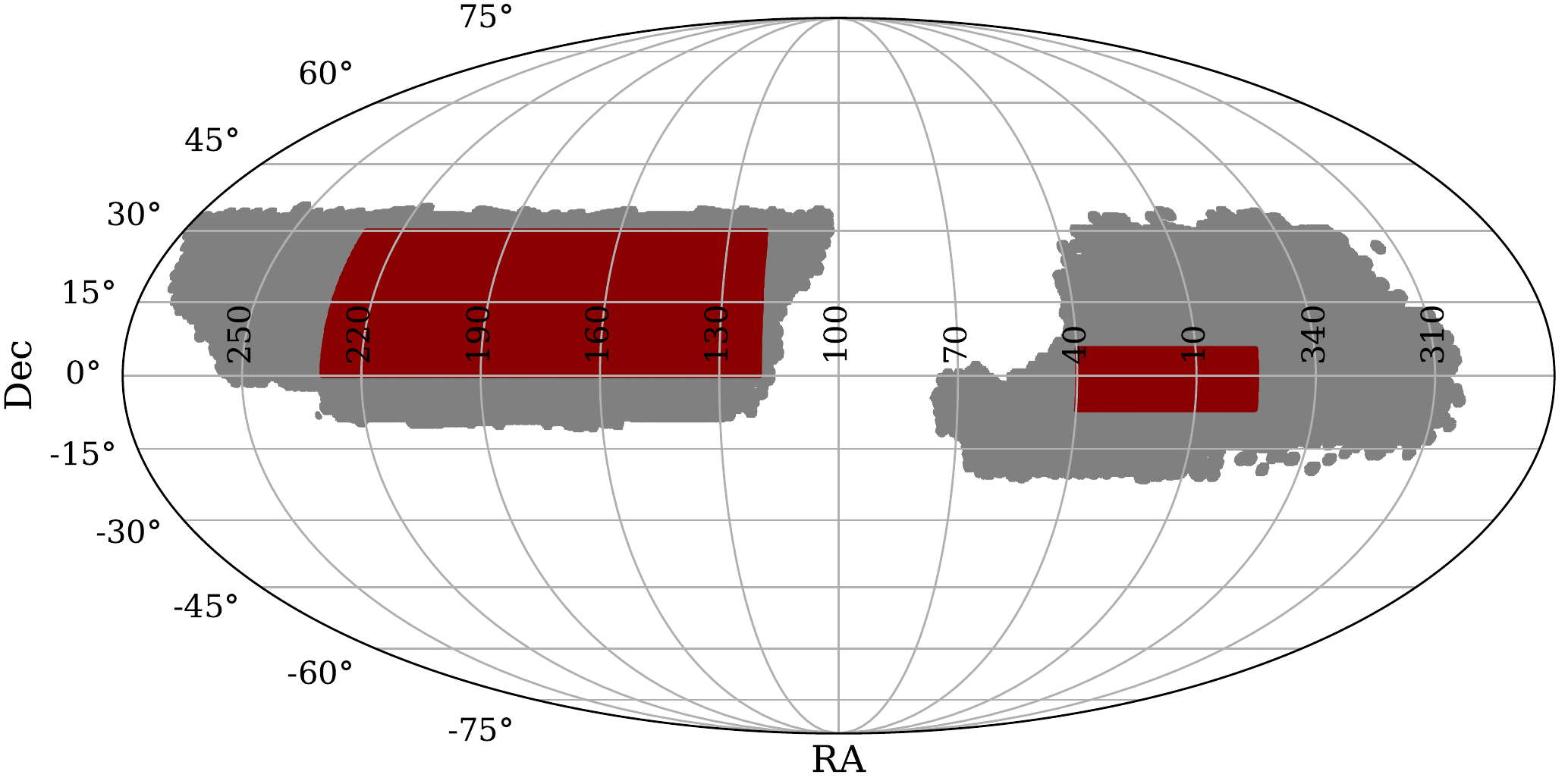}
    \caption{Two patches over which we calculate the counts-in-cells distributions and moments. These  were chosen by eye as survey regions with no tattered edges and relatively few holes.}
\label{fig:cic_patches}
\end{figure}

We calculate the counts-in-cells statistics over two large fields, one in each galactic hemisphere, with effective areas $S_{\text{eff},N} = 3300$ deg$^2$ and $S_{\text{eff},S} = 562.5$ deg$^2$ (Figure~\ref{fig:cic_patches}). We select these fields to be regular in shape and relatively smooth, avoiding areas that are tattered or full of holes. We calculate the probability distribution $P(N)$ for each field, then measure the weighted average and standard error of the factorial moments $F_{p}$ \citep{Wolk++13}:
\begin{align*}
\bar{F}_{p} &= \frac{ \sum\limits_{i=N,S}^{}S_{\text{eff},i}F_{p,i}}{ \sum\limits_{i=N,S}^{}S_{\text{eff},i}} \\
(\Delta F_p)^2 &= \frac{ \sum\limits_{i=N,S}^{}S_{\text{eff},i}(F_{p,i} - \bar{F}_{p})^2}{
  \sum\limits_{i=N,S}^{}S_{\text{eff},i}} 
\end{align*}
Following the reasoning of \cite{Wolk++13}, we do not perform the complex error propagation from factorial moments to correlation functions, as the error estimate is already a crude approximation limited by the small number of fields. 

\begin{figure}
	\centering
	\includegraphics[width=\columnwidth]{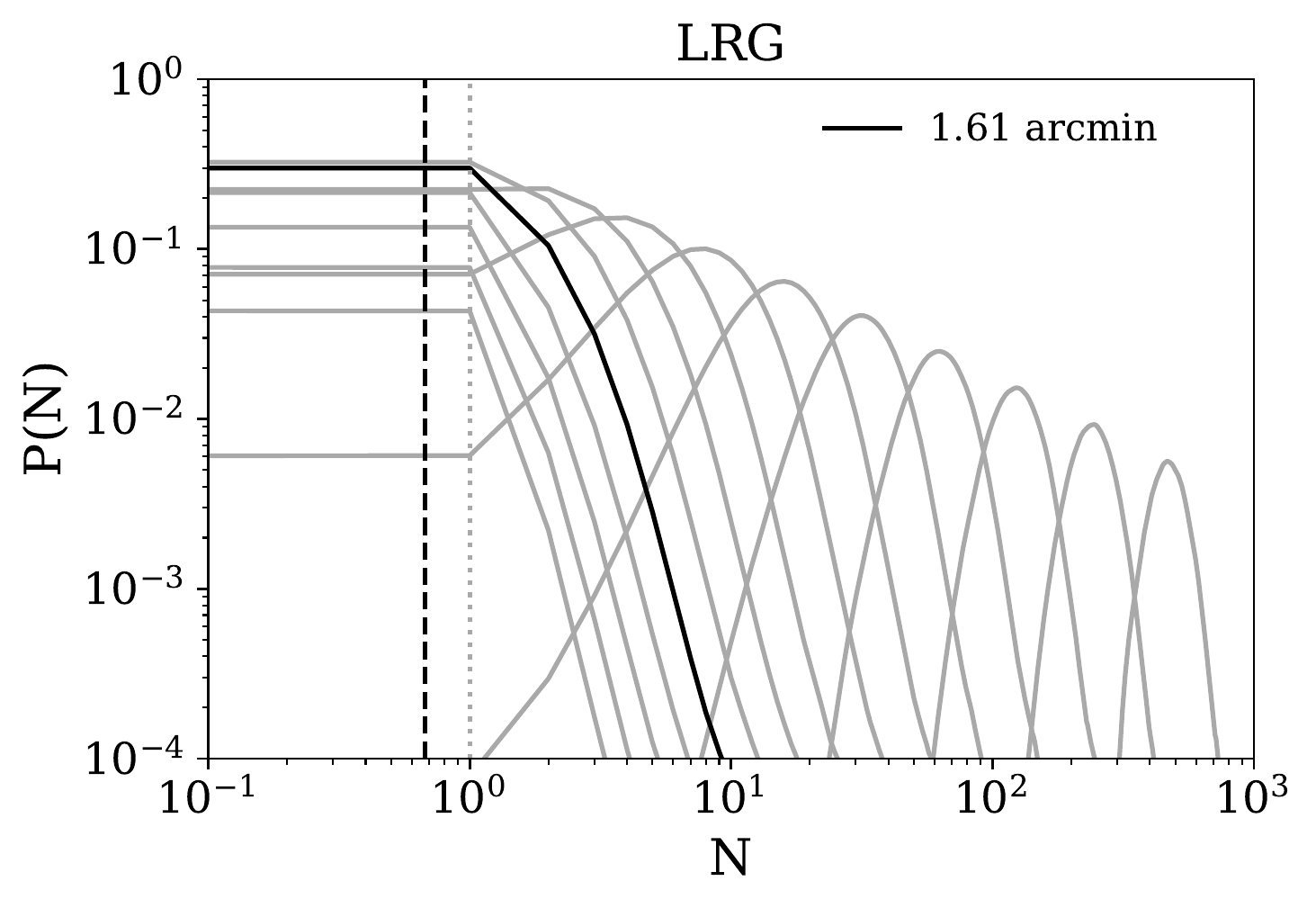}
    \includegraphics[width=\columnwidth]{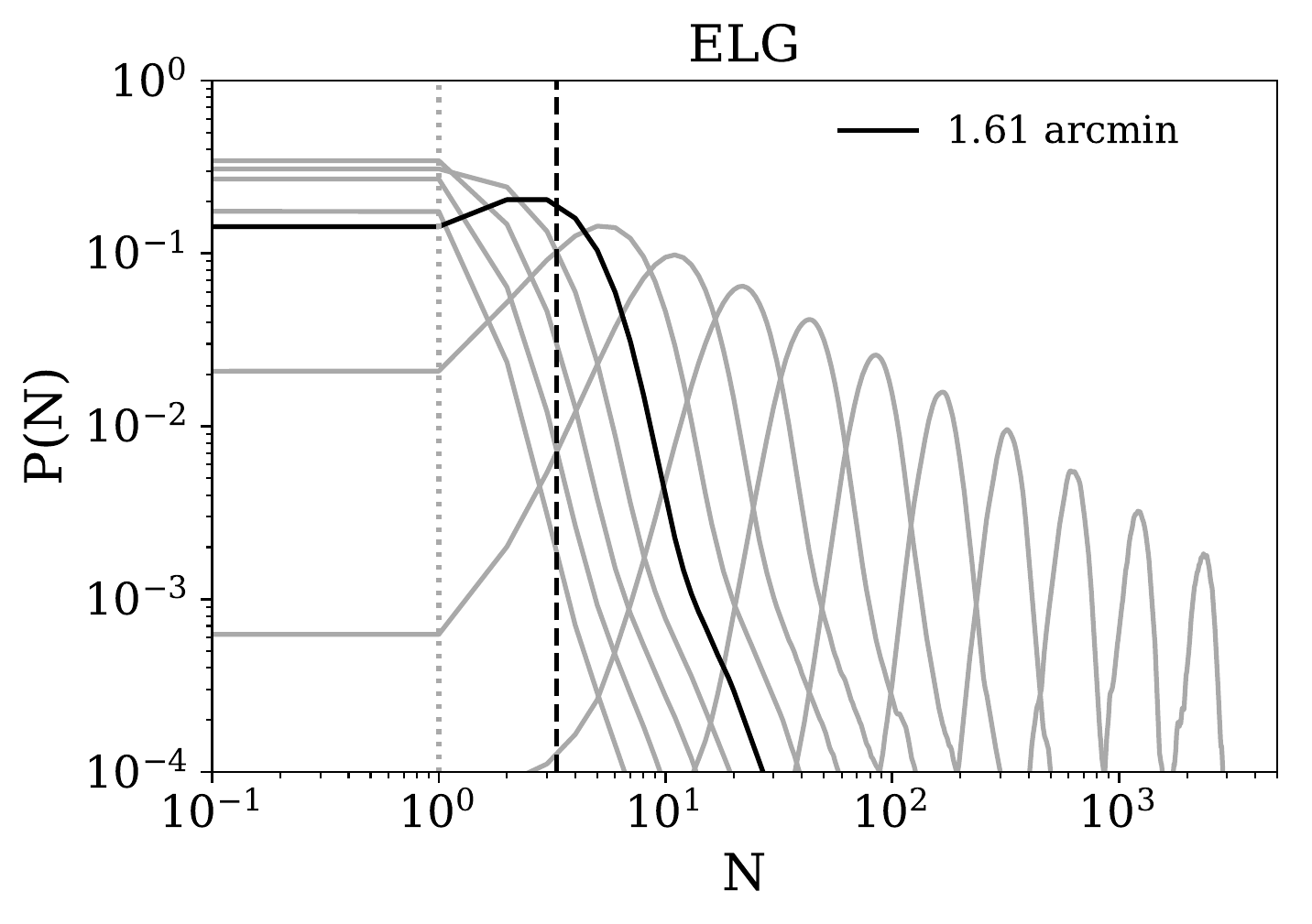}	
    \includegraphics[width=\columnwidth]{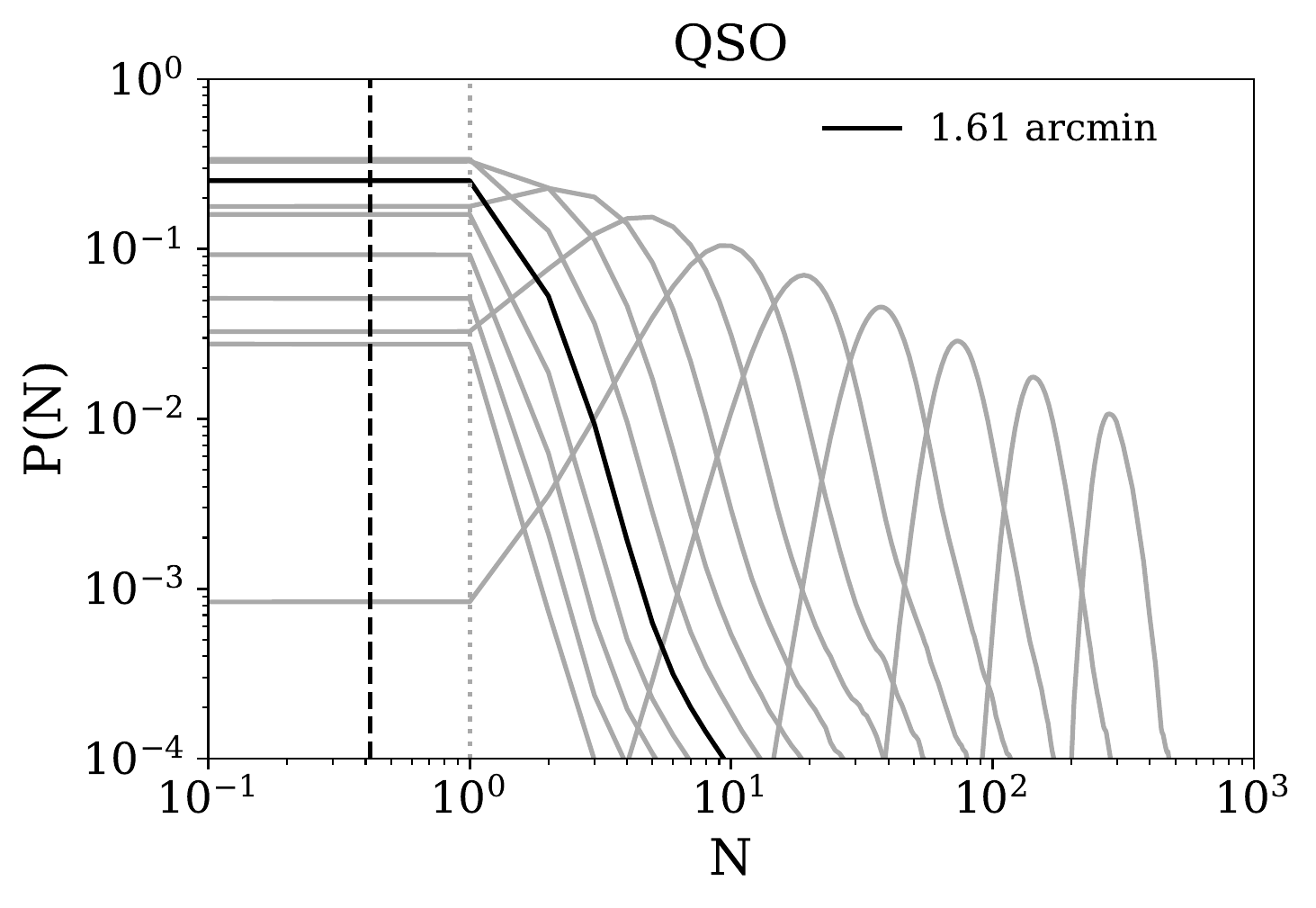}	
	\caption{$P(N)$ vs. $N$ for 15 logarithmically spaced cell widths from $\theta = 0.01^{\circ}$ to $\theta = 1^{\circ}$. The highlighted cell has width $0.268^{\circ} \approx 1.61$ arcmin, close to the fiber patrol radius of $1.4$ arcmin, and the dashed vertical lines correspond to the average target density times the cell area. The dotted vertical line at $N=1$ marks the limit where shot noise dominates ($\bar{N} < 1$).}
\label{fig:cic_pn}
\end{figure}

Figure~\ref{fig:cic_pn} shows the probability distributions $P(N)$ evaluated in square cells with different widths from $\theta = 0.01^\circ$ to $\theta = 1^\circ$ for each target class, with a cell width close to the fiber patrol radius ($1.4^\prime \approx 0.023^{\circ}$) highlighted. The dashed vertical line drawn for this highlighted cell represents the expected number of targets calculated from multiplying the mean target density with the cell area. Note that even for ELGs, the first few angular bins are shot-noise dominated ($\bar{N} < 1$ so most cells contain one or zero targets).

Table~\ref{table:cic_all} presents the following quantities for each of the three main target classes: 
\begin{itemize}[leftmargin=0.4cm]
    \item mean $\bar{N} \equiv \langle N \rangle$
    \item variance $\sigma^2 \equiv \langle (N - \bar{N})^2 \rangle$
    \item skewness $\langle (N - \bar{N})^3 \rangle/\sigma^3$
    \item kurtosis $\langle (N - \bar{N})^4 \rangle/\sigma^4$ 
    \item cell-averaged angular correlation functions $\overline{w}_2$, $\overline{w}_3$, $\overline{w}_4$
\end{itemize}


From the above quantities, other quantities of interest can be determined, such as the  hierarchical moments $S_p = \overline{w}_p / \overline{w}_2^{p-1}$ (\citealt{SzapudiSzalay93b}, \citealt{ColombiSzapudi01}), fitted power law parameters for $\overline{w}_2$ (see e.g. \citealt{BlakeWall02}), etc. More directly, these quantities can be used for the training and testing of mock catalogs.

\begin{table*}
\begin{tabular}{ccccccccc}
 & Width & $\bar{N}$ & $\sigma^2$ & $\frac{\langle (N - \bar{N})^3 \rangle}{\sigma^3}$ & $\frac{\langle (N - \bar{N})^4 \rangle}{\sigma^4}$ & $\overline{w}_2$ & $\overline{w}_3$ & $\overline{w}_4$ \\
\hline
\textbf{LRG} & $0.010^{\circ}$ & $0.0490 \pm 0.0019$ & $0.0550 \pm 0.0023$ & $7.50 \pm 0.23$ & $260 \pm 35$ & $2.410 \pm 0.014$ & $250 \pm 18$ & $88000 \pm 8100$ \\
 & $0.014^{\circ}$ & $0.0950 \pm 0.0036$ & $0.1120 \pm 0.0047$ & $7.30 \pm 0.35$ & $340 \pm 49$ & $1.870 \pm 0.022$ & $146.0 \pm 9.4$ & $40000 \pm 3400$ \\
 & $0.019^{\circ}$ & $0.1840 \pm 0.0069$ & $0.2320 \pm 0.0097$ & $7.50 \pm 0.45$ & $410 \pm 58$ & $1.420 \pm 0.025$ & $81.0 \pm 4.3$ & $16000 \pm 1200$ \\
 & $0.027^{\circ}$ & $0.350 \pm 0.013$ & $0.490 \pm 0.020$ & $8.10 \pm 0.48$ & $470 \pm 55$ & $1.060 \pm 0.025$ & $45.0 \pm 1.5$ & $6200 \pm 300$ \\
 & $0.037^{\circ}$ & $0.680 \pm 0.026$ & $1.050 \pm 0.042$ & $9.10 \pm 0.45$ & $550 \pm 44$ & $0.780 \pm 0.024$ & $24.90 \pm 0.21$ & $2520 \pm 21$ \\
 & $0.052^{\circ}$ & $1.320 \pm 0.050$ & $2.290 \pm 0.086$ & $10.00 \pm 0.38$ & $620 \pm 29$ & $0.560 \pm 0.020$ & $13.10 \pm 0.15$ & $990 \pm 26$ \\
 & $0.072^{\circ}$ & $2.550 \pm 0.096$ & $5.10 \pm 0.18$ & $10.10 \pm 0.22$ & $617.0 \pm 4.3$ & $0.390 \pm 0.015$ & $6.30 \pm 0.21$ & $360 \pm 28$ \\
 & $0.100^{\circ}$ & $4.90 \pm 0.19$ & $11.40 \pm 0.36$ & $9.200 \pm 0.090$ & $500 \pm 22$ & $0.270 \pm 0.012$ & $2.80 \pm 0.13$ & $110 \pm 12$ \\
 & $0.139^{\circ}$ & $9.50 \pm 0.36$ & $26.00 \pm 0.75$ & $7.550 \pm 0.016$ & $330 \pm 20$ & $0.1820 \pm 0.0086$ & $1.090 \pm 0.066$ & $27.0 \pm 3.5$ \\
 & $0.193^{\circ}$ & $18.40 \pm 0.70$ & $61.0 \pm 1.6$ & $5.710 \pm 0.016$ & $190 \pm 12$ & $0.1260 \pm 0.0063$ & $0.410 \pm 0.028$ & $5.80 \pm 0.83$ \\
 & $0.268^{\circ}$ & $35.0 \pm 1.3$ & $145.0 \pm 3.2$ & $4.070 \pm 0.049$ & $95.0 \pm 6.9$ & $0.0880 \pm 0.0047$ & $0.150 \pm 0.012$ & $1.20 \pm 0.19$ \\
 & $0.373^{\circ}$ & $69.0 \pm 2.6$ & $359.0 \pm 6.0$ & $2.790 \pm 0.093$ & $45.0 \pm 4.2$ & $0.0620 \pm 0.0036$ & $0.0560 \pm 0.0060$ & $0.250 \pm 0.044$ \\
 & $0.518^{\circ}$ & $132.0 \pm 5.1$ & $906.0 \pm 8.7$ & $1.90 \pm 0.13$ & $21.0 \pm 2.6$ & $0.0440 \pm 0.0029$ & $0.0210 \pm 0.0031$ & $0.050 \pm 0.011$ \\
 & $0.720^{\circ}$ & $256.0 \pm 9.8$ & $2338.0 \pm 5.3$ & $1.30 \pm 0.14$ & $11.0 \pm 1.6$ & $0.0320 \pm 0.0024$ & $0.0080 \pm 0.0016$ & $0.0100 \pm 0.0028$ \\
 & $1.000^{\circ}$ & $490 \pm 19$ & $6110 \pm 99$ & $0.90 \pm 0.13$ & $6.0 \pm 1.1$ & $0.0230 \pm 0.0020$ & $0.00360 \pm 0.00083$ & $0.00230 \pm 0.00080$ \\
\cmidrule{1-9}
\textbf{ELG} & $0.010^{\circ}$ & $0.2350 \pm 0.0030$ & $0.290 \pm 0.010$ & $5.0 \pm 1.2$ & $90 \pm 47$ & $1.10 \pm 0.27$ & $30 \pm 15$ & $2000 \pm 1100$ \\
 & $0.014^{\circ}$ & $0.4530 \pm 0.0058$ & $0.630 \pm 0.034$ & $5.0 \pm 1.5$ & $110 \pm 56$ & $0.90 \pm 0.21$ & $19.0 \pm 9.4$ & $800 \pm 520$ \\
 & $0.019^{\circ}$ & $0.870 \pm 0.011$ & $1.40 \pm 0.10$ & $6.0 \pm 1.8$ & $120 \pm 61$ & $0.70 \pm 0.16$ & $12.0 \pm 5.7$ & $400 \pm 230$ \\
 & $0.027^{\circ}$ & $1.690 \pm 0.022$ & $3.20 \pm 0.30$ & $7.0 \pm 2.0$ & $140 \pm 67$ & $0.50 \pm 0.12$ & $7.0 \pm 3.4$ & $200 \pm 100$ \\
 & $0.037^{\circ}$ & $3.260 \pm 0.042$ & $7.30 \pm 0.79$ & $7.0 \pm 2.2$ & $160 \pm 75$ & $0.380 \pm 0.088$ & $4.0 \pm 1.9$ & $80 \pm 49$ \\
 & $0.052^{\circ}$ & $6.290 \pm 0.080$ & $17.0 \pm 1.9$ & $7.0 \pm 2.2$ & $150 \pm 72$ & $0.260 \pm 0.055$ & $1.90 \pm 0.92$ & $30 \pm 18$ \\
 & $0.072^{\circ}$ & $12.10 \pm 0.15$ & $37.0 \pm 3.7$ & $6.0 \pm 1.8$ & $110 \pm 54$ & $0.170 \pm 0.030$ & $0.80 \pm 0.35$ & $7.0 \pm 4.6$ \\
 & $0.100^{\circ}$ & $23.40 \pm 0.30$ & $83.0 \pm 5.6$ & $5.0 \pm 1.2$ & $70 \pm 30$ & $0.110 \pm 0.013$ & $0.30 \pm 0.11$ & $1.60 \pm 0.92$ \\
 & $0.139^{\circ}$ & $45.30 \pm 0.57$ & $192.0 \pm 3.3$ & $4.00 \pm 0.50$ & $50 \pm 11$ & $0.0720 \pm 0.0036$ & $0.110 \pm 0.021$ & $0.40 \pm 0.12$ \\
 & $0.193^{\circ}$ & $87.0 \pm 1.1$ & $460 \pm 24$ & $3.40 \pm 0.25$ & $34.0 \pm 3.4$ & $0.0490 \pm 0.0018$ & $0.0480 \pm 0.0058$ & $0.110 \pm 0.019$ \\
 & $0.268^{\circ}$ & $169.0 \pm 2.1$ & $1200 \pm 170$ & $3.0 \pm 1.0$ & $30 \pm 15$ & $0.0350 \pm 0.0049$ & $0.030 \pm 0.013$ & $0.050 \pm 0.038$ \\
 & $0.373^{\circ}$ & $326.0 \pm 3.9$ & $3100 \pm 760$ & $3.0 \pm 1.7$ & $30 \pm 24$ & $0.0260 \pm 0.0065$ & $0.020 \pm 0.014$ & $0.030 \pm 0.034$ \\
 & $0.518^{\circ}$ & $628.0 \pm 7.3$ & $9000 \pm 3000$ & $3.0 \pm 2.1$ & $30 \pm 28$ & $0.0200 \pm 0.0071$ & $0.010 \pm 0.013$ & $0.020 \pm 0.027$ \\
 & $0.720^{\circ}$ & $1210 \pm 14$ & $30000 \pm 11000$ & $3.0 \pm 2.3$ & $30 \pm 29$ & $0.0170 \pm 0.0071$ & $0.010 \pm 0.010$ & $0.020 \pm 0.020$ \\
 & $1.000^{\circ}$ & $2340 \pm 25$ & $80000 \pm 39000$ & $2.0 \pm 2.2$ & $20 \pm 25$ & $0.0140 \pm 0.0068$ & $0.0070 \pm 0.0079$ & $0.010 \pm 0.012$ \\
\cmidrule{1-9}
\textbf{QSO} & $0.010^{\circ}$ & $0.02990 \pm 0.00011$ & $0.03700 \pm 0.00054$ & $18.0 \pm 1.6$ & $1500 \pm 290$ & $7.90 \pm 0.42$ & $2900 \pm 440$ & $1900000 \pm 420000$ \\
 & $0.014^{\circ}$ & $0.05780 \pm 0.00022$ & $0.0800 \pm 0.0018$ & $25.0 \pm 3.2$ & $3000 \pm 650$ & $6.70 \pm 0.43$ & $2300 \pm 410$ & $1500000 \pm 360000$ \\
 & $0.019^{\circ}$ & $0.11160 \pm 0.00042$ & $0.1820 \pm 0.0064$ & $36.0 \pm 5.7$ & $5000 \pm 1300$ & $5.70 \pm 0.44$ & $1800 \pm 370$ & $1100000 \pm 300000$ \\
 & $0.027^{\circ}$ & $0.21550 \pm 0.00081$ & $0.440 \pm 0.022$ & $49.0 \pm 8.9$ & $9000 \pm 2300$ & $4.70 \pm 0.44$ & $1300 \pm 310$ & $800000 \pm 230000$ \\
 & $0.037^{\circ}$ & $0.4160 \pm 0.0016$ & $1.080 \pm 0.077$ & $60 \pm 12$ & $12000 \pm 3400$ & $3.80 \pm 0.41$ & $900 \pm 240$ & $500000 \pm 150000$ \\
 & $0.052^{\circ}$ & $0.8030 \pm 0.0030$ & $2.70 \pm 0.25$ & $70 \pm 15$ & $13000 \pm 4200$ & $2.90 \pm 0.37$ & $600 \pm 170$ & $240000 \pm 85000$ \\
 & $0.072^{\circ}$ & $1.5510 \pm 0.0058$ & $6.70 \pm 0.76$ & $70 \pm 17$ & $12000 \pm 4200$ & $2.10 \pm 0.30$ & $300 \pm 100$ & $100000 \pm 38000$ \\
 & $0.100^{\circ}$ & $2.990 \pm 0.011$ & $16.0 \pm 2.1$ & $70 \pm 16$ & $10000 \pm 3300$ & $1.50 \pm 0.22$ & $170 \pm 54$ & $40000 \pm 13000$ \\
 & $0.139^{\circ}$ & $5.780 \pm 0.021$ & $39.0 \pm 5.2$ & $60 \pm 13$ & $7000 \pm 2300$ & $1.00 \pm 0.15$ & $80 \pm 24$ & $11000 \pm 4000$ \\
 & $0.193^{\circ}$ & $11.160 \pm 0.040$ & $90 \pm 12$ & $50 \pm 10$ & $5000 \pm 1500$ & $0.640 \pm 0.094$ & $34.0 \pm 9.9$ & $3000 \pm 1100$ \\
 & $0.268^{\circ}$ & $21.550 \pm 0.077$ & $210 \pm 27$ & $42.0 \pm 7.5$ & $3400 \pm 930$ & $0.400 \pm 0.055$ & $13.0 \pm 3.6$ & $700 \pm 250$ \\
 & $0.373^{\circ}$ & $41.60 \pm 0.15$ & $450 \pm 59$ & $32.0 \pm 5.4$ & $2000 \pm 510$ & $0.240 \pm 0.033$ & $4.0 \pm 1.2$ & $140 \pm 49$ \\
 & $0.518^{\circ}$ & $80.30 \pm 0.29$ & $1000 \pm 130$ & $23.0 \pm 3.7$ & $1000 \pm 260$ & $0.140 \pm 0.019$ & $1.40 \pm 0.38$ & $25.0 \pm 8.5$ \\
 & $0.720^{\circ}$ & $155.20 \pm 0.56$ & $2100 \pm 270$ & $16.0 \pm 2.5$ & $500 \pm 130$ & $0.080 \pm 0.011$ & $0.40 \pm 0.11$ & $4.0 \pm 1.4$ \\
 & $1.000^{\circ}$ & $300.0 \pm 1.1$ & $4500 \pm 570$ & $10.0 \pm 1.5$ & $230 \pm 54$ & $0.0460 \pm 0.0061$ & $0.120 \pm 0.031$ & $0.60 \pm 0.20$ \\
\hline
\end{tabular} 
\caption{Mean, variance, skewness, kurtosis, and cell-averaged $n$-point angular correlation functions for each of the three main DESI target classes measured in square cells.}
\label{table:cic_all}
\end{table*}




\subsection{Clustering as a function of magnitude}
\label{sec:characterization/magdep}
Angular clustering is expected to scale with sample depth \citep{Peebles80}, so analyzing $w(\theta)$ as a function of magnitude provides another test for the presence of systematics.

\begin{figure}
\centering
\includegraphics[width=0.99\linewidth]{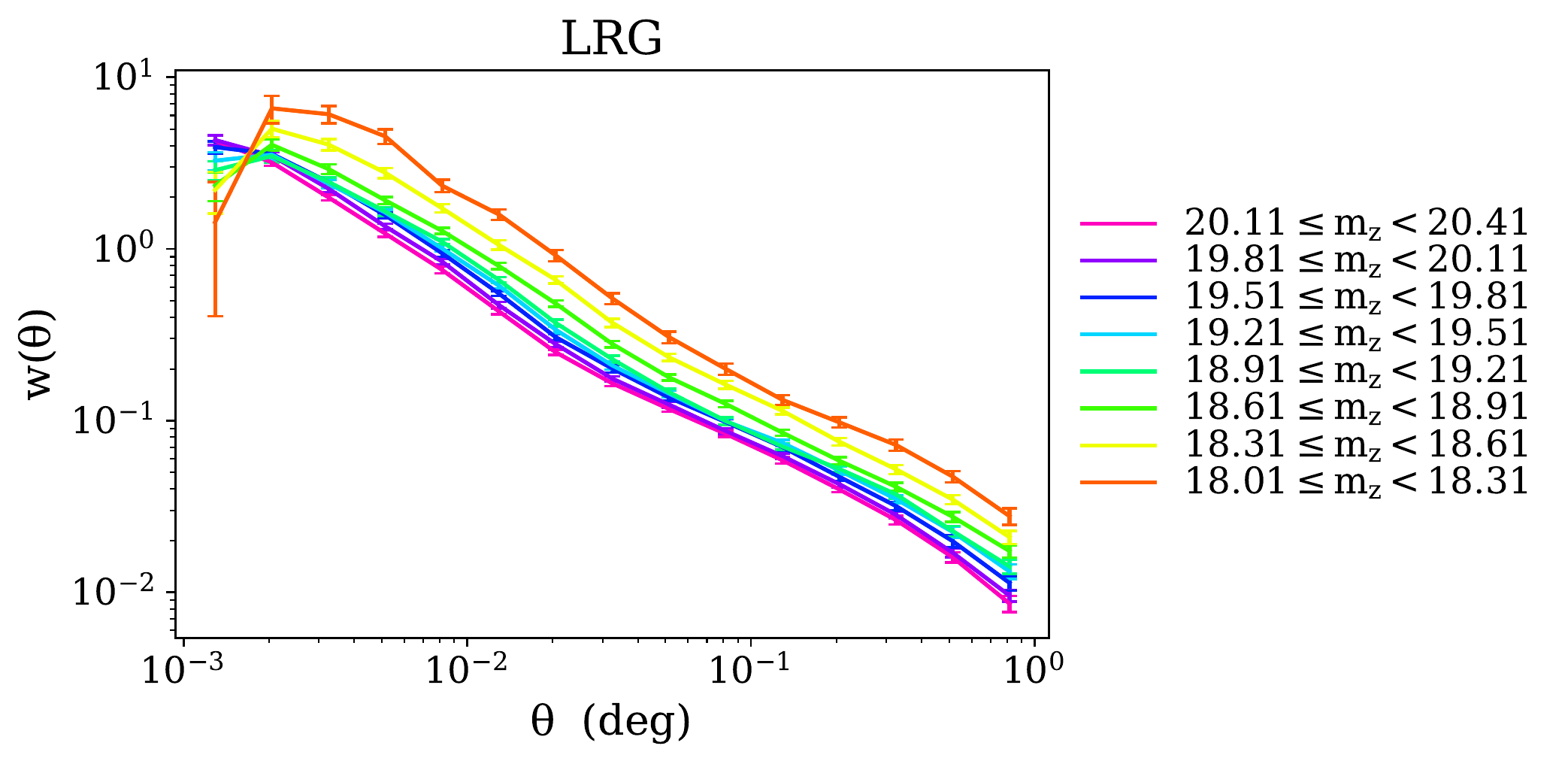}
\vspace{0.1cm}
\includegraphics[width=0.99\linewidth, trim=0 0 0 0.8cm, clip]{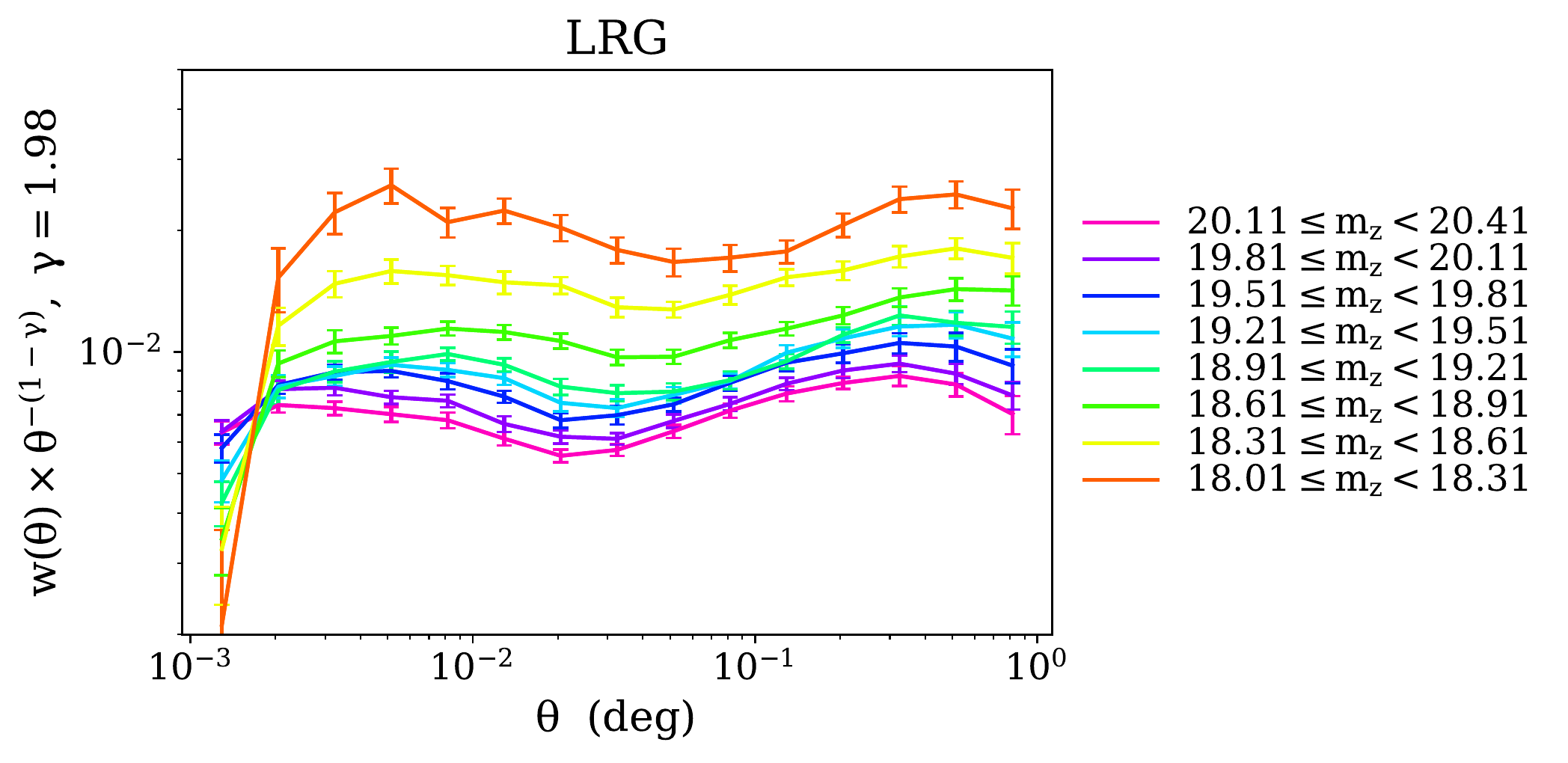}
\caption{Angular correlation functions for LRGs in eight magnitude bins (upper plot), with the angular dependence scaled out for a fixed slope of $\gamma = 1.98$ (lower plot), which is the slope determined from fitting over the full LRG sample in Table~\ref{table:fiducial_powerlaw}.}
\label{fig:lrg-magbin-wtheta}
\end{figure}

\begin{figure}
\centering
\includegraphics[width=0.99\linewidth]{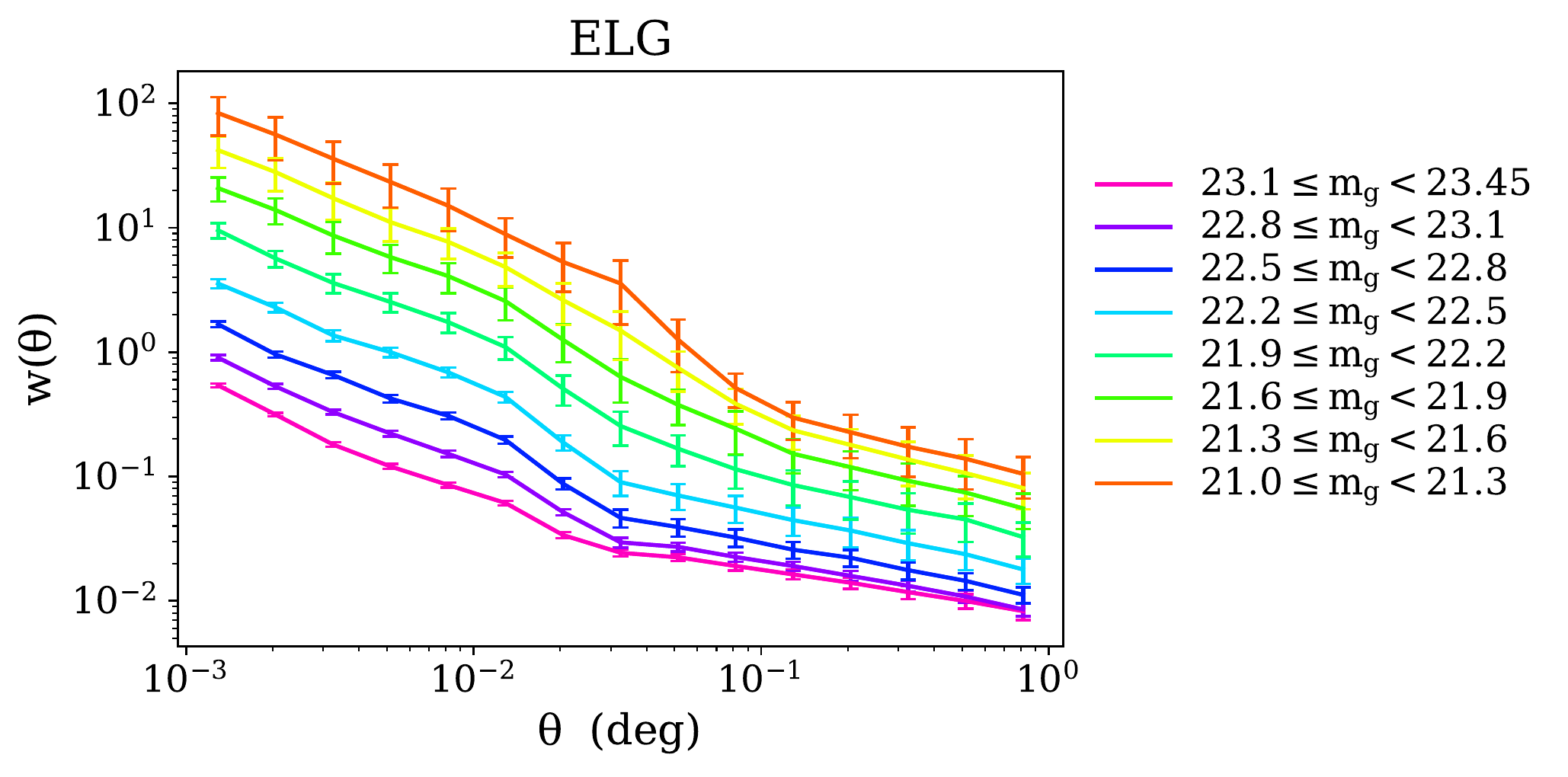}
\vspace{0.1cm}
\includegraphics[width=0.99\linewidth, trim=0 0 0 0.8cm, clip]{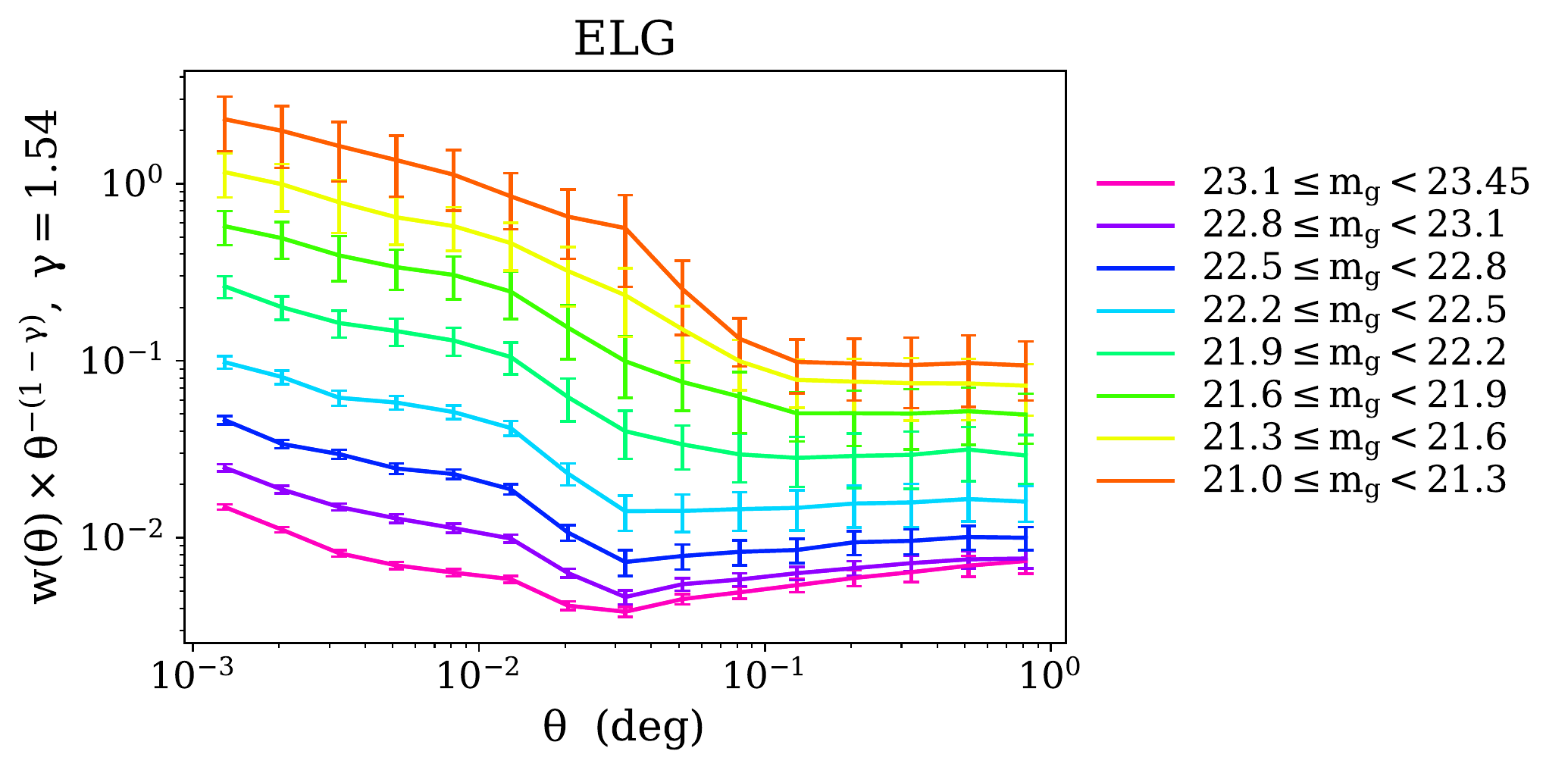}
\caption{Angular correlation functions for ELGs in eight magnitude bins (upper plot), with the angular dependence scaled out for a fixed slope of $\gamma = 1.54$ (lower plot), which is the slope determined from fitting over the full ELG sample in Table~\ref{table:fiducial_powerlaw}.}
\label{fig:elg-magbin-wtheta}
\end{figure}

We divide the LRG and ELG samples into eight disjoint, equally wide magnitude slices, from the bright limit ($m_z = 18.01$ for LRGs, $m_g = 21$ for ELGs) to the faint limit ($m_z = 20.41$ for LRGs, $m_g = 23.4$ for ELGs) of the target selection. For each bin, we evaluate $w(\theta)$\footnote{Note, we do not re-evaluate the photometric weights for each magnitude bin but instead apply the same weights to all bins.}. The results are shown in Figures~\ref{fig:lrg-magbin-wtheta} and Figures~\ref{fig:elg-magbin-wtheta}, with the upper plots showing $w(\theta)$ for each slice, and the lower plots showing the same functions but with the angular dependence divided out using a representative value of $\gamma$ determined from fitting the full sample (Table~\ref{table:fiducial_powerlaw}). 

Qualitatively, the results match expectation: the brighter subsamples have larger clustering amplitudes and break from the power law form at larger scales. We also note that, for both LRGs and ELGs, the minimum inflection slides to smaller scales at fainter magnitudes, again consistent with a 1-halo to 2-halo transition; fainter bins are at higher redshift (so the characteristic scale of the transition will shift to smaller angles due to the larger angular diameter distance) and/or lower luminosity (thus the 1-halo term will be weaker, as less luminous galaxies reside in less massive halos).

The best fit values for the clustering amplitudes $A_w$ and slopes $\gamma$ are reported in Tables~\ref{table:lrg-magbinned} and \ref{table:elg-magbinned}. We fit $A_w$ with the fixed representative value of $\gamma$, and also fit $A_w$ and $\gamma$ simultaneously, finding similar results in either case. We also perform the fits over different sets of angular scales, starting with a minimum cutoff of $\theta = 0.005$ for LRGs to avoid scales where the power law model appears to break down, and obtaining fits for $\theta < 0.05^{\circ}$ and $0.05^{\circ} < \theta < 1^{\circ}$ separately as well as for the full range.

\begin{table*}
\centering
\begin{tabular*}{0.7\textwidth}{*{7}{c}}
\toprule
\multicolumn{7}{c}{LRG} \\
$\theta_{\text{min}}$, $\theta_{\text{max}}$ & $m_z$ bin & med $m_z$  & \# objects & $\log_{10}A_w$ ($\gamma = 1.98$) & $\log_{10}A_w$ & $\gamma$ \\
\midrule
0.005$^\circ$, 0.05$^\circ$ & 20.11, 20.41 & 20.26 & 868849 & $-3.94 \ ^{-3.92}_{-3.96}$ & $-4.01 \ ^{-3.88}_{-4.15}$ & $2.00 \ ^{2.07}_{1.93}$ \\
 & 19.81, 20.11 & 19.96 & 790503 & $-3.90 \ ^{-3.88}_{-3.92}$ & $-3.97 \ ^{-3.82}_{-4.12}$ & $2.00 \ ^{2.08}_{1.92}$ \\
 & 19.51, 19.81 & 19.66 & 692920 & $-3.83 \ ^{-3.81}_{-3.86}$ & $-3.91 \ ^{-3.72}_{-4.10}$ & $2.00 \ ^{2.09}_{1.91}$ \\
 & 19.21, 19.51 & 19.37 & 596349 & $-3.80 \ ^{-3.78}_{-3.82}$ & $-3.87 \ ^{-3.71}_{-4.04}$ & $2.00 \ ^{2.08}_{1.92}$ \\
 & 18.91, 19.21 & 19.06 & 518696 & $-3.77 \ ^{-3.75}_{-3.79}$ & $-3.84 \ ^{-3.68}_{-4.01}$ & $2.00 \ ^{2.08}_{1.92}$ \\
 & 18.61, 18.91 & 18.78 & 434355 & $-3.69 \ ^{-3.68}_{-3.70}$ & $-3.76 \ ^{-3.66}_{-3.86}$ & $2.00 \ ^{2.04}_{1.96}$ \\
 & 18.31, 18.61 & 18.48 & 257434 & $-3.56 \ ^{-3.55}_{-3.58}$ & $-3.63 \ ^{-3.50}_{-3.76}$ & $2.00 \ ^{2.06}_{1.94}$ \\
 & 18.01, 18.31 & 18.19 & 137380 & $-3.41 \ ^{-3.39}_{-3.43}$ & $-3.48 \ ^{-3.25}_{-3.72}$ & $2.00 \ ^{2.10}_{1.90}$ \\
0.05$^\circ$, 1$^\circ$ & 20.11, 20.41 & 20.26 & 868849 & $-3.85 \ ^{-3.83}_{-3.87}$ & $-3.56 \ ^{-3.50}_{-3.61}$ & $1.87 \ ^{1.91}_{1.83}$ \\
 & 19.81, 20.11 & 19.96 & 790503 & $-3.82 \ ^{-3.80}_{-3.85}$ & $-3.56 \ ^{-3.50}_{-3.61}$ & $1.88 \ ^{1.92}_{1.84}$ \\
 & 19.51, 19.81 & 19.66 & 692920 & $-3.78 \ ^{-3.76}_{-3.81}$ & $-3.43 \ ^{-3.39}_{-3.48}$ & $1.85 \ ^{1.88}_{1.82}$ \\
 & 19.21, 19.51 & 19.37 & 596349 & $-3.75 \ ^{-3.72}_{-3.78}$ & $-3.31 \ ^{-3.26}_{-3.35}$ & $1.81 \ ^{1.85}_{1.78}$ \\
 & 18.91, 19.21 & 19.06 & 518696 & $-3.74 \ ^{-3.71}_{-3.77}$ & $-3.27 \ ^{-3.22}_{-3.31}$ & $1.80 \ ^{1.83}_{1.77}$ \\
 & 18.61, 18.91 & 18.78 & 434355 & $-3.67 \ ^{-3.65}_{-3.69}$ & $-3.25 \ ^{-3.23}_{-3.27}$ & $1.82 \ ^{1.83}_{1.81}$ \\
 & 18.31, 18.61 & 18.48 & 257434 & $-3.55 \ ^{-3.53}_{-3.57}$ & $-3.20 \ ^{-3.17}_{-3.23}$ & $1.85 \ ^{1.86}_{1.83}$ \\
 & 18.01, 18.31 & 18.19 & 137380 & $-3.43 \ ^{-3.41}_{-3.46}$ & $-3.01 \ ^{-2.96}_{-3.06}$ & $1.82 \ ^{1.85}_{1.78}$ \\
0.005$^\circ$, 1$^\circ$ & 20.11, 20.41 & 20.26 & 868849 & $-3.90 \ ^{-3.88}_{-3.92}$ & $-3.65 \ ^{-3.60}_{-3.70}$ & $1.90 \ ^{1.93}_{1.87}$ \\
 & 19.81, 20.11 & 19.96 & 790503 & $-3.87 \ ^{-3.85}_{-3.88}$ & $-3.68 \ ^{-3.63}_{-3.74}$ & $1.92 \ ^{1.95}_{1.89}$ \\
 & 19.51, 19.81 & 19.66 & 692920 & $-3.81 \ ^{-3.80}_{-3.83}$ & $-3.67 \ ^{-3.62}_{-3.73}$ & $1.94 \ ^{1.97}_{1.91}$ \\
 & 19.21, 19.51 & 19.37 & 596349 & $-3.78 \ ^{-3.76}_{-3.80}$ & $-3.61 \ ^{-3.56}_{-3.67}$ & $1.93 \ ^{1.96}_{1.90}$ \\
 & 18.91, 19.21 & 19.06 & 518696 & $-3.76 \ ^{-3.74}_{-3.77}$ & $-3.62 \ ^{-3.56}_{-3.67}$ & $1.94 \ ^{1.96}_{1.91}$ \\
 & 18.61, 18.91 & 18.78 & 434355 & $-3.68 \ ^{-3.67}_{-3.69}$ & $-3.55 \ ^{-3.51}_{-3.59}$ & $1.94 \ ^{1.96}_{1.92}$ \\
 & 18.31, 18.61 & 18.48 & 257434 & $-3.55 \ ^{-3.54}_{-3.57}$ & $-3.45 \ ^{-3.40}_{-3.49}$ & $1.94 \ ^{1.97}_{1.92}$ \\
 & 18.01, 18.31 & 18.19 & 137380 & $-3.43 \ ^{-3.41}_{-3.44}$ & $-3.40 \ ^{-3.33}_{-3.47}$ & $1.97 \ ^{2.00}_{1.94}$ \\
\bottomrule
\end{tabular*}
\caption{Best fit parameters from modelling the angular clustering of LRGs in $z$-band magnitude bins using Equation~\ref{eqn:limber-pl}. The clustering amplitude $A_w$ is reported for a fixed slope of $\gamma = 1.98$ (taken from the fit over the full LRG sample; see Table~\ref{table:fiducial_powerlaw}), as well as the results of fitting amplitude and slope simultaneously. $\theta_{\text{min}}$ and $\theta_{\text{max}}$ are the angular scales fit over, and $m_z$ is in AB magnitudes.}
\label{table:lrg-magbinned} 
\end{table*}

\begin{table*}
\centering
\begin{tabular*}{0.7\textwidth}{*{7}{c}}
\toprule
\multicolumn{7}{c}{ELG} \\
$\theta_{\text{min}}$, $\theta_{\text{max}}$ & $m_z$ bin & med $m_z$  & \# objects & $\log_{10}A_w$ ($\gamma = 1.54$) & $\log_{10}A_w$ & $\gamma$ \\
\midrule
0$^\circ$, 0.05$^\circ$ & 23.1, 23.4 & 23.27 & 8530522 & $-3.15 \ ^{-3.08}_{-3.22}$ & $-4.10 \ ^{-3.98}_{-4.22}$ & $1.80 \ ^{1.87}_{1.73}$ \\
 & 22.8, 23.1 & 22.97 & 4768510 & $-2.98 \ ^{-2.91}_{-3.07}$ & $-4.66 \ ^{-4.59}_{-4.72}$ & $1.99 \ ^{2.02}_{1.95}$ \\
 & 22.5, 22.8 & 22.68 & 2462312 & $-2.67 \ ^{-2.59}_{-2.76}$ & $-4.43 \ ^{-4.32}_{-4.54}$ & $2.00 \ ^{2.05}_{1.95}$ \\
 & 22.2, 22.5 & 22.38 & 1224671 & $-2.36 \ ^{-2.27}_{-2.47}$ & $-4.09 \ ^{-3.94}_{-4.24}$ & $2.00 \ ^{2.06}_{1.94}$ \\
 & 21.9, 22.2 & 22.07 & 614394 & $-1.97 \ ^{-1.88}_{-2.09}$ & $-3.68 \ ^{-3.53}_{-3.84}$ & $2.00 \ ^{2.05}_{1.95}$ \\
 & 21.6, 21.9 & 21.77 & 334669 & $-1.74 \ ^{-1.61}_{-1.90}$ & $-3.31 \ ^{-3.16}_{-3.47}$ & $2.00 \ ^{2.05}_{1.95}$ \\
 & 21.3, 21.6 & 21.47 & 205434 & $-1.74 \ ^{-1.45}_{-2.28}$ & $-3.00 \ ^{-2.90}_{-3.11}$ & $2.00 \ ^{2.03}_{1.97}$ \\
 & 21.0, 21.3 & 21.16 & 138431 & $-1.74 \ ^{-1.17}_{-2.31}$ & $-2.68 \ ^{-2.62}_{-2.75}$ & $2.00 \ ^{2.01}_{1.98}$ \\
0.05$^\circ$, 1$^\circ$ & 23.1, 23.4 & 23.27 & 8530522 & $-3.23 \ ^{-3.21}_{-3.26}$ & $-2.93 \ ^{-2.87}_{-2.99}$ & $1.43 \ ^{1.48}_{1.39}$ \\
 & 22.8, 23.1 & 22.97 & 4768510 & $-3.15 \ ^{-3.13}_{-3.17}$ & $-2.80 \ ^{-2.79}_{-2.82}$ & $1.41 \ ^{1.42}_{1.40}$ \\
 & 22.5, 22.8 & 22.68 & 2462312 & $-2.99 \ ^{-2.98}_{-3.01}$ & $-2.76 \ ^{-2.74}_{-2.78}$ & $1.45 \ ^{1.46}_{1.44}$ \\
 & 22.2, 22.5 & 22.38 & 1224671 & $-2.77 \ ^{-2.76}_{-2.78}$ & $-2.64 \ ^{-2.62}_{-2.65}$ & $1.49 \ ^{1.50}_{1.48}$ \\
 & 21.9, 22.2 & 22.07 & 614394 & $-2.47 \ ^{-2.46}_{-2.48}$ & $-2.53 \ ^{-2.49}_{-2.58}$ & $1.56 \ ^{1.59}_{1.54}$ \\
 & 21.6, 21.9 & 21.77 & 334669 & $-2.22 \ ^{-2.19}_{-2.25}$ & $-2.51 \ ^{-2.42}_{-2.60}$ & $1.66 \ ^{1.71}_{1.61}$ \\
 & 21.3, 21.6 & 21.47 & 205434 & $-2.04 \ ^{-2.01}_{-2.08}$ & $-2.47 \ ^{-2.33}_{-2.61}$ & $1.72 \ ^{1.80}_{1.65}$ \\
 & 21.0, 21.3 & 21.16 & 138431 & $-1.93 \ ^{-1.89}_{-1.97}$ & $-2.37 \ ^{-2.17}_{-2.57}$ & $1.72 \ ^{1.83}_{1.62}$ \\
0$^\circ$, 1$^\circ$ & 23.1, 23.4 & 23.27 & 8530522 & $-3.17 \ ^{-3.12}_{-3.21}$ & $-4.08 \ ^{-3.99}_{-4.17}$ & $1.79 \ ^{1.85}_{1.74}$ \\
 & 22.8, 23.1 & 22.97 & 4768510 & $-3.05 \ ^{-2.99}_{-3.11}$ & $-4.16 \ ^{-4.06}_{-4.25}$ & $1.87 \ ^{1.92}_{1.81}$ \\
 & 22.5, 22.8 & 22.68 & 2462312 & $-2.80 \ ^{-2.73}_{-2.87}$ & $-4.09 \ ^{-3.99}_{-4.19}$ & $1.92 \ ^{1.96}_{1.87}$ \\
 & 22.2, 22.5 & 22.38 & 1224671 & $-2.53 \ ^{-2.45}_{-2.63}$ & $-4.00 \ ^{-3.89}_{-4.11}$ & $1.98 \ ^{2.02}_{1.94}$ \\
 & 21.9, 22.2 & 22.07 & 614394 & $-2.32 \ ^{-2.22}_{-2.43}$ & $-3.69 \ ^{-3.59}_{-3.79}$ & $2.00 \ ^{2.04}_{1.96}$ \\
 & 21.6, 21.9 & 21.77 & 334669 & $-2.13 \ ^{-2.04}_{-2.24}$ & $-3.35 \ ^{-3.25}_{-3.44}$ & $2.00 \ ^{2.03}_{1.97}$ \\
 & 21.3, 21.6 & 21.47 & 205434 & $-1.98 \ ^{-1.88}_{-2.09}$ & $-3.11 \ ^{-2.99}_{-3.23}$ & $2.00 \ ^{2.04}_{1.96}$ \\
 & 21.0, 21.3 & 21.16 & 138431 & $-1.89 \ ^{-1.80}_{-2.01}$ & $-2.99 \ ^{-2.76}_{-3.22}$ & $2.00 \ ^{2.09}_{1.91}$ \\
\bottomrule
\end{tabular*}
\caption{Best fit parameters from modelling the angular clustering of ELGs in $g$-band magnitude bins using Equation~\ref{eqn:limber-pl}. The clustering amplitude $A_w$ is reported for a fixed slope of $\gamma = 1.54$ (taken from the fit over the full ELG sample; see Table~\ref{table:fiducial_powerlaw}), as well as the results of fitting amplitude and slope simultaneously. $\theta_{\text{min}}$ and $\theta_{\text{max}}$ are the angular scales fit over, and $m_g$ is in AB magnitudes.}
\label{table:elg-magbinned} 
\end{table*}

\section{SPECTROSCOPIC CROSS-CORRELATIONS}
\label{sec:characterization_cross}

\subsection{Clustering as a function of redshift}
\label{sec:characterization/dndz}
\subsubsection{External catalogs}

\begin{figure}
\includegraphics[width=\columnwidth]{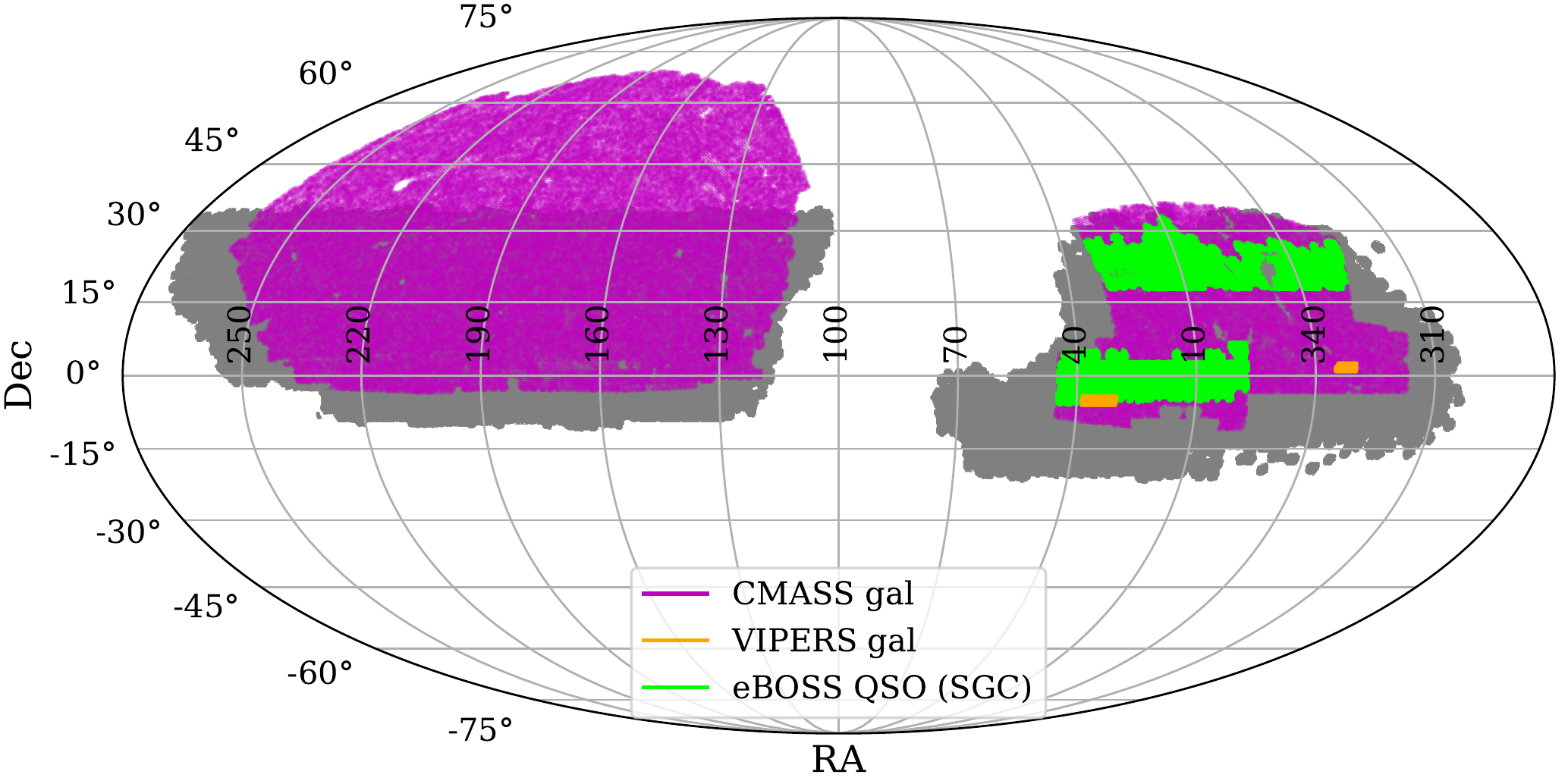}
\caption{Visualizing the overlap between the DECaLS DR7 footprint (gray) and the footprints of the external catalogs used for cross-correlations. Positions are mapped using Mollweide projection.}
\label{fig:spec_overlapping_area}
\end{figure}

\begin{figure}
\includegraphics[width=\columnwidth]{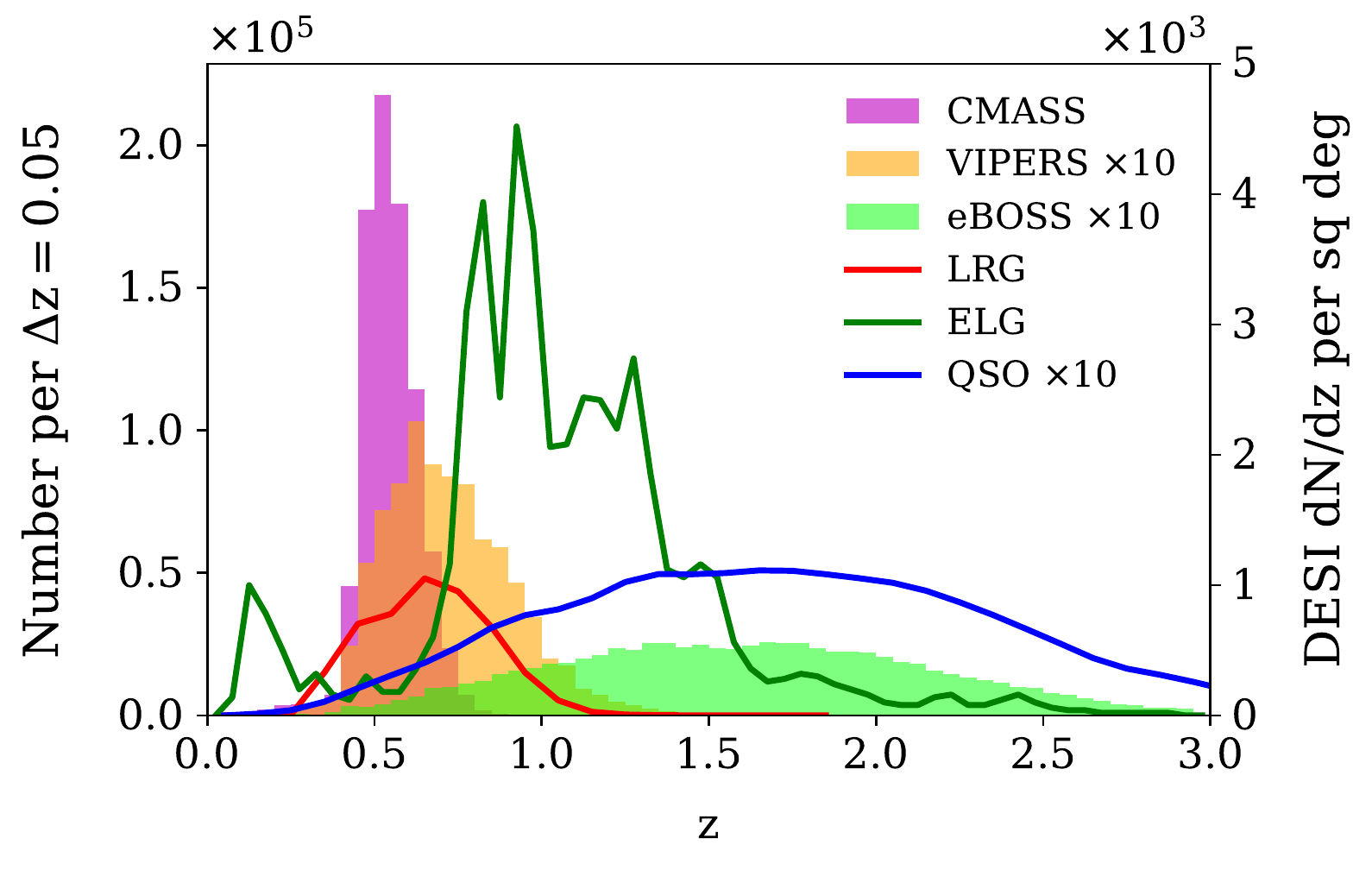}
\caption{Visualizing the redshift ranges for DESI targets compared to catalogs from spectroscopic surveys which overlap the DECaLS footprint. The solid lines correspond to the expected $dN/dz$ per square degree of the DESI target classes, while the histograms are from the external catalogs.}
\label{fig:spec_overlapping_z} 
\end{figure}

To probe the clustering as a function of redshift through cross-correlations, as described in Sections~\ref{sec:methods/xcorrs} and~\ref{sec:methods/dndz}, we make use of several external spectroscopic catalogs. We use the CMASS galaxy sample from DR12 of the Baryon Oscillation Spectroscopic Survey (BOSS; \citealt{Eisenstein11}; \citealt{BOSS13}), which selects higher redshift galaxies at $0.4 < z < 0.8$ and has significant angular overlap with the DECaLS footprint. We also use galaxies from the the final data release of the VIMOS Public Extragalactic Redshift Survey\footnote{\url{http://vipers.inaf.it/}} (VIPERS; \citealt{VIPERS18}). VIPERS extends over two narrow CFHTLS fields, \texttt{W1} and \texttt{W4}, with a combined area of approximately 23.5 deg$^2$, and has nearly $90,000$ redshifts out to $z \sim 1$. Finally, we use the main sample of QSOs from eBOSS DR14 \citep{Dawson++16}, which overlaps with the DECaLS footprint in the south galactic cap. Figure~\ref{fig:spec_overlapping_area} shows where the footprints of these surveys intersect with DECaLS DR7, and Figure~\ref{fig:spec_overlapping_z} demonstrates how their redshift distributions span the expected redshift ranges of the DESI targets. 

\subsubsection{Projected real-space cross-correlation functions}

\begin{figure}
\includegraphics[width=\columnwidth]{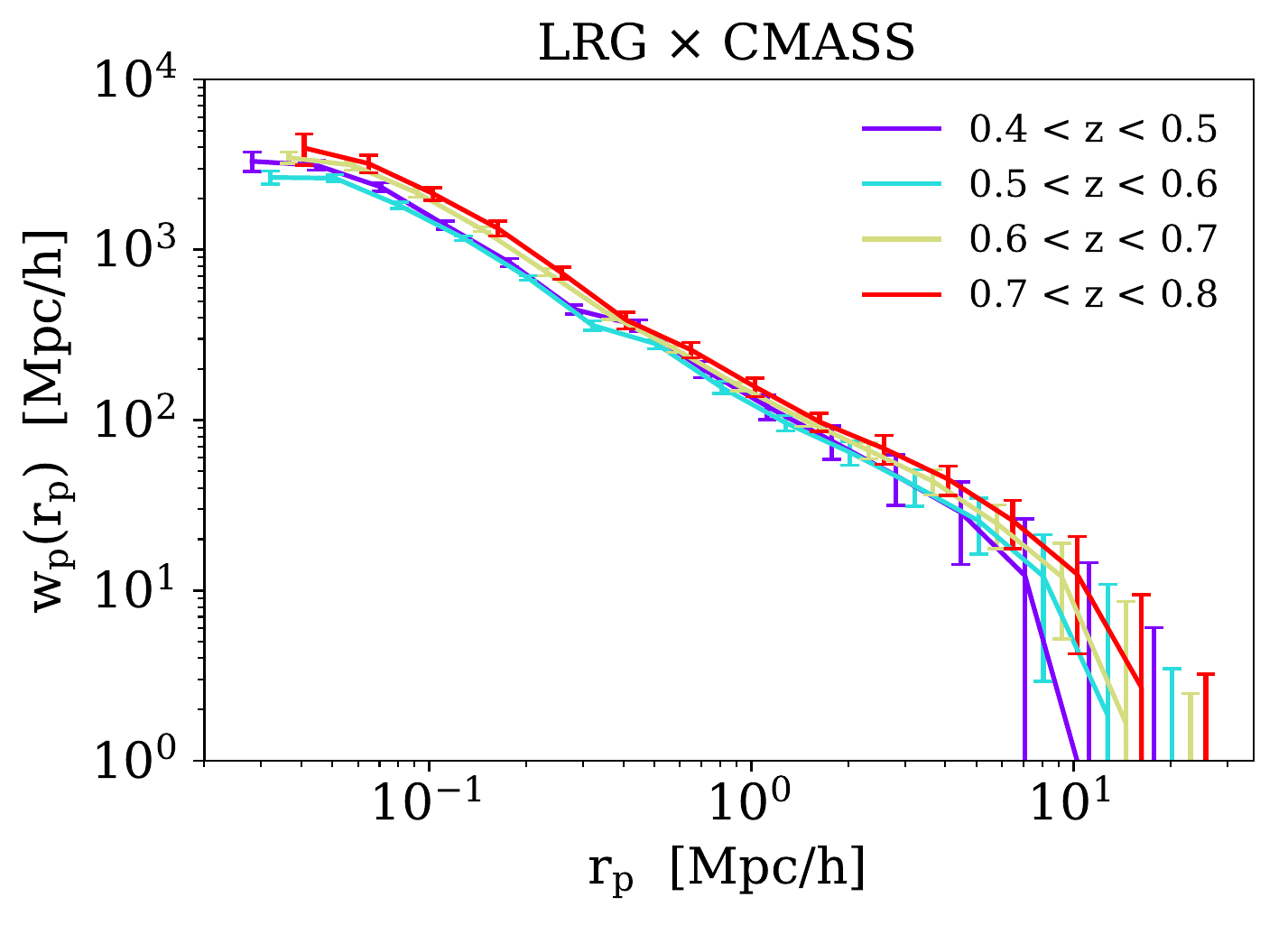}
\includegraphics[width=\columnwidth]{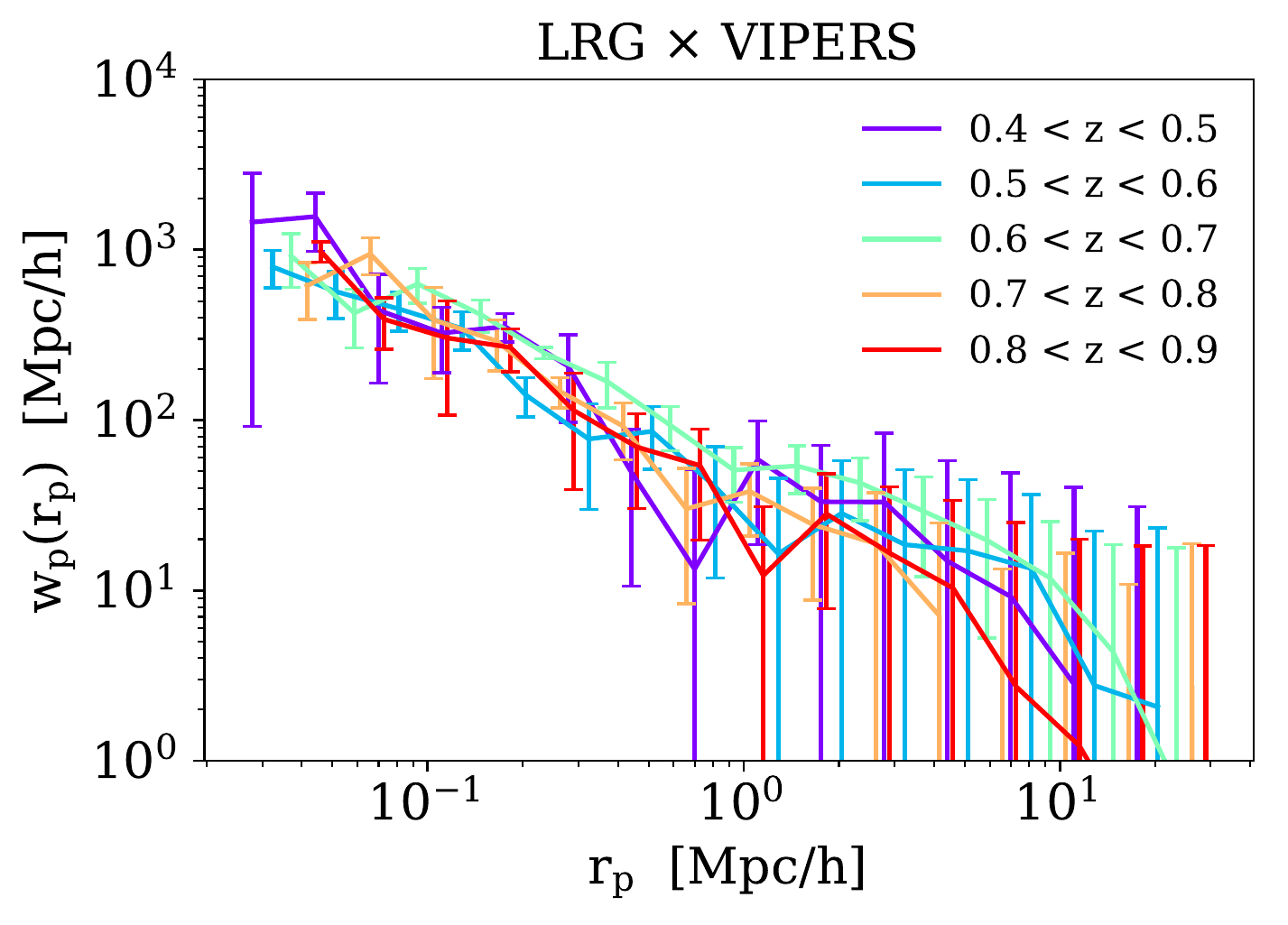}
\includegraphics[width=\columnwidth]{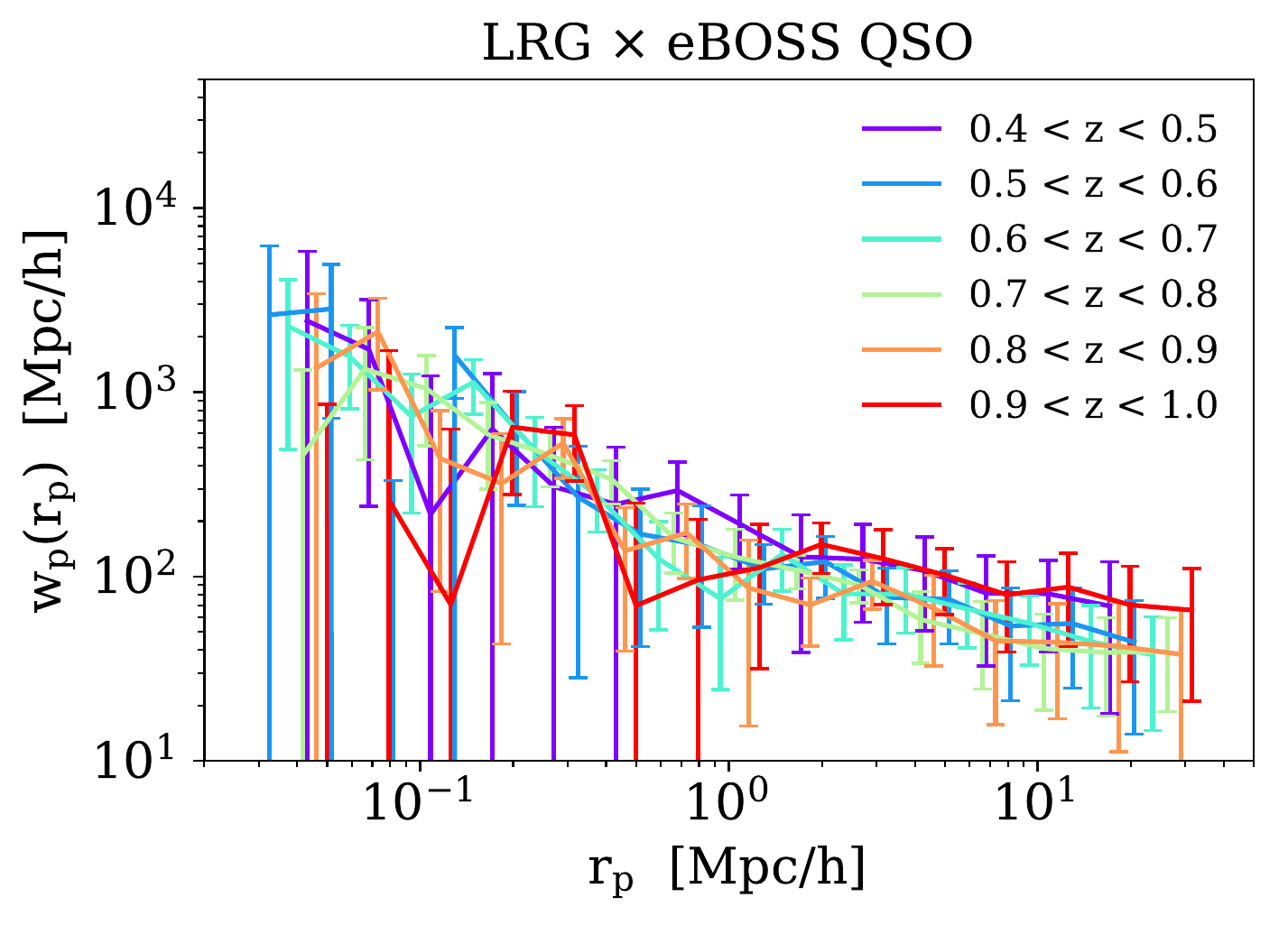}
\caption{Projected real-space cross-correlations between LRGs and three external samples with spectroscopic redshifts: CMASS galaxies, VIPERS galaxies, and eBOSS QSOs. Error bars are from bootstrapping.}
\label{fig:LRG_xcorrs}
\end{figure}

We present the real-space projected cross-correlation functions (derived in Section~\ref{sec:methods/xcorrs}) for LRGs in Figure~\ref{fig:LRG_xcorrs}, using CMASS galaxies, VIPERS galaxies, and eBOSS QSOs in bins of width $\delta z = 0.1$. Some noisy redshift bins are omitted from the plots. The error bars are from bootstrapping on the area, and therefore are likely overestimated for VIPERS, which has very small fields. 

\begin{figure}
\includegraphics[width=\columnwidth]{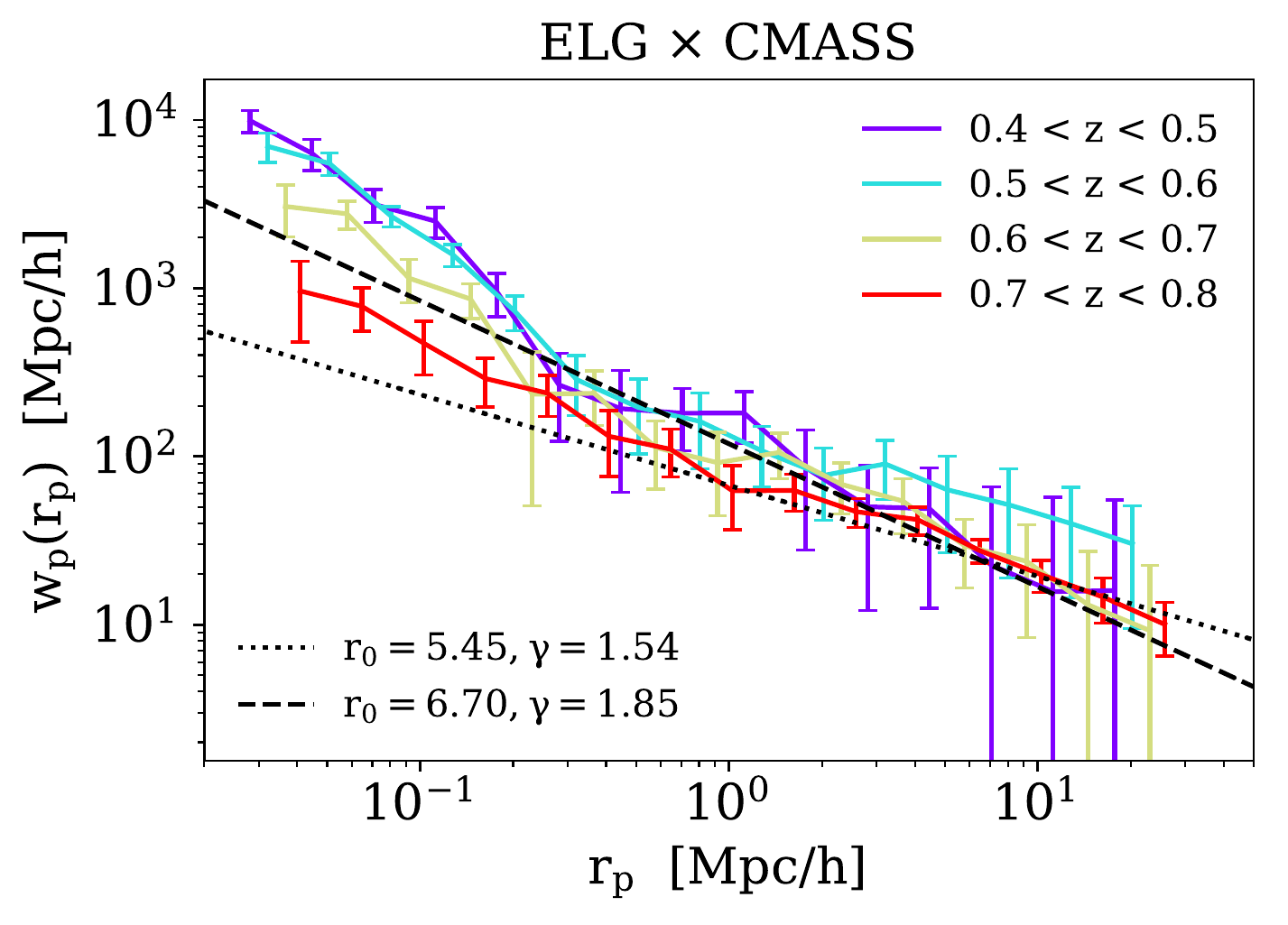}
\caption{Projected real-space cross-correlations between ELGs and CMASS galaxies. Error bars are from bootstrapping. Dashed and dotted lines are the power-law fits of the ELG autocorrelation from Table~\ref{table:fiducial_powerlaw}.}
\label{fig:ELG_xcorrs}
\end{figure}

For ELGs, we initially find that the cross-correlations with CMASS galaxies flatten above $\theta \sim 0.01^{\circ}$ for all redshift bins. However, using the CMASS systematics weights in concert with our own photometric weights eliminates this effect, indicating a correlation between systematics in the two catalogs, likely the anti-correlation with stars found in both DESI ELGs and CMASS samples. The projected real-space cross-correlations are plotted in Figure~\ref{fig:ELG_xcorrs}, along with the power-law predictions from the ELG $w(\theta)$ fits in Table~\ref{table:fiducial_powerlaw}, which we translate into $w_p(r_p)$ using Equation~\ref{eqn:wp}. We thus have consistency between ELG $w(\theta)$ in Figure~\ref{fig:final-wtheta}, $C_{\ell}$ in Figure~\ref{fig:elg_cls}, and $w_p(r_p)$ in Figure~\ref{fig:ELG_xcorrs}. We also note that the break at small scales becomes less pronounced at higher redshift, as the 2-halo term becomes more dominant.

More puzzlingly, ELGs appear to show no correlation with eBOSS QSOs over the overlapping redshift range. This null signal is consistent within error bars across all redshift bins, and remains null even when switching to brighter ELG subsamples. At present, we wish to avoid speculating on why there is no cross-correlation between ELGs and eBOSS QSOs, as a full investigation with survey validation data and spectra is expected to paint a much clearer picture. Given the reasonable ELG autocorrelation and ELG $\times$ CMASS cross-correlation, we do not believe this is indicative of catastrophic failure in the ELG sample. Finally, we note that the QSO cross-correlations are too noise-dominated to obtain a meaningful signal.

\subsubsection{Clustering $dN/dz$}

For ELGs and QSOs, issues with cross-correlation measurements discussed in the previous section prevent us from obtaining meaningful $dN/dz$ over the full redshift ranges of the targets. We therefore focus on LRGs and defer further investigation of ELGs and QSOs to a future work. 

\begin{figure}
\includegraphics[width=0.99\linewidth]{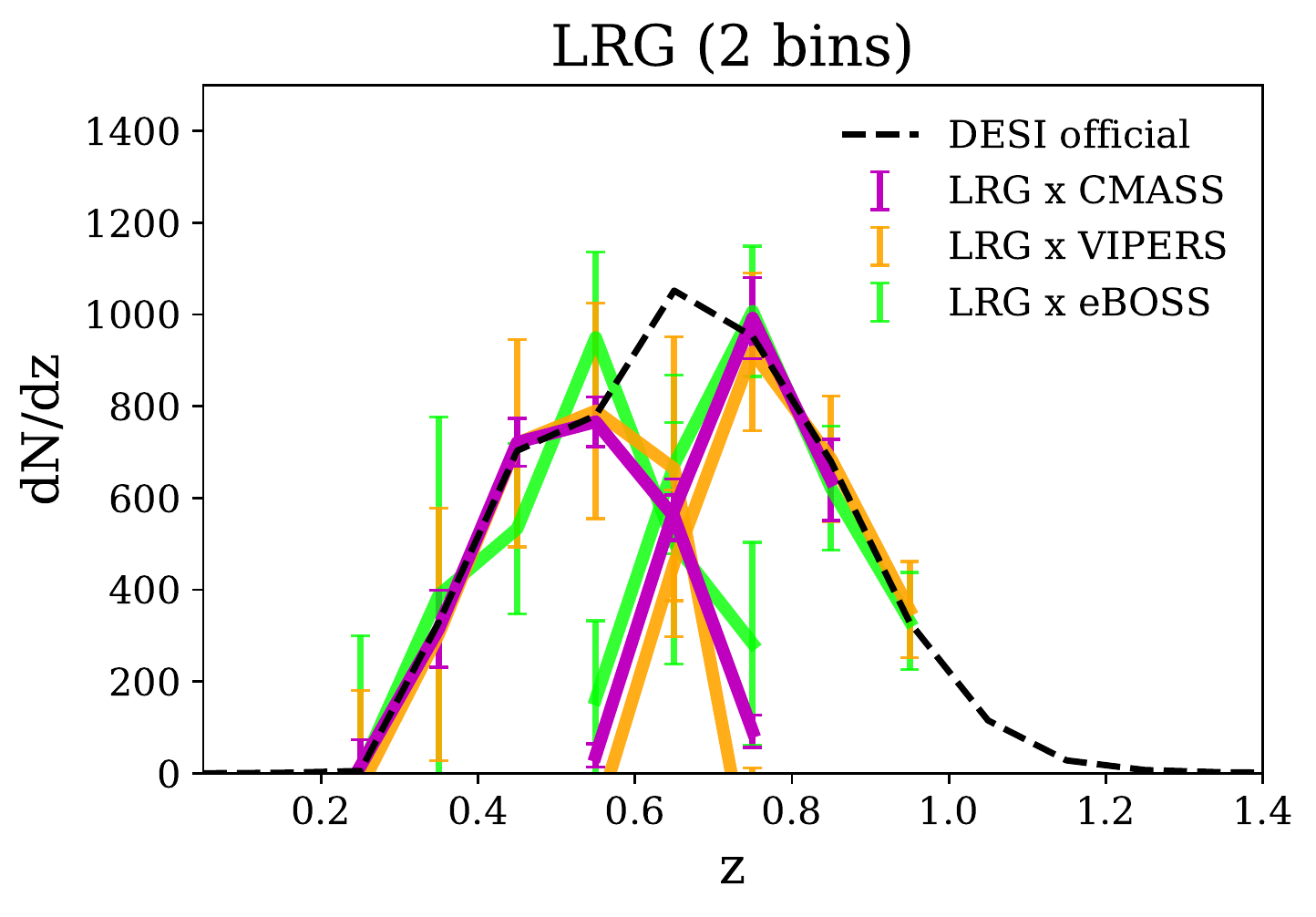}
\includegraphics[width=0.99\linewidth]{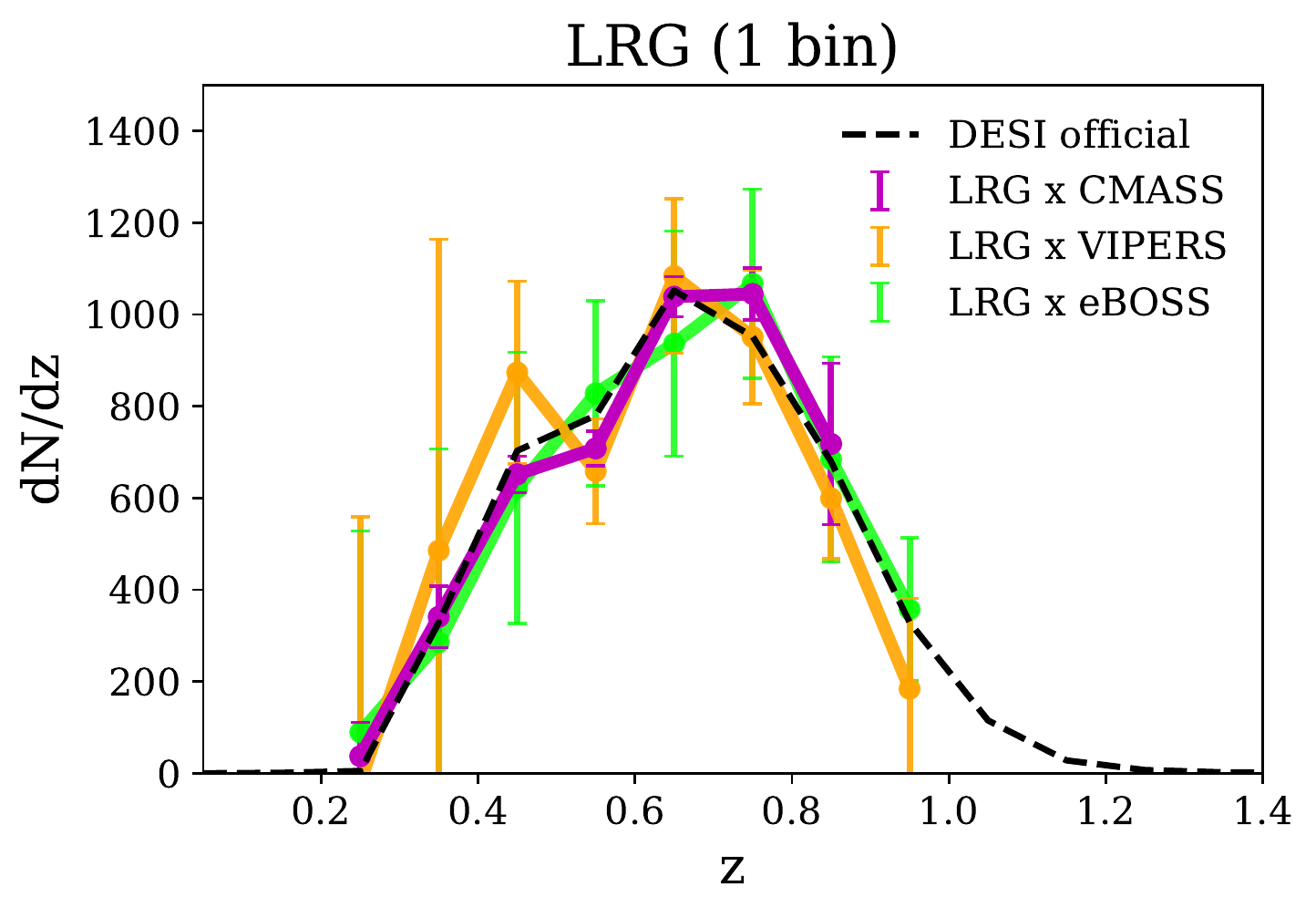}
\caption{The clustering-based $dN/dz$ for LRGs derived from cross-correlations with CMASS galaxies (magenta), VIPERS galaxies (orange), and eBOSS QSOs (lime), with the expected $dN/dz$ plotted as a dashed line. The upper plot shows $dN/dz$ calculated using two $r - W1$ color bins, a proxy for a $z \sim 0.6$ cut in order to reduce the impact of bias evolution, while the lower plot is determined using the full sample. Error bars are from propagating bootstrap errors from the cross-correlations through the $dN/dz$ calculation.}
\label{fig:lrg-dndz}
\end{figure}

Using the method outlined in Section~\ref{sec:methods/dndz}, we integrate over each set of cross-correlations in the overlapping redshift ranges to piece together the shape of the LRG $dN/dz$. We choose the minimum and maximum physical scales of integration in such a way as to reduce the propagated errors on $dN/dz$; for LRG $\times$ CMASS and LRG $\times$ VIPERS, we use $s_{\text{min}} = 0.05$ $h^{-1}$ Mpc, $s_{\text{max}} = 5$ $h^{-1}$ Mpc, whereas for LRG $\times$ eBOSS we use $s_{\text{min}} = 0.2$ $h^{-1}$ Mpc, $s_{\text{max}} = 10$ $h^{-1}$ Mpc.

To minimize the potential impact of bias evolution in the photometric sample, we first divide it by color before cross-correlating each subsample separately, as discussed by \cite{Menard13}, \cite{Schmidt13}, \cite{Rahman15}, \cite{Gatti18},. In Figure 1 of \cite{Prakash16}, the photometric redshifts of eBOSS LRGs are plotted in color-space, with a transition from mostly $ z < 0.6 $ objects to mostly $ 0.6 < z < 1.0 $ objects occurring when $ r - W1 $ is in the range between $2$ and $3$. Motivated by this, we select a roughly median value of $ r - W1 = 2.6 $ to create two similarly sized LRG subsamples. We cross-correlate these two subsamples separately with the three external catalogs, with the combined results shown in Figure~\ref{fig:lrg-dndz} and compared to the results derived without binning the sample.  

We find that the clustering $dN/dz$ from all three cross-correlations match very well with the fiducial FDR $dN/dz$.\footnote{We note that the fiducial redshift distributions in Figure~\ref{fig:spec_overlapping_z} are of the targets selected from imaging, including contaminants.} Along with the excellent agreement between measured and fiducial bias found in Section~\ref{sec:characterization/cls}, this suggests that the LRG sample will be able to fully meet the cosmology goals of the collaboration. Additionally, the upper panel of Figure~\ref{fig:lrg-dndz} confirms that the color cut at $r-W1 = 2.6$ effectively splits the LRG sample into high and low redshift subsamples with an approximate boundary at $z \sim 0.65$.

\FloatBarrier

\subsection{Clustering as a function of luminosity}
\label{sec:characterization/lumdep}
\begin{figure}
\centering
\includegraphics[width=0.99\linewidth]{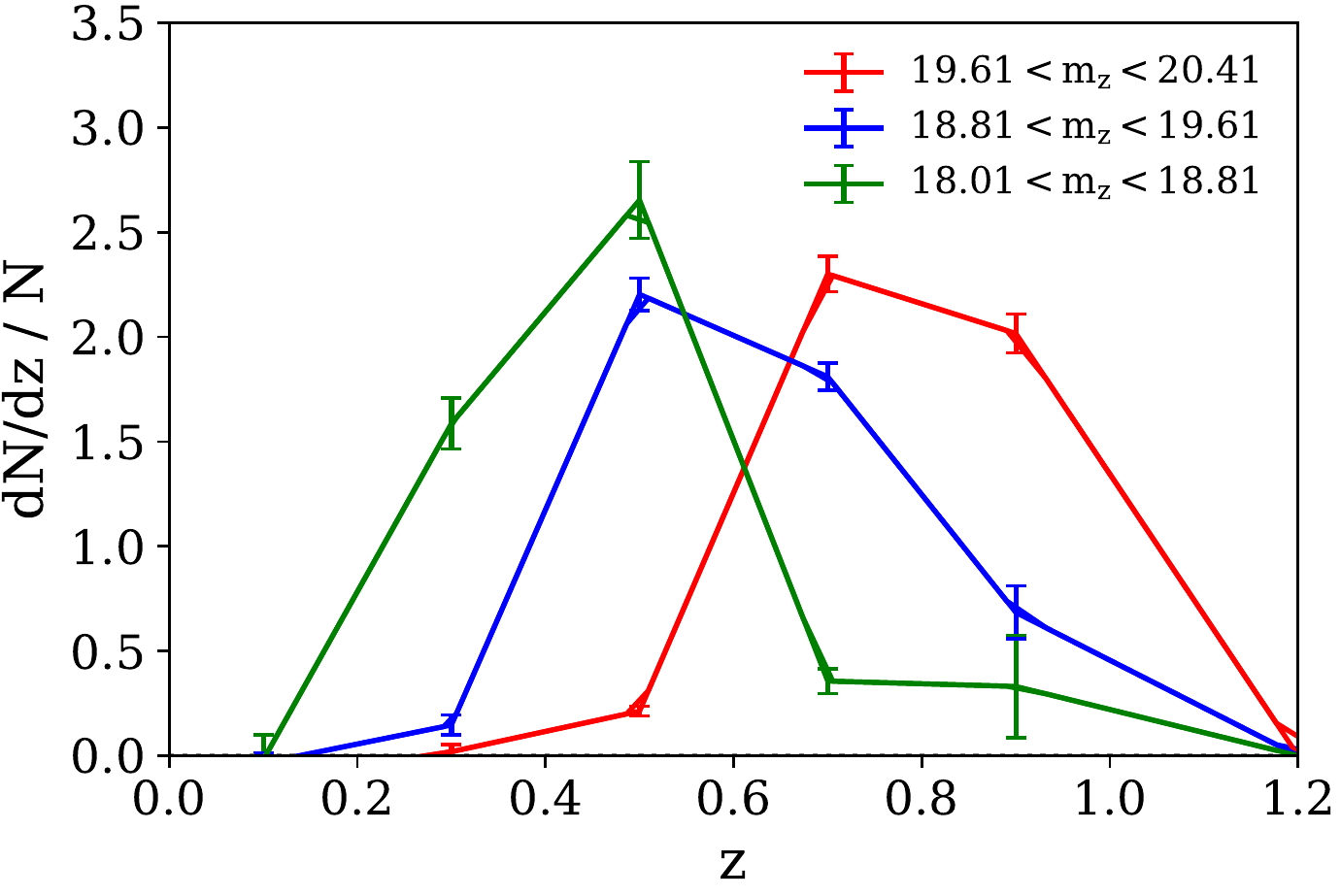}
\caption{Normalized $dN/dz$ for each of three broad magnitude bins, derived from cross-correlations with CMASS galaxies in broad redshift bins of $\delta z = 0.2$. Error bars are from propagating bootstrap errors from the cross-correlations through the $dN/dz$ calculation.}
\label{fig:lrg-3bin-dndz}
\end{figure}

By cross-correlating magnitude binned LRGs with redshift binned spectroscopic catalogs, we can also probe the luminosity dependence of the sample. We begin by dividing the LRGs into three broader magnitude bins from $m_z=18.01$ to $m_z=20.41$, the bright and faint limits, respectively, of the target selection. To improve signal-to-noise, we also double the widths of the redshift bins to $\delta z = 0.2$ and focus on the cross-correlations with CMASS galaxies, which involve the smallest error bars. Through these cross-correlations, we can crudely reconstruct $dN/dz$ for each of the magnitude bins, shown in Figure~\ref{fig:lrg-3bin-dndz}. The behavior is as expected; brighter objects are at lower mean redshift, with the redshift distributions generally appearing as deeper and deeper copies of each other.

This result allows us to convert angular cross-correlations into projected real-space cross-correlations, as detailed in Section~\ref{sec:methods/xcorrs}, giving us $w_p(r_p)$ in three broad luminosity bins for a given redshift bin. We select a bin near the middle of the CMASS redshift range\footnotemark, $0.4 < z < 0.6$, and fit the clustering for each corresponding luminosity bin to a power law, $w_p(r_p) = A r_p^{1-\gamma}$. The results are given in Table~\ref{table:lumbinned}. 

\footnotetext{Since CMASS galaxies are selected by color cuts, as are our LRGs, objects at the edges of the redshift range (particularly the low end, where interlopers become more probable) may be physically different from objects in the middle, with different clustering.}

\begin{table}
\centering
\begin{tabular*}{0.4\textwidth}{*{3}{c}}
\toprule
\multicolumn{3}{c}{LRG $\times$ CMASS, $\ 0.4 < z < 0.6$} \\
\toprule
$m_z$ bin & $M_z(\bar{z}=0.5)$ & $\log_{10}A$ ($\gamma = 1.98$) \\
\midrule
19.61, 20.41 & -21.87, -21.07 & $-3.40 \ ^{-3.36}_{-3.44}$ \\
18.81, 19.61 & -22.67, -21.87 & $-2.51 \ ^{-2.49}_{-2.53}$ \\
18.01, 18.81 & -23.47, -22.67 & $-2.56^{-2.54}_{-2.59}$ \\
\bottomrule
\end{tabular*}
\caption{Fits to the luminosity binned LRG-CMASS cross-correlations assuming $w_p(r_p) = A r_p^{1-\gamma}$. The first column specifies the LRG apparent magnitude bins, while the second column calculates the corresponding absolute magnitude bin at the midpoint of the CMASS redshift bin $0.4 < z < 0.6$.}
\label{table:lumbinned} 
\end{table}


\FloatBarrier
\section{SUMMARY AND CONCLUSIONS}
\label{sec:conclusions}
In order to fully realize the statistical power of the DESI experiment, it is vital to assess the quality of the imaging data and target definitions, and to account for any non-cosmological sources of spatial fluctuations in the galaxy catalogs that could bias the cosmological analyses. In the first part of this paper, we diagnose causes of systematic errors in the clustering of DESI main targets selected from imaging and present masks and photometric weights aimed at reducing these effects. The masks and weights will be used in construction of cosmological clustering samples. Our key results are summarized below:

\begin{itemize}[leftmargin=0.4cm]

\item We find that obscuration due to bright stars in the foreground creates significant variations in density, particularly for QSOs. Implementing aggressive masks around Tycho-2 and WISE stars, which remove $7$-$8\%$ of the usable sky area for galaxies and $28\%$ of the usable sky area for QSOs, dramatically improves the agreement of the angular correlation functions with cosmological predictions.

\item We determine that a general mask around large galaxies and other extended sources is not indicated; however, by visual inspection, we discover that failing to mask around the Coma cluster and M3 will create a significantly overestimated clustering signal for QSOs in the north galactic cap. Additionally, the images of a small number of very bright stars are plagued by complex patterns of reflected pupil ghosts, which create structures in the ELG density beyond the scope of even our highly conservative bright star masks.

\item We find that masked LRGs exhibit only minor density variations as a function of potential systematics such as stellar density, extinction, airmass, seeing, sky brightness, and exposure time. By contrast, masked ELG and QSO densities still fluctuate significantly, with the most dominant systematics, stellar density and extinction, affecting densities by as much as $10\%$. We find that ELGs are anti-correlated with stars and extinction, while QSOs are positively correlated. Photometric weights calculated by performing multilinear regression on these trends significantly ameliorates them.

\item We perform angular cross-correlations between the targets and stars, and again find that LRGs are uncorrelated, ELGs are anti-correlated, and QSOs are positively correlated. Dividing the stars into three galactic latitude bins, we use QSO-star cross-correlations and star autocorrelations to estimate the stellar contamination fraction in the QSO sample as a function of galactic latitude.

\end{itemize}

We stress that this process has been highly iterative, with our efforts continuously informing the evolution of DESI's imaging data reduction pipelines and target selection algorithms. The Legacy Survey Data Release 9, which is currently being processed and will be used for DESI target selection, has made several algorithmic upgrades motivated in part by the feedback in this study. These include changes to the pixel-level flat-field response functions and an improved modeling of sky subtraction. The latter improvements are most pronounced for ELG targets that are faint relative to the sky and therefore have their target densities modulated by errors in the sky modeling. The new sky model also helps with issues of scattered light around bright stars that affect the selection of QSO targets. In addition, the imaging team has added more aggressive foreground masking and has flagged other bad data from the list of problematic regions identified in this study (such as the ghost pupils around certain visibly bright stars, and the M3/NGC contaminants in the north). Similarly, the target selection algorithms have been iteratively updated many times in response to our findings.

In addition to being a crucial first step towards constraining cosmology with DESI clustering measurements, our findings have important implications for other ongoing and future imaging surveys. As multi-epoch surveys become deeper and more sensitive, they will be increasingly limited by systematic uncertainties from instrument calibration, survey characteristics, and observing conditions. Our framework for identifying and mitigating the effects of such systematics, such as our new approach to quantifying contamination due to stars, is therefore highly relevant and widely applicable to future imaging surveys.

After applying masks and weights, we devote the second part of this paper to modeling the properties of the samples, providing the first large-scale clustering analysis of DESI targets. Modeling the samples is an important first step for doing cosmology with DESI, and our clustering results will also aid in the creation and validation of accurate mock catalogs. Additionally, we present several new methodologies, including the technique of probing the luminosity dependent clustering by cross-correlating magnitude-binned photometric samples with redshift-binned spectroscopic samples. These methods can be applied to other clustering studies with deep photometric data, for instance in future studies with data from the Large Synoptic Survey Telescope \citep{LSST09}. Our main results are outlined below:

\begin{itemize}[leftmargin=0.4cm]

\item We present the average densities before and after the corrections have been applied, finding that all three target densities are in reasonably good agreement with expectation after masking and weighting.

\item We model the angular correlation functions of the samples, assuming power law spatial correlation functions. For LRGs, we recover values which agree very well with earlier studies. For ELGs, we see a broken power law, with different slopes for $\theta < 0.01^{\circ}$ and $\theta > 0.05^{\circ}$ which agree reasonably well with similar studies. For QSOs, we obtain a highly inflated value for the clustering amplitude, indicating that substantial contamination remains.

\item We compare the observed angular power spectra to theory to determine the linear large-scale bias, and also probe the scale dependence of the bias in the weakly nonlinear regime. For LRGs, we find a value of $b_0$ that agrees very well with the DESI FDR prediction. For ELGs, we find a value of $b_0$ that is higher than the conservative lower limit given by the FDR but is similar to values from the literature and is self-consistent with our angular and real-space clustering measurements. By contrast, the observed angular power spectrum for QSOs is a poor fit to theory, with all scales seemingly affected by non-cosmological signals.

\item We use cross-correlations with external spectroscopy to determine real-space projected cross-correlation functions in redshift bins, through which we also derive clustering $dN/dz$. For LRGs, the clustering as a function of redshift behaves as expected, and we see an excellent match with the expected $dN/dz$ from target selection. For ELGs, the redshift-binned cross-correlations with CMASS are consistent with expectation, but cross-correlations with eBOSS QSOs show no significant correlation. For QSOs, the cross-correlations are not currently clean enough to meaningfully model $dN/dz$.

\item The clustering of LRGs and ELGs as a function of magnitude also behaves as predicted, with clustering amplitude scaling with depth. We provide fits to the angular correlation functions in magnitude bins. We also cross-correlate magnitude binned LRGs with redshift binned CMASS galaxies to probe luminosity dependent clustering. 

\item We present counts-in-cells moments and cell-averaged higher order correlation functions to further facilitate mock calibration and validation.

\end{itemize}

Overall, our results suggest that the quality of the imaging and the selection of targets are suitable for achieving the ambitious scientific objectives of the DESI collaboration. With imaging surveys completed and spectroscopic first light announced in October 2019, the commissioning phase is on track for completion in January 2020. After a survey validation period in the spring, the 5-year survey is expected to begin in the summer of 2020. We look forward to the exciting and impactful new science that DESI will enable in the coming decade.


\section*{Acknowledgements}
The authors thank Stephen Bailey, Daniel Eisenstein, Shirley Ho, Dustin Lang, Jeffrey Newman, Anand Raichoor, Ashley Ross, Hee-Jong Seo, Mike Wilson, Christophe Y\`{e}che, and Pauline Zarrouk for many useful discussions. E.K. and M.W. are supported by the U.S. Department of Energy, Office of Science, Office of High Energy Physics under Award No. DE-SC0017860. DS, JG, ML and JM are supported by  the Director, Office of Science, Office of High Energy Physics of the U.S. Department of Energy under Contract No. DE-AC02-05CH11231, and by the National Energy Research Scientific Computing Center (NERSC), a DOE Office of Science User Facility under the same contract.  This work also made extensive use of the NASA Astrophysics Data System and of the \texttt{astro-ph} preprint archive at \texttt{arXiv.org}. Additional support for DESI is provided by the U.S. National Science Foundation, Division of Astronomical Sciences under Contract No. AST-0950945 to the National Optical Astronomy Observatory; the Science and Technologies Facilities Council of the United Kingdom; the Gordon and Betty Moore Foundation; the Heising-Simons Foundation; the National Council of Science and Technology of Mexico; and by the DESI Member Institutions. The authors are honored to be permitted to conduct astronomical research on Iolkam Duag (Kitt Peak), a mountain with particular significance to the Tohono O'odham Nation. 


\bibliographystyle{mnras}
\bibliography{mybib}

\appendix

\bsp	
\label{lastpage}
\end{document}